\documentclass[twocolumn,apj]{openjournal}

\usepackage[breaklinks,colorlinks,citecolor=blue]{hyperref}

\newcommand{\be}{\begin{equation}}
\newcommand{\ee}{\end{equation}}
\newcommand{\bea}{\begin{eqnarray}}
\newcommand{\eea}{\end{eqnarray}}
\newcommand{\Dd}{\mathrm{d}}

\usepackage[utf8]{inputenc}
\usepackage{graphicx}
\usepackage{listings}

\usepackage{tikz}
\usepackage[misc]{ifsym}

\usepackage{color}
\definecolor{dkgreen}{rgb}{0,0.6,0}
\definecolor{gray}{rgb}{0.5,0.5,0.5}
\definecolor{mauve}{rgb}{0.58,0,0.82}

\begin{document}

\title{\bf A Beginner's Guide to Working with Astronomical Data}
\author{Markus P{\"o}ssel\\ \small Haus der Astronomie and Max Planck Institute for Astronomy}

\maketitle

{\small

\tableofcontents
}

\section{Introduction}

Compare professional astronomy today with how it was 50 years ago, and you will recognise some continuity --- but also a number of fundamental changes. Perhaps the key change is that astronomy has come to rely almost completely on digital data. Modern telescopes with their CCD cameras produce digital images and, with the help of suitable dispersive elements, digital astronomical spectra. An in-depth analysis of a particularly well-studied object will be able to make use of digital images and spectra taken at different wavelengths --- some taken by ground-based telescopes and some, like extreme ultraviolet or X rays, which can only be provided by space telescopes.

Furthermore, in-depth studies of selected objects are only part of what modern astronomy has to offer. We also live in an era of extensive surveys: large-scale undertakings to photograph, or take spectra of, wider regions of the sky in a systematic way. These surveys not only produce many images and spectra, but also extensive {\em catalogues} of the objects observed, listing various of their properties. With such catalogues comes the ability to make statistical deductions about astronomical objects: If you want to know whether, say, elliptical galaxies are, in general, brighter than spiral galaxies (to pick an artificially simple example), you, consult a suitable galaxy catalogue and look up the brightness values for a large number of elliptical and a large number of spiral galaxies. 

Modern surveys produce considerable amounts of data. For home use, we have become used to Gigabytes (1 Gigabyte =1000 Megabyte) and Terabytes (TB; 1 Terabyte = 1000 Gigabyte): a DVD holds 4.7 Gigabytes, and hard drives now routinely hold a Terabyte or more. The ESO Survey Telescope VISTA produces about 150 TB worth of data per year, and the Large Synoptic Survey Telescope (LSST) currently under construction is predicted to produce 500 TB worth of image data {\em per month}.

Then, there a increasingly large and detailed simulations. Take the IllustrisTNG simulation, which follows the evolution of a large portion of the universe from shortly after the Big Bang to the present. The smallest but most detailed of the TNG runs, TNG50, follows the fate of a cube that, in the present universe, has a side-length of 50~Mpc. Within this volume, matter is represented by 10 billion point particles representing Dark Matter and 10 billion point particles representing gas. The two simulation runs TNG300 and TNG100 which were made available to the public in December 2018 sum up to more than one Petabyte of data (PB; 1 Petabyte = 1000 Terabytes).

Increase the amount of data, and at some point it will become impractical to download a complete data set onto your own computer for analysis. This is where data base operations become important: the data is stored in dedicated data centres, and is accessible online; in order to work with the data, you use the Internet to send specific queries (''Give me the list of all galaxies on the Southern hemisphere which are brighter than X''). In this way, the only data you download is the data you specifically need for your research. The next step is not far off: when even those pre-selected data sets become too cumbersome to handle, researchers can run their analysis programs remotely on the dedicated servers where the data is stored. Infrastructure to allow just this, notably JuPyter notebooks, are becoming increasingly common.

All this implies that digital data analysis skills are part of the essential skill sets of modern astronomers. Some of the skills needed for a given research project will be very specific, involving custom software to be used for a very particular kind of analysis, or custom software to be written by the researcher herself. These special skills must be learned and honed on the job. But there are other skills which are more elementary and more general. Teaching a selection of those skills is the purpose of this text. It was originally written for interns at Haus der Astronomie in Heidelberg, in particular for participants of our International Summer Internship Program\footnote{\url{http://www.haus-der-astronomie.de/en/what-we-do/internships/summer-internship}} aimed at students in the final years of high school, or for students who have just finished high school and are about to start college. 

The text is meant to give a first introduction to working with astronomical data. It does not cover the more detailed astronomical use cases, but instead is meant to help students familiarise themselves with the basic tools needed for such work, and learn to apply basic techniques and tools that are fairly universal.

\subsection{Types of data}

When it comes to data from observational astronomy, most data sets we will be dealing with fall into one of the following categories:
\begin{itemize}
\item {\bf Image data} --- in its simplest form, an image is a two-dimensional array of pixels, where each pixel value denotes a brightness value. In an ordinary color image, each pixel will have three brightness values, denoting the contribution from red, green, and blue (RGB). Since astronomers use many specialist filters beyond these three colors, astronomical ''color'' images can have even more color values per pixel. Astronomical images usually show a region of the night sky. 
\item {\bf Spectra} --- simple spectra show us how the energy of the light emitted by an object is distributed among the different possible wavelengths. Such simple spectra are one-dimensional: for each wavelength value, we know the contribution of light from that particular wavelength region.
\item {\bf Data cubes} --- think of these as an enhanced version of astronomical images. An example is a data cube from what is known as integral field spectroscopy (IFS); such a data cube is like a two-dimensional image, but now each pixel contains not a brightness value, but a whole spectrum received from the region of the sky within that pixel. Since we have a one-dimensional spectrum for each pixel of a two-dimensional image, that gives us in effect a three-dimensional object: a data cube
\item {\bf Catalog data} --- on a higher level of analysis, astronomers make catalogues of the properties of different types of astronomical objects. A star catalog, for instance, could list position, proper motion, parallax, brightness (in various wavelength bands), and effective temperature for each of a specific selection of stars.
\end{itemize}
This list is not complete --- for instance, in interferometric imaging, when you are trying to reconstruct an image by combining coherently the measurements of different telescopes (``aperture synthesis''), your raw data will be time-stamped data from the single telescopes, and the initial processing will involve cross-correlating the data between those telescopes. But while the list is not complete, it should cover the great majority of current astronomical use cases.

We will take a first look at examples for each data type in section \ref{DataBasics}. These different data usually come with {\bf meta-data}, that is, descriptive information about the data. Astronomical images typically include information about the circumstances of when and how the image was taken (what telescope, what time, what exposure time, what pointing?), and about where in the sky the object in question is located (in the shape of data allowing the user to relate image pixels to an astronomical coordinate system).

For simulations, the situation is more diverse, but there are two fundamental schemes:
\begin{itemize}
\item {\bf N-body simulations} --- here, matter is represented by point particles. A point particle can represent different kinds of objects: it could be a lump of gas, or a star, or a group of a few $10^4$ or $10^5$ stars in a galaxy, or a lump of dark matter. But every particle has a position in space, and as the simulation runs, the particle positions change.
\item {\bf Grid-based simulations} --- here, space is divided into basic cells, for instance a space-filling set of small cubes. For each cell, basic properties (such as density, temperature, quantities of the different types of matter present) are tracked; those values change as the simulation runs. In more complex simulations, the grid itself can also change in ways that are adapted to making the simulation more efficient (``adaptive grid'').
\end{itemize}

\subsection{Types of tools}

When it comes to the tools for working with these various kinds of data, we can distinguish two types. 
\begin{itemize}
\item {\bf Application software} is software written for a specific set of tasks. In everyday electronic life, an image viewer allows us to inspect image data, an image manipulation program such as Adobe Photoshop or The Gimp allows us to change images in specific ways, and Microsoft Office Excel is a common way of dealing with catalog-data in the form of spread-sheets. 
\end{itemize}
Good application software has the advantage of being comparatively easy to operate --- in line with modern usage, such software offers you a menu structure for the selection of task, and a graphical and interactive interface. Also, while application software performs only a limited selection of tasks, good application software performs those tasks rather well, having been written by specialists who have a lot of experience with the kind of task in question.

Increasingly, web applications are playing an important role in everyday life --- applications that you run in your web browser, with a grey area between more complex applications and dynamically generated web pages. Dynamic web content does play an important role in astronomy, as well: Many researchers in astronomy will begin their day by looking over the new research articles posted as e-prints at [\href{https://arxiv.org/list/astro-ph/recent}{https://arxiv.org/list/astro-ph/recent}], use NASA's Astrophysical Data Service ADS [\href{https://ui.adsabs.harvard.edu/}{https://ui.adsabs.harvard.edu/}] to look up specific research articles, find information about specific objects in the CDS's SIMBAD astronomical data base [\href{http://simbad.u-strasbg.fr/simbad/}{http://simbad.u-strasbg.fr/simbad/}] or the NASA/IPAC extragalactic data base NED [\href{https://ned.ipac.caltech.edu/}{https://ned.ipac.caltech.edu/}] --- and, along the rest of us, use Google as a search tool or read up on helpful answers on sites such as Stack Overflow.

We will use some application software in the following, namely SAOImage DS9 for images and TOPCAT for operations involving tables. But in astronomical research, application software is usually not enough. For simple image operations you might get by with firing up the DS9 software, for instance. But at some point, sooner rather than later, you will want to do something more specialised, and more automatised, than application software can provide. Similarly, for data analysis. 
In some of the simplest cases, you might get away with loading the catalogue in Microsoft Excel and start analyzing your data in there. But in all other cases, including almost all of the interesting ones, your analysis will need a little more flexibility. That is when, again, you start writing a bit of code that helps you choose the right entries from the catalogues, and to produce helpful diagrams -- plots and histograms -- that allow you to make sense of your data. Then it becomes time to make use of a different kind of tool:
\begin{itemize}
\item A {\bf programming language} is a tool for writing your own custom applications. 
\end{itemize}
When you are using such a programming language for data analysis, you are in effect writing yourself a custom application that can be used for the specific analysis problem you need to solve. This approach has the advantage that you have (nearly) full control over what you will be doing what your data. 

It would not be an effective use of your time if you were to re-invent the wheel by using a programming language for writing your own routines for standard tasks such as opening different kinds of data files, or standard analysis operations. Instead, you should make use of useful collections of routine operations that have been written by helpful other people. In the ecosystems of programming languages, these usually come in the form of {\bf libraries} or {\bf modules}: chunks of codes that are easily included in your own program, and give you pre-programmed functionality you can use for your own specific purposes. That way, you need not write everything from scratch. But you will still need to program in order to string these tools together to do your bidding. 

What qualifies as a routine operation will depend on context, of course. Specialised astronomical modules provide you with tools for higher-level operations that are routine in astronomy, but not elsewhere. An ephemeris module will help you find the position of Solar System Bodies, for instance. Some routines may be adapted to a specific telescope, allowing you to reduce and analyze that telescope's data. While you are still learning, you will want to avoid some of those higer-level modules and re-invent at least some of the wheels in question, since writing a routine for completing some specified astronomical task is a good way of understanding what that particular task, and the astronomy behind it. When have become more advanced, ready-made modules represent a different problem: In research, it is important that you understand the different steps in any analysis you are doing. Using a module which is a ``black box'' for you represents a step in your analysis that you do not fully understand. In those cases, it is even more important than usual for your analysis to include suitable cross-checks and tests to ensure that it is indeed doing what you intend it to do.

\subsection{Concepts and operations}

In general, when working with astronomical data analysis, you need not understand all areas and aspects of a programming language. But there are certain concepts, and certain types of operations/manipulations, which constitute the basic working knowledge of virtually every astronomer working with data.

This starts with {\bf basic mathematical operations}. When you go from the magnitude to the flux emitted by an astronomical object, you will need the ``x to the power of n'' operation; on the way back, the logarithm. Whenever you perform calculations with your data, you will need the appropriate operations.

Data points come in sets: the pixel data for an image corresponds to a two-dimensional arrangement, while a list of properties for astronomical objects will correspond to a one-dimensional chain of values. Programming languages feature suitable {\bf data structures} for this kind of connected data, such as {\bf lists}, {\bf arrays}, {\bf tuples}, or different kinds of {\bf table} (the meaning of those words can differ somewhat between one programming language and the next). 

Knowledge of these data types and the various ways of manipulating them is a must, along with knowledge of more basic types such as {\bf strings}, {\bf integers} or {\bf floating point numbers} --- and of course the basic concept of storing values in a variable in the first place!

For operations on our data, we need {\bf control structures}. If we want to perform a certain operation on every element of a list, for instance, we will need something like a {\bf for loop}. In order to distinguish between different cases --- a structure that allows us to apply a certain combination of operations to every element of a list. 

There will also be situations where we might want to perform a certain operation on some elements of the list, but not on specific other elements --- to accomplish this, we need {\bf if clauses}, that is, tools that tell our program to apply certain operations only if specific conditions are met, but not otherwise. 

In addition to this kind of general knowledge, which is required when learning pretty much any general programming language, astronomers should have at their disposal a set of programming tools for more specific tasks --- which often equates with familiarity with particular libraries or modules. Often, we want to {\bf visualize} our data, so knowledge of how to create various kinds of {\bf plots}, {\bf diagrams} or {\bf histograms} (both one-dimensional histograms and two-dimensional density plots) is essential.

Last but not least, how do we get data into our program, and our results out again? If we have obtained the data by downloading a file, we will need to know about proper {\bf input/output operations} (in short, i/o). For certain data formats, such as the ubiquitous FITS image files that are the usual format for astronomical images, or for astronomical tables in FITS or VOTable format, there are special functions that read the data in a way that makes it particularly easy to start working with them. 

When we do not download the data in the form of files, but instead access astronomical data bases, there is an additional issue. We need to tell the data base which specific set of data we would like to access. In order to do so, we must submit a {\bf data base query}, or {\bf query} for short, to the data base: a formalized request for data, written in a specific query language.  A number of astronomical data bases are organised in the shape of a {\bf Virtual Observatory} (VO) --- data bases that conform to certain common standards to enable easy access for all astronomers. The query language for the VO is the {\bf Astronomical Data Query Language (ADQL)}, which is similar to the more general Structured Query Language (SQL, pronounced either ''S--Q--L'' or ''sequel''). Queries in this language are useful both in the context of an application software like TOPCAT, where they can used in the framework of the {\bf Table Access Protocol} to download a specific subset of data from an online data base via the Internet, or as part of a Python program.

There is another aspect of all this, which would require a tutorial of its own for proper treatment: data can be {\bf generated} by software, too. Astronomy isn't only about observing. In the end, we want to understand the objects we observe. That involves creating simplified models for these objects. If a star is (put simply) a gigantic ball of plasma, held together by its own gravity and heated up by nuclear fusion reactions in its core, then if we create a {\bf simulation} of such an object, using our knowledge of the laws of physics, the resulting model should have similar properties to a real star (as we can check using observations).

Simulations, too, require coding. In physics, only the simplest situations can be described ``analytically'', that is, writing down what happens in terms of simple functions such as $\sin(x), \cos(x)$, polynomials and the like. For more complicated situations, you will need to simulate what happens numerically: starting with the initial situation, and then letting the computer reconstruct, time step by time step, what happens. We will encounter a very simple simulation in section \ref{Simulation}.

\subsection{Software/language choices}

Every text on data processing has the same problem: For specific applications, there is usually more than one application software, and of course there are numerous programming languages. In teaching about data processing, one should include specific examples, and students should work through such examples themselves, gaining hands-on experience with all that data processing involves. If the author chooses to present these examples in one specific programming language, at least some students will later, when they are working on a specific project, need to re-learn a different programming language.

This is not as bad as it sounds, though. Most programming languages, and most applications, share similar concepts and functions. Once you have learned about those in the framework of one specific programming language, or application software, switching to another language or software will be much easier than starting from scratch. Thus, learning what this tutorial has to offer is definitely not a waste of time, even if it should turn out that later on you will need to adapt to other software.

In this tutorial, I have chosen some common application software for simple operations: {\bf SAOImage DS9} (DS9 for short) is a comparatively simple image viewer that also allows some basic manipulation of astronomical images. {\bf TOPCAT} is a standard tool for accessing data from the Virtual Observatory. The programming language used for more complex tasks is {\bf Python}, which is widely used in astronomy. This wide use has a great advantage: astronomers have been writing helpful astronomy-specific libraries and modules for Python, and are actively maintaining them. If you're starting a career in astronomy, chances are that you will do your basic programming in Python.

All that said, let's get started. To get our bearings, we start with something simple: before we delve into astronomical Python and start coding ourselves, let us begin with two simple use cases for application software: In section \ref{DS9}, we will look at astronomical images and combine red, green and blue filter images into a color image. In section
\ref{TOPCAT}, we will look at some basic table operations with TOPCAT.

\section{Data basics: images, spectra, tables}
\label{DataBasics}

In astronomy, just as in other sciences, we are not interested in data for data's sake. We want to do astrophysics: we want to use data to further our understanding of the universe. Before we look at specific tools, and learn how to use them, let us consider some of the properties of astronomical data, as well as some of the specific ways of extracting information from them.

\subsection{Images: Colour, brightness, pixels}
\label{ImagesBasics}

\begin{figure}[htbp]
\begin{center}
\includegraphics[width=\linewidth]{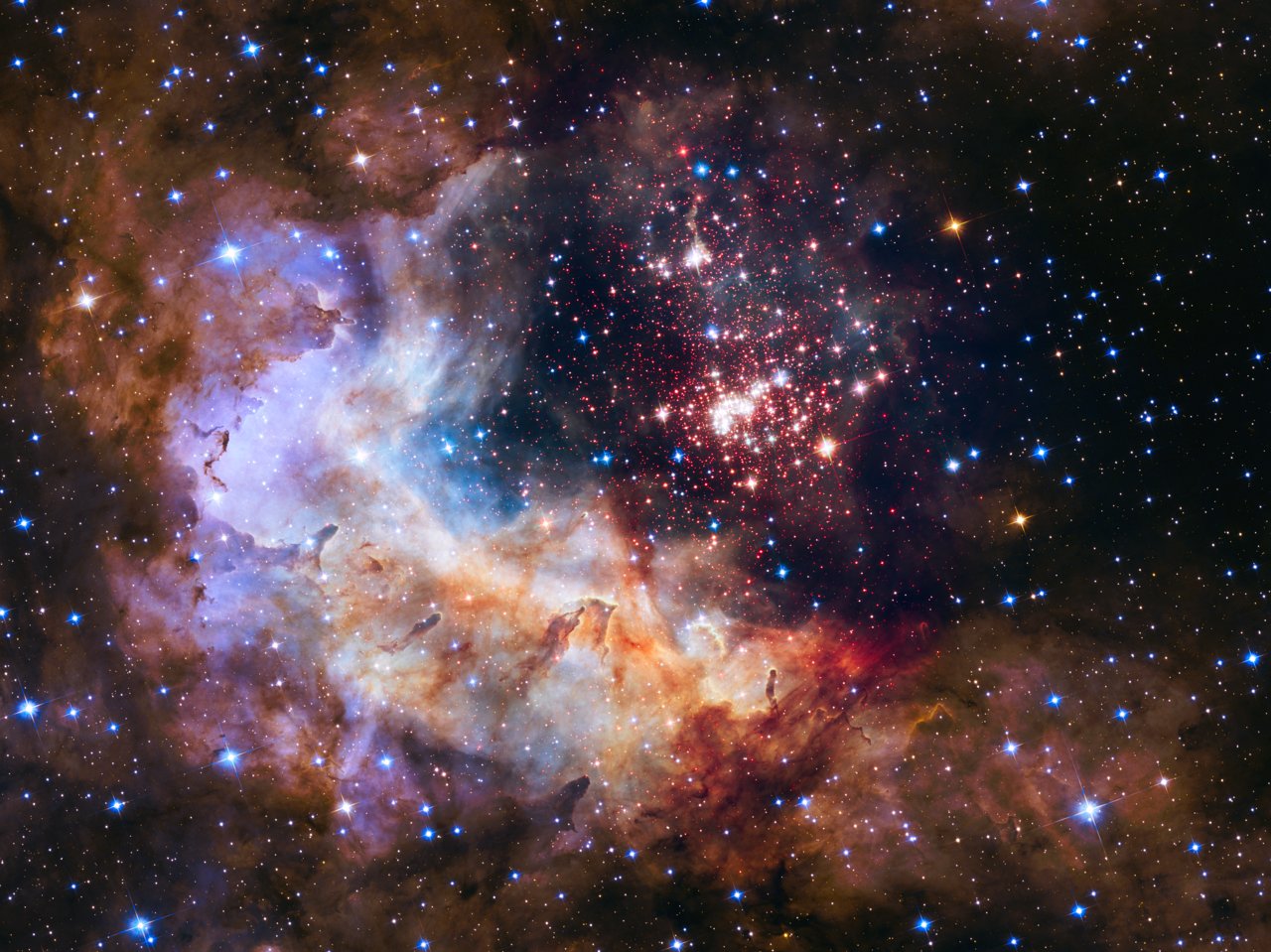}
\caption{Image of the open star cluster Westerlund 2, taken by the Hubble Space Telescope. The image data was downloaded from \href{https://www.spacetelescope.org/images/heic1509a/}{spacetelescope.org}. Image credit: NASA, ESA, the Hubble Heritage Team (STScI/AURA), A. Nota (ESA/STScI), and the Westerlund 2 Science Team }
\label{Westerlund2}
\end{center}
\end{figure}

Astronomers take images of astronomical objects, using telescopes and suitable instruments attached to those telescopes. Public versions of such images can be stunningly beautiful, and contribute significantly to the fascination of the general public with astronomy. The underlying science images are commonly stored in a format known as FITS, which stands for the ``Flexible Image Transport System'' --- a flexible file format that astronomers have been using for images, spectra, data tables and more since the 1980s. When you encounter a professional astronomical image, it is likely to be in that particular format, with file extensions ``.fits'' or ``.fit'' on an older Windows machine. Figure \ref{Westerlund2} shows one of the iconic images from the Hubble Space Telescope, namely the open cluster Westerlund 2. 

Using this image as an example, we can demonstrate a number of properties of astronomical image data. Phenomenologically, the image contains two types of information: the stars we see in the image are much too small for even Hubble to see any of their structure. They appear as {\bf point sources}. In addition, we have  {\bf extended sources}, in this case a region of ionized hydrogen (HII, in astronomical parlance), which are, as the name says, extended areas with varying brightness and colour.

\begin{figure}[htbp]
\begin{center}
\includegraphics[width=0.2\textwidth]{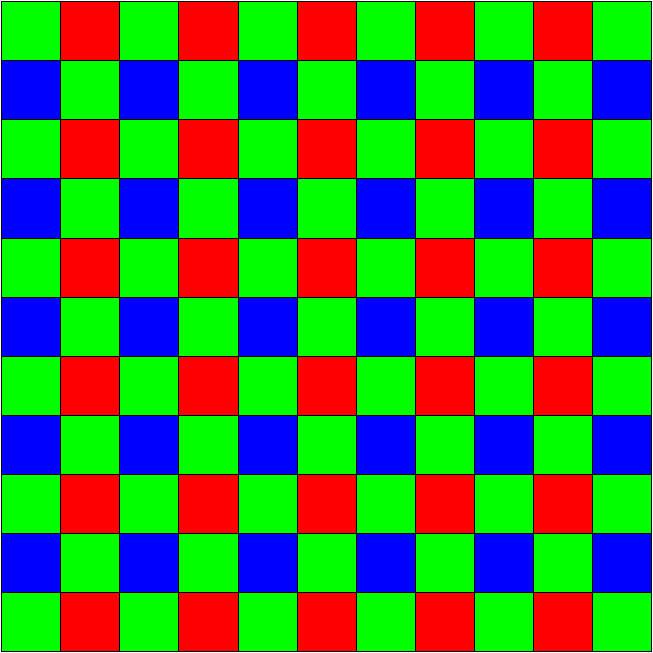}
\caption{Part of a Bayer mask pattern: an array of filters installed in front of detector pixels to enable the quick creation of a colour image }
\label{BayerMask}
\end{center}
\end{figure}

While we tend to think of astronomical images as a rendition of ``what's up there in the sky,'' there are several aspects in which such images are {\em not} faithful renditions --- and those aspects are crucial for understanding astronomical image data. Let us start with the {\bf colours}. Professional astronomical images are black-and-white images. One reason for this is that digital cameras are, at their most basic, black-and-white. For each pixel, they can only record how much light has fallen onto the collecting area for that pixel (more specifically: how many photons have fallen). Consumer cameras as in your smartphone or your digital camera are only able to produce colour images because they have an array of filters installed in front of the array of light-detecting sensor pixels. A common pattern is the Bayer mask, part of which is shown in figure \ref{BayerMask}.

When such a consumer camera has taken an image, the colour information is interpolated, and a colour image is displayed and saved. (If you have a camera that can save images in some kind of ``raw'' mode, you can see the not-yet-interpolated pattern.)

For astronomical images, a fixed filter mask is impractical for several reasons. Astronomers would like to get the full resolution for their images, so reducing resolution in each color band by having information about the green brightness in every second pixel only, and about red and blue in every fourth pixel, is a drawback. Colour interpolation means that some of the colour information gets lost.  Also, astronomers use a bewildering array of possible filters, not just those corresponding to red, green, and blue (R, G, B) --- some of those filters capture a wider wavelength range, while narrow-band filters might capture just a particular spectral line. Astronomers need the flexibility of putting these different filters in front of their camera. On the plus side, most astronomical objects change only very slowly. Using different filters in succession, taking an image first through one filter, then through the next, is perfectly feasible; those images will all show the target object in effectively the same state. (On the rare occasions where speed is of the essence, astronomers will use something like parallel cameras observing through different filters. A case in point are gamma ray burst afterglows, which fade on a time scale of several minutes. The GROND instrument at the 2.2m ESO/MPG telescope at La Silla observatory was optimised for observing such afterglows through seven different filters, simultaneously.)

From these separate images, we can reconstruct colour images. (One simple way of doing this will be shown in section \ref{MakingColourImage}.) If your aim is to produce a ``pretty picture,'' there are several filter combinations that will serve. Perhaps the best-known is the combination of images taken with the filters B, V, and R respectively from the Johnson-Morgan photometric system to represent the colours blue, green, and red. 

\begin{figure}[htbp]
\begin{center}
\includegraphics[width=\linewidth]{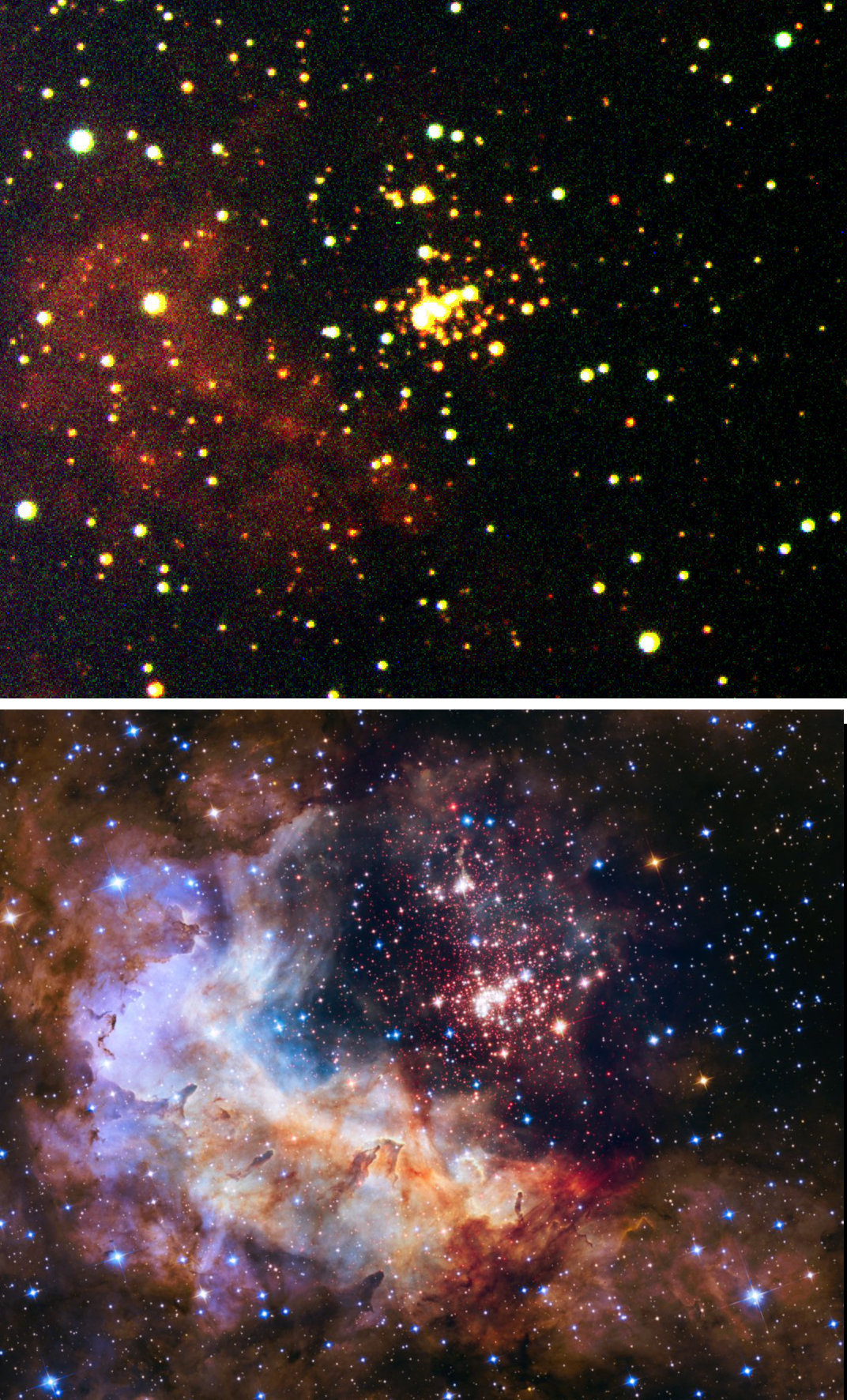}
\caption{Two images of Westerlund 2. Image credit top: 2 m Faulkes Telescope operated by Las Cumbres Observatory at Siding Spring under license \href{https://creativecommons.org/licenses/by-nc/2.0/deed.en_US}{CC BY-2.0}. Bottom image: NASA, ESA, the Hubble Heritage Team (STScI/AURA), A. Nota (ESA/STScI), and the Westerlund 2 Science Team
}
\label{WesterlundComparison}
\end{center}
\end{figure}

But most images taken with the Hubble Space Telescope do not include B, V and R versions. Colour images composed from them will be false-colour, and look markedly different from the colours we see around us. That is OK as long as you use the colours only to help you discern structural details, but you should be aware that the result is not a faithful rendition of astronomical colours. (There are heated discussions among amateur astronomers\footnote{These days, with sophisticated technology available, the so-called amateur astronomers are working very, very professionally indeed, and produce spectacular images.}
about what constitutes faithful colours; some amateurs strive for the optimum of faithful colour rendition. Their images are probably the closest you can get these days to faithful colouring.) Figure \ref{WesterlundComparison} shows two images of the same region side by side: the Westerlund 2 image already shown in figure \ref{Westerlund2}, and an image of the same region, taken in April 2017 with the 2 m Faulkes Telescope operated by Las Cumbres Observatory at Siding Spring in Australia. 
The left-hand image was reconstructed from three images taken through B, V, and R filters. The red colour of the gas cloud is probably fairly realistic. The Hubble image no the right, while much higher resolved and full of structural details, doesn't show faithful colours. Few Hubble images do (which does not take away from, but possibly adds to their power to fascinate viewers).

Another aspect to keep in mind is the brightness scale of the image. FITS images typically have 16-bit brightness values, that is, each pixel can take on values from 0 to $2^{16} = 65~536$ (apart from a possible offset value); alternatively, brightness values within that range are stored as floating-point numbers. Even in the integer case the best current computer screens cannot faithfully display that dynamic range. And even if they could, the result of simply displaying pixel brightness in a linear fashion would not let you see the interesting details, and understand your image. 

\begin{figure}[htbp]
\begin{center}
\includegraphics[width=\linewidth]{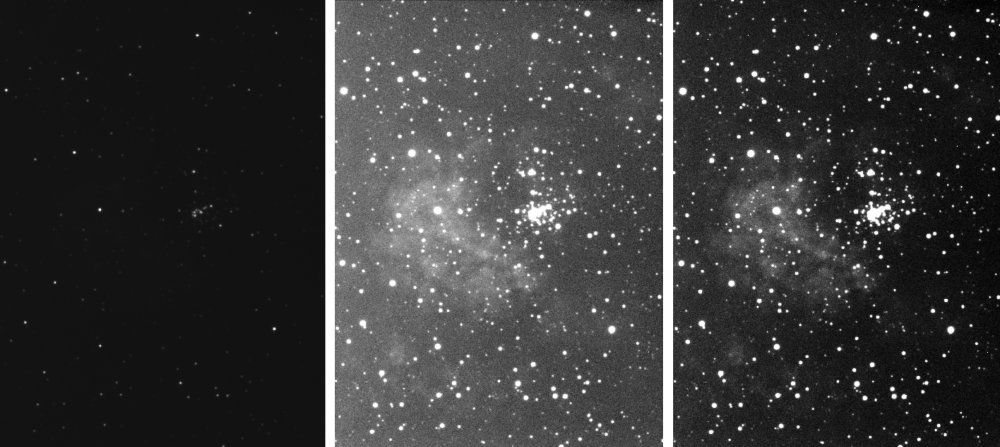}
\caption{Part of the Westlund 2 image taken through an R (red) filter in April 2017 with the 2 m Faulkes Telescope operated by Las Cumbres Observatory at Siding Spring in Australia with different scaling. Left: Linear scaling from 0 to 65536. Center: Linear scaling from 4572 to 6002. Right: Square scaling from 4572 to 6002}
\label{ContrastComparison}
\end{center}
\end{figure}

Published astronomical images will always\footnote{With possible exceptions when an author is trying to make a point about scaling and astronomical images.} have some kind of {\bf (brightness) scaling} applied to them. Figure \ref{ContrastComparison} again shows the 2 m Faulkes Telescope image of Westerlund 2, this time only the one taken through the red R filter. The leftmost version has linear scaling (pixel value linearly related to brightness shown in the image), with the smallest pixel value taken as black and the brightest taken as white. The only structure visible corresponds to the locations of the brightest stars. No cloud structure is visible in this rendition. The center version is still using a linear map, but now every pixel value smaller than 4573 is displayed as black, and every pixel value larger than 6001 is displayed as white, and brightness values between 4572 and 6002 are displayed as the various shades of grey in between.
By concentrating on this narrow range of brightness values, we not only see more stars, but also some of the structure of the cluster's hydrogen clouds. But at this scaling, the background sky looks rather bright. If we insert a square function --- where the displayed brightness is proportional to the square of the pixel brightness value --- we obtain a clearer differentiation between the dark background and the brighter areas corresponding to the cloud. This is shown in the version on the right.

Choosing a good scaling is not an exact science, but a matter of artisanship: a good scaling will serve to illustrate the structures that are scientifically interesting. But you should always keep in mind that there were choices involved in creating the image. 

\begin{figure}[htbp]
\begin{center}
\includegraphics[width=\linewidth]{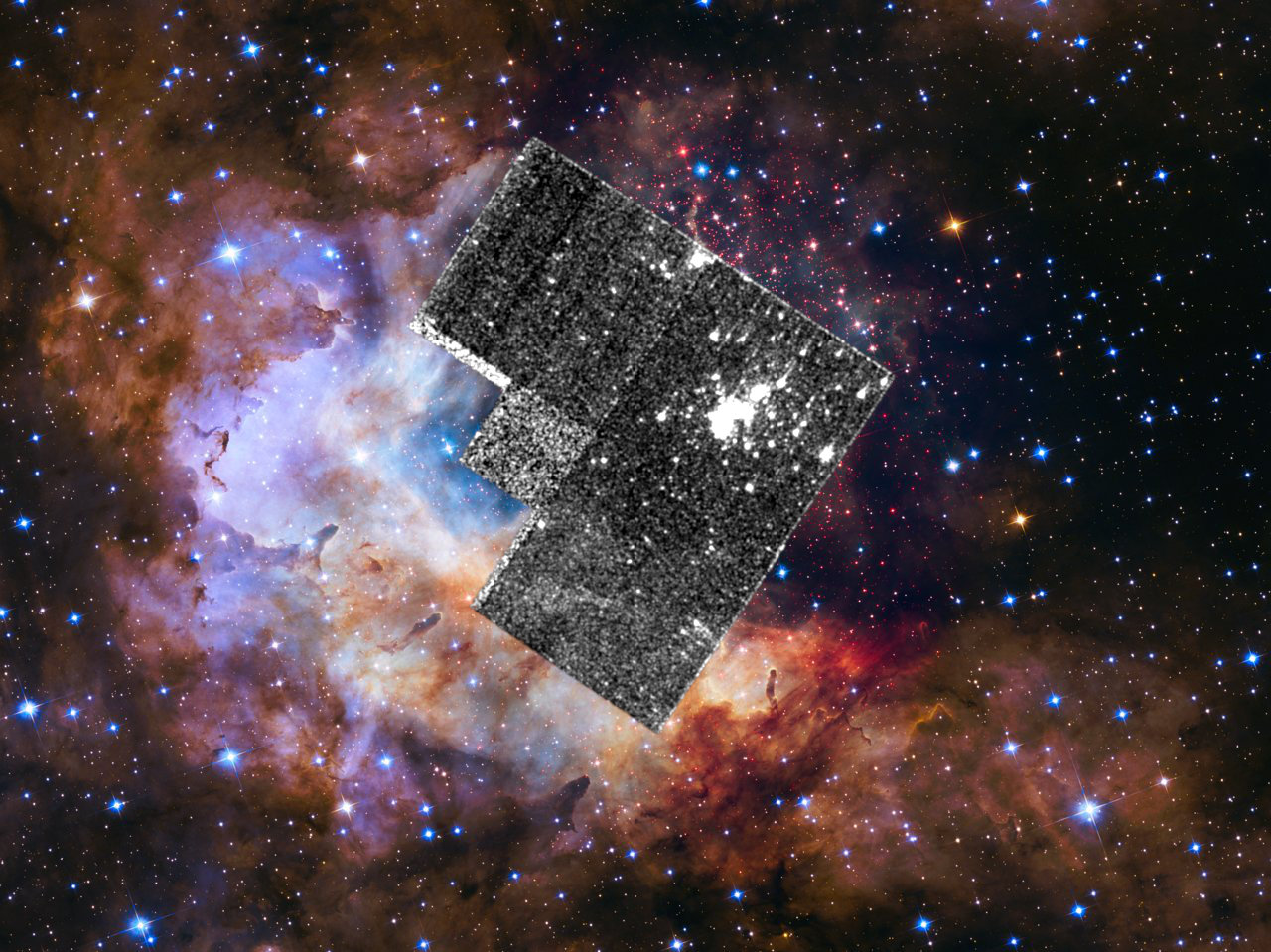}
\caption{Image credit: NASA, ESA, the Hubble Heritage Team (STScI/AURA), A. Nota (ESA/STScI), and the Westerlund 2 Science Team}
\label{WesterlundInset}
\end{center}
\end{figure}

A number of images are really {\bf mosaics}, where several images have been stitched together to form a larger picture. The beautiful Hubble version of Westerlund 2 in figure \ref{Westerlund2} is a case in point, as it is a composite image using observations with Hubble's Advanced Camera for Surveys (ACS) and its Wide Field and Planetary Camera 3 (WFPC3). Figure \ref{WesterlundInset} shows
a sample WFPC3 image (although 
probably not one used in the final composite\footnote{
I didn't find any of the original WFPC3 images in the MAST archive; all I could find were already (smaller) composites.}) 
pasted into the final colour image to give you an idea of the footprint of the WFPC3. In fact, the WFPC3 inset is already a blend of four images, created from the three image chips (CCDs) of the Wide Field Camera (the three larger squares) and the image chip of the Planetary Camera (smaller square nestled into the corner formed by the other three).

\begin{figure}[htbp]
\begin{center}
\includegraphics[width=\linewidth]{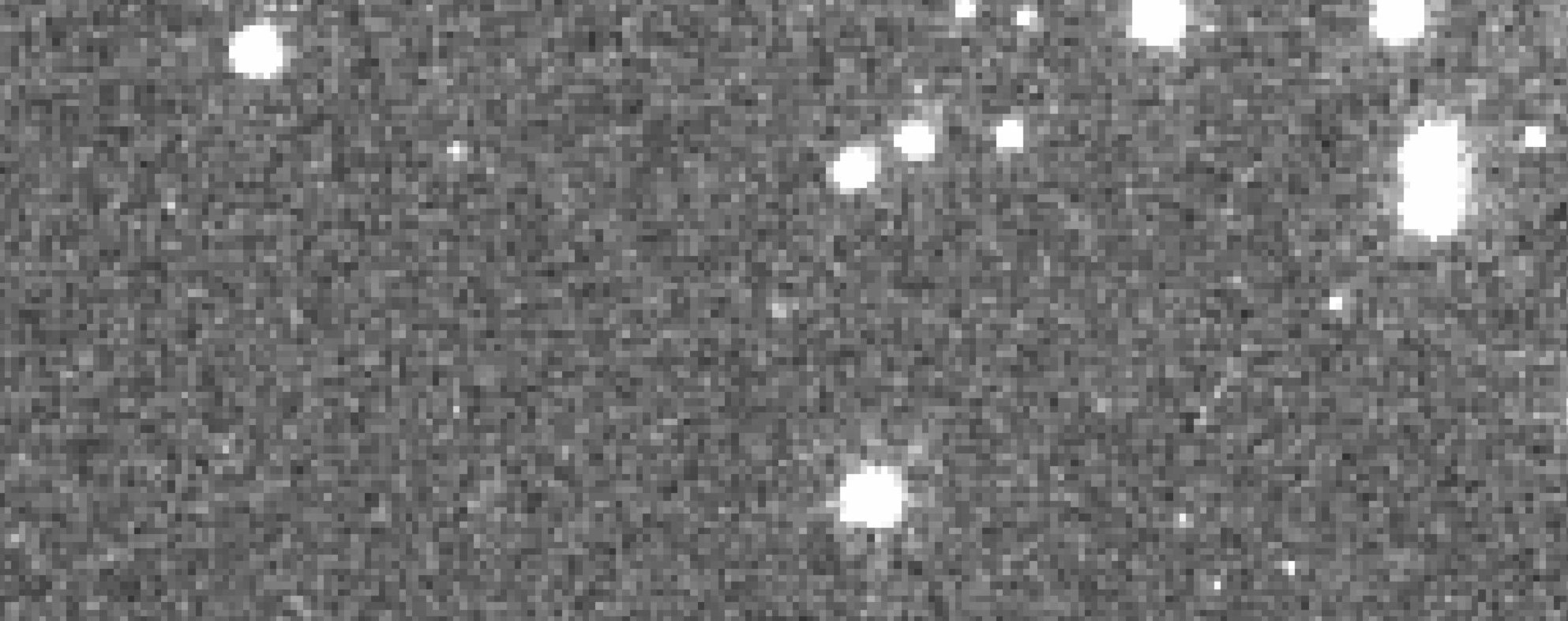}
\caption{Zoom in on the WFPC2 image of Westerlund 2. Image credit: NASA, ESA}
\label{Westerlund2Zoom}
\end{center}
\end{figure}

As a next step, let's zoom in into the WFPC3 images, more concretely: into one of the Wide Field Camera squares. The result can be seen in figure \ref{Westerlund2Zoom}.
There are several points of note. The first is that the image is made up of discrete square fields: pixels. That is no surprise if you have ever looked very, very closely at digital photographs. It also means that, at the lowest level, working with digital images means working with pixels, and with the brightness value associated with each pixel.

\begin{figure}[htbp]
\begin{center}
\includegraphics[width=0.35\linewidth]{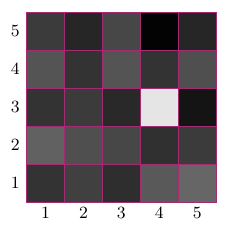}
\caption{Pixels and pixel coordinates}
\label{PixelExample}
\end{center}
\end{figure}

Pixel positions are described by a pair of (integer) coordinates for each pixel. A schematic example is shown in figure \ref{PixelExample}.

The brightest pixel in the $5\times 5$ array would have the coordinates $(4,3)$, since it is in the fourth column from the left, and in the third row from the bottom. (Beware, other conventions exist! Some count rows from the top. Some start the count at number 0, not number 1.)

\subsection{Images: PSF and noise}

\begin{figure}[htbp]
\begin{center}
\includegraphics[width=\linewidth]{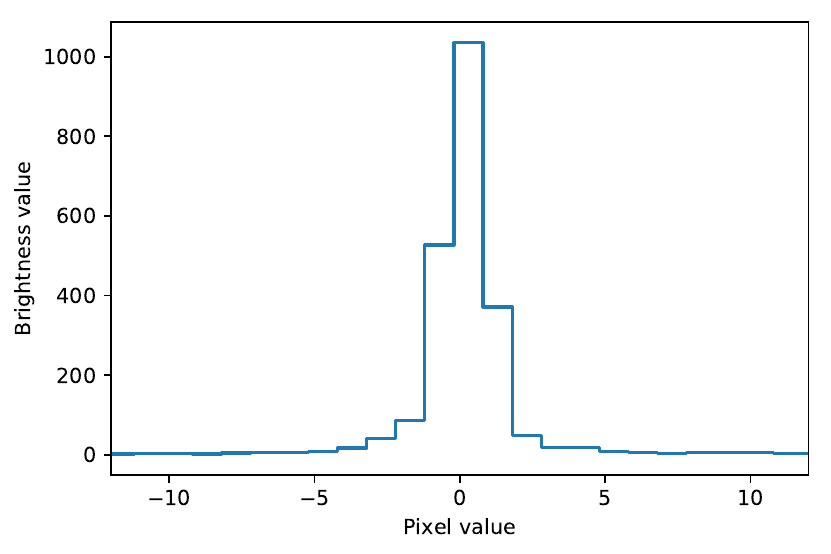}
\caption{Brightness profile of the bright star near the bottom center of figure \ref{Westerlund2Zoom} }
\label{Westerlund2PSF}
\end{center}
\end{figure}

Back to figure \ref{Westerlund2Zoom}. The disk-shaped bright objects in the image are stars. Here's the thing: Westerlund 2 is at a distance of about 20~000 light-years from us. At that distance, even an especially large super giant with 1500 solar radii should subtend an angle of a mere $0.002''$. Each pixel in the WFPC2 image has a side length of $0.1''$. If our image were a faithful map showing the exact direction whence light reaches us from the sky, even these largest known stars would fall within a single pixel. Instead, they and the much more common markedly smaller stars are smeared out and appear in the image as disks. Figure \ref{Westerlund2PSF} shows a brightness profile of the star at the bottom center of figure \ref{Westerlund2Zoom}. The disk is a few pixel wide.
Why the disk? Why not a single pixel? The answer, as you probably know, is that light is a wave phenomenon, and that a wave passing through an opening --- in this case, the aperture of the telescope --- is diffracted. The result is a diffraction pattern that makes the image of a point source a disk (if you were to look very closely, a disk with concentric rings around it). The larger the telescope aperture, the smaller the disk --- which is one key reason to build ever larger telescopes: ever better resolution for the resulting images. The function that defines the brightness distribution that results when a telescope instrument produces an image of a point source is called the {\em point-spread function}, abbreviated PSF.

\begin{figure}[htbp]
\begin{center}
\includegraphics[width=\linewidth]{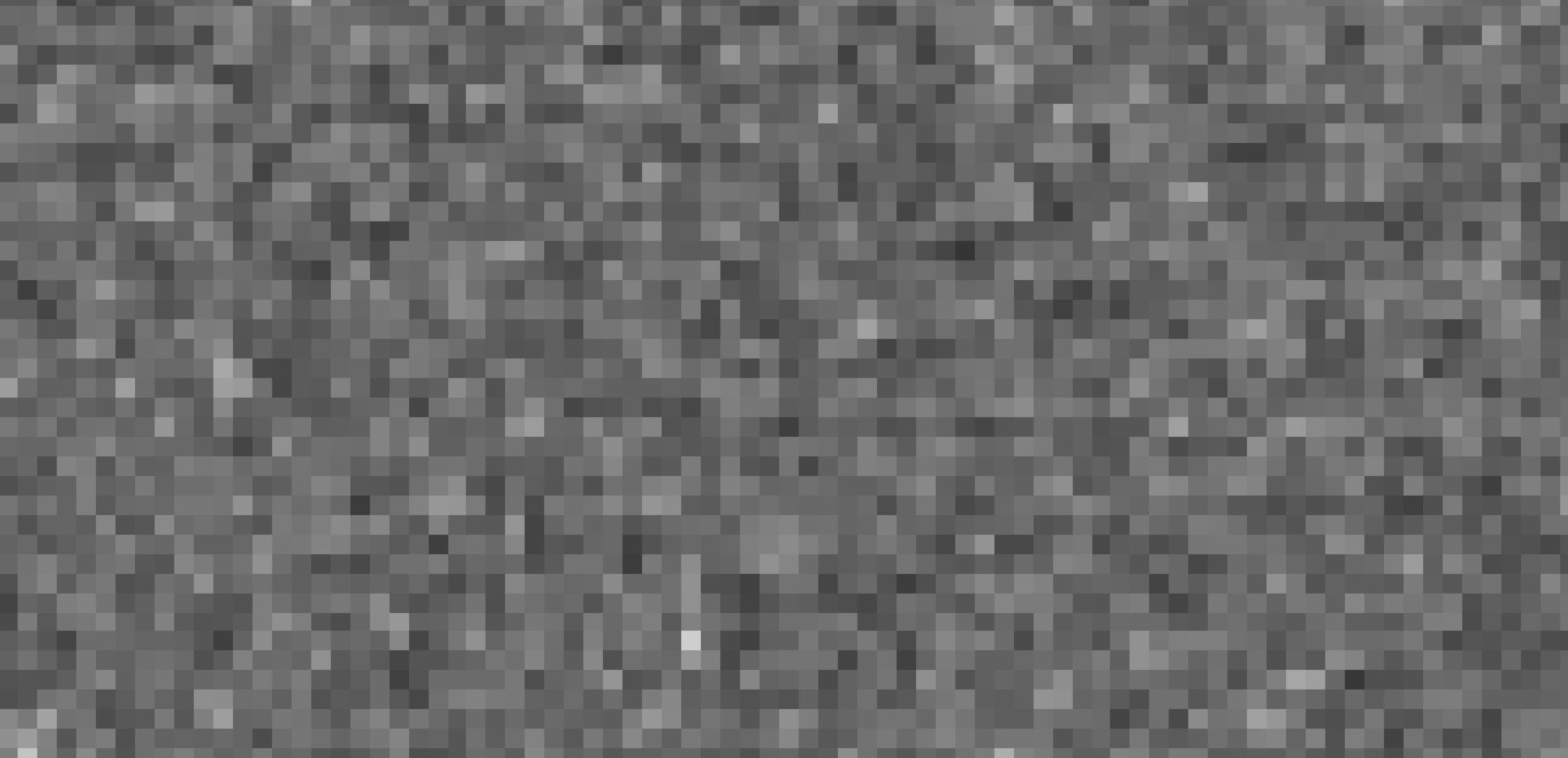}
\caption{Zooming in on part of the background of figure \ref{Westerlund2Zoom}}
\label{Westerlund2bg}
\end{center}
\end{figure}

Last but not least, look at the part of figure \ref{Westerlund2Zoom} that is not stars, but background. The background is not uniformly black, but mottled grey --- a section, shown at even larger magnification, can be seen in figure \ref{Westerlund2bg}.
There are several reasons that some of the pixels are brighter, others less bright. One is the presence of distant, unresolved astronomical objects. But those cannot explain the small-scale variation from pixel to pixel --- remember that even a point source would appear as a smeared-out disk! Instead, the variability is {\bf noise} --- a spurious addition that tells us nothing about the astronomical light sources out there. 

The most fundamental effect is one of statistics: light reaches our detectors in the form of photons, of light particles. The intensity of light reaching us from a specific source determines the {\em probability} of a photon arriving within a certain time interval. But the arrival itself is a random event. (You probably know a similar situation: radioactive decay, where the decay probability per unit time is constant, but each decay will still occur at an unpredictable random time.) This randomness translates into pixel brightness fluctuations. The relative size of these fluctuations shrinks as the total number of photons collected grows. This is the other key reason why astronomers want large telescopes (and in most cases still need long exposure times, in addition): the more light they can collect from a distant object, the smaller the relative fluctuations, the higher the signal-to-noise ratio, and thus the clearer the image of those distant structures.

There are other kinds of noise. If you inspect raw, unprocessed images taken with the Hubble Space Telescopes, you will find one kind that is typical for space telescopes: traces left by cosmic particles depositing their energy in the detector, leading to either longish streaks or more sharply defined dots, depending on the direction the particle was travelling. 
\begin{figure}[htbp]
\begin{center}
\includegraphics[width=\linewidth]{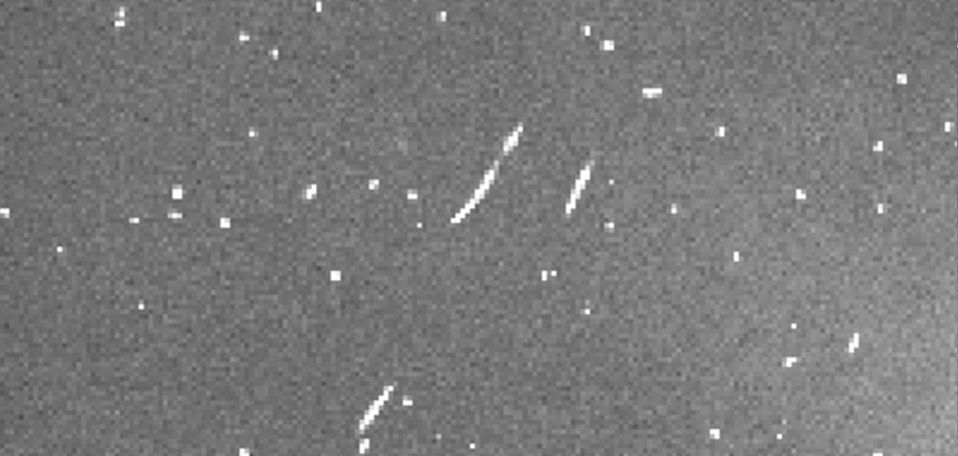}
\caption{Traces of cosmic ray particles in an image taken with the Hubble Space telescope. Image credit: NASA and ESA}
\label{CosmicRayTraces}
\end{center}
\end{figure}
Figure \ref{CosmicRayTraces} shows an example (albeit from Hubble targeting a different object, the Eagle nebula). Also, there is noise from the electronic devices involved (although cooling key electronic elements down can reduce that kind of noise considerably).

\subsection{Images: Noise and flatfielding}

When astronomers prepare data for the extraction of astronomical information, a process commonly called {\bf data reduction}, there are several typical steps they take in order to reduce both the noise produced in their instrument and the instrument's idiosyncrasies when it comes to recording brightness.\footnote{In preparing this section, I have profited from two online sources: The lecture notes by S. Littlefair for the course {\em PHY217 -- Observational Techniques for Astronomers} he taught in 2014, \href{http://slittlefair.staff.shef.ac.uk/teaching/phy217/}{http://slittlefair.staff.shef.ac.uk/teaching/phy217/}, last access 2019-11-01,
and parts of the e-book by R. A. Jansen, {\em Astronomy with Charged [sic] Coupled Devices} (2006), \href{http://www.public.asu.edu/~rjansen/ast598/2006ACCD.ebook...1J.pdf}{http://www.public.asu.edu/\~rjansen/ast598/2006ACCD.ebook...1J.pdf}, last access 2019-11-01.
} 

The image chips consist of little pixel elements; light falling onto a pixel sets free some electrons. In a CMOS, each pixel also contains the electronics, including a little amplifier, to read out a signal that indicates the number of electrons, and thus the amount of light. In a CCD camera, the read-out process is more involved, and involves herding electrons to the end of each pixel row, then moving them to an amplifier. In both cases, ideally, the number of electrons will be in direct proportion to the number of photons that have fallen onto that pixel during the exposure time. And in the end, those electrons are dumped onto a capacitor, whose voltage is measured. Since the voltage across a capacitor is proportional to the accumulated charge, the result gives us a measure of the number of electrons, and thus of the amount of light we have captured. The analog voltage value is fed into an analog-to-digital converter (ADC) which converts the voltage value into an integer digital number, corresponding to analog-to-digital-units (ADUs) or, for short, to counts associated with that pixel.

The conversion factor from the number of electrons to ADUs is called the {\bf gain}, typically given in units of $e^-$/ADU. Since the ADU values are digital, a gain higher than 1 will introduce {\bf quantization noise} through rounding errors --- for a gain of 6 $e^-$/ADU, the value 1 ADU could stand for anything between 1 and 5 electrons. A common range is for the digital numbers used to store the count value to have 16 bits, and for each pixel to have count numbers between 0 and 65~535.

There is also the question of choosing the zero level for our voltage --- that is, the level that will correspond to the number 0 in our digitalized documentation for each pixel. When no light falls onto the pixel, there is still bound to be some level of noise, corresponding to a voltage value which is called the {\em bias}. But mapping the {\em average} bias, call it $V_{0,avg}$, to the zero value of the ADU is usually not a good idea. Our measurements of that zero-light voltage are going to fluctuate slightly, its values sometimes a little higher, sometimes a little lower. Frequently, we will want to average images, either pixel by pixel or after aligning the astronomical objects in those images. When taking such averages, above-average and below-average bias levels should cancel out. But if we set zero ADU to correspond to the average bias, then all fluctuations below that level will be mapped to zero as well, and cannot cancel the fluctuations that are higher than average.
You can take care of the bias by taking a {\bf bias frame} --- an image taken with your camera shutter closed (so no external light falls onto your chip) and an exposure time as close to zero as your camera can manage. Subtract that bias frame --- or, again, the average of several such frames --- from your science image, and you've taken care of the bias. 

Depending on the type of camera you are using, and on the type of observations, you might need to perform different kinds of subtraction. For older CCD cameras in particular, you might need to take a {\bf dark frame}, recording the contribution of electric currents due to thermal fluctuations, and making a suitable correction. Under excellent conditions, and in particular for infrared observations, you might need to take a {\bf sky frame} to subtract the sky brightness, including effects like airglow, faint radiation emitted by the Earth's atmosphere itself. 

\begin{figure}[htbp]
\begin{center}
\includegraphics[width=0.9\linewidth]{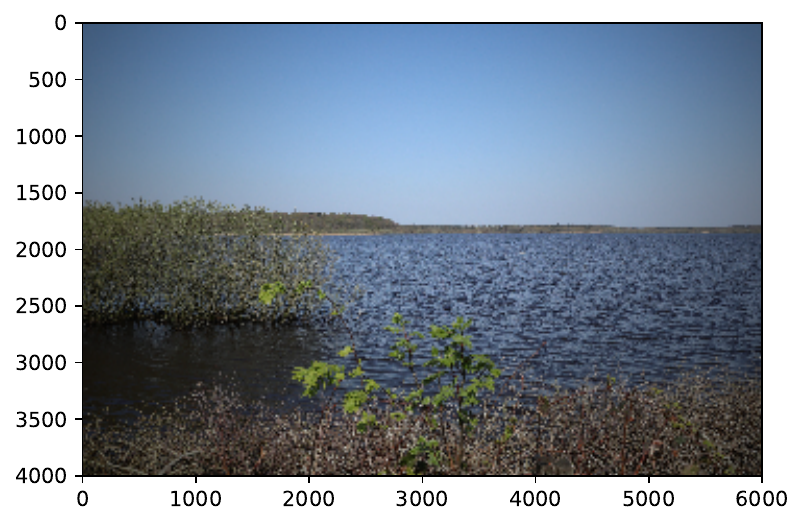}
\caption{Holiday snap with cxslearly visible vignetting}
\label{SampleVignette}
\end{center}
\end{figure}

The corrections mentioned so far --- bias frame, dark frame, sky frame --- all involved erroneous additions to pixel values, and the correction required the subtraction of suitable correction frames. Another, different type of correction is necessary because the sky is bound to get mapped to your digital image unevenly --- some parts of it too bright, others not quite bright enough. I will illustrate this using a non-astronomical image, since the effect is more obvious in an everyday setting. Fig.~\ref{SampleVignette} shows a holiday snapshot, a lake in Northern Germany. 

But you can see that the brightness is not as you would expect: near the edges, the image becomes darker. This is not a property of Northern-German lake landscape; it is a property of the camera I have used to record the image, an effect known as {\em vignetting}.\footnote{Actually, I have added that effect afterwards, artificially, but let's pretend it is real. Real vignettes look very similar.} In a telescope image, you might also see dark, ring-shaped smudges reminiscent of coffee stains; those are caused by dust flecks on the telescope mirror.

Systematic brightness distortions like this can be removed as follows. In our example in Fig.~\ref{SampleVignette}, the brightness of each pixel is a combination of the brightness of the part of the object (in this case, landscape) we have recorded on the one hand, and the sensitivity (or lack thereof) of our telescope-instrument combination in that specific image region on the other.

\begin{figure}[htbp]
\begin{center}
\includegraphics[width=0.9\linewidth]{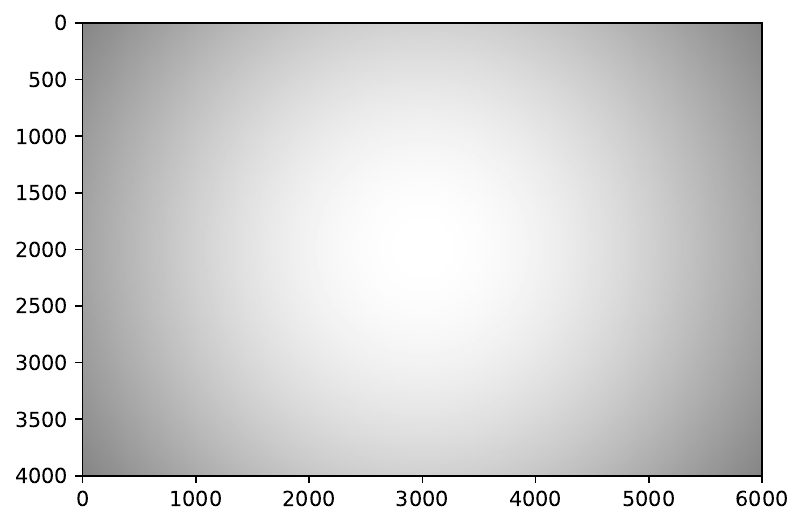}
\caption{Flatfield image taken with the same telescope-instrument combination as the holiday snap, with clearly visible vignetting}
\label{SampleVignetteFF}
\end{center}
\end{figure}

We can reconstruct at least the relative sensitivity of our telescope-instrument combination by taking an image of a scenery with completely uniform brightness. In that case, all brightness variations are, by definition, not due to our target scenery (which is completely uniform), but due to sensitivity variations. Such an image can be seen in Fig.~\ref{SampleVignetteFF}, and is called a {\bf flat field (image)}. In astronomy, such images are produced either by pointing the telescope at a uniformly lit canvas within the dome, or else by waiting for dusk (or dawn) and pointing the telescope at the sky, whose scattered light will be sufficiently uniform across the typical narrow fields of view of astronomical telescopes (``sky flats,'' specifically ''dusk flats'' or ''dawn flats'').
For the flatfield image, too, one can produce dark frames (taken at the same exposure as the flatfield image) and subtract them, resulting in a suitably corrected {\bf master flat}. That master flat encodes the sensitivity for each image pixel. With this information, we can correct for the sensitivity variations as follows. Assume that a pixel in the master flat is twice as bright as a second pixel. That would mean our setup is only half as sensitive for the second pixel than for the first. But if we were to take an image, and then to multiply the brightness value for the second pixel with the factor two, we would have compensated for the different sensitivity levels.

\begin{figure}[htbp]
\begin{center}
\includegraphics[width=0.9\linewidth]{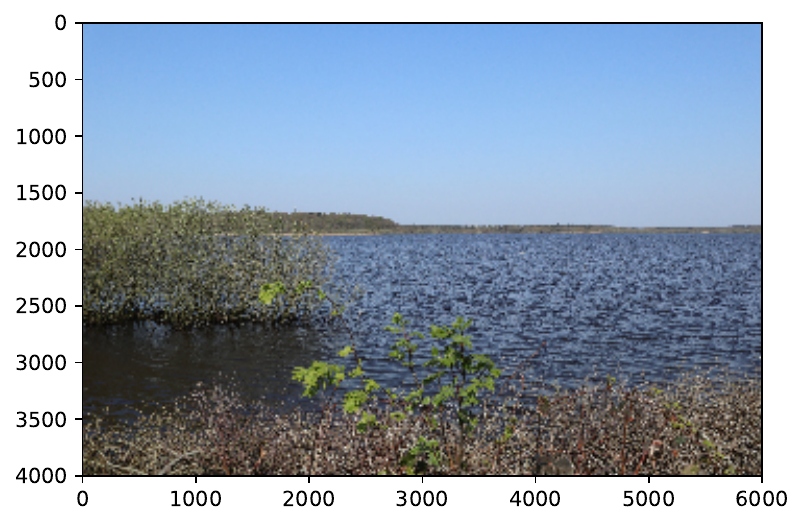}
\caption{Flatfield-corrected holiday snap}
\label{SampleVignetteCorr}
\end{center}
\end{figure}

More generally, we can restore proper relative brightnesses of all our pixels by dividing our science frame by our master flat, pixel by pixel. The result for our holiday snap can be seen in Fig.~\ref{SampleVignetteCorr}.
The compensation is not perfect. For instance, if fewer photons have reached a certain detector pixel, then the statistical noise will be somewhat larger for that pixel. Dividing the pixel value by a factor, as one does in flatfielding, will not get rid of that additional noise. 

All in all, we have learned the basics of how astronomers are reducing their image data --- compensating for noise that is added to each pixel by subtracting suitable compensation terms (bias frame, dark frame, sky frame), and afterwards compensating for sensitivity variations by dividing by a suitable compensation term (flatfield image).

To sum up the last few sections: The digital astronomical images used by astronomers are made of pixels; what we see is in part determined by the properties of our target object, but in part by the properties of the optical system used (telescope plus instrument and their PSF), and in part by noise. The ``elementary images'' are black-and-white, and usually taken through a specific filter. When such images are displayed, additional decisions were involved about how to represent brightness. Published images frequently use colour to convey additional information --- although in most cases, these are false colour images, which do not reproduce the colour of the object we would perceive could we view it directly. Astronomers employ dark frames and flatfielding to reduce certain types of noise, and of sensitivity variations. Sometimes, images are fit together to form a larger mosaic.

When working with images, we need to keep all this in mind --- after all, we want to use the information contained in the image to make deductions about the astronomical objects observed. To do that, we need to know which aspects of the image really {\em do} contain information about the object --- and not information about the telescope-instrument combination, or photon statistics.

\subsection{Images: astronomical information}

So far, we have talked mostly about image artefacts --- what makes an image different from the real thing. Time to talk about the physics behind it all: What information is contained in astronomical images? The information important for classical astronomy, for a start: Images contain {\bf position information} about astronomical objects, information about where exactly an object is located in the sky (for object whose position does not change in the usual coordinate system), or about how its position changes over time. 

In the era of classical astronomy, this was the main purpose of observatories: determining the positions of stars in the sky, as an aid to celestial navigation. This also included measurements that allowed for precise time-keeping: until the advent of stable quartz clocks in the mid-20th century, documenting the periodic changes in the night sky, in particular the diurnal motion during one (sidereal) day, was the most accurate time-keeping method.

In modern astronomy, determining stellar positions remains an important sub-field, which for the last few years has been dominated by ESA's astrometry satellite Gaia. Accurate catalogues of stellar positions not only provide a framework for localising astronomical objects in general. Via the {\em parallax effect}, they also provide information about the distances of stars in our cosmic neighbourhood, which in turn is a prerequisite for farther-reaching methods of astronomical distance determination. Knowledge of astronomical distances is crucial for making deductions about object's luminosity. (In principle, an object that appears to be bright in the night sky could have a rather faint luminosity, but appear bright since it is very close to us, or else have a really high luminosity while being rather more distant.)

Then, there is {\bf photometry}, that is, determining the (apparent) brightness of astronomical objects. If we have chosen proper exposure times, and made all necessary corrections, we could deduce the brightness of a star from the sum of the values of the pixels associated with that star. In practice, it's difficult to separate star pixels from non-star pixels, but there is a simpler way known as {\bf aperture photometry}: 
\begin{figure}[htbp]
\begin{center}
\includegraphics[width=\linewidth]{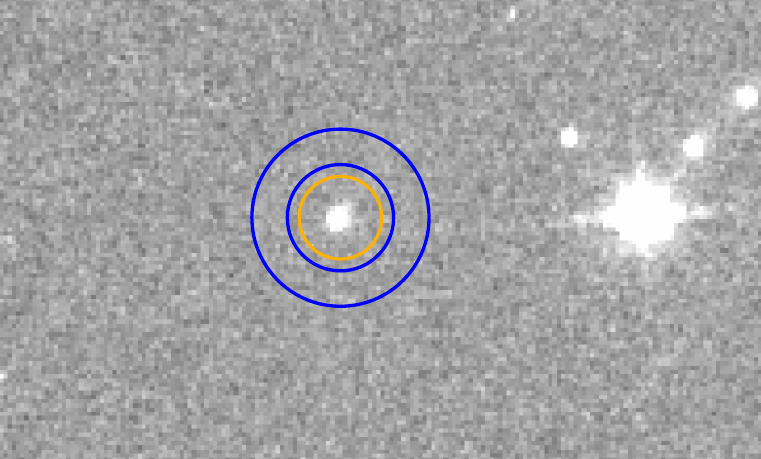}
\caption{Aperture photometry in a part of the Hubble Space Telescope image of Westerlund 2. Image credit: NASA, ESA, the Hubble Heritage Team (STScI/AURA), A. Nota (ESA/STScI), and the Westerlund 2 Science Team}
\label{ApPhot}
\end{center}
\end{figure}
Define a circular region that contains the PSF of the star completely (the yellow circle in Fig.~\ref{ApPhot}). At some small distance outside, define an annular region (bounded by the two blue circles in Fig.~\ref{ApPhot}). Assuming that the central circle contains only the star we are interested in, and the annular region no discernible star at all, we can argue as follows: If we sum up the pixel values within our central circle, we get the light from the star plus background light. We can estimate the background light as follows: On average, the background brightness should be the same in the central circle and in the surrounding annulus. In the annulus, we can determine the average brightness by summing up the pixel values in the annulus and then dividing by the number of pixels (equivalently, by the area of the annulus). Multiply this average by the number of pixels in the central circle (equivalently, by the area of the circle), and the result will be the brightness contribution from the background within the central circle. Subtract this contribution from the total sum of the pixel values within the central circle, and what is left is a measure of the brightness of the star. Note that photometric measurements are sometimes made with the telescope slightly out of focus, distributing the object's light over a greater number of detector pixels for greater accuracy.

In astronomical practice, stars are point-like objects. For extended objects, we can measure a {\em surface brightness}, given in brightness per angular area in the sky, and we can measure how that brightness varies from location to location. Such brightness maps contain information about the amount of material we are seeing. The situation is more complicated when densities are so high that some of the matter obscures our view of whatever matter lies behind (that is, if the matter in question is ``optically thick''). In the simplest case, we can see all of the light from the matter of, say, a nebula (the nebula is ``optically thin''), and the brightness in a certain area of the sky allows us to estimate how many atoms we are seeing in that area --- a {\bf column density} since we cannot deduce the three-dimensional structure, only the number of atoms within that column of the three-dimensional object which gets projected to the sky-region in question.

Brightness measurements will only ever cover some limited region of the electromagnetic spectrum. Some of the limitation comes about by the kind of telescope we use. An ordinary optical telescope will be able to receive visible light, near-infrared light, and ultraviolet light (which, for ground-based telescopes, is somewhat pointless since almost all UV light is filtered out in the Earth's atmosphere). But its camera would not be able to detect, say, mid-infrared light, let alone X-rays. In practice, as we have already seen, astronomers voluntarily restrict themselves to even narrower portions of the spectrum, by using suitable filters. This allows for quantitative description of the colors of astronomical objects. An object that is bright when viewed through a blue filter, but dim when viewed through a red filter, will be blueish in color. 

What we have called the brightness so far, summing up pixel values in our image, is proportional to the number of photons from a certain source (or a certain area of the sky) entering our telescope during the exposure time. Since the exposure time is the same for all the objects in our image, the ratio of brightness values for two such objects is equal to the ratio of the energy per unit time (in the given filter band) we receive from those two objects. 

Also, since in both case we are using the same telescope, and hence the same collecting area, the ratio is equal to the ratio of the energy per unit time per unit area (again in the given filter band) for those objects, or using the appropriate technical term: the ratio of their {\bf fluxes}. 

Flux ratios are how astronomers traditionally compare the apparent brightness of celestial objects --- except that, to ensure some degree of backwards-compatibility with the naked-eye-based, 2000-year-old Ancient Greek magnitude system (as one does), those ratios are measured on a logarithmic scale. Specifically, if $F_1$ and $F_2$ are the flux values for light we receive from two objects 1 and 2 in a specific filter band, then their {\bf apparent magnitudes} in that band are defined as
\be
m_1-m_2 = -2.5\cdot\log\left(\frac{F_1}{F_2}\right).
\label{AstroMagnitudeFormula}
\ee
A reference point for the magnitude system is chosen by setting a value for the magnitude of a specific star in a specific filter band; for instance, in the V filter band that roughly corresponds to a green filter, the star Vega was originally chosen as a zero point, although his modern visual magnitude is $m_V = +0.03$. 

Note the minus sign --- magnitude values are larger for {\em fainter} stars. With the naked eye, under good conditions, you can observe stars with magnitudes up to about $m_V=6.5.$ For extended objects, we would need to document the energy per unit time and unit collecting area that reaches us from a given solid angle in the sky. There, the {\bf intensity}, as energy received per unit time per unit area per unit solid angle is the appropriate descriptive quantity.

Brightness, of course, can change. Different types of variable stars, for instance, can be distinguished by the shapes of their light curves, which document how their brightness changes over time. The transit method for detecting exoplanets also relies on light-curve measurements. 

Last but not least, images also contain spatial, 3D information about the objects under study. Typically, that information is projected onto the sky --- we do not see the full three-dimensional structure of, say, a gas cloud; instead, we have one particular fixed perspective on that cloud. Interpreting what we see typically involves models for the physical, three-dimensional structure, whose predictions can then be compared to what we actually observe.

\subsection{Spectra}
\label{Spectra}

\begin{figure}[htbp]
\begin{center}
\includegraphics[width=\linewidth]{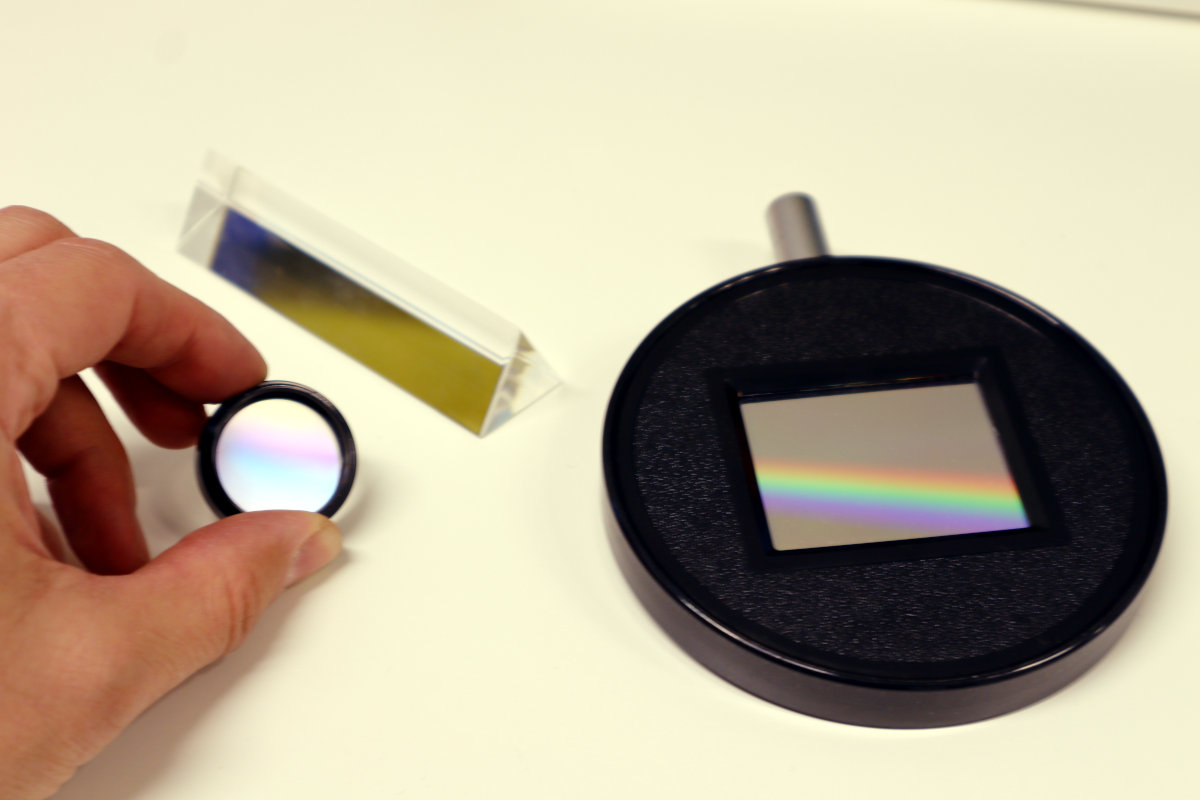}
\caption{The three basic types of dispersive element, from left to right: transmission grating, prism, and reflective grating}
\label{DispersiveExample}
\end{center}
\end{figure}

\begin{figure}[htbp]
\begin{center}
\includegraphics[width=\linewidth]{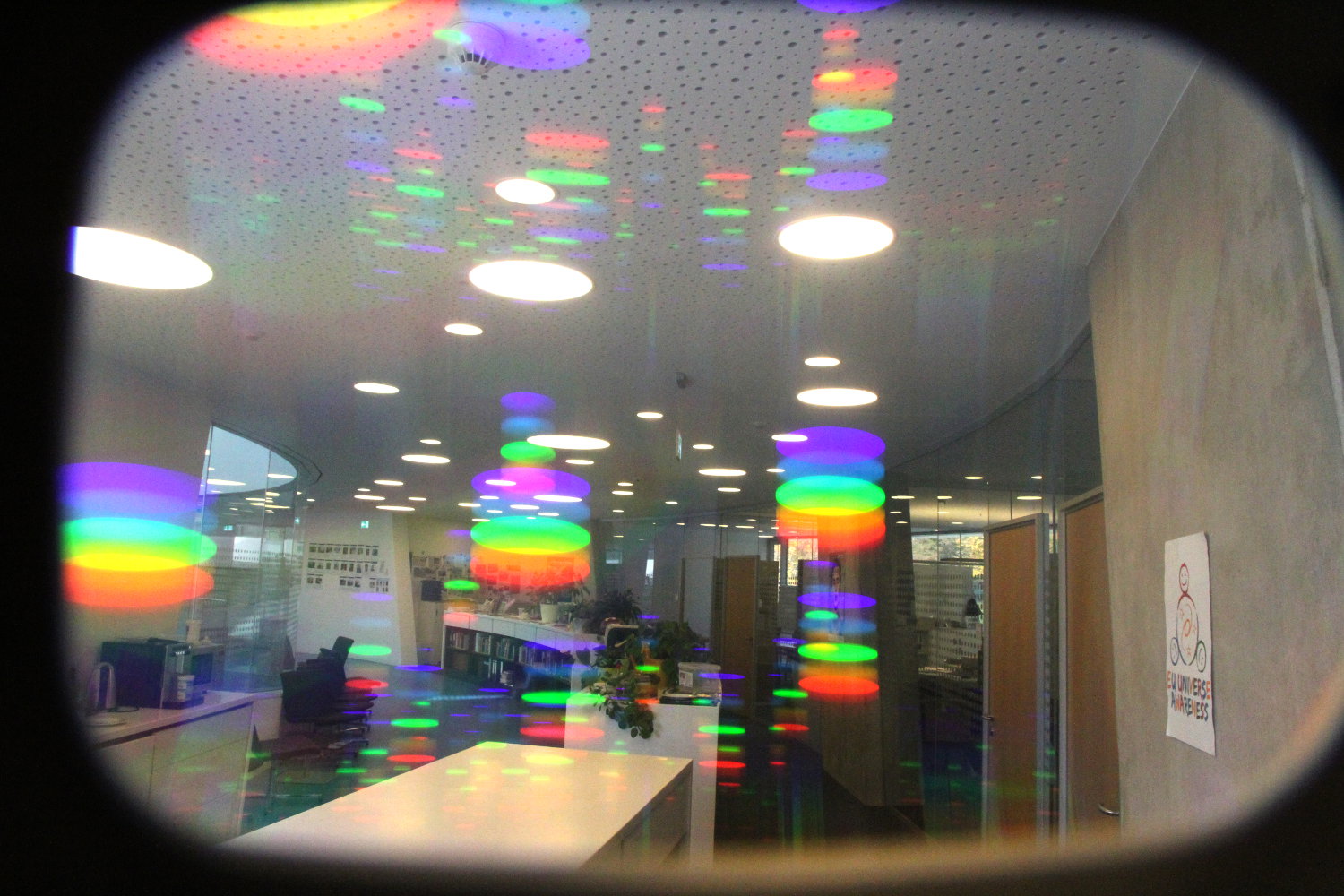}
\caption{Light sources, imaged through a dispersive element (``spectral glasses'')}
\label{SpectraDispersive}
\end{center}
\end{figure}

On to a central kind of data set in astrophysics: spectra! A spectrograph contains a {\bf dispersive element} (or even more than one), which splits the incoming light into its rainbow colours or, in physics terminology, into its different wavelengths. Examples for the three basic types of dispersive element an be seen in Fig.~\ref{DispersiveExample}.
Figure \ref{SpectraDispersive} shows ceiling lamps, imaged through a dispersive grid, namely through ``spectral glasses'' that can be used to demonstrate dispersion effects.
As you can see in the figure, though, the spectral decomposition makes for a hodgepodge of effects. The coloured images of the lamps, each corresponding to a spectral line, overlap, creating a mix of spatial information and wavelength information that is not easily disentangled.
\begin{figure}[htbp]
\begin{center}
\includegraphics[width=\linewidth]{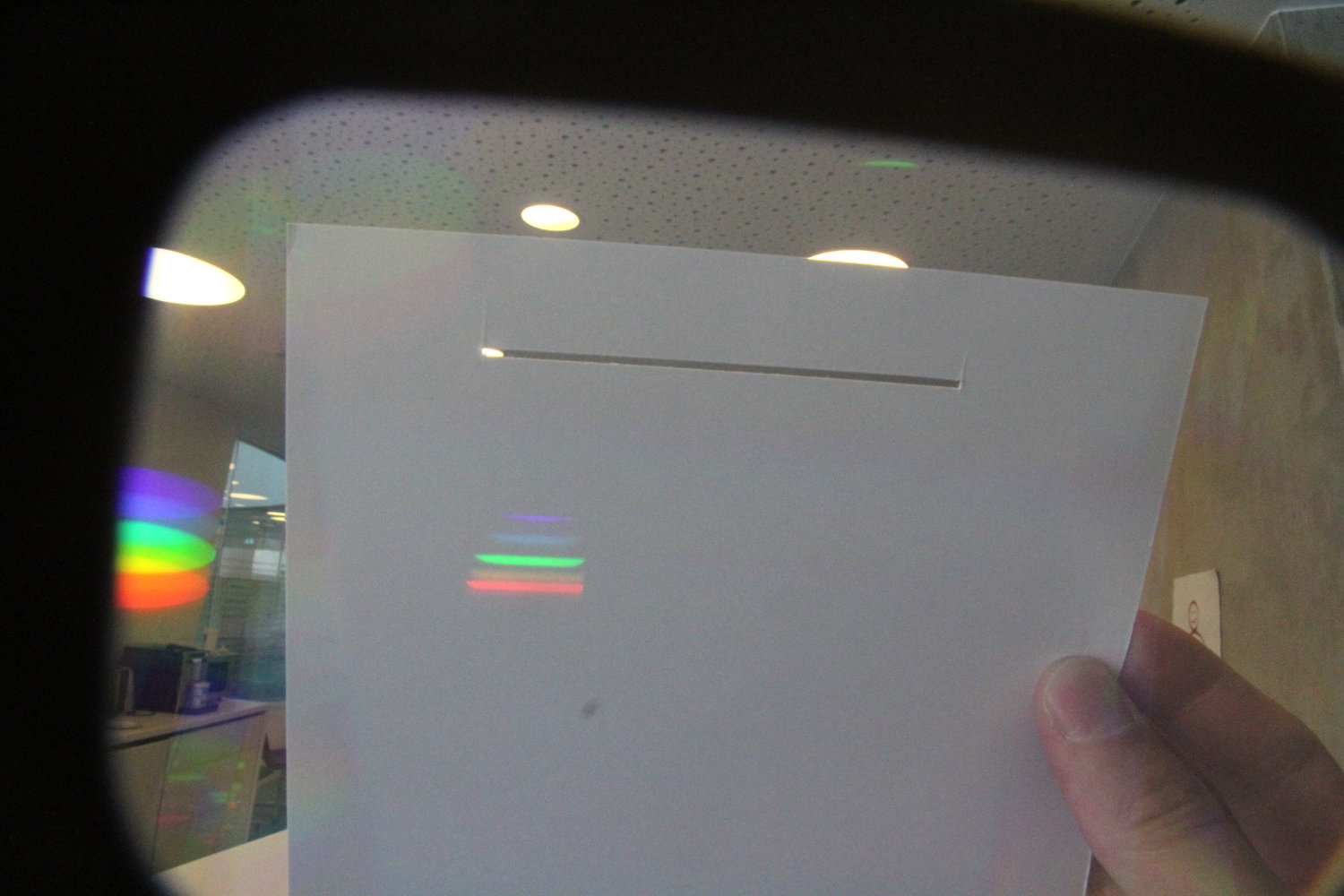}
\caption{One of the light sources masked by a slit, and imaged through the same dispersive element}
\label{SpectraSlit}
\end{center}
\end{figure}
A common solution is shown in figure \ref{SpectraSlit}: introduce a slit mask, with the slit oriented at right angles to the direction of spectral dispersion. If you keep the slit small, the overlaps between the successive images of the slit will be small as well. In the case of separate spectral emission lines, as in this example, you can then see the lines clearly separated. 

\begin{figure}[htbp]
\begin{center}
\includegraphics[width=\linewidth]{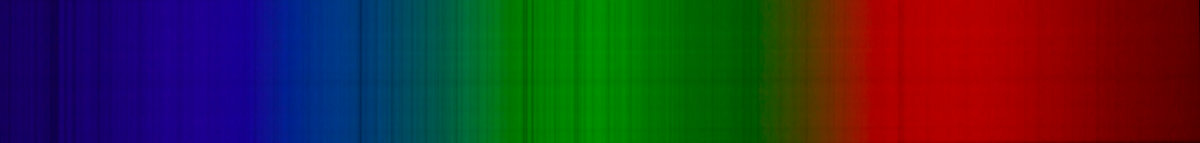}
\caption{Spectrum of sunlight, reflected by the Moon}
\label{SampleSpectrum}
\end{center}
\end{figure}

When you see the image of a spectrum that is a broad colourful band (possibly with bright emission lines, or dark absorption lines), you see a succession of such slit images, each copy indicating the brightness at that particular wavelength. One such image can be seen in figure \ref{SampleSpectrum}: a spectrum of sunlight reflected by the Moon, produced with a Baader DADOS spectrograph.	
The image was taken for demonstration purposes with a consumer digital camera, and thus shows colours --- any professional image of a spectrum would, of course, be in black and white; the colour carries no additional information, as the position of a spectral feature along the horizontal axis already defines its wavelength, and hence its colour.

\begin{figure}[htbp]
\begin{center}
\includegraphics[width=\linewidth]{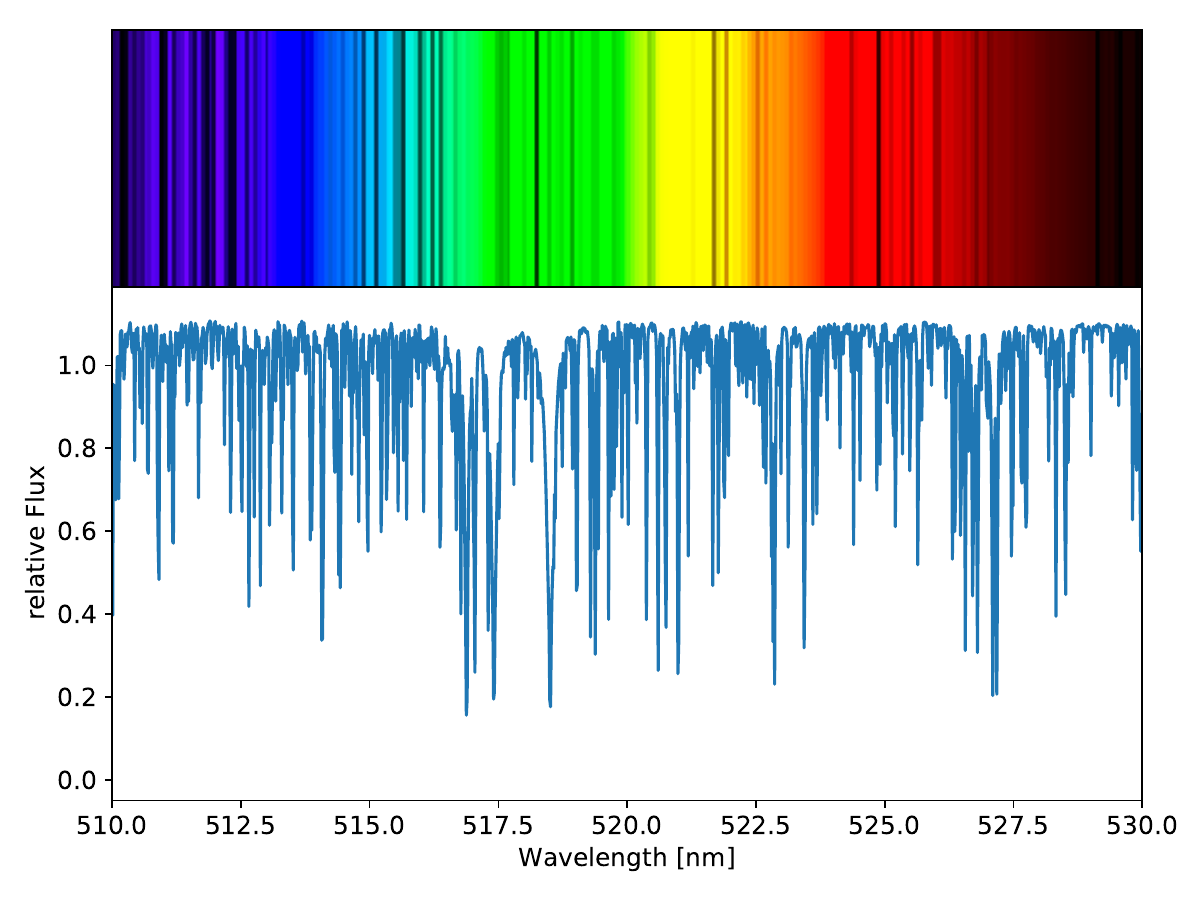}
\caption{Solar spectrum as a curve (bottom) and the reconstructed version of a slit spectrum image (top). Data from the IAG Solar Flux Atlas, Reiners et al. 2016 }
\label{SpectralCurve}
\end{center}
\end{figure}

\begin{figure}[htbp]
\begin{center}
\includegraphics[width=\linewidth]{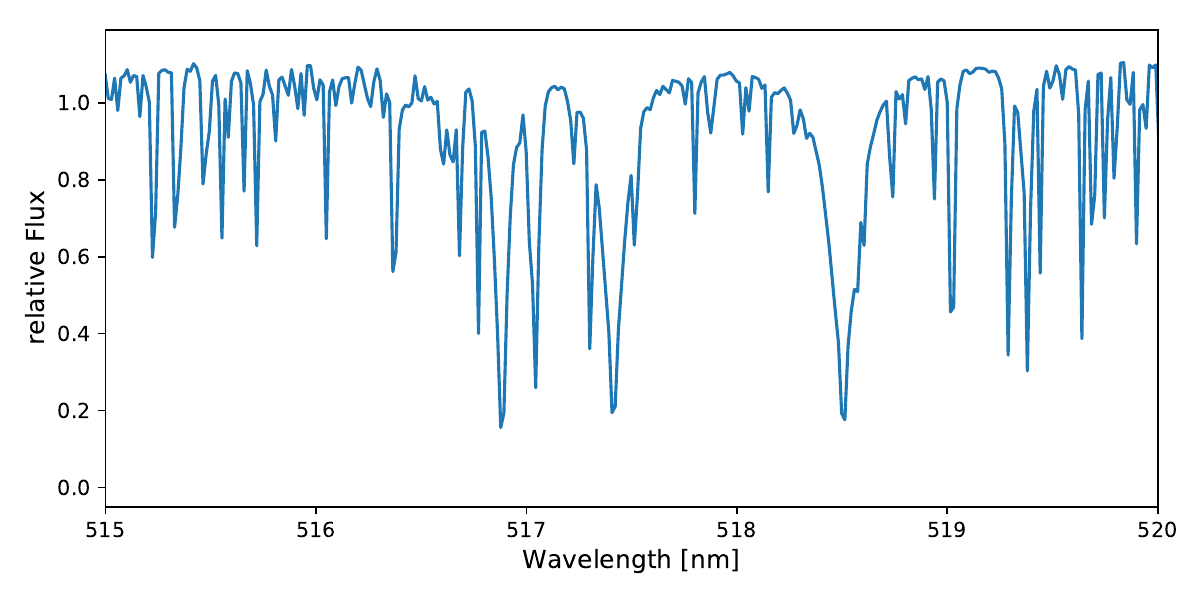}
\caption{Narrow region from a solar spectrum. Data from the IAG Solar Flux Atlas, Reiners et al. 2016}
\label{SpectralCurveDetail}
\end{center}
\end{figure}

The information contained in a spectrum is one-dimensional --- for each wavelength, we have a quantity indicating how much light is emitted in that particular wavelength region. Thus, it is natural to plot spectral data as a curve. The top part of figure \ref{SpectralCurve} shows an artificial color image of a Solar spectrum, complete with dark Fraunhofer lines. The bottom plots the spectrum as a curve, with wavelength plotted along the x axis and relative flux on the y axis. The requisite data is taken from IAG solar flux atlas\footnote{
Reiners et al. 2016,\newline [\href{http://adsabs.harvard.edu/abs/2016A&A...587A..65R}{http://adsabs.harvard.edu/abs/2016A\&A...587A..65R}].}
The curve is quite complex, with a forest of absorption lines --- narrowly defined minima --- one next to the other. If we zoom in by plotting a much smaller wavelength interval, as in figure \ref{SpectralCurveDetail}, you can see the the lines themselves have characteristic shapes. 

When working with raw data from spectra, certain reduction steps need to be taken. Some of those are similar to the reduction of image data: Flatfielding is again needed to compensate for differing sensitivity of the instrument in different parts of the spectrum. This is more difficult for a spectrum than for an image, since for a true flat field, you would need a perfectly flat spectrum. Instead, any well-known, preferably smooth calibration spectrum can be used to deduce the varying sensitivity. Dark frames again can be used to take into account that the electronics of the detector will produce some spurious brightness in the image, which needs to be subtracted.

Wavelength calibration is another necessity. After all, the spectral spread has a specific meaning --- light is separated according to wavelength (or frequency). In order to map specific wavelengths to the direction along which dispersion takes place, astronomers often employ specific calibration lamps, which contain a gas or a mixture of gases that produce a hopefully dense array of known emission lines. For a simple amateur spectrograph, you might use Neon for the purpose; at the professional level, you might for instance find a mixture of Thorium, Argon, and Neon. Sometimes, the calibration lines will be recorded separately; in other cases, they are recorded concurrently with the astronomical observation to allow for a direct comparison. A special case of the latter, unavoidable for ground-based telescopes, are {\bf telluric lines} --- absorption lines created not in outer space, but by light absorption in the Earth's atmosphere. Such telluric lines can be used for calibration, as well.

\begin{figure}[htbp]
\begin{center}
\includegraphics[width=\linewidth]{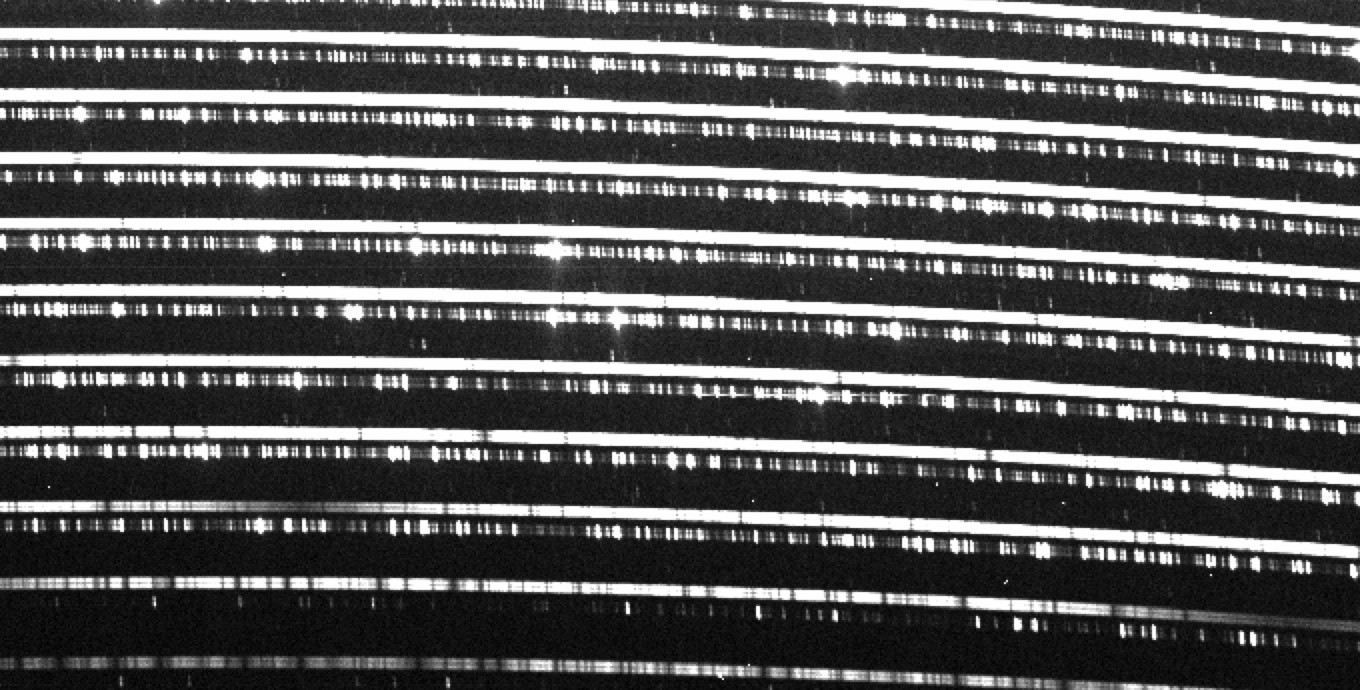}
\caption{Region within a raw image of a spectrum of the star HIP66974 and a calibration lamp, taken with ESO's FEROS spectrograph in June 2015 (Data set SAF+FEROS.2015-06-13T23:16:46.772). Retrieved from ESO's Science Archive on 24 April 2019 }
\label{FEROSSpectrum}
\end{center}
\end{figure}

Image distortions can make spectral data reduction particularly challenging. A particularly complex case are Echelle spectrographs, where two kinds of spectral dispersion are combined: A grating will, in fact, produce several different spectra, called ``spectra of different order''. Higher orders tend to overlap each other, but the different overlapping partial spectra can be separated by a second stage of dispersion in the direction orthogonal to the initial dispersion. The result are different rows of partial spectra, allowing astronomers to capture an entire high-resolution spectrum on a standard, two-dimensional camera chip. The raw image of one such Echelle spectrum, taken with the FEROS spectrograph at the MPG/ESO 2.2-metre telescope at ESO's La Silla observatory, can be seen in Fig.~\ref{FEROSSpectrum}.
This is just a small region, about 10\%, of a much larger image. As you can see, the horizontal, curved stripes always come in pairs: the stripe on top is mostly white with some dark absorption lines (which increase lower in the image), while the lower stripe consists of a fairly dense forest of emission lines. The upper stripe of each pair is the science image, in this case of the star HIP66974, a star with the same spectral type (G2V) as the Sun and thus a fairly similar spectrum. 

The lower stripe in each instance is the calibration lamp --- hence the many emission lines, each marking a well-known reference wavelength. Reducing this spectrum would mean to map the different stripes to their proper wavelength regions (using the calibration lines), unbending the curved image, and properly calibrating the brightness over the different parts of the image.

Once the spectrum is reduced, or if one is working with a reduced spectrum in the first place, the spectrum as a whole and in particular the spectral lines contain a wealth of information about the object in question. Systematic Doppler shifts in the spectral lines indicate whether or not the light source is moving towards us or away from us. Doppler shifts that change periodically over time contain information about objects orbiting each other, from double stars to exoplanets detected by the radial velocity method. Simple data analysis in these cases proceeds by fitting the individual spectral  lines, finding their central wavelength, and tracing the changes of that wavelength over time. 

The shape and relative depth of spectral lines of a star contains information about the star's metallicity, that is, the fraction of elements heavier than Helium contained in the star's atmosphere, about the surface gravity and about the effective temperature. Some specific spectral lines corresponding to radioactive elements can be used to reconstruct the age of stars, and have been used to find the oldest stars in existence. 

The simplest part of such an analysis is about identifying the lines corresponding to specific chemical elements; these lines show which elements are present in the star's atmosphere. 

Some such lines are indicated in figure \ref{SpectralCurve}. In all of these cases, analysis usually proceeds by creating spectra based on suitable models and comparing those with the actual observations, finding the best fit. 

\subsection{Data cubes}

So far, we have talked about two-dimensional images (where the two dimensions correspond to an area on the night sky) and one-dimensional spectra (where the one dimension corresponds to wavelengths). Data cubes are the combination of this: We have a two-dimensional image of the night sky, but at each pixel location, we have not only a single brightness value, but instead a whole spectrum. 

With two plus one dimensions, we are effectively looking not at a two-dimensional rectangle, but a three-dimensional cube. A data cube does not contain {\em all} the information reaching us from a certain region of the sky at a certain time (polarisation information is missing), but it comes impressively close. 

One way of obtaining such a data cube is with {\bf Integral Field Spectroscopy} ---  for instance: splitting an image into comparatively large ``pixels,'' each of which is channeled into a glas fibre which transmits its light to a spectrograph, where the spectrum is then recorded. 

Another natural way of recording such a data cube is in interferometry, where the light from several telescopes is combined in a coherent way, making use of the wave nature of light. In reconstructing images from interferometric measurements, one can distinguish (to a certain degree) between contributions with different wavelengths; in effect, this allows for the reconstruction of a three-dimensional data cube.

Human beings are not equipped for really three-dimensional vision. What we call three-dimensional vision is really just seeing surfaces within (sparsely populated) three-dimensional space. We cannot see all the points within a three-dimensional data cube at once. 
\begin{figure}[htbp]
\begin{center}
\includegraphics[width=\linewidth]{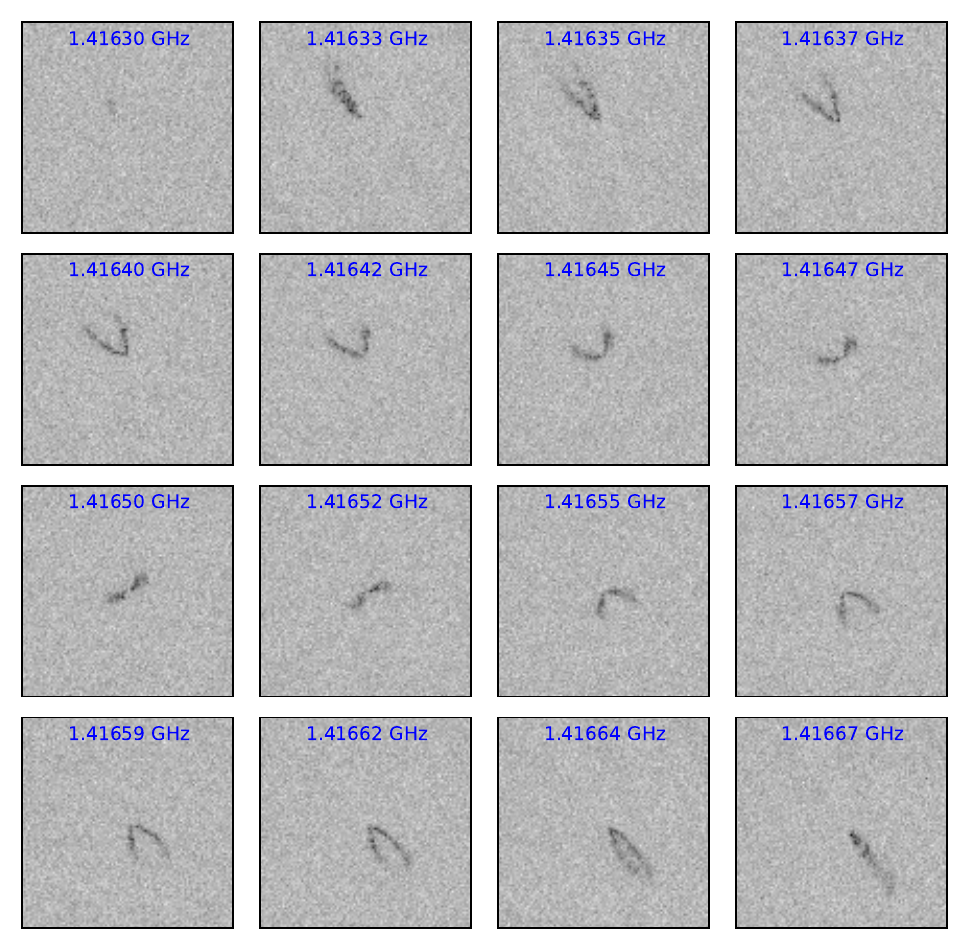}
\caption{16 of the 72 channels recorded for the galaxy NGC 3198, from the THINGS survey (\href{https://ui.adsabs.harvard.edu/abs/2008AJ....136.2563W/abstract}{Walter et al. 2008}),
[\href{http://www.mpia.de/THINGS/}{http://www.mpia.de/THINGS/}] }
\label{THINGSchannelmap}
\end{center}
\end{figure}
Fig.~\ref{THINGSchannelmap} shows one solution: showing separate images for different regions within the spectrum. In this particular case, the ``channel map'' shows 16 of the 72 frequency bins around the 21 cm hydrogen line that is characteristic for atomic hydrogen (that is, hydrogen atoms; not bound into hydrogen molecules, not ionized to form a plasma, just simple atoms). 

One of the most interesting applications of data cube data is to extract the information they contain about the large-scale motion of matter. Fig.~\ref{THINGSchannelmap} shows 21 cm radiation emitted by hydrogen atoms, but some of that radiation is shifted to lower and some to higher frequencies --- why? Because some of the atoms are moving towards us, others away from us, and their 21 cm radiation undergoes a corresponding Doppler shift. 

\begin{figure}[htbp]
\begin{center}
\includegraphics[width=\linewidth]{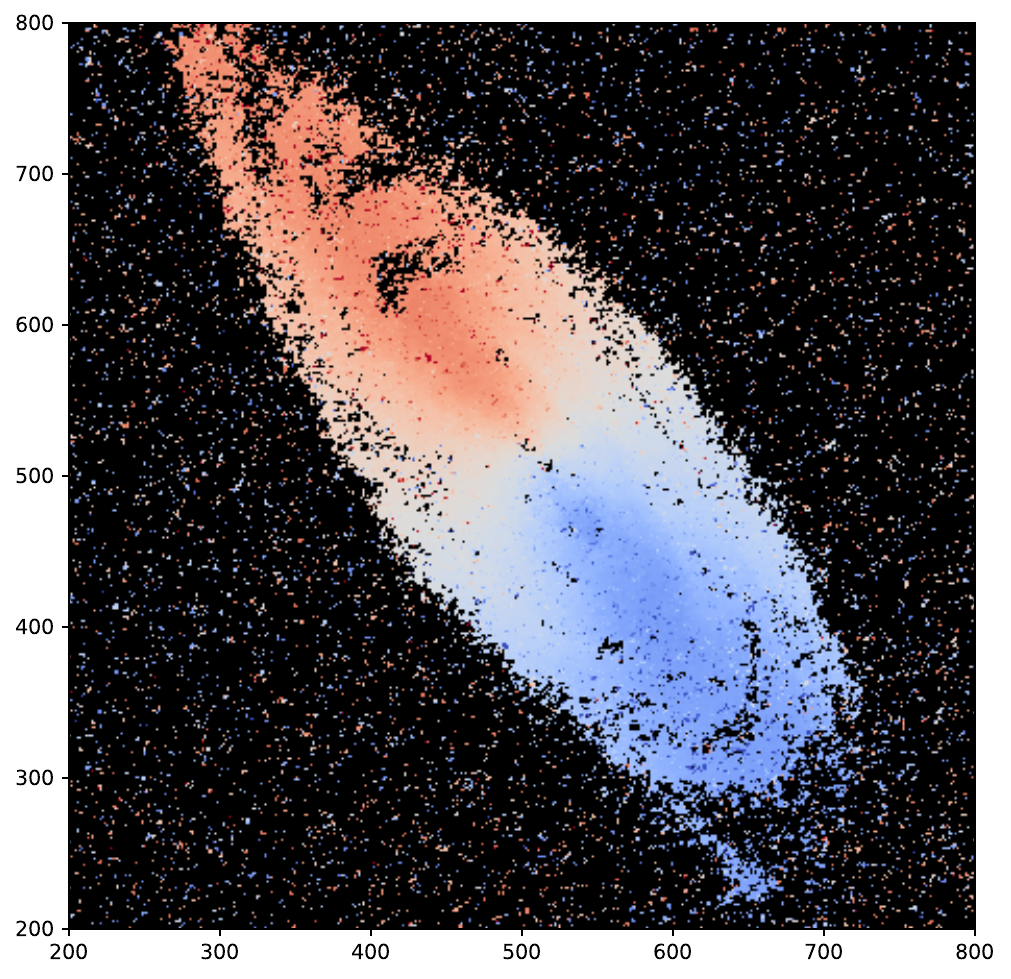}
\caption{Velocity map for the galaxy NGC 3198, from the THINGS survey (\href{https://ui.adsabs.harvard.edu/abs/2008AJ....136.2563W/abstract}{Walter et al. 2008}),
[\href{http://www.mpia.de/THINGS/}{http://www.mpia.de/THINGS/}] }
\label{ThingsFirstMoment}
\end{center}
\end{figure}

The data cube contains information on the radial velocity of the gas we see in the different frequency channels. We can combine that information to make a color picture whose color encodes the average radial velocity of gas in each region of the image, giving what is called a {\bf first moment map}. That picture is shown in Fig.~\ref{ThingsFirstMoment}. 
Reddish regions are moving away from us, blueish regions towards us. The combined picture is that of a disk galaxy whose stars are rotating as a whole, one side of the disk coming towards us and the other moving away from us in that coordinated motion. 

Alternatively, instead of taking the average, we could compute the standard deviation of the velocity values associated with each pixel. That would give us an estimate not of the bulge motion of gas in that region, but of the diversity of motion, the spread of radial velocities.

In a similar manner, we can use data cube information to map all those quantities that can be derived from spectra --- the presence of specific elements and thus the chemical composition, on larger scales the prevalence of different kinds of stars, and more.

\subsection{High-level data: catalogues and tables}
\label{HighLevel}

Once astronomers have derived observational or physical quantities from their observations and measurements --- deduced the temperature of stars from their spectra, or their luminosity from their apparent brightness and some measure of their distance --- they can compile catalogues containing such higher-level, derived physical information. Analysing this kind of high-level data is broadly similar to statistical analysis in other fields, and uses the same general tools.

To begin with, a catalogue is no more than a list. Conventionally, each row in that list represents a separate object, and each column represents a property. If you just look at the numbers in a big list, you are sure to miss the forest for the trees. In order to extract trends, distribution, systematic correlations, functional relationships from the data, in astronomy: in order to understand what physical laws and evolutionary pathways has made objects the way they are, we need employ proper tools. Statistical analysis is a wide field, and in this basic introduction, we will only look at the most basic of descriptive tools. 

A very basic tool is a {\bf histogram}, which allows us to see the distribution of values for a single quantity within our sample (e.g. among the objects of our catalogue). The basic principle is simple: within the range $x_{min} \le x\le x_{max}$ spanned by the values for the quantity $x$, we define $N$ bins of equal size, each with a lower boundary $x_i$ and an upper boundary $x_{i+1}$. A value $x$ falls into the bin with index $i$ if
$x_i< x\le x_{i+1}$. We then draw the bins as rectangles, whose height is proportional to the number of values which fell into the bin. This gives us a measure of how prevalent (or not) specific values are.

Histograms can also be logarithmic. In that case, we divide the range of the quantity we intend to map into bins of equal {\em logarithmic} size. For instance, if one bin contains stars that are between $1=10^0$ times and $10=10^1$ times as luminous as the Sun, the next bin would be from $10^1$ to $10^2$ solar luminosities $L_{\odot}$, and the following one from $10^2$ to $10^3\:L_{\odot}$. Those bins are of equal size when it comes to their exponents. Such logarithmic binning is useful for physical properties that are spread across a wide spectrum of scales. 

\begin{figure}[htbp]
\begin{center}
\includegraphics[width=0.8\linewidth]{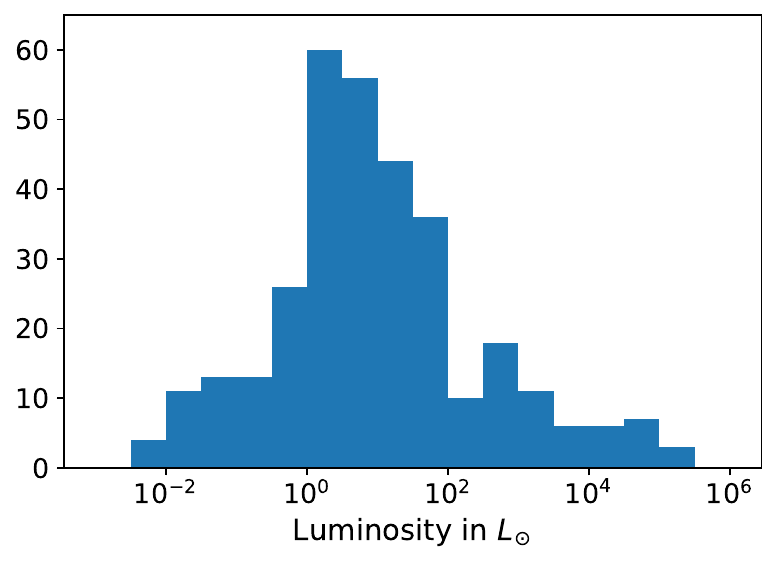}
\caption{Automatically-generated histogram}
\label{HistogrammLeuchtkraft}
\end{center}
\end{figure}

Here is a simple example for a histogram with logarithmic bins. Fig.~\ref{HistogrammStrichliste} shows the basic preparations for manually drawing a histogram: a tally sheet for putting the stars in a certain data set into their proper bins. The visual appearance of the tally sheet already constitutes a simple histogram-like representation, although in an unusual sideways orientation.
\begin{figure}[htbp]
\begin{center}
\bgroup
\footnotesize
\def\arraystretch{1.0}
\begin{tabular}{r@{ to  }r|l|r}
$10^{-3}$&$3\cdot 10^{-3}\; L_{\odot}$ &  & 0 \\\hline
$3\cdot 10^{-3}$&$ 10^{-2}\; L_{\odot}$  &\StrokeThree  & 3  \\\hline
$10^{-2}$&$3\cdot 10^{-2}\; L_{\odot}$  & \StrokeFive \StrokeFive & 10\\\hline
$3\cdot 10^{-2}$&$ 10^{-1}\; L_{\odot}$ &\StrokeFive \StrokeFive \StrokeThree  & 13 \\\hline
$10^{-1}$&$3\cdot 10^{-1}\; L_{\odot}$  & \StrokeFive \StrokeFive \StrokeTwo & 12 \\\hline
$3\cdot 10^{-1}$&$10^{0\phantom{-}}\; L_{\odot}$  & \StrokeFive \StrokeFive \StrokeFive \StrokeFive \StrokeFive \StrokeOne  & 26 \\\hline
$10^{0\phantom{-}}$&$3\cdot 10^{0\phantom{-}}\; L_{\odot}$ & \StrokeFive \StrokeFive \StrokeFive \StrokeFive \StrokeFive \StrokeFive \StrokeFive \StrokeFive \StrokeFive \StrokeFive \StrokeFive \StrokeFive  & 60 \\\hline
$3\cdot 10^{0\phantom{-}}$&$ 10^{1\phantom{-}}\; L_{\odot}$  &\StrokeFive \StrokeFive \StrokeFive \StrokeFive \StrokeFive \StrokeFive \StrokeFive \StrokeFive \StrokeFive \StrokeFive \StrokeFive   & 55 \\\hline
$10^{1\phantom{-}}$&$3\cdot 10^{1\phantom{-}}\; L_{\odot}$  &  \StrokeFive \StrokeFive \StrokeFive \StrokeFive \StrokeFive \StrokeFive \StrokeFive \StrokeFive \StrokeFour& 44 \\\hline
$3\cdot 10^{1\phantom{-}}$&$ 10^{2\phantom{-}}\; L_{\odot}$  &\StrokeFive \StrokeFive \StrokeFive \StrokeFive \StrokeFive \StrokeFive \StrokeFour &34  \\\hline
$10^{2\phantom{-}}$&$3\cdot 10^{2\phantom{-}}\; L_{\odot}$  &\StrokeFive \StrokeFive & 10 \\\hline
$3\cdot 10^{2\phantom{-}}$&$ 10^{3\phantom{-}}\; L_{\odot}$  &\StrokeFive \StrokeFive  \StrokeFive \StrokeThree  & 18 \\\hline
$10^{3\phantom{-}}$&$3\cdot 10^{3\phantom{-}}\; L_{\odot}$  & \StrokeFive \StrokeFive \StrokeOne &11 \\\hline
$3\cdot 10^{3\phantom{-}}$&$ 10^{4\phantom{-}}\; L_{\odot}$  & \StrokeFive \StrokeOne  & 6\\\hline
$10^{4\phantom{-}}$&$3\cdot10^{4\phantom{-}}\; L_{\odot}$  & \StrokeFive & 5\\\hline
$3\cdot 10^{4\phantom{-}}$&$ 10^{5\phantom{-}}\; L_{\odot}$  & \StrokeFive \StrokeOne & 6 \\\hline
$10^{5\phantom{-}}$&$3\cdot 10^{5\phantom{-}}\; L_{\odot}$  &  \StrokeThree & 3\\\hline
$3\cdot 10^{5\phantom{-}}$&$ 10^{6\phantom{-}}\; L_{\odot}$  &   & 0 \\\hline
\end{tabular}
\egroup
\caption{Tally sheet in preparation for a hand-crafted histogram}
\label{HistogrammStrichliste}
\end{center}
\end{figure}
Hand-crafted histograms have gone the way of so many other hand-crafted things. The automatically-generated version for the same histogram can be seen in Fig.~\ref{HistogrammLeuchtkraft}.
We will look at ways of generating such histograms in sections \ref{TOPCAThistogram} (TOPCAT) and \ref{HistogramsPython} (Python). 

The data set is DEBCat,\footnote{Southworth 2014, [\href{https://arxiv.org/abs/1411.1219}{https://arxiv.org/abs/1411.1219}]. The data and supplemental information are available online at [\href{http://www.astro.keele.ac.uk/jkt/debcat/}{http://www.astro.keele.ac.uk/jkt/debcat/}].} a collection of more than a hundred well-studied transiting double stars. In that particular set-up, astronomers have sufficient information to be able to reconstruct the stars key physical properties like mass, radius, luminosity and temperature, and derive properties such as a star's average density. 

Let us use some of those properties to look at another very common tools are diagrams populated with data points in which we plot one quantity against another. Each axis stands for a specific physical quantity (commonly scaled either linearly or logarithmically), and each data point corresponds to a pair of values, one for each of the axis quantities.

\begin{figure}[htbp]
\begin{center}
\includegraphics[width=\linewidth]{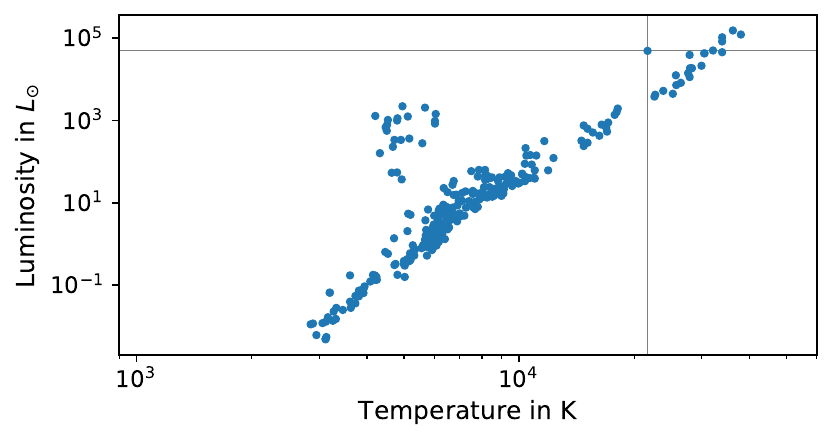}
\caption{Hertzsprung-Russell diagram using physical quantities (temperature and luminosity) instead of spectral classes and luminosity classes. Data from DEBCat}
\label{HRDiagram}
\end{center}
\end{figure}

A famous example, with two logarithmically scaled axes, is the Hertzsprung-Russell diagram, a version of which is shown in Fig.~\ref{HRDiagram}.
The values corresponding to each data point are read off in the usual way, as shown by the two grey auxiliary lines: from the dot, go horizontally to the vertical axis; read off the value indicated at the intersection point, in this case a luminosity of about $5\cdot 10^4\:L_{\odot}$ and a temperature of somewhat more than $20~000\:\mbox{K}$. 

In the diagram, you can clearly discern a linear structure going from bottom left to top right, and a cloud of dots hovering to the upper left of that linear structure. Astronomers call the linear structure the {\em main sequence}, and the stars in it the {\em main sequence stars}. 

\begin{figure}[htbp]
\begin{center}
\includegraphics[width=\linewidth]{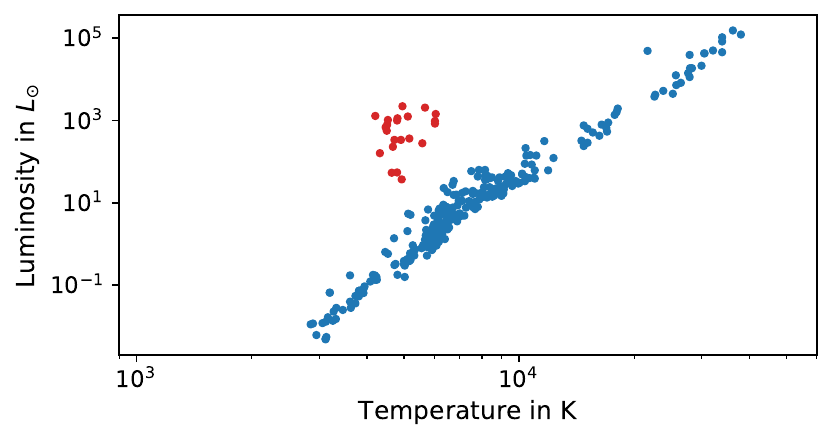}
\caption{Hertzsprung-Russell diagram using physical quantities (temperature and luminosity) instead of spectral classes and luminosity classes. Data from DEBCat}
\label{HRDiagramRedBlue}
\end{center}
\end{figure}

\begin{figure}[htbp]
\begin{center}
\includegraphics[width=0.8\linewidth]{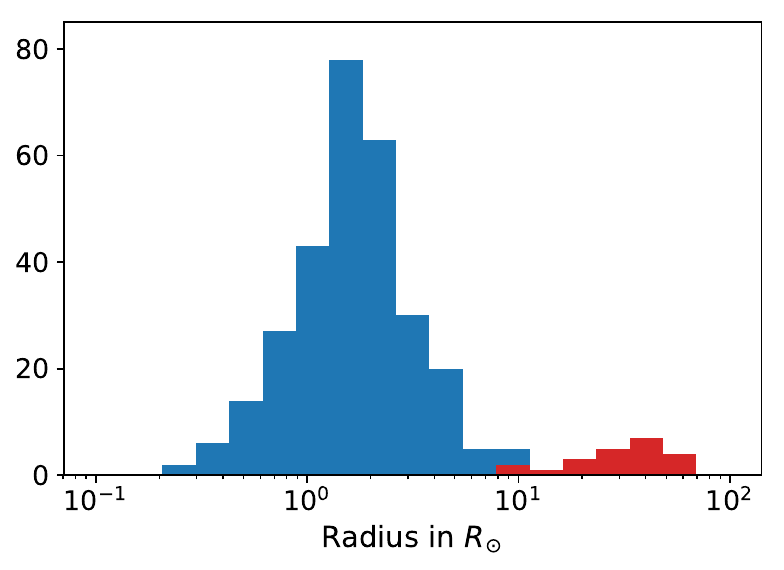}
\caption{Separate histograms for the blue and red data points from Fig.~{HRDiagramRedBlue} }
\label{RadiiBlueRed}
\end{center}
\end{figure}

There is a way of include additional information in histograms and 2-dimensional diagrams: use colour! In the simplest case, we can use colour to distinguish between different populations of data points. For instance, let us colour the points in the upper-left cloud on the Hertzsprung-Russell diagram red, as Fig.~\ref{HRDiagramRedBlue}.
So far, the colouring hasn't brought us any great advantage. The cloud was apart from the rest before, and it is apart from the rest now. But let us carry this color scheme over to a histogram, for instance, plotting histograms for the red and the blue dots side by side, using the same bins. The result for a histogram of radii is shown in Fig.~\ref{RadiiBlueRed}.
Now we see that the stars corresponding to those red dots have considerably larger radii than their main sequence counter parts. The size distributions are clearly separate. Those red-dot stars are veritable giants! We know from the Hertzsprung-Russell diagram that their temperatures are somewhere between 4000 and 6000 K, going from reddish to yellowish. So these red giant stars were named with excellent reason.

\begin{figure}[htbp]
\begin{center}
\includegraphics[width=\linewidth]{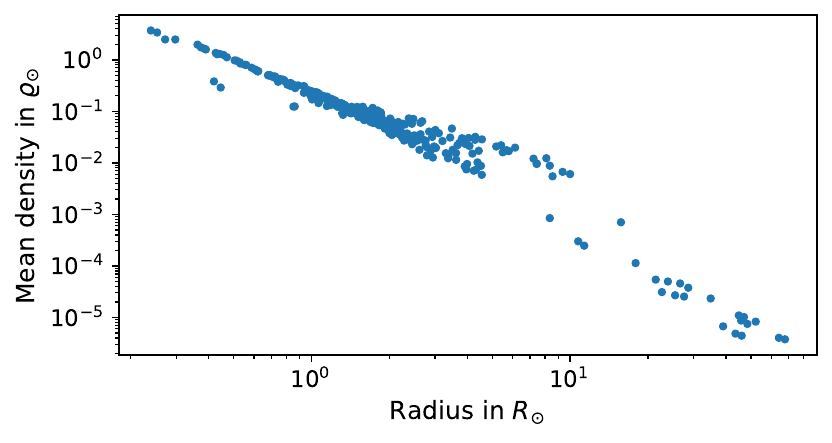}
\caption{Plotting radius against density, both in solar units. Data from DEBCat}
\label{RadDensMono}
\end{center}
\end{figure}

\begin{figure}[htbp]
\begin{center}
\includegraphics[width=\linewidth]{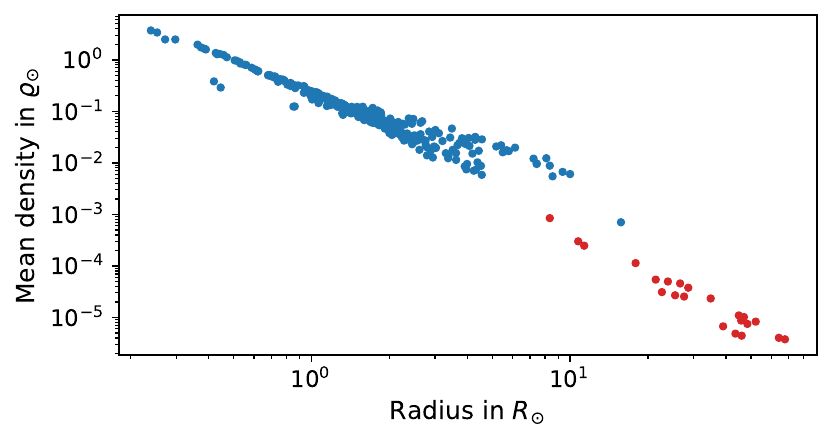}
\caption{Plotting radius against density, both in solar units. Color marks red giants vs.\ main sequence stars. Data from DEBCat}
\label{RadDensColor}
\end{center}
\end{figure}

Let's look at a diagram plotting, say, radius against density, as in Fig.~\ref{RadDensMono}.
In that diagram, it is not clear which are the main sequence stars and which are the red giants. With the red-blue distinction, the situation becomes clear, as shown in Fig.~~\ref{RadDensColor}.
Red giants are not only generally larger in radius than main sequence stars, they are also considerably less dense. Our data points us in the right direction: in the modern view, main sequence stars go through a red giant phase after they exhausted the hydrogen fusion fuel in their cores, their atmospheres swelling up and cooling down in the process, leading to a large, reddish star with drastically reduced mean density.

\begin{figure}[htbp]
\begin{center}
\includegraphics[width=0.9\linewidth]{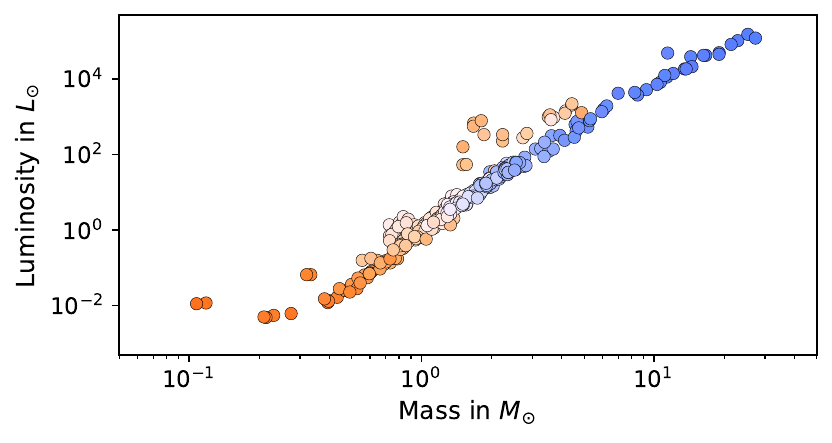}
\caption{Mass-luminosity diagram with data points plotted in the color corresponding to a star's temperature. Data from DEBCat}
\label{MassLumColor}
\end{center}
\end{figure}

\begin{figure}[htbp]
\begin{center}
\includegraphics[width=\linewidth]{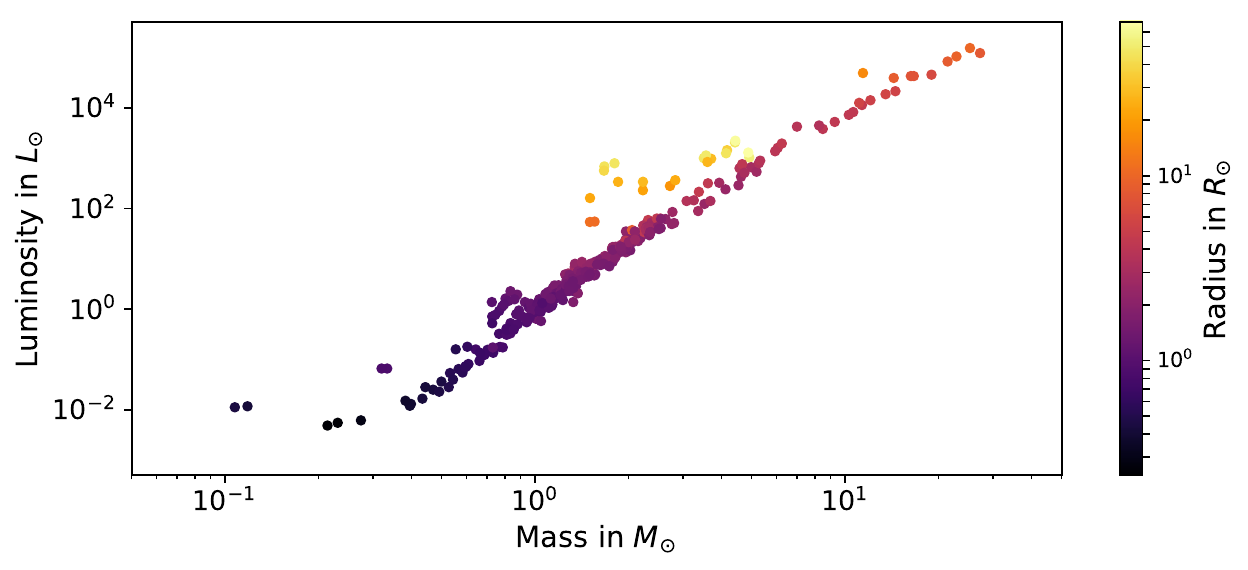}
\caption{Mass-luminosity diagram with data points plotted using an artificial color map that indicates the stars' radii. Data from DEBCat}
\label{MassLumRadiusColor}
\end{center}
\end{figure}

We can use colour more quantitatively than just to express class membership in a two-colour scheme. Colour can add an (imperfect) third dimension to our diagrams. In the version of a mass-luminosity diagram shown in Fig.~\ref{MassLumColor}, each data point has the proper star color corresponding to it's temperature (as determined from the star's spectral properties\footnote{Information about this kind of color mapping can be found on \href{http://www.vendian.org/mncharity/dir3/starcolor/}{http://www.vendian.org/mncharity/dir3/starcolor/}}). This color-coding immediately allows you to identify the red giants, see that their mass range is a subset of the mass range of the main sequence stars, but that the red giants are larger and more reddish.
Color scales can also be artificial, and different color maps are available for the purpose.  In Fig.~\ref{MassLumRadiusColor}, the color now indicates the radius of each star, with the scale shown by the colorbar on the right.
Clearly, stellar radii grow along the main sequence, but the red giants, in their little cloud of data points above the main sequence, are larger still.

A good color map can make your diagram much easier to understand; a bad one can be confusing. Also, you should take into account accessibility issues. Your color maps should be accessible even to people with certain forms of colour-blindness.\footnote{Some information on this kind of accessibility can be found on [\href{https://betterfigures.org/2015/06/23/picking-a-colour-scale-for-scientific-graphics/}{https://betterfigures.org/2015/06/23/picking-a-colour-scale-for-scientific-graphics/}]} Unless you are plotting a spectrum, avoid the rainbow colour map. Instead, consider color maps that have been designed to be accessible for those with colour blindness, as well as to print well in black and white --- for instance the Viridis family of colour maps.

Patterns in such diagrams indicate interesting relationships. Is there a linear relation --- do data points for certain physical quantities fall on a straight line in a linear diagram? Or is there a power law at work, $y\sim x^a,$ in which case the data points would fall on a straight line in a log-log diagram? In this way, diagrams can help us find systematic relations between our data. 

This is not as straightforward as it sounds, of course. In a two-dimensional diagram, we can plot at most three different quantities (if we make clever use of a color map). We could try all different pairs of physical quantities relevant for the situation we are looking at, and might get lucky in finding interesting relationships in that way. A three-dimensional diagram is possible, but would need to be interactive so we can view it from all different sides to get a feeling for the three-dimensional structures. Of course, the basic physical quantities can be combined to yield compound quantities. Complex relationships between quantities, longer polynomials involving several quantities for instance, or differential/integral relations, are much less straightforward to read off such diagrams.

The typical way of extracting information about systematic relationships from data is to {\bf fit a function} to the data. Assume that we have data points $(x_i,y_i)$ for $i=1,\ldots,N$, each representing a pair of quantities. A common measure for how well those data points satisfy a general relationship $y=f(x)$ is as follows.

If the relationship were to hold perfectly, then we would have $y_i=f(x_i)$. In real life, functional relationships are not that perfect. Even in cases where the relationship $y=f(x)$ is the basis for our set of data points, measurement errors will lead to deviations. Moreover, exact relationships are rare; the much more common case is that the relationship is approximate, and that data points scatter around the curve $y=f(x)$.

For a single data point, the quantity
\be
\Delta y_i = y_i-f(x_i) 
\ee
is a measure of the deviation of the data point from the relationship. What is the best way of summarising the deviations for our data set as a whole? We can say what is definitely {\em not} a good measure: taking the sum of all the $\Delta y_i$, since deviations may be positive and negative, and the sizeable deviations associated with different data points could cancel each other out, skewing the result --- we could even get an overall measure of zero, indicating no deviation, in a situation where the $\Delta y_i$ are huge, but cancel pair-wise!

To avoid this, we could take the sum of the absolute values $|\Delta y_i|$ but as we shall see later on, it is useful for the measure we choose to be differentiable. That why a better choice is the sum over the squares of the deviations, 
\be
S\equiv \sum_{i=1}^N [ y_i-f(x_i) ]^2.
\label{SquareSum}
\ee
Commonly, we have an idea for the basic properties of the function $f(x)$, but not about the explicit form of the function. For instance, we might have reason to believe the function to be linear (since that is what an x-y diagram suggests), $f(x) = ax+b,$ but do not know the values for $a$ and $b$. 

In such a case, we can use the quantity (\ref{SquareSum}) to find the best fit. Let us make explicit that the function depends on the parameters $a,b$, namely as $f(x,a,b) = ax+b.$ For our set of data points and for any given pair of values $a,b$, we have
\be
S(a,b)\equiv \sum_{i=1}^N [ y_i-f(x_i,a,b) ]^2.
\label{SquareSumParam}
\ee
Fitting the function to the data involves minimising $S(a,b)$; since the expression in question is the sum of the squares of the deviations, this is known as the {\bf least-squares method}. For linear functions, there is even an analytic solution: At the minimum of $S(a,b)$ as a function of $a$, the derivative of $S(a,b)$ with respect to $a$ must be zero (this is why it was useful to choose $S(a,b)$ to be readily differentiable). The same goes for $b$; impose both of those conditions, and you can find $a$ and $b$ directly. In the general case, no such analytical solutions are possible, and the best fit is found numerically.

There are a number of ways to go from here, some more advanced, some less. In fitting a function to data points, you can {\bf give different weight} to different contributions to (\ref{SquareSumParam}) with the measurement errors for each data point; in this way, those data points that are less-well known will also contribute less to the choice of parameters. 

Then, there is the problem of {\bf outliers}, that is, lonely data points that are far from the rest, and probably not because of their physical properties but because of measurement errors. Astronomers also make use of {\bf Bayesian techniques} in order to estimate the parameter values best fitted to their model, or to compare different models with each other. All of these issues are beyond the scope of at least this version of my basic introduction.\footnote{Some additional information can be found in the classic paper by Hogg, Bovy and Lang 2010, [\href{https://arxiv.org/abs/1008.4686}{https://arxiv.org/abs/1008.4686}].} 

The problem of making sense of multidimensional data is of interest far beyond astronomy, and a main task of what has become known as {\bf data science}. Astronomers apply numerous tools that have much more general application in order to solve their data science problems. 

For astronomers, this is not a matter of merely applying well-established methods and tools to new data sets. Instead, the way that astronomers look at their data, and make their deductions, is continually evolving. It is certainly more common for astronomers to derive new results by utilizing new data, but it is equally possible to make new deductions from an existing data set, by applying new methods.

As an example of a comparatively recent development, {\bf machine learning} has begun to play a role within astronomy. Machine learning comprises a certain subset of algorithms that can be used to find pattern in data, often involving a training phase during which the software learns about classifying certain kinds of data before moving on to new classifications.\footnote{Some information can be found in Ntampaka et al. 2019 [\href{https://arxiv.org/abs/1902.10159}{https://arxiv.org/abs/1902.10159}]. }

After this general overview, let us consider astronomical data sets, and explore ways of viewing or analyzing them. We start with astronomical images, and a simple application software for viewing them.

\section{SAOImage DS9 and astronomical images}
\label{DS9}
\begin{figure}[htbp]
\begin{center}
\includegraphics[width=\linewidth]{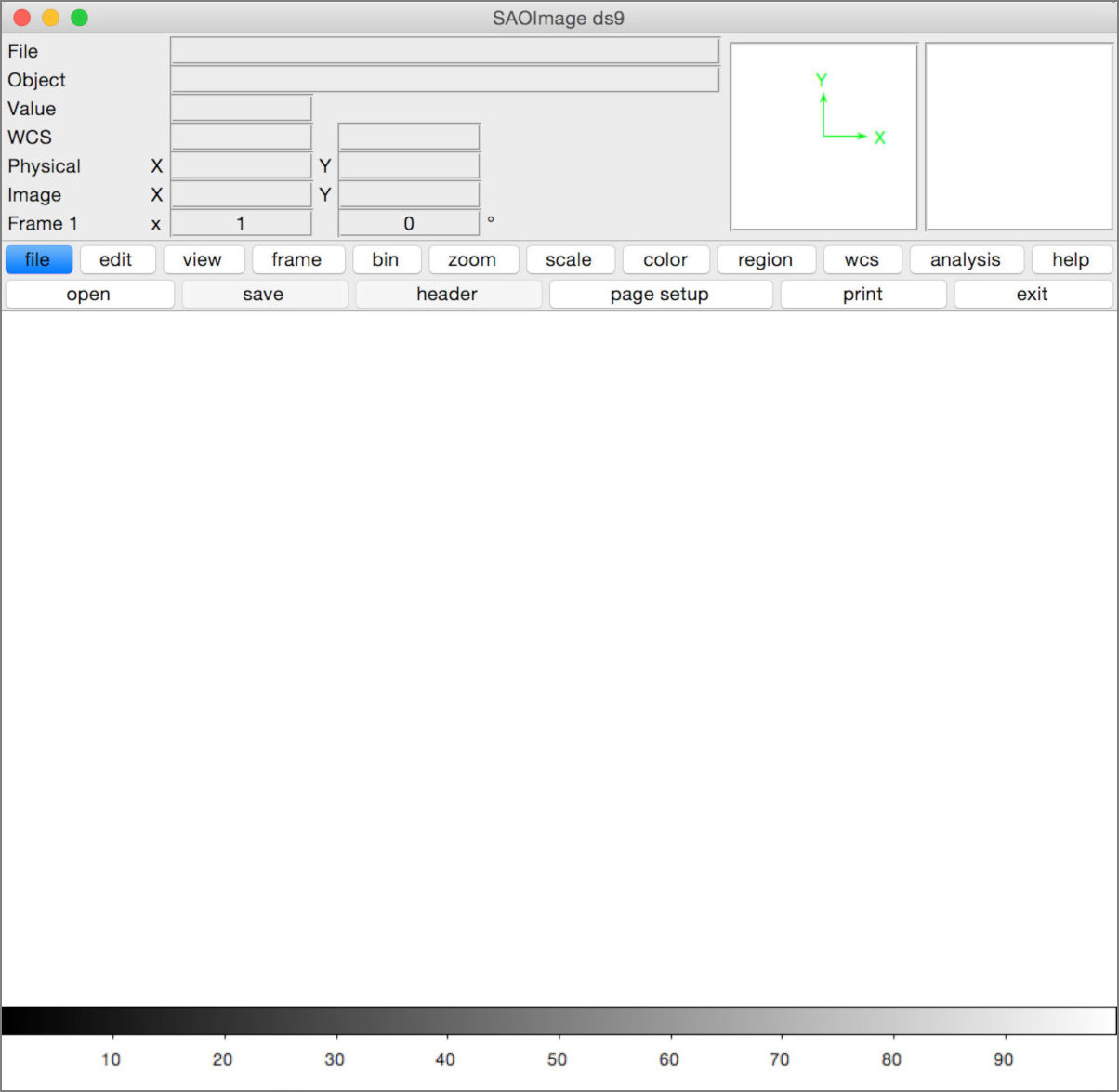}
\caption{The initial screen of DS9 right after the software has started}
\label{SAOOpen}
\end{center}
\end{figure}

When astronomy is in the public eye, a large portion of the attention goes to spectacular astronomical images. In this section, we will take a closer look at the scientific versions of such images. The application software we will use in this section is called SAOImage DS9. It was developed at the Smithsonian Astrophysical Observatory (SAO)  and is available for download from [\href{http://ds9.si.edu/}{http://ds9.si.edu/}]. When you start DS9, it will look as in Fig.~\ref{SAOOpen}.
Top left, there are fields with information about the image(s); top right, two small windows we will talk about later. Below, there is a horizontal menu with two rows. I will call the top row the ``main menu.'' Whenever you click on a field in the top row, the bottom row will display a set of associated commands. I will call this bottom row the ``secondary menu''. Below the secondary menu is the main image window, with a color bar (currently greyscale) below that. There is also a horizontal menu on top --- on a Mac at the top of the screen, on Windows or on Linux at the top of the window. I will call this the ``top menu.'' This is mostly a duplicate of the menu with the two rows within the window, but it does provide some additional options, and will occasionally be needed. Should operating systems change in the future, the menu placement may change as well, but currently, the positioning is as I have described. If DS9 opens additional windows, those can come with their own top menu, different from that of the main window. Details of what the menu items are called might vary slightly from version to version; I have used version 7.5.

\subsection{Loading a Hubble image}
\label{LoadHubble}

Let's load an astronomical image file. We get our file from the {\bf Hubble legacy archive} at the Space Telescope Science Institute (STScI) in Baltimore, which operates the Hubble Space Telescopes and other space telescopes. The legacy archive is where all the older Hubble images are stored. It can be found at the URL [\href{https://hla.stsci.edu/hlaview.html}{https://hla.stsci.edu/hlaview.html}]. There is a helpful search field. Let's search for M 16, the Eagle Nebula, by entering ``M16'' in the search field and pressing ``Search''.

\begin{figure}[htbp]
\begin{center}
\includegraphics[width=\linewidth]{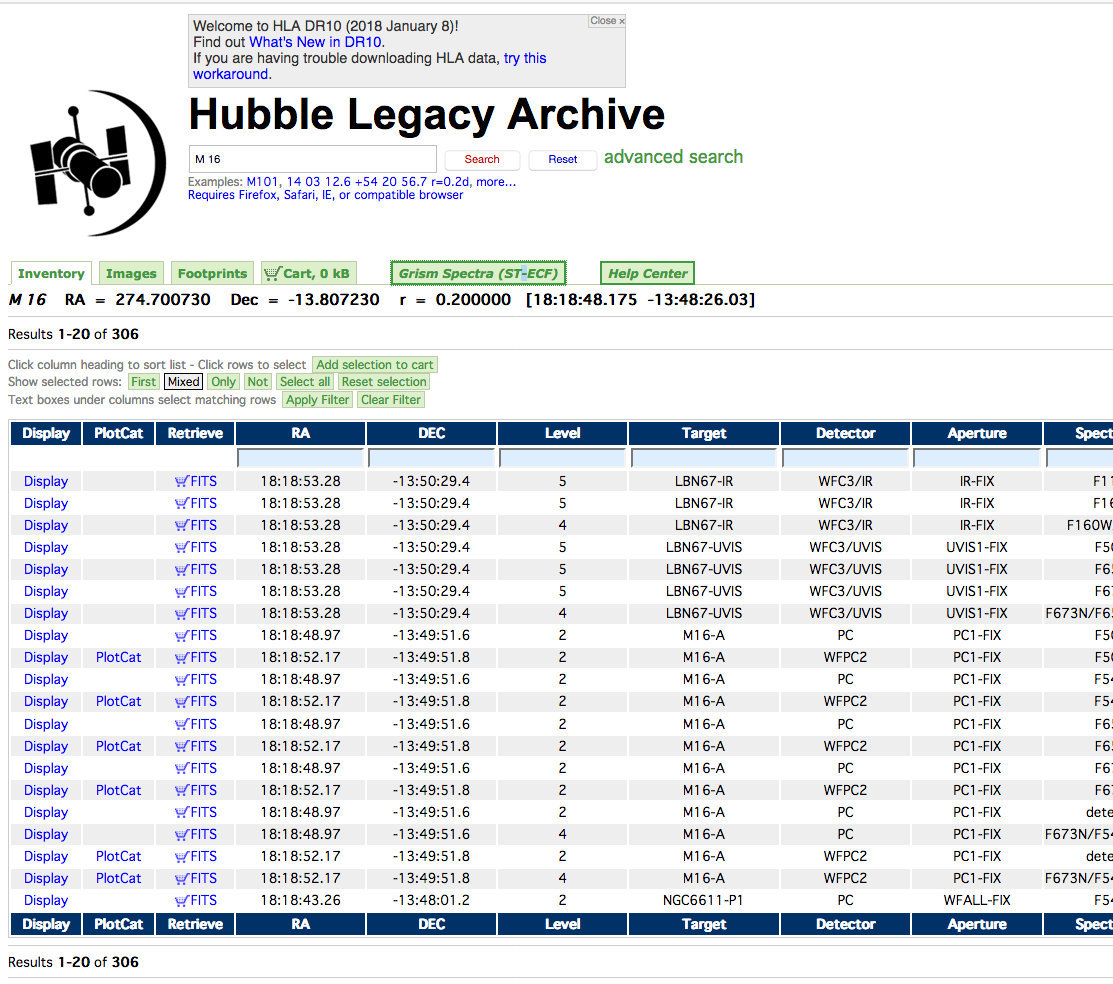}
\caption{Search results from the Hubble Legacy Archive}
\label{HLSIntro}
\end{center}
\end{figure}

You should get a very wide result screen, the leftmost bit of which is shown in Fig.~\ref{HLSIntro}.
You will probably need to scroll right to see the columns 13 and 14 we are interested in. In column 14, called ``Dataset'', look for the data set names
\begin{itemize}
\item hst\_05773\_05\_wfpc2\_f502n\_wf
\item hst\_05773\_05\_wfpc2\_f656n\_wf
\item hst\_05773\_05\_wfpc2\_f673n\_wf
\end{itemize}
At the time of this writing, it is fairly easy to find these images. If you don't, you can try something different: In the search field on [\href{https://hla.stsci.edu/hlaview.html}{https://hla.stsci.edu/hlaview.html}], click on ``Advanced search''. In the main field, enter ``M16'', and in the proposal ID field, enter ``05773''. The result will be a much shorter list, including the images listed above. Even in the far future, when the interface might have changed, searching for the proposal ID in addition to the object name should return a list that includes those images --- even future archives should be ``legacy-proof.''

The images in question were all taken on April 1, 1995, and belong to one of the most iconic Hubble images: the pillars of creation. Each file should be around 53 MB in size. You can download the files by either clicking the little shopping cart icon for each image and then going to the shopping cart tab, or by right clicking on the shopping cart icon and choosing ``save link as''.

If you know astronomical abbreviations, you will be able to make some initial sense out of these dataset names: hst, for instance is bound to mean that we are downloading data from the Hubble Space Telescope. WFPC2 is the ``Wide-Field and Planetary Camera 2" on that telescope, wf says that we are downloading the wide-field camera images. f502, f656 and f673 denote different filters which have been placed before the camera for these respective images. We will combine the three images into a colour image --- but it is going to be a false-color images, since those three filters do not correspond to red, green, and blue!

Last but not least, when you have downloaded the images, you will notice that the filename extension indicates that these are FITS files, in the most common format used for scientific images in astronomy; the filename extension is either ``fits'' or possibly if you are on an older Windows machine, ``fit''. 

\subsection{A first look at the Eagle Nebula M16}

\begin{figure}[htbp]
\begin{center}
\includegraphics[width=\linewidth]{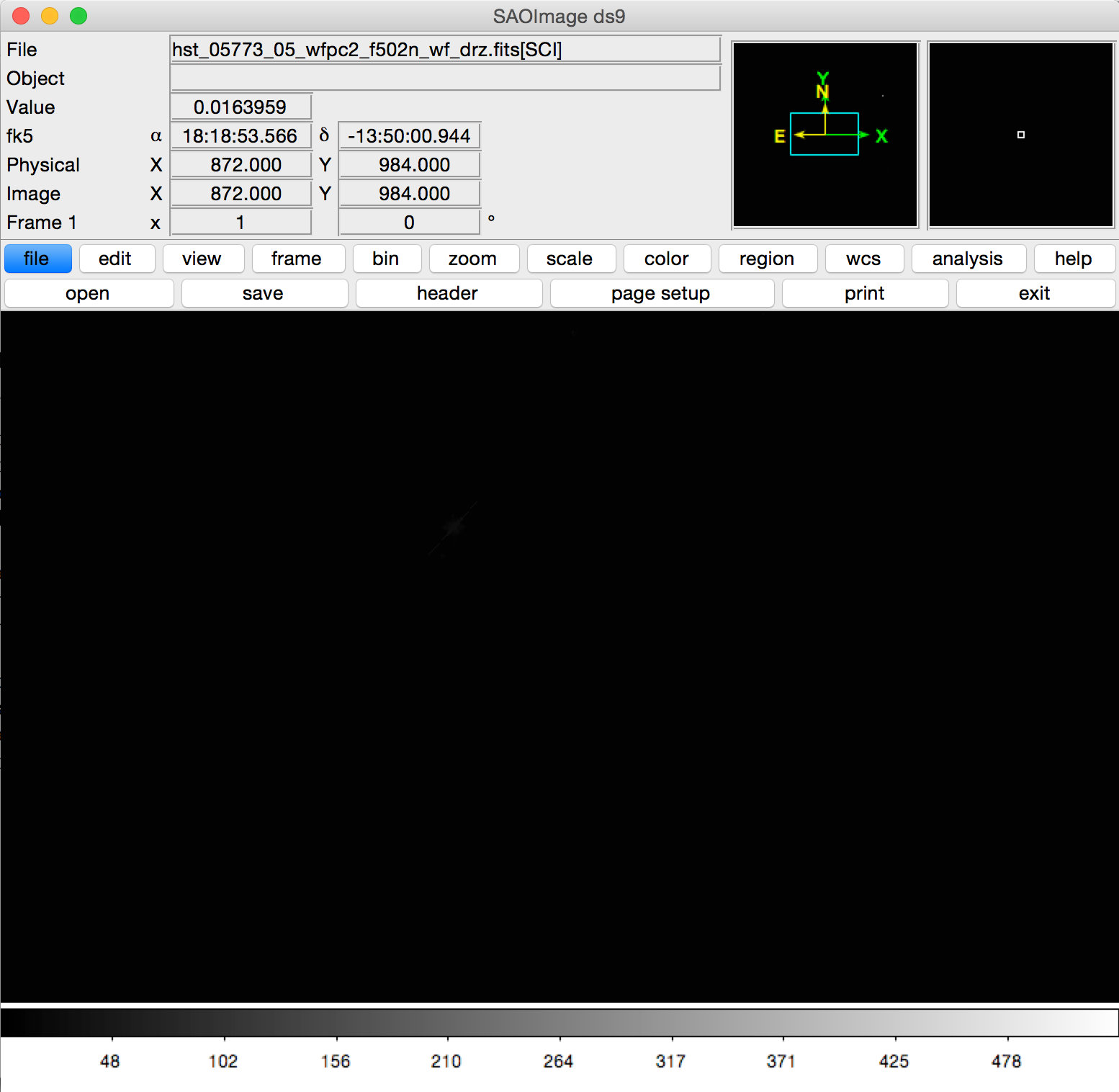}
\caption{Opening an HST Eagle Nebula image in DS9}
\label{EagleOpen1}
\end{center}
\end{figure}

Now that we have the files safely stored away, we can open them using DS9. To this end, go to the main menu row and click on ``file'' (which is probably highlighted to begin with); from the secondary menu that appears directly below, choose ``open''. In the usual pop-up choose-a-file window, I'll choose the first of the Hubble files we downloaded, hst\_05773\_05\_wfpc2\_f502n\_wf\_drz.fits. Once the file is open, the DS9 window should look as in Fig.~\ref{EagleOpen1}.

The image is disappointingly black. We need to find a better brightness scale to see what is going on. Astronomical images typically capture an amazing dynamic range, that is, an amazing range of different brightness values for each pixel (concretely, 65536 different brightness values per pixel, compared with the 256 of a typical RGB pixel). Displaying such an image on a computer monitor, or printing it, can never do this range full justice. Instead, we need to pick and choose --- which part of the brightness range do we want to display, and which way of compressing the brightness scale shows us the most information about the image?

There is no single right way of doing this, and there is no standard way that will guarantee the best results for all possible astronomical pictures. Instead, this is a matter of combining experience (your own and that of others!) with some experimentation to arrive at a result that works for you. You should always be aware that such a result is not a naked view of the astronomical data --- what you see is determined both by the astronomical data and by the choices you have made in displaying that data. (Also note that these display options do not change the image itself; the image file itself is unchanged, and you are only changing the way you are viewing the data.)

\begin{figure}[htbp]
\begin{center}
\includegraphics[width=\linewidth]{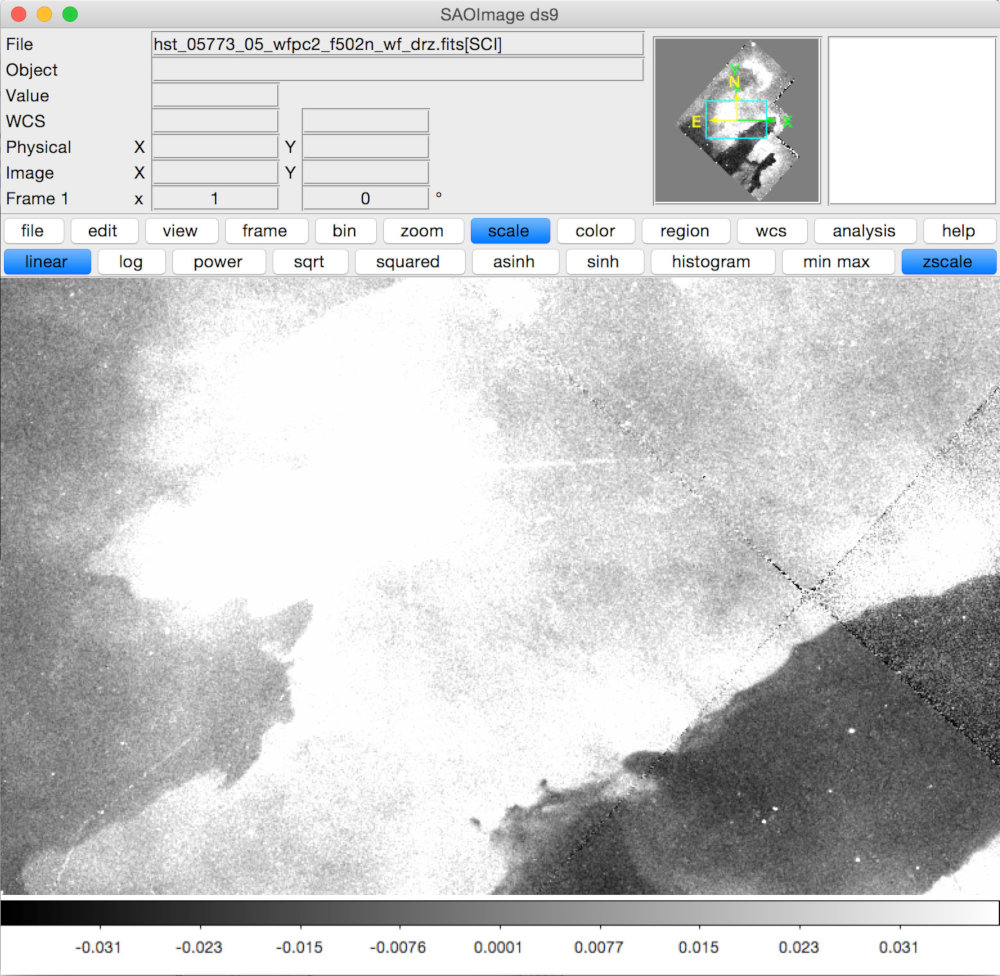}
\caption{HST Eagle nebula image in DS9 with zscale applied}
\label{Eagle2}
\end{center}
\end{figure}

To experiment a bit, go to the main menu and choose ``scale''. From the secondary menu that will come up, choose ``zscale''. Where previously (in ``minmax'' mode) the image had been displayed with the minimum pixel brightness set to black, and the maximum pixel brightness to white, the colors are now mapped to values closer to the median pixel brightness. As a result, your image should look something like this in Fig.~\ref{Eagle2}.
This still looks fairly raw, and rather different from the pretty astronomy pictures you see in the media. But at least we can discern some structures. The main window below the two rows of menus only shows part of the image. In the overview window (second to right, on top) you can see the whole of the image. The cyan frame in the overview window marks the part of the image that is visible in the main window. Drag it around (left-click the mouse and drag) to explore other parts of the image. Alternatively, you can go to ``zoom'' in the main menu and choose one of the options in the secondary menu to see the image as a whole (``zoom fit''), or in more detail (e.g. ``zoom 2'').

In the ``scale'' menu, you can also choose another compression method instead of linear --- see how it affects the view! (Again, the image itself is not changed by your choice.) Also, instead of the grayscale display, you can go to the main menu point ``color'' and select another color map. 

\subsection{Coordinates: Navigating the image}

\begin{figure}[htbp]
\begin{center}
\includegraphics[width=\linewidth]{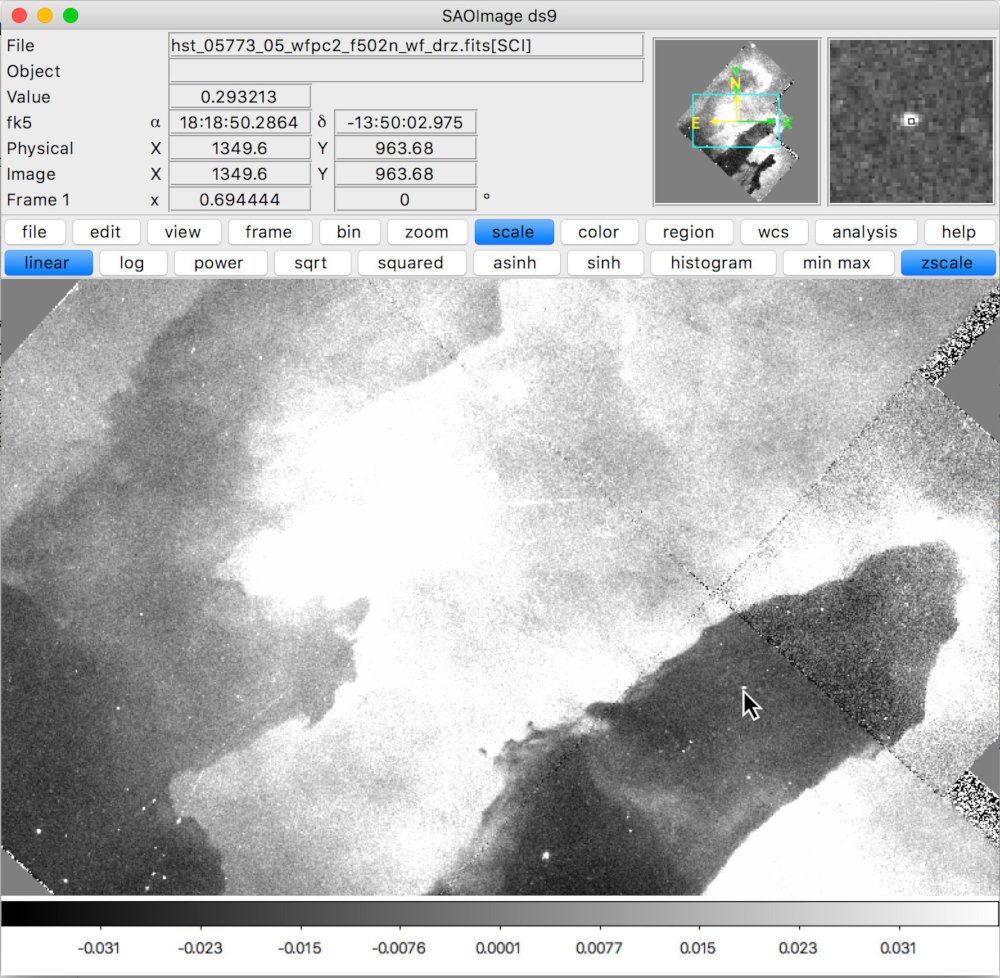}
\caption{HST Eagle nebula image with cursor on a star}
\label{Eagle3}
\end{center}
\end{figure}

When your cursor is on the main image, what you will see will be something like in Fig.~\ref{Eagle3}.
In this image, the tip of the cursor is placed on a star. The detailed image (inset image top right) shows the star's little disk, with DS9's own representation of the cursor on top. Let us take a closer look at the information on the top left. ``File'' is simply the file name. If the header identifies the object by name in a suitable way, that is what would be displayed in ``Object.'' The ``Value'' field gives you the value of that particular pixel. 

Below, we have the sky position information for the cursor, given in the ``World Coordinate System'' (WCS). ``fk5'' tells you that the sky coordinate system is defined using the reference stars of the Fifth Fundamental Catalogue (Fundamentalkatalog 5), which was published in 1988 by Astronomisches Recheninstitut Heidelberg (ARI, now a part of Heidelberg University). The coordinates themselves are those of the equatorial system, which is analogous to latitude and longitude on Earth: longitude corresponds to the {\em right ascension} $\alpha$ (sometimes abbreviated to RA, ra, or R.A.), latitude corresponds to the {\em declination} $\delta$ (sometimes abbreviated to Dec, dec, or DE). 

Both RA and Dec are given here in the standard sexagesimal notation. Right ascension is given in hours (in hour example: 18), minutes (18) and seconds (50.2864), written as 18:18:50.2864. (In other contexts, it might be written as $18^h\:18^m\:50^s.2864$, but the meaning is always the same: Each object lies on a particular meridian (that is, on part of a great circle through the two poles of the coordinate system). The value of the right ascension indicates where that meridian intersects the celestial equator, measured eastwards along the celestial equator, starting from the vernal point --- the point where the Sun crosses the celestial equator northwards, at the time of the vernal equinox (at or around March 20 each year). For this measurement, 24 hours correspond to the full circle. The integer hour value is followed by the number of full minutes, with 60 minutes in a full hour, as expected; the third position gives the seconds, again with 60 seconds in a minute.

Declination is given in degrees northwards (positive sign) or southwards (negative sign) from the celestial equator, which itself is at $\delta=0^{\circ}$. In the image, the notation is again sexagesimal, with a southward (because of the minus sign) $13$ degrees, 50 arc minutes and 2.975 arc seconds written as 13:50:2.975. (In another context, you might see this written as $13^{\circ}50'2''.975.$)

If you go to ``WCS'' in the top menu, you can change the notation from sexagesimal to degrees; the latter will show both right ascension and declination as a decimal number denoting degrees, in our example $274.7095$ for the right ascension, and $-13.9341$ for the declination (in both cases with a few more significant digits).

Just like in the usual geographical coordinate system, a difference of one degree in declination corresponds to the same length, wherever we are on the celestial sphere, just like a difference of one degree in latitude does. Differences in right ascension, on the other hand, correspond to smaller angular distances the closer you go to the celestial poles, analogous to what happens with geographic longitude. If you want to move the cursor around a bit to estimate the angular scale of your image (how many pixels corresponding to, say, 10 arc seconds), use declination for the purpose, not right ascension. 

The "Image" coordinates below denote the X and Y coordinate of a pixel within the given image. In a FITS file, the pixel in the bottom left corner has the coordinate $(1,1)$. If you zoom in, you will see that DS9 assigns this coordinate value to the center of the pixel. The lower left corner of the lower left corner has the pixel coordinates $(0.5,0.5)$. In our example the ``Physical'' coordinate is the same as the image coordinate. If the image you are looking at is only a part of a larger image, you are likely to find the physical and the image coordinates differ: The physical coordinates would still be those of the original image, $(1,1)$ the coordinates of the bottom left pixel of the CDD camera. The image coordinates would be those of the smaller image, $(1,1)$ the coordinates of {\em its} bottom left pixel.

How does DS9 know which sky coordinates to map to the pixel coordinates? That (meta-)information is contained in a special part of any FITS file: the header. 

\subsection{Meta-Information: the FITS header}

Astronomical images can only be interpreted correctly if you know the conditions under which they were produced. What filter was used? How long was the image exposed? When and where was it taken? FITS files include this meta-information in a dedicated section called the {\bf header}. We can inspect the FITS header of our image by going on ``file'' in the main menu, and choosing ``header''.

\begin{figure}[htbp]
\begin{center}
\includegraphics[width=\linewidth]{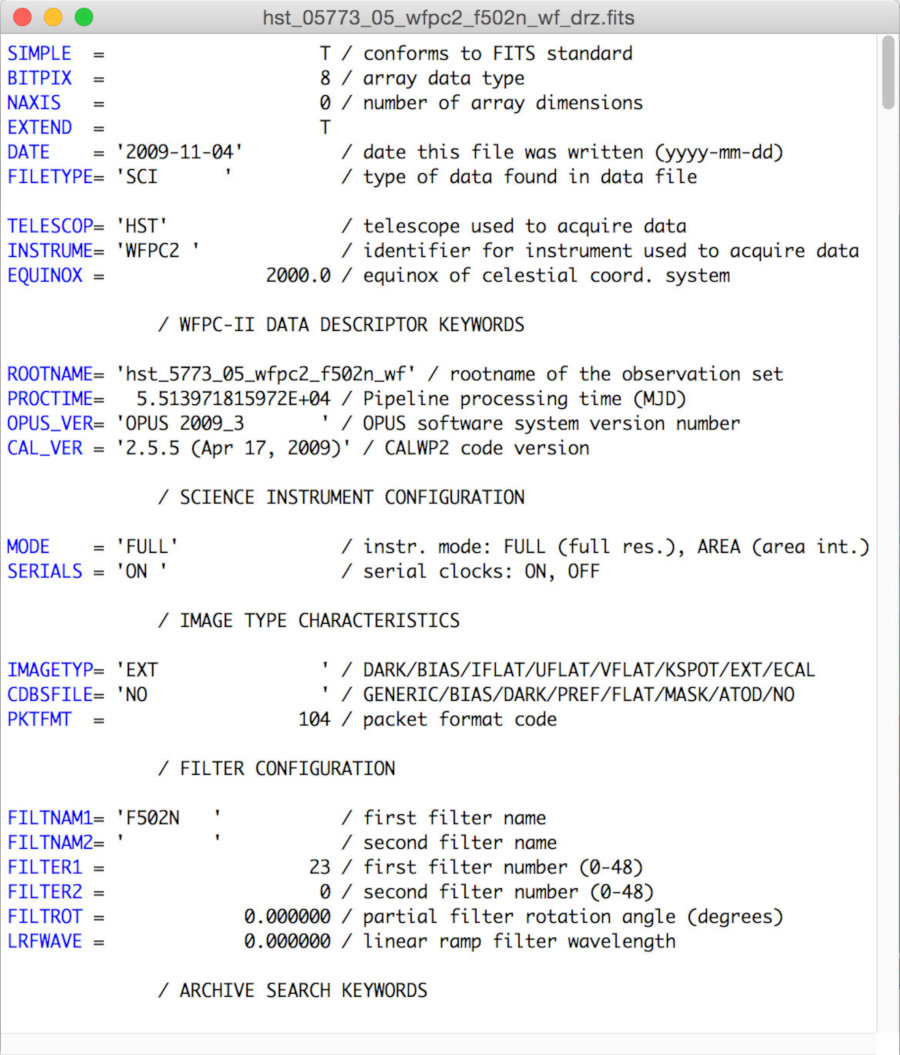}
\caption{First header of the HST Eagle nebula image}
\label{EagleHeader1}
\end{center}
\end{figure}

Once you do this, you are given two choices; the second one is the file name with [SCI] appended. Start by looking at the first header, which looks like in Fig.~\ref{EagleHeader1}.
You don't need to understand this in detail (I certainly don't!), but some bits are fairly clear: among other things, the header lists the file name convention, telescope and instrument name, date when this was processed, and the filter F502N that was used to take the image. Information is encoded in a two-part way: each particular chunk of information has a keyword (here shown in blue, on the left) and an assigned value. The two are linked by the equal sign. Often, this is followed by a slash, after which there is a comment with a description of the keyword's meaning.

If you scroll down, you can find different kinds of information: the position of the Sun at the moment of observation is encoded there, the angle between the Moon's position and the pointing direction of the telescope, the observation start time and end time. Some of those keywords are specific to the telescope, instrument, organization or project in question. Others are more general. Under EXPTIME, you will commonly find the exposure time in seconds, and under DATE-OBS the date of the start of the observation in year--month--day format. TIME-OBS gives the time of the start of the observations; the time zone is Universal Time (UT). The multiple lines marked HISTORY typically contain information about how the image has already been processed, which files were used as flatfield or dark frame, and which software was used for the processing.

\begin{figure}[htbp]
\begin{center}
\includegraphics[width=\linewidth]{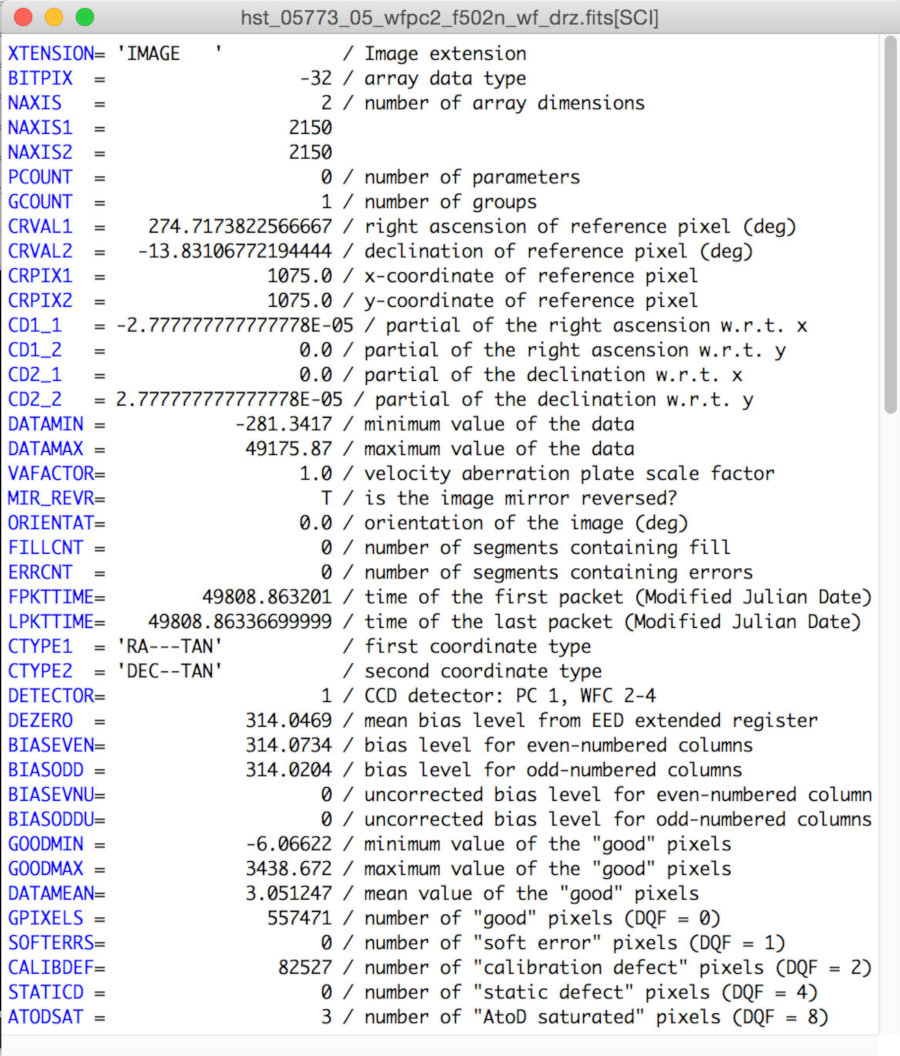}
\caption{Second header of the HST Eagle nebula image}
\label{EagleHeader2}
\end{center}
\end{figure}

The [SCI] version is more image-specific, as you can see in Fig.~\ref{EagleHeader2}.
It gives you lots of details about the instrument and telescope properties during these particular observations, and important information about the context of the data. Here, too, some keywords will be specific to the project in question, and others more universal. NAXIS will tell you the number of axes you are dealing with, in this case two of them. NAXIS1 and NAXIS2 will give you the width and height of the image, in pixels. BUNIT will give you the units of the pixel values. In this case, it is a generic ``COUNTS/S'', counts per second, in other cases it might be Jansky per beam or similar units.

The CRVAL1, CRVAL2, CRPIX1, CRPIX2 and the CD1\_1, CD1\_2, CD2\_1 and CD2\_2 contain the information that allows you to calculate, for each pixel, the values of the two equatorial coordinates: right ascension and declination. The CD values correspond to a matrix mapping the two coordinate systems to each other. Often, the axes of the WCS and the image coordinate systems are parallel to each other. (You can check this in the overview window in the upper right corner, second from the right, where the North and East directions are shown as N and E, and the image coordinate directions as X and Y.) In that case, CD1\_2 and CD2\_1 are zero, and for the usual square pixels, CD1\_1 and CD2\_2 are equal (up to a sign) and denote the image scale. In our case, CD2\_2 = 2.777777777777778E-05 denotes the degrees that correspond to the width of a single pixel; 
$2.\overline{7}\cdot 10^{-5}$ degrees per pixel corresponds to $0.1$ arc seconds per pixel. CD1\_1 has an extra minus sign because the X direction and the East direction are anti-parallel in this image.

\subsection{Making a colour image}
\label{MakingColourImage}

As a nod towards pretty astronomical pictures, let us combine partial images into an RGB color image. Recall from section \ref{ImagesBasics} that astronomical cameras take black-and-white images, each through a previously chosen filter (or, in some cases, no filter at all). These different filter images can be combined afterwards to produce an RGB color image. 

In most cases, the result will be a false-color image, however, since the astronomical filters used will not correspond to the proper red, green and blue filters. So let's make a color image! Since we have played with various switches in DS9, we should reset; the easiest way is to quit and restart the software.

\begin{figure}[htbp]
\begin{center}
\includegraphics[width=0.25\linewidth]{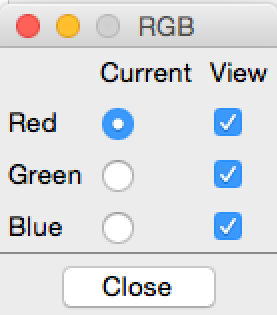}
\caption{The RGB popup menu}
\label{RGBMenu}
\end{center}
\end{figure}

Once DS9 is up and running again, create a color frame by clicking ``frame'' in the main menu, and then ``rgb'' in the secondary menu. A little extra menu will pop up, which can be seen in Fig.~\ref{RGBMenu}.

With the ``red'' option checked, go to ``file'' in the main menu and ``open'' in the submenu and load hst\_05773\_05\_wfpc2\_f673n\_wf\_drz.fits. Next, check green and load hst\_05773\_05\_wfpc2\_f656n\_wf\_drz.fits. You should already see a superposition of the red and green parts at this stage. Finally,  check blue and load hst\_05773\_05\_wfpc2\_f502n\_wf\_drz.fits. There is your composite color image, but it's looking rather dark. 

With the small RGB menu window active, go to the top menu (at the top of your screen on a Mac, and at the top of your window frame in Windows). Under the top menu point ``Lock'', choose ``scale''. That way,  when you change the scale, the change will affect all the three images equally. 

\begin{figure}[htbp]
\begin{center}
\includegraphics[width=\linewidth]{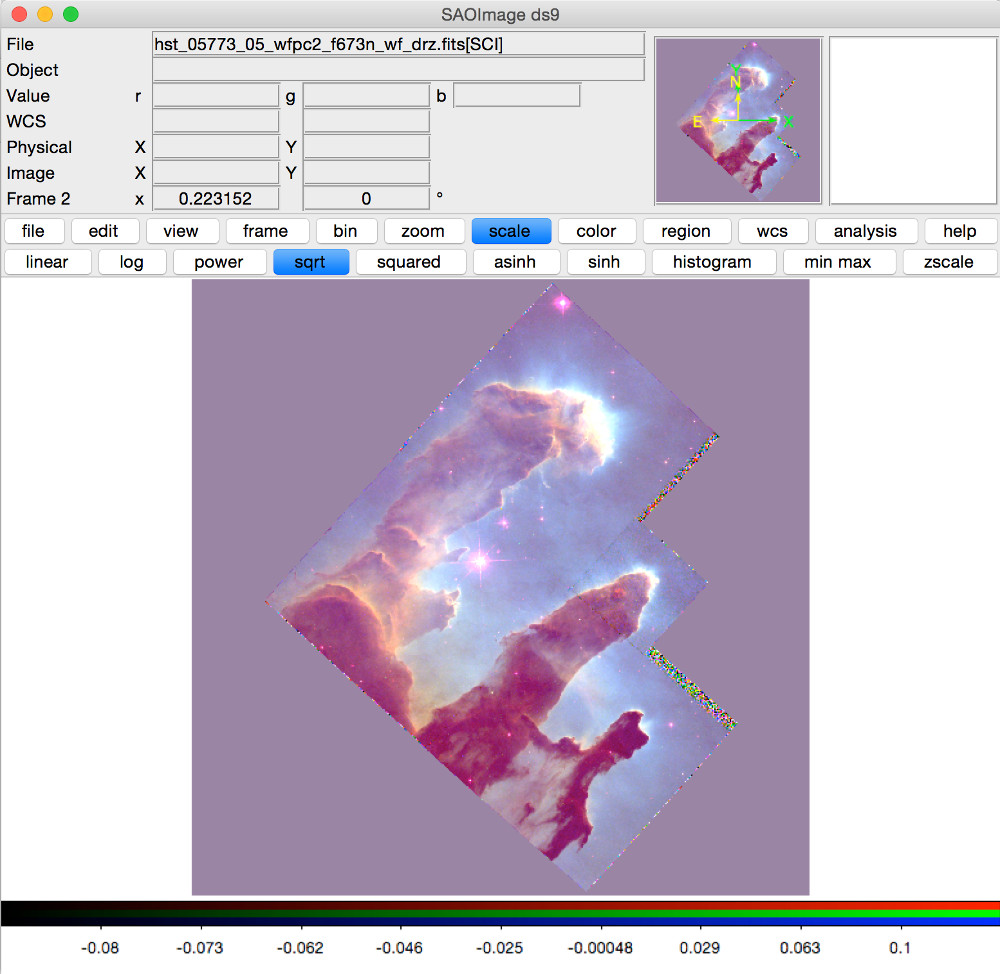}
\caption{Colour image of the ``pillars of creation'' with DS9}
\label{Pillars1}
\end{center}
\end{figure}

By setting the scale to ``sqrt'' (a form of compression) and the scale limits to 99\% (only possible under ``scale'' in the top menu) I get an image that's pretty close to the usual appearance of the pillars of creation, cf. Fig.~\ref{Pillars1}.
The Hubble Space Telescope images you can find in the media have had significant extra work applied to them, up to and including a person going over the image region by region, removing remaining impurities by hand, and cleaning up the boundaries.

\subsection{Catalogs}
\label{DS9DR9}

In the end, we're not here for the pretty pictures. We're here for the science. And since today's science builds upon what was done before, we will look at ways o accessing the information that is already out there about the region whose image we are looking at. We will work with an image from data release 9 (DR9) of the Sloan Digital Sky Survey (SDSS). The original SDSS was the first large-scale digital survey, a pioneering project that provided astronomers with large amounts of data that was of consistently high quality. Were, before, astronomers had counted themselves lucky if they could do a statistical analysis with a few hundred galaxies, the original SDSS gave them data on more than 900.000! The DR9 is part of the third incarnation of the survey, SDSS-III.

I have chosen an image fairly randomly by going to \href{https://dr9.sdss.org/fields/}{https://dr9.sdss.org/fields/} and entering the RA 20.0 and the Dec 20.0 in the ``Search by Object Coordinates'' and hitting ``Submit''. On the results page, click on the link ``g-band FITS'' to download the file shot through the SDSS g filter. Unzip the file to obtain the unadorned FITS file named ``frame-g-007923-5-0307.fits'' which is 12.4 MB in size. Open that image with DS9.  In order to look at the image, choose a linear zscale.

\begin{figure}[htbp]
\begin{center}
\includegraphics[width=\linewidth]{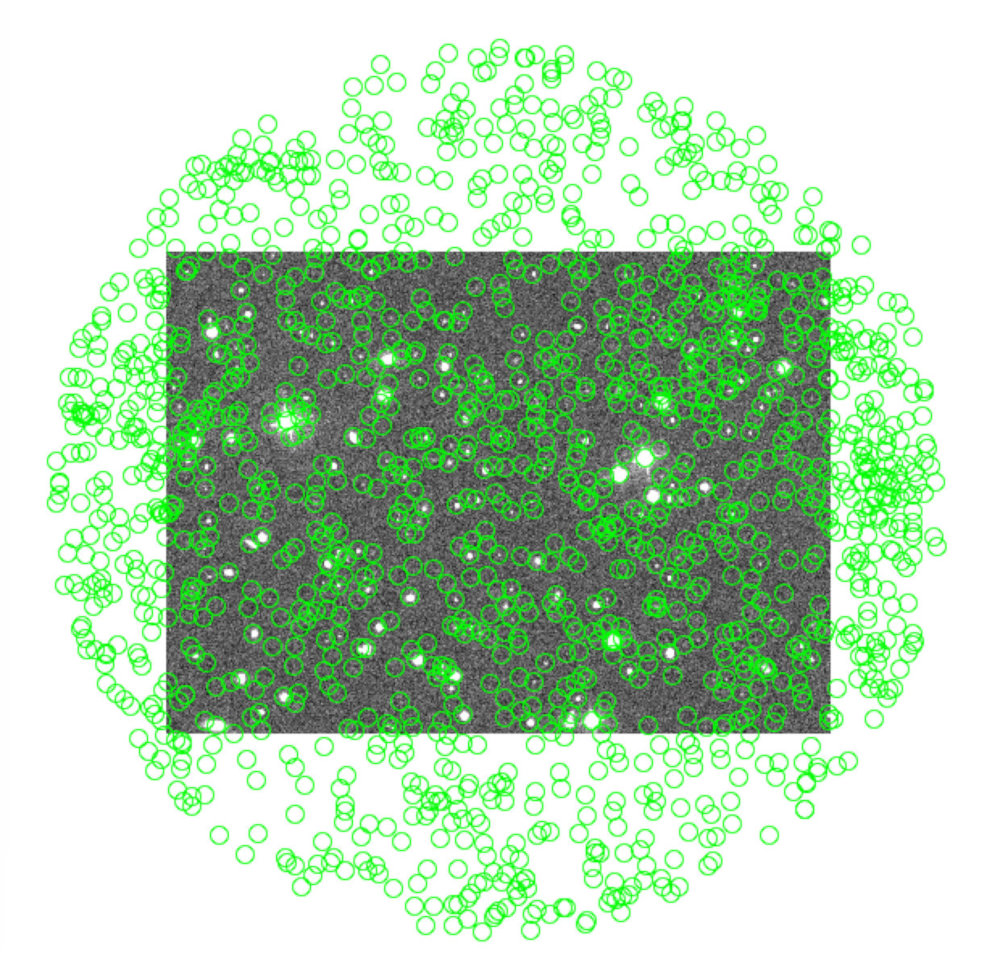}
\caption{Zoomed-out version of our SDSS image, with SDSS DR9 catalog stars circled in green}
\label{SDSS9src}
\end{center}
\end{figure}
For an image with accurate positioning data (contained, as we have seen, in the FITS header), DS9 can show the positions of the known stars. To display them, go to the main window's top menu. From the dropdown menu ``Analysis,'' go to the ``Catalogues'' dropdown sub-menu, from there to ``Optical'' and there, choose ``SDSS Release 9,'' which is the catalog associated with the image's data release. If you zoom out sufficiently far, your main image will look as in Fig.~\ref{SDSS9src}. The image, and the surrounding area, are covered in small green circles, and each circle marks a star (or possibly other object) that is listed in the SDSS DR9 catalogue.

\begin{figure}[htbp]
\begin{center}
\includegraphics[width=\linewidth]{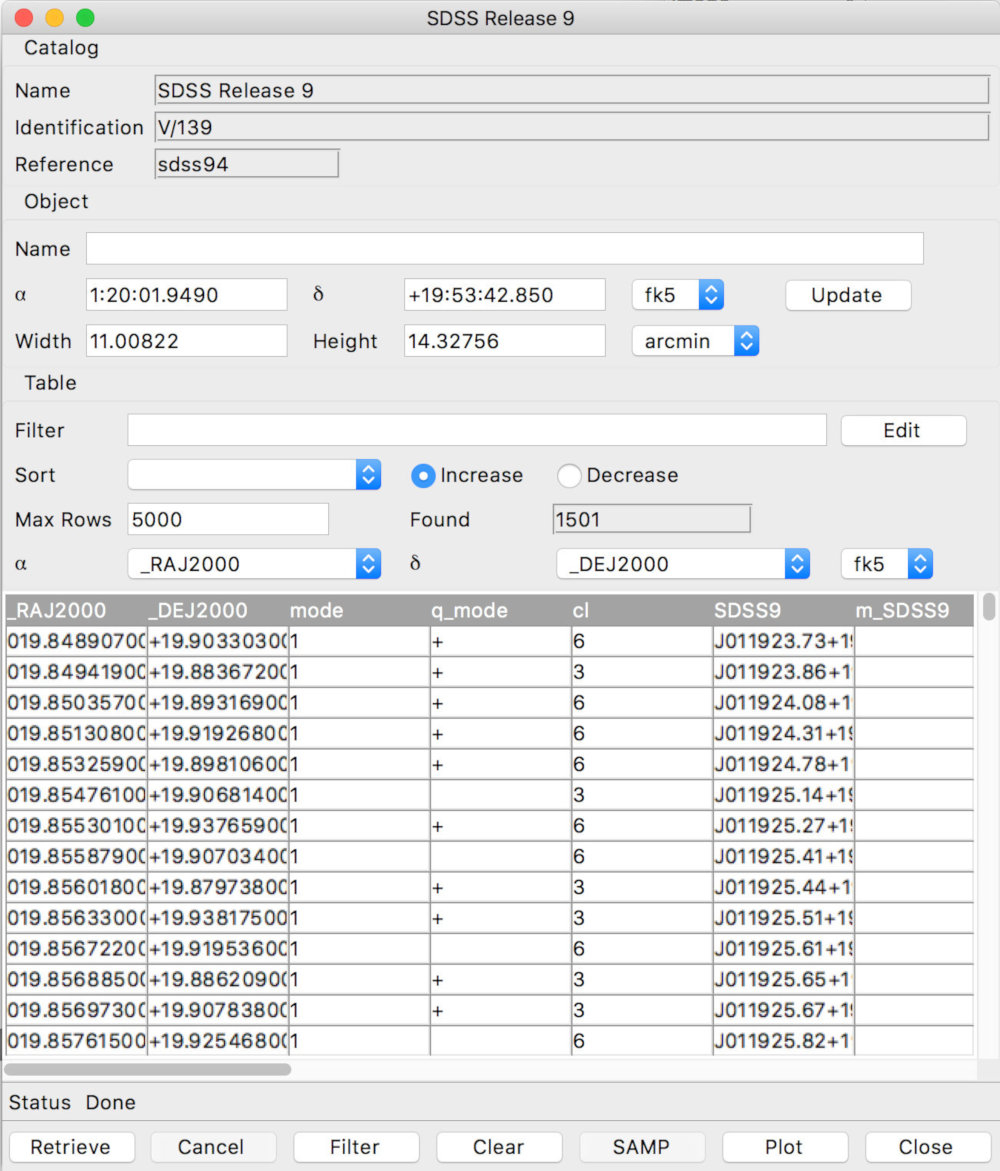}
\caption{Catalog window for the SDSS DR9 stars within our window}
\label{CatalogWindow}
\end{center}
\end{figure}

At the same time, DS9 opens up a catalog window, as shown in Fig.~\ref{CatalogWindow}.
Near the bottom of this window is a list of objects. Each object corresponds to one of the green circles in the image. Use the scroll bar to look at the list's different columns, and you will see that, in addition to the object's position (RA and Dec), there is information about the object's brightness: the different apparent magnitudes, as measured using the SDSS filter set {\em ugriz}, corresponding to specific filters centered in the near-UV (u filter, magnitude is given in the umag column), the blue-green part of the spectrum (gmag), the red region (rmag), the border region between red and near-infrared (imag) and an infrared filter beyond that (zmag).

If you click on a line in that table, corresponding to the entry for a specific object, the object's marker circle will blink red a few times in the main window, and centre the view on that object In that way, you can zoom in on specific objects and have a closer look at them. Conversely, if the main window is active and you go to ``Edit'' in the top menu and, in that submenu, choose ``Catalog'', then you can click on any of the little catalog marker circles in the image, and in the catalog window, the corresponding row will be visible and highlighted. (Choose ``None'' in the ``Edit'' dropdown menu to go back to the normal editing mode.)

The filter field in the catalog window allows you to filter out those parts of the catalog that do not meet your criteria. For instance, entering ``\$gmag $< 17$'' (where gmag is the name of the column, to which you have added a dollar sign \$, which denotes in some programming languages that this is a variable) and clicking on the ``Filter'' button near the bottom of the window will only keep those objects whose g-magnitude is smaller (brighter!) than $m_g=17$. You can connect several such conditions with \&\& for a logical ``and'' or $\Vert$ for logical ``or'', e.g. 
``$\mbox{\$gmag} < 17\;\; \Vert \;\; \mbox{\$umag }< 16$'' if any object with either $m_g<17$ or $m_u<16$ is fine.

The sort functionality of the catalog window provides another possibility for finding your way around the catalog data. To the right of the ``Sort'' marker is a menu that lets you choose any of the columns of the catalog. To the right, by checking a box, you can choose whether to sort the catalog using values in that particular column in ascending or descending order (increase or decrease).

\subsection{Photometry with regions and statistics}
\label{DS9Photometry}

Let us continue with the SDSS image we had opened in the previous section. If you haven't done so, choose the SDSS DR9 catalog. Sort the catalog by gmag in increasing order so you can pick out specific values for gmag. Go to the star with gmag 19.659. It's at around RA 20.0714 and Dec $+19.9777$, corresponding to the pixel coordinates $X=1819,\; Y=1215$.

\begin{figure}[htbp]
\begin{center}
\includegraphics[width=\linewidth]{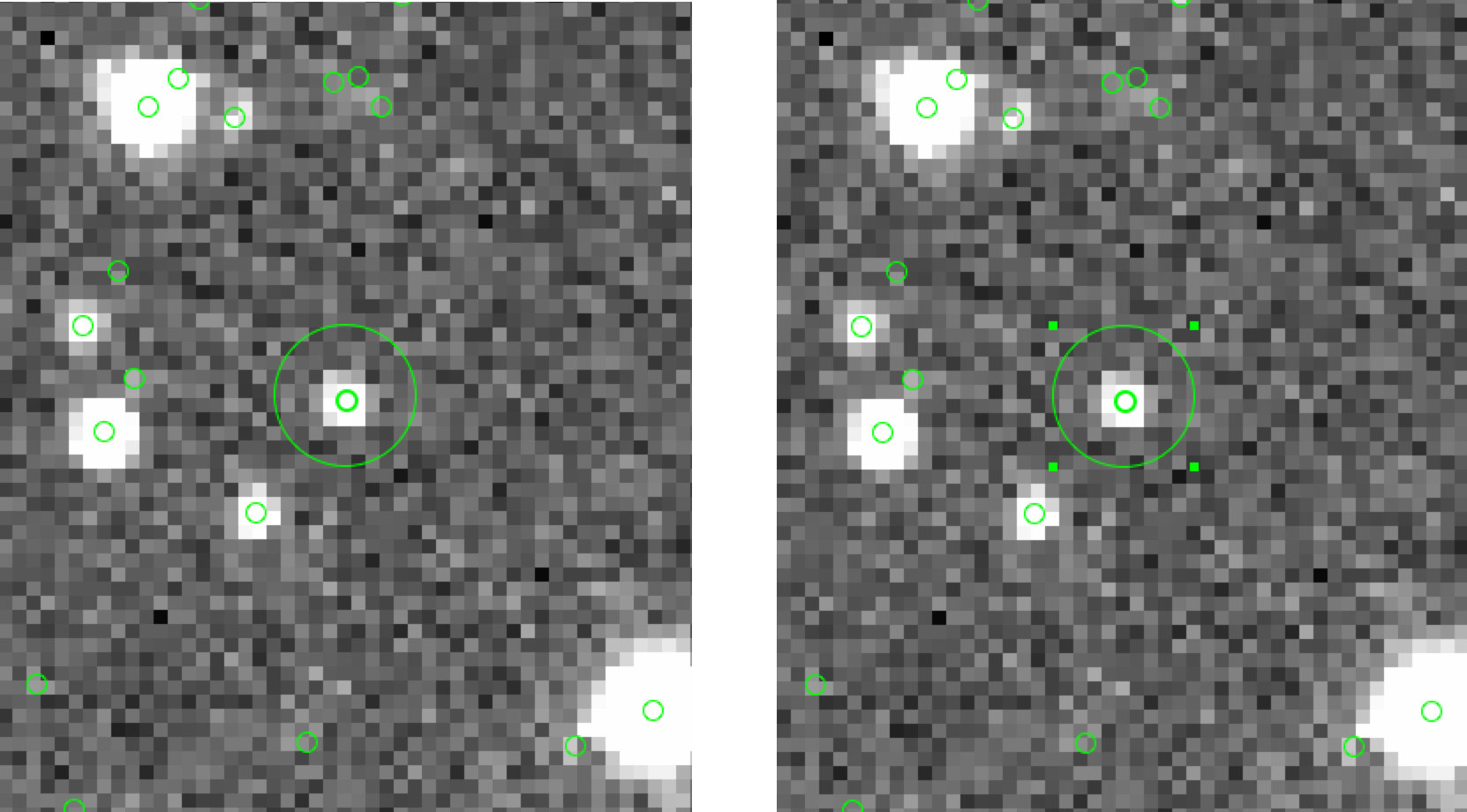}
\caption{A circular region, not selected (left) and selected (right). The much smaller green circles are the catalog indicators and not related to the region itself}
\label{DS9Selected}
\end{center}
\end{figure}

Next, we are going to define a circular region around this star. First, I zoom in considerably, going to ``Zoom 4'' and clicking ``Zoom in'' once. In order to do add the region, go to the top menu and, from the dropdown menu ``Edit'', choose the item ``Region.'' Now we are in region mode. Still in the top menu, from the dropdown menu ``Region'', go to the sub-menu ``Shapes'' and select ``Circle''. Now, click on the star we have chosen. A larger green circle outline appears, marking the region we have chosen, cf. the left part of Fig.~\ref{DS9Selected}.  Click within that circle, and the region is selected, indicated by the four dots framing it, as in the right part of Fig.~\ref{DS9Selected}. Now you can pull the circle around by clicking, holding and dragging. Centre it on the star.

\begin{figure}[htbp]
\begin{center}
\includegraphics[width=\linewidth]{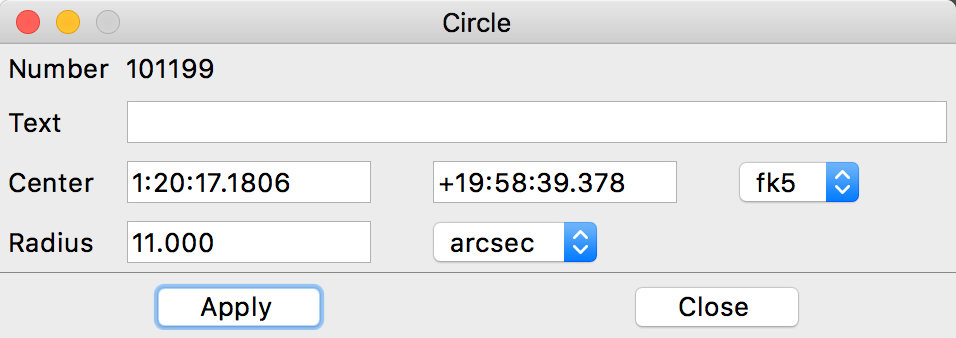}
\caption{The extra window for our circular region}
\label{RegionWindow}
\end{center}
\end{figure}

By double clicking on the circle (while still in the ``Region'' edit mode), you can bring up an extra window describing the region, cf. Fig.~\ref{RegionWindow}. Let us call this the ``region window,'' for short.
The window allows you to read off the basic properties of your region --- in the case of a circle, center and radius. You can also change the properties by hand, giving them specific values. We will change the radius of our circle, making it as large as possible while keeping it sufficiently small not to include any other stars than that in the centre. Trial-and-error suggests to me that around 11 arc seconds is a good value in this particular case --- we just about avoid the next star to the lower left of our region. To change the radius value to 11 arc seconds, first change the unit of the radius to ``arcsec'' by choosing that value from the dropdown menu that initially reads ``Degrees''. Input the value 11 for the radius. Press ``Apply'' so that your choice is applied to the region. You will probably see the region circle get a little larger in consequence.

\begin{figure}[htbp]
\begin{center}
\includegraphics[width=\linewidth]{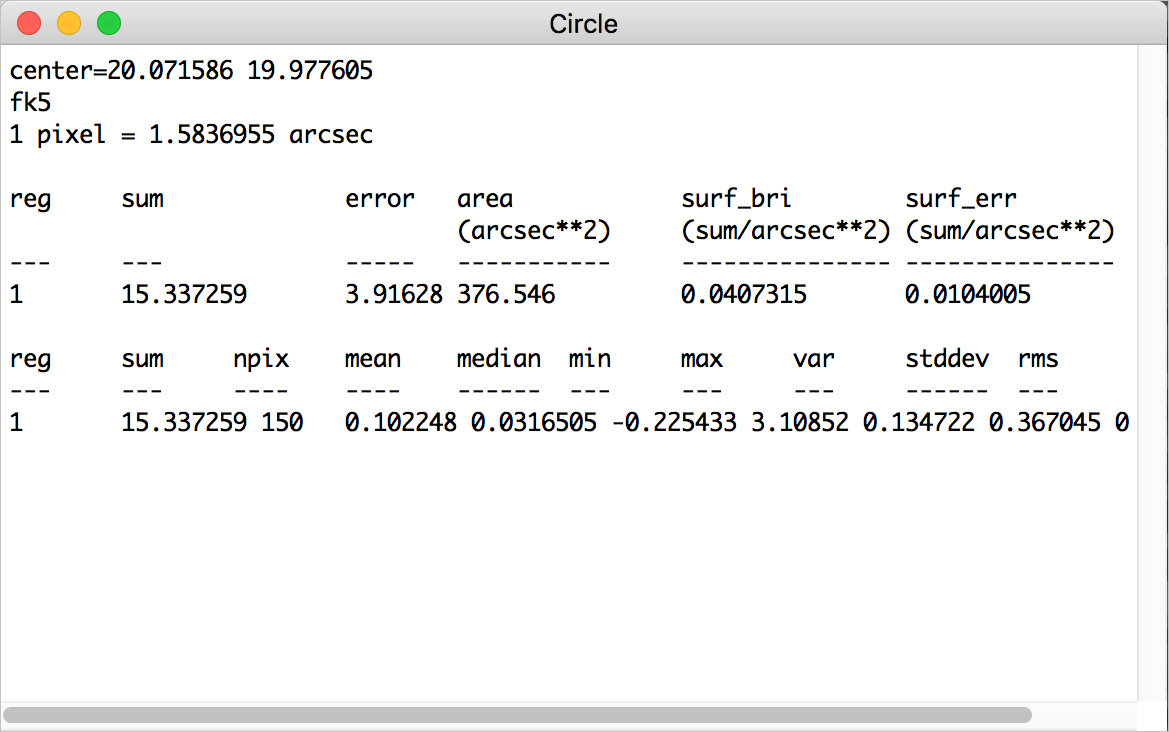}
\caption{Statistics window for our region}
\label{StatisticsWindow}
\end{center}
\end{figure}

As long as the small region window is active, you can go to the top menu, go to the dropdown menu ``Analysis'', and choose ``Statistics''. Yet another window will pop up, this one with information about the pixels in the selected region, cf. Fig.~\ref{StatisticsWindow}.
This window provides us with several interesting bits of information. It shows us the center of the circle, and also the pixel scale, mapping image pixels to arc second intervals on the sky. A bit lower, we are provided with the sum of the pixel values within the region, in this case $C_1=15.34$, and its measurement error, the region's area in square arc seconds, namely $A_1=376.55$, as well as the resulting surface brightness.

The bottom line shows you additional statistical descriptors for the pixel values within the region, namely mean, median, minimum, maximum, variance, standard deviation and root-mean-square (rms).

\begin{figure}[htbp]
\begin{center}
\includegraphics[width=\linewidth]{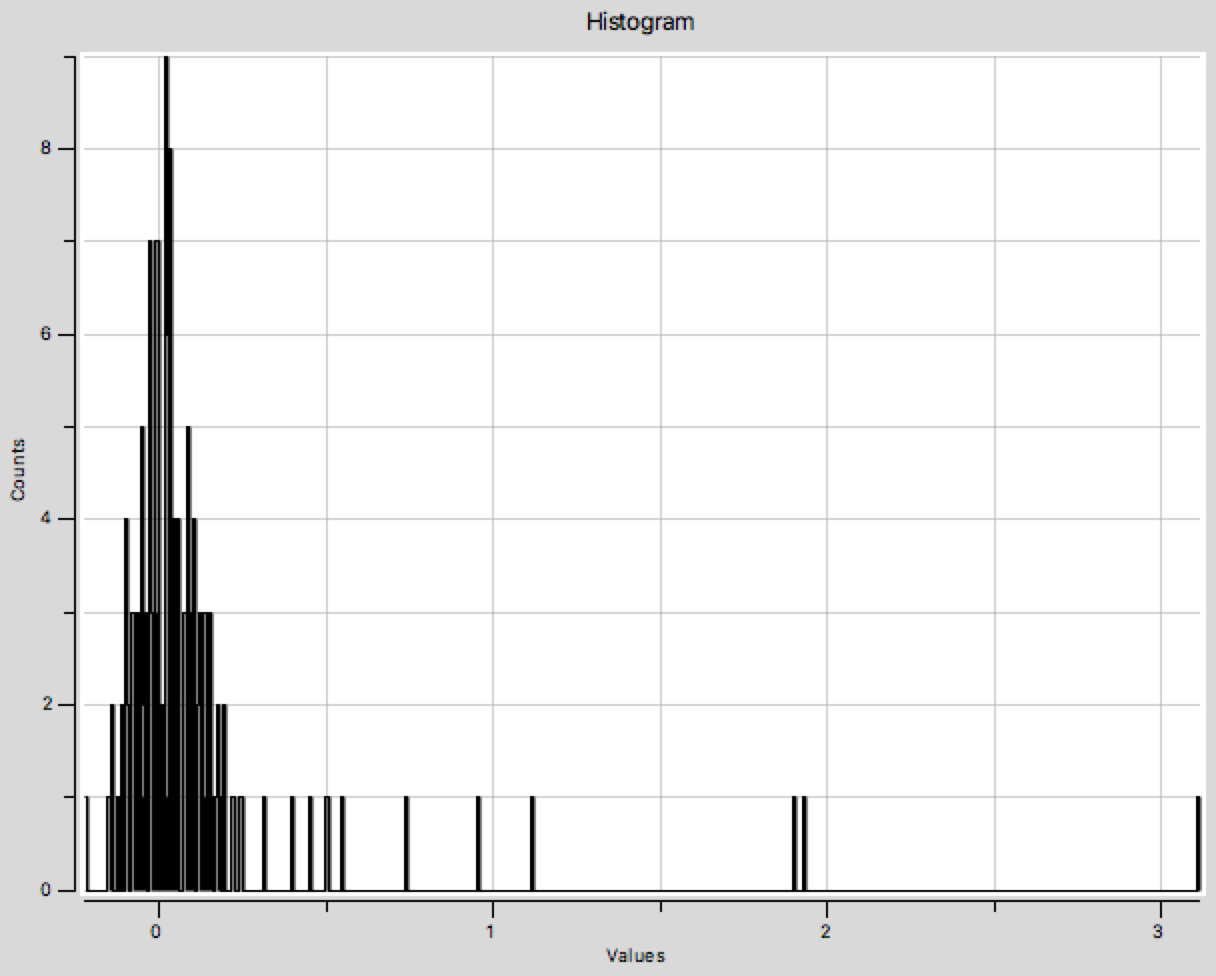}
\caption{Histogram for the pixel values in our selected region}
\label{DS9Histo}
\end{center}
\end{figure}

You can also get DS9 to display a histogram of pixel values. Activate the region window and, in the top menu, from the dropdown menu ``Analysis,'' choose ``histogram''. In our case, most of the values appear to be centered around zero, cf. Fig.~\ref{DS9Histo}. That appears to be the background plus noise.

Let us use the statistics function for regions to perform a simple aperture photometry measurement. In most FITS files of astronomical images, a pixel value is proportional to the light falling onto that particular portion of the detector (although possibly there may be an overall offset added in), that is, proportional to the flux density of the incoming light. (When comparing different images, you need to be a bit more circumspect --- for instance, if those images have different exposure times, and the pixel values correspond to the energy that was collected, you will need to rescale in order to make a valid comparison between an object in one and an object in the other image.) In the case of our SDSS, the BUNIT keyword tells us we are dealing with a property that is proportional to a value in Jansky, and thus a flux. 

Let us make a relative brightness measurement within the image, as follows. First, a second measurement on our first selected star, this time with a smaller circle that just about takes in the star and its immediate neighbourhood; I choose a radius of 8 arc seconds, and get a sum of pixel values $C_2=14.17$ and area $A_2=200.82$. The brightness density of the annulus bounded by the two circles, in pixel values per square arc second, is
\be
c_{bg} = \frac{C_1-C_2}{A_1-A_2}. 
\ee
After all, $C_1-C_2$ is the sum of the pixel values in the bigger circle minus the sum in the smaller circle; their difference is the sum of pixel values in the annulus. Dividing this by the area of the annulus $A_1-A_2$ in square arc seconds gives us the brightness density. In our case, we have $c_{bg}=0.006658$ per square second.

The amount of light reaching us from within the smaller, second circle is the light we receive from the star plus the light reaching us from the background sources within that circle. We assume that the background contribution per pixel from the little circle is about the same as for the surrounding annulus, which would add up to a total background contribution of  $c_{bg}\cdot A_2$. Subtracting this from the sum $C_2$, we get an estimate for the amount of light $l$ received from the star, namely
\be
l = C_2 - \frac{C_1-C_2}{A_1-A_2}\cdot A_2.
\label{BrightnessFormula}
\ee
In our particular case, $l_A=12.83$. Note that subtracting the background contribution also subtracts any overall offset values, so $l$ should indeed be proportional to the amount of light that was received from this star.

Let us repeat this measurement with another star. Going down the list, I select the one with g magnitude 17.529, near the bottom of the image frame, at $X=1123,\;Y=34$. Repeating the same measurements with a larger circle 13 arc seconds in radius and a smaller circle with a radius of 8 arc seconds, I obtain $C_1=108.36$, $A_1=534.70$ square arc seconds, $C_2=107.57$, $A_2=200.82$, and applying formula (\ref{BrightnessFormula}), I obtain $l_B=107.09$.

The flux ratio between the two is
\be
\frac{l_B}{l_A} = 8.35
\ee
so in this sense, and in this filter band, the second star is a bit over 8 times brighter than the first.

We can cross-check that against the published g magnitudes. After all, given that the collecting area and the exposure time are the same in both cases, the ratio of our values $l_B$ and $l_A$ should be the same as the ratio of the intensities of the two stars. Inserting this into equation (\ref{AstroMagnitudeFormula}) for the astronomical magnitudes, we find that we should have
\be
m_A-m_B = -2.5\cdot\log\left(\frac{l_A}{l_B}\right) = 2.30.
\ee
The difference in the stars' catalog g magnitudes (gmag) is 2.13. Even with our simple measurements, we have reproduced the relative brightness of those stars with a deviation of 0.17 mag, corresponding to an error in their relative flux of 17\%. Acceptable for a relatively crude, off-the-cuff measurement.

Given suitable data, we could measure changes in brightness in the same way: Compare a non-variable star (or an ensemble of several such stars) with a star whose brightness varies over time, and document the varying star's light curve. Given suitably precise measurements, you can even detect a transiting exoplanet in this way, although this will typically require considerable accuracy --- 1\% in some cases, much higher accuracy in others.

\subsection{Profiles}

DS9 also gives you the tools to create brightness profiles from astronomical images. We will take an example image from the THINGS survey by Fabian Walter et al.,\footnote{A description of the THINGS survey can be found in Walter et al. 2008, \href{https://arxiv.org/abs/0810.2125}{https://arxiv.org/abs/0810.2125}} the data of which is available for download from the THINGS project website at [\href{http://www.mpia.de/THINGS/Data.html}{http://www.mpia.de/THINGS/Data.html}]. THINGS stands for ``The HI Nearby Galaxy Survey'' and provides observations of nearby galaxies in and around the HI 21 cm line, undertaken with the Very Large Array (VLA) in Socorro, New Mexico.

For the following, please download the ``ro'' version of the ``moment 1'' file for the galaxy NGC 3198, which is named ``NGC\_3198\_RO\_MOM1\_THINGS.FITS''. This is a radial velocity map, with each pixel showing the average speed at which atomic hydrogen gas in that region of the galaxy moves away from us or towards us. Open the file with DS9, and from the FITS header, you will see that the units for each pixel are ``METR/SEC,'' meter per second.

Let us go into region mode again,  going to the top menu, there to the dropdown menu ``Edit'', and in that menu, on ``Region.'' We choose a particular type of region: in the top menu, open the dropdown menu ``Region'' and in the sub-menu ``Shape,'' choose ``Projection''.

\begin{figure}[htbp]
\begin{center}
\includegraphics[width=\linewidth]{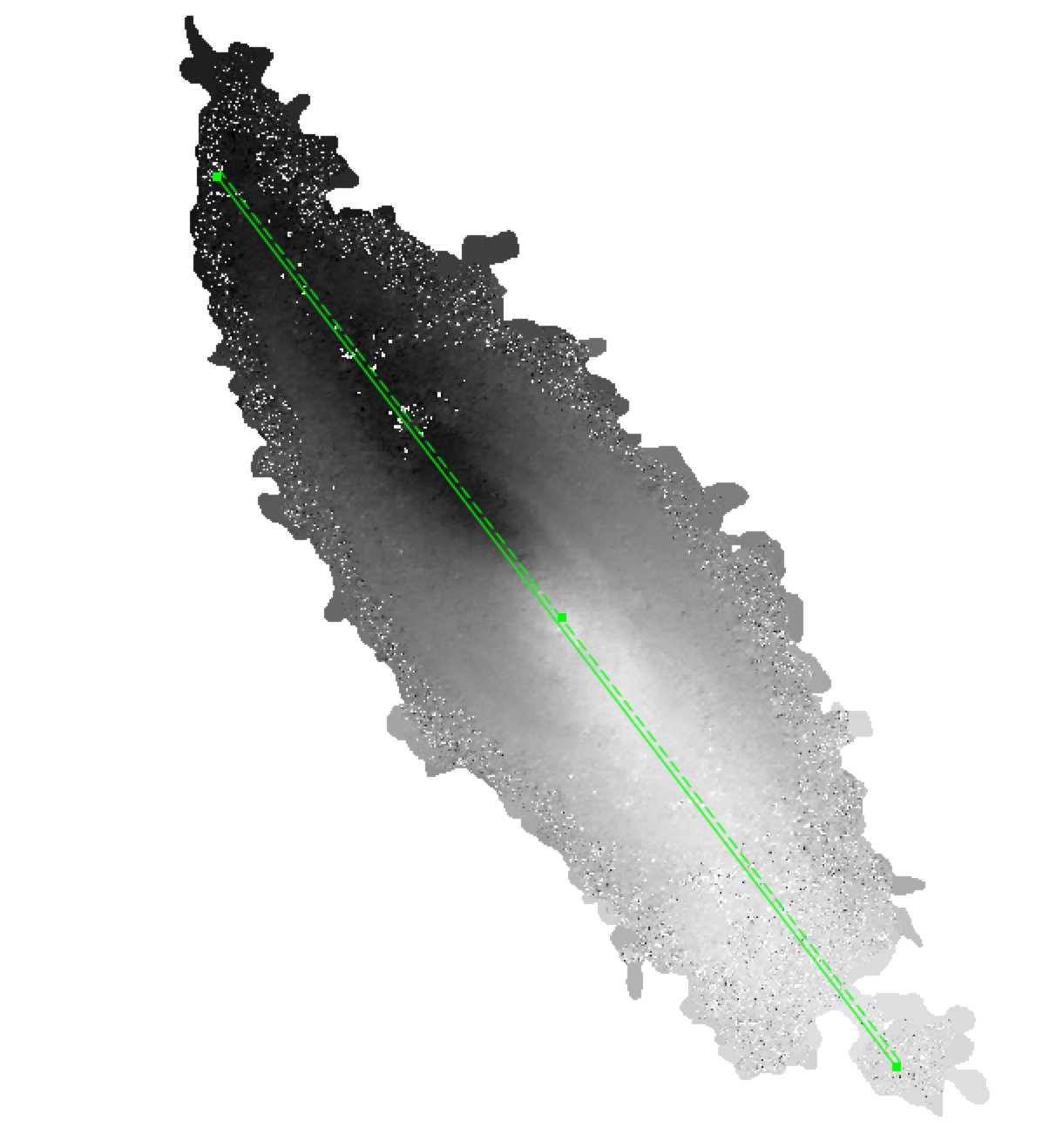}
\caption{A profile shape laid along the major axis of the galaxy NGC 3198's THINGS first moment map.}
\label{ProfileNGC}
\end{center}
\end{figure}
Now, click on one point of the image, hold your mouse button and drag to another point. DS9 will join the two points with a slim rectangle and, at the same time, open a new window that shows the profile of pixel values along that slim rectangle. Pixel values are averaged over the direction perpendicular to the profile; in that way, you can iron out some of the local variations (including noise) and get an accurate representation of trends on larger scales.\footnote{If you want the sum of the pixel values and not the average, you can change that in the top menu associated with the region window --- under ``Analysis'', you find a ``Method'' sub-menus where you can set a check mark at ``Average'' (default) or ``Sum''.} If the profile window has gotten lost, select the region and double-click the region window; in the top menu, under ``Analysis'', uncheck and re-check the ``Plot 2D'', and the profile window should re-appear.

\begin{figure}[htbp]
\begin{center}
\includegraphics[width=\linewidth]{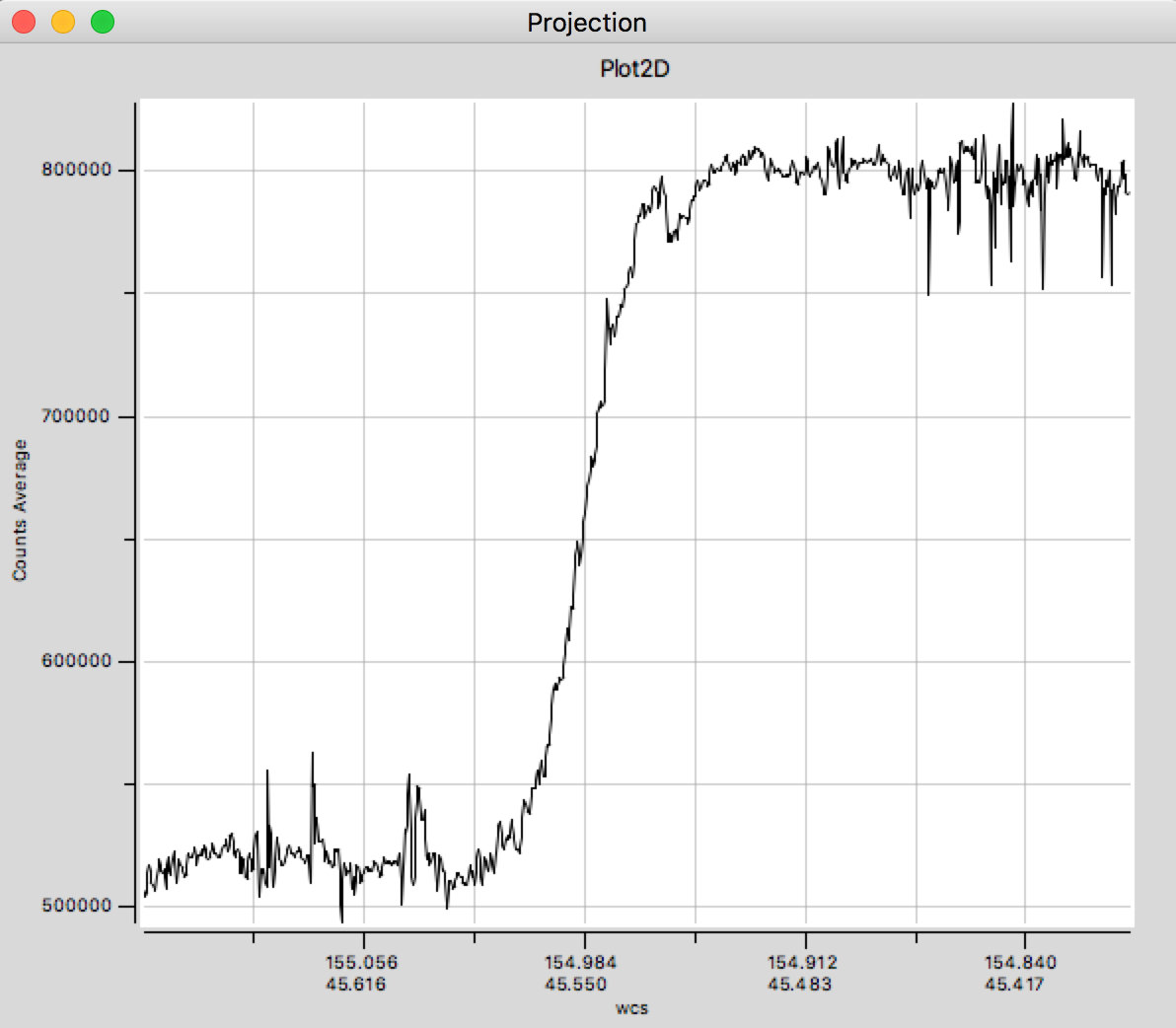}
\caption{Radial velocity profile of NGC 3198 based on the first moment map.}
\label{ProfileNGCCurve}
\end{center}
\end{figure}

Click on that rectangle while still in region mode, and little dots appear --- you can drag the outer dots around to change the position, orientation and length of the projection area, and the centre dot to change the width. In our example, I have laid the profile, 6 pixels wide along the major axis of the apparent ellipse formed by the galaxy NGC 3198 in the sky, see Fig.~\ref{ProfileNGC}. The corresponding profile curve can be seen in Fig.~\ref{ProfileNGCCurve}. 

When the profile curve window is active, you can go to the item ``File'' in the top menu and, from the dropdown menu, choose ``Statistic''. That will give you minimum, maximum, mean and a few other statistical descriptors about the data. 

In our case, we can see that the radial velocity grows anti-symmetrically with distance from the galaxy's center. The mean radial velocity is at 670 km/s, which is the velocity at which the galaxy as a whole recedes from us. The curve shows how different parts of the galaxy move towards us or away from us as the galaxy rotates. (In order to determine the rotation speed of the galaxy, we would need to de-project the radial velocities, though, as the disk of the galaxy is tilted relative to our line of sight.) 

The fact that the curve flattens out at larger radii is a piece of the puzzle that is the evidence for dark matter: Going by the galaxy's visible matter (in the shape of stars and gas), one would expect the rotation speed to decrease beyond a certain distance from the centre. The fact that it doesn't points to a surrounding halo of additional, non-luminous matter. Comparing the rotation curve measurement with the expected result based on the galaxy's brightness, it is possible to quantify how much dark matter is needed to keep the galaxy together. Rotation curves like this, such as the ones measured by Vera Rubin and Kent Ford in the 1970s, are an important piece of the puzzle that has led astronomers to postulate the existence of dark matter, which does not absorb or emit light or other forms of electromagnetic radiation.

\section{TOPCAT and table data}
\label{TOPCAT}

Next, let us look at application software for dealing with higher-level astronomical data, such as catalog data or, more generally, tables. The software is called TOPCAT, which stands for ``Tool for OPerations on Catalogues And Tables.'' It was developed by the astronomer Mark Taylor at the University of Bristol, and its tagline is ``does what you want with tables''. TOPCAT can be downloaded under this link [\href{http://www.star.bris.ac.uk/~mbt/topcat/}{http://www.star.bris.ac.uk/\~{}mbt/topcat/}]. It can do many, many things with tables, including creating diagrams, histograms, all-sky plots and the like. We will only scratch the surface of TOPCAT's functionality. You can find more comprehensive tutorials on the TOPCAT page, notably here: [\href{http://www.star.bristol.ac.uk/~mbt/topcat/#further}{http://www.star.bristol.ac.uk/\~{}mbt/topcat/\#further}].

TOPCAT is based on Java, so if you haven't got a suitably recent version of the Java Runtime Environment (JRE) on your computer, you will need to download it from the Oracle website at [\href{http://www.oracle.com/technetwork/java/index.html}{http://www.oracle.com/technetwork/java/index.html}]. (Fortunately, installing Java is comparatively pain-free!)

\subsection{Opening a table file}
\label{TOPCATOpenTable}

\begin{figure}[htbp]
\begin{center}
\includegraphics[width=\linewidth]{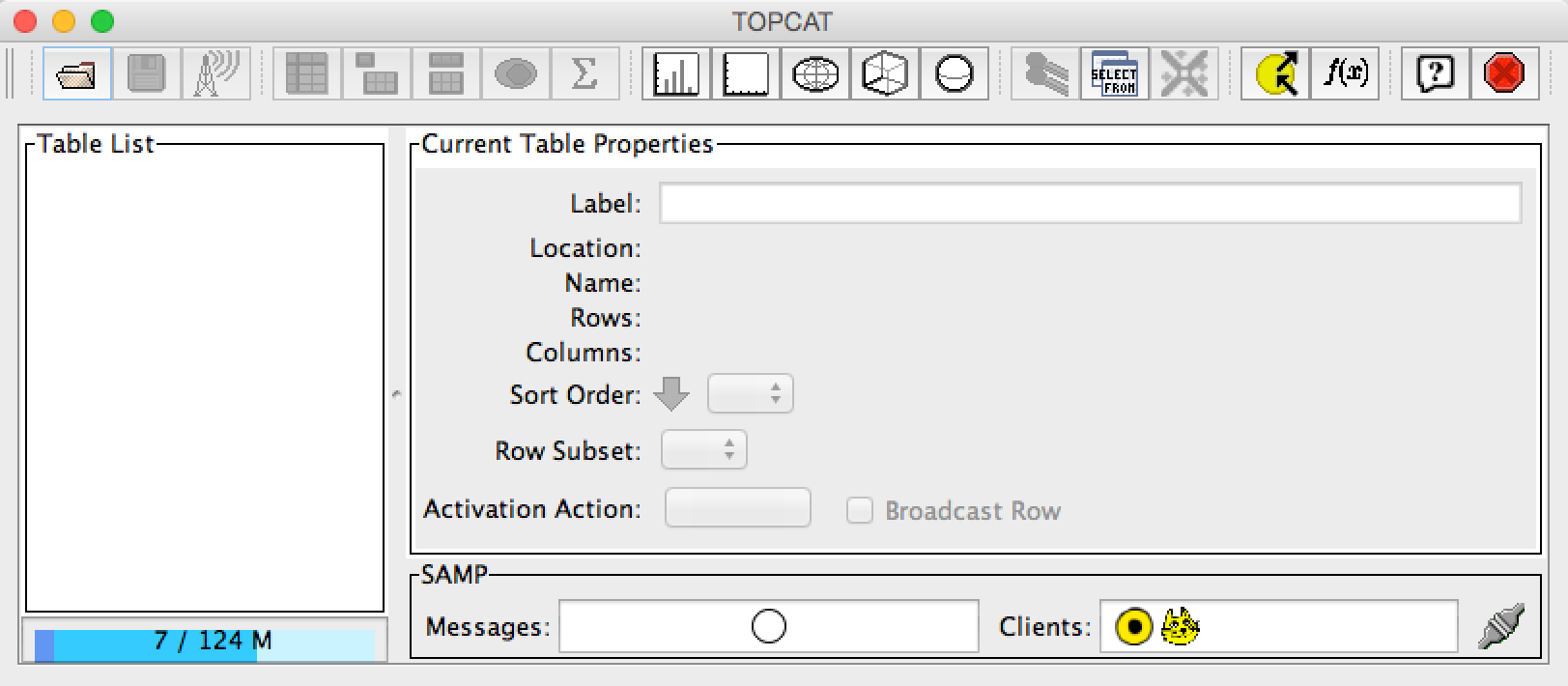}
\caption{The TOPCAT main window}
\label{TOPCATMain}
\end{center}
\end{figure}

When you open TOPCAT, it looks as shown in Fig.~\ref{TOPCATMain}. There will be an additional row of menu keywords --- on the Windows and Linux versions at the time of this writing (Windows), this row can be found across the top of the window, in the current Mac operating system (I'm on macOS 10.12.6) at the top of the screen. Again, as operating systems evolve over time, the position of this menu might change. I will call it the top menu, for short.

\begin{figure}[htbp]
\begin{center}
\begin{tabular}{p{2cm}p{2cm}p{2cm}}
\hspace*{0.75em}\includegraphics[width=0.05\textwidth]{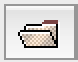}
& \includegraphics[width=0.05\textwidth]{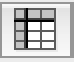}
& \includegraphics[width=0.05\textwidth]{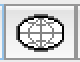}
\end{tabular}\\
\begin{tabular}{p{2cm}p{2cm}p{2cm}}
\scriptsize Load Table &\scriptsize  View Data
& \scriptsize  Sky Plot 
\end{tabular}\\[0.5em]
 \begin{tabular}{p{2cm}p{2cm}p{2cm}}
\hspace*{0.75em}\includegraphics[width=0.05\textwidth]{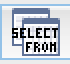}
& \includegraphics[width=0.05\textwidth]{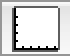}
& \includegraphics[width=0.05\textwidth]{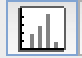}
\end{tabular}\\
\begin{tabular}{p{2cm}p{2cm}p{2cm}}
\hspace*{0.75em}\scriptsize \hspace*{0.5em}TAP
&\scriptsize Plane Plot
& \scriptsize Histogram
\end{tabular}
\caption{Some of the TOPCAT icons}
\label{TOPCATIcons}
\end{center}
\end{figure}

TOPCAT opens additional windows for different functionality, but what Fig.~\ref{TOPCATMain} shows is the TOPCAT base window. If you close that window, you will close TOPCAT.
Near the top, there is a row of icons; Fig.~\ref{TOPCATIcons} shows a few that we will need in the following. A mouse-over will also tell you what each icon does.
Let us load an example table from a file. Specifically, go to the data page of the Galaxy Zoo project, which can be found at [\href{https://data.galaxyzoo.org/}{https://data.galaxyzoo.org/}]. Specifically, we will look at the Galaxy zoo data release 2, table 5 ``Table 5 -- Main sample, spectroscopic redshifts'' near the middle of the page. You can see there are several possibilities for download: a csv file, a fits file and a VO table file. Please download the fits file.\footnote{Should the galaxy zoo pages have changed drastically since the time this text was written, you can find a local version of the same table at \href{http://www.haus-der-astronomie.de/working-with-astro-data}{http://www.haus-der-astronomie.de/working-with-astro-data} } The meaning of the columns can be found under this link: [\href{https://data.galaxyzoo.org/data/gz2/zoo2MainSpecz.txt}{https://data.galaxyzoo.org/data/gz2/zoo2MainSpecz.txt}]. The scientific paper describing the data release is \href{https://arxiv.org/abs/1308.3496v2}{Willett et al. 2013, [https://arxiv.org/abs/1308.3496v2]}.  The file itself is zipped (that is, compressed so as to make its file size smaller). TOPCAT should be smart enough to unzip it on its own.\footnote{If it doesn't, then, on Windows, you will need to use software such as 7--zip to unpack the file. On a Linux distribution, running the gunzip command in the command line should work (``gunzip zoo2MainSpecz.fits.gz''). On a Mac, double clicking on the file in the finder should do the trick.}

In order to open the file, click on the ``Open Table'' symbol
in the TOPCAT base window (cf. Fig.~\ref{TOPCATIcons}), or go to the ``File'' tab in the top menu and choose ``Open Table'' there. In most cases, we can leave the ``Format'' selection in the ``Open Table'' window on ``(auto)'' and TOPCAT will find out on its own the type of the table file, and open the file accordingly. If you should get an error when attempting to open the file, you can try to specify the table type explicitly using the ``Format'' dropdown menu. 

\begin{figure}[htbp]
\begin{center}
\includegraphics[width=\linewidth]{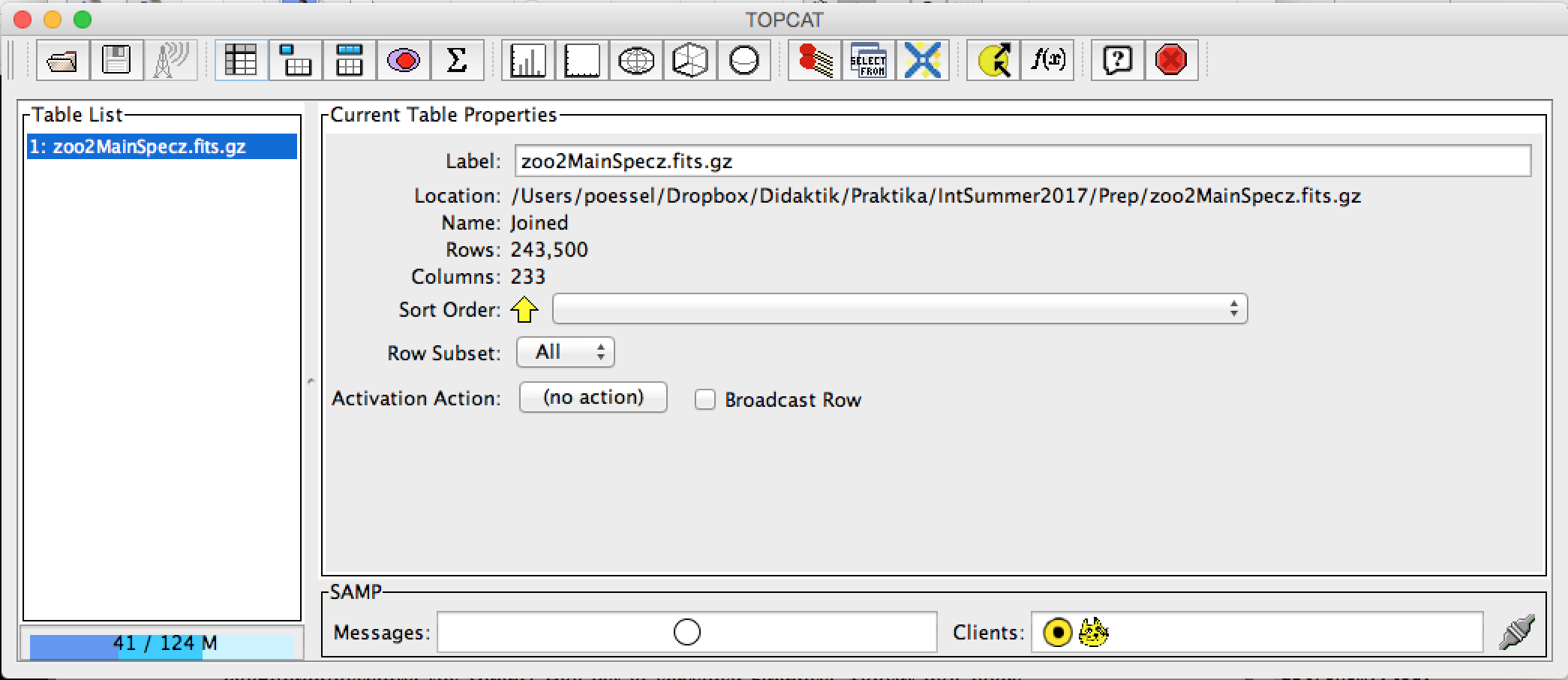}
\caption{TOPCAT after having loaded our file}
\label{TOPCATloaded}
\end{center}
\end{figure}

Open the ``Filestore Browser'', select the file choose zoo2MainSpecz.fits.gz, the Galaxy Zoo list of classified galaxies, and click on ``OK'' to load it. Topcat now looks as in Fig.~\ref{TOPCATloaded}.

\begin{figure}[htbp]
\begin{center}
\includegraphics[width=\linewidth]{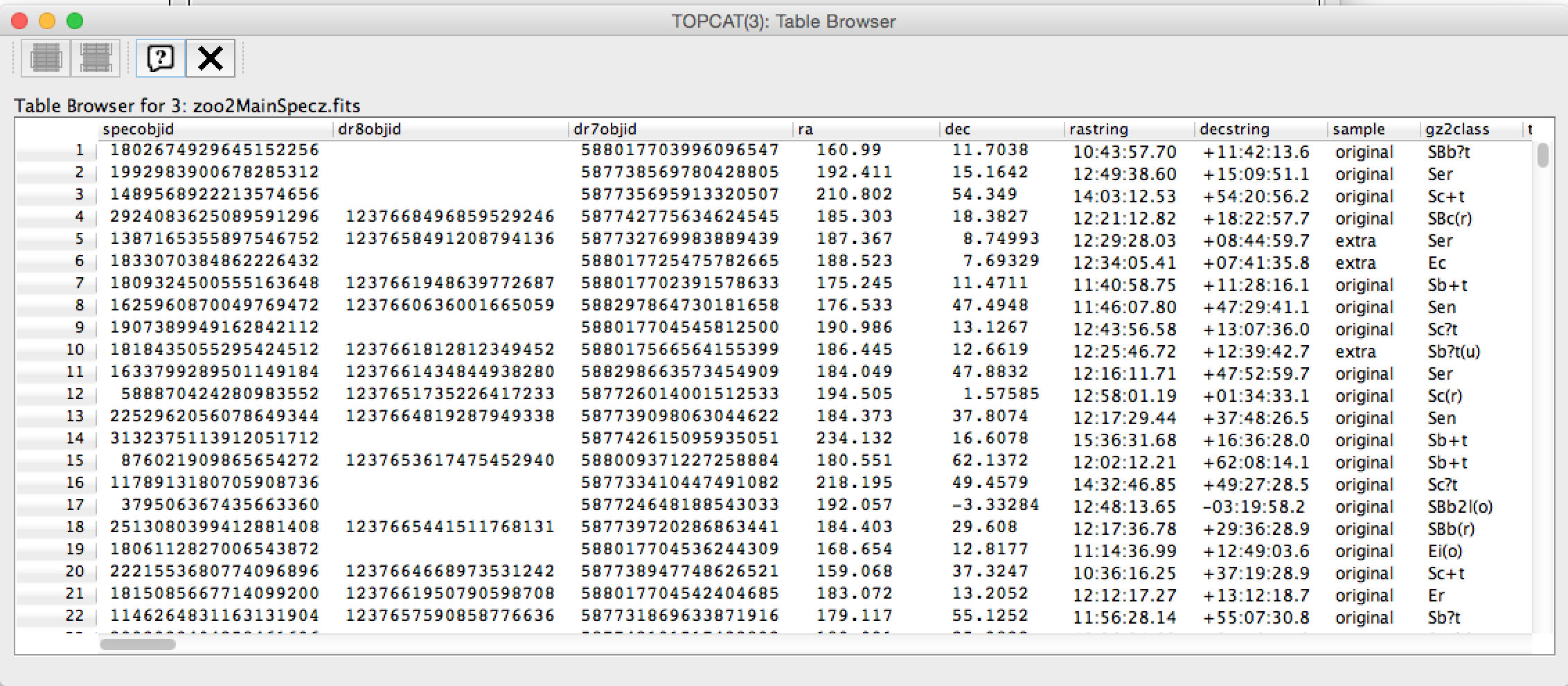}
\caption{The TOPCAT main window with table data loaded}
\label{TOPCATTableData}
\end{center}
\end{figure}

The name of the table file, the number of rows and the number of columns are displayed, and some of the icons that were initially greyed out (because they only work once a table is loaded) are now accessible.  By clicking on the ``View table data'' icon (cf. Fig.~\ref{TOPCATIcons}) or by going to the top menu, choosing the ``Views'' drop-down menu and clicking ``Table Data'', you can get an overview of your table data, column names included, in a new window, as in Fig.~\ref{TOPCATTableData}.

Click the icons next to ``View table data'' and you will be given either the general table data (such as the number of columns) or an overview of the columns and their properties, respectively. The column overview includes a description (where available), and the type of variable stored. 

\subsection{Making a sky plot}

TOPCAT provides several possibilities of viewing your data. Let's begin with functionality that is specific to astronomy. When you are given a catalog of astronomical objects, you might be interested in seeing where on the celestial sphere these objects are located. Are you looking at a full-sky survey, or a catalog that spans only one particular region? Let us ask this question for the Galaxy Zoo list of galaxies we have loaded into TOPCAT. Where on the celestial sphere are these galaxies located?

\begin{figure}[htbp]
\begin{center}
\includegraphics[width=\linewidth]{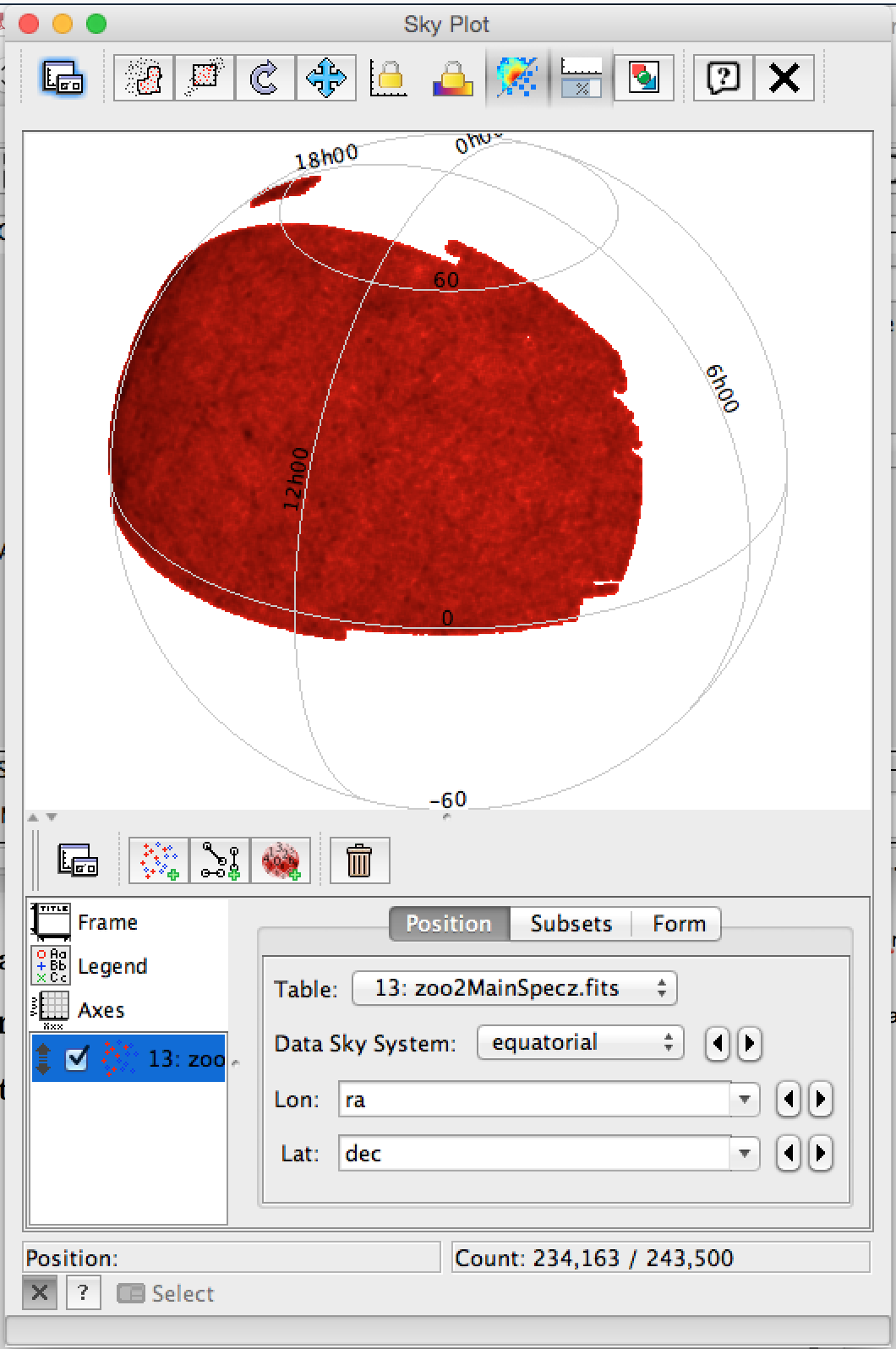}
\caption{The TOPCAT Skyplot window}
\label{TOPCATSkyplot}
\end{center}
\end{figure}

To this end, go to the main window (where we clicked the ``view data'' icon previously) and click the sky plot icon (cf. Fig.~\ref{TOPCATIcons}). Alternatively, in the top horizontal menu, choose Graphics $\to$ Sky Plot. Standard catalog data will include columns giving right ascension and declination, usually indicated by abbreviations like RA or Dec or similar, of the objects in question. TOPCAT is usually smart enough to find those columns, and plot your objects' positions in the sky. The result will be displayed in an extra window, as shown in Fig.~\ref{TOPCATSkyplot}.
You can click and hold with your left mouse button on the sky sphere, dragging it around. That way, you can see that the SDSS did a detailed study of a larger patch of the sky, but also studied selected strips of sky in nearly the opposite direction. The physical motivation for this: If you look only in one direction, you might miss that the universe has completely different properties in another direction. Checking up on at least some sample regions elsewhere in the sky is not foolproof, but can at least go some way towards showing you that the overall properties of the universe indeed do not depend on your direction of view (in other words, that the universe is isotropic).

The Sky Plot window comes with its own variety of options for customisation. For instance, by clicking on ``Axes'' in the bottom left field, you can easily vary at least some aspects of the visualisation. If you click on ``Axes'', the options field directly to the right will change. In the tab ``Projection'' you can change the projection from the default ``sin'' (which shows you a projection of the celestial sphere) to the world-map like ``aitoff'' or to the rectilinear coordinate plot ``car''. Under the tab ``grid'', you can shift the slider ``Grid Crowding'' to make the grid of coordinate lines denser or less dense. Have a look around, and try some of the options.

Before we explore TOPCAT a bit further, let us look at other ways of obtaining a table to work with: not as a file to be loaded, but directly from an astronomical data service: via the Virtual Observatory.

\subsection{Virtual Observatory (VO) services}
\label{TOPCATVO}

Astronomical data bases are getting larger and larger, and downloading all the data beforehand becomes more and more awkward as the file size increases. Fortunately, there are ways of searching online data bases for exactly the data you need for your project --- that way, you only handle the data you really need. For tables, the necessary search can be handled by the ''table access protocol'' TAP, using the astronomical data base query language ADQL (which is very similar to SQL). Is TOPCAT still open? If not, open it please.

\begin{figure}[htbp]
\begin{center}
\includegraphics[width=\linewidth]{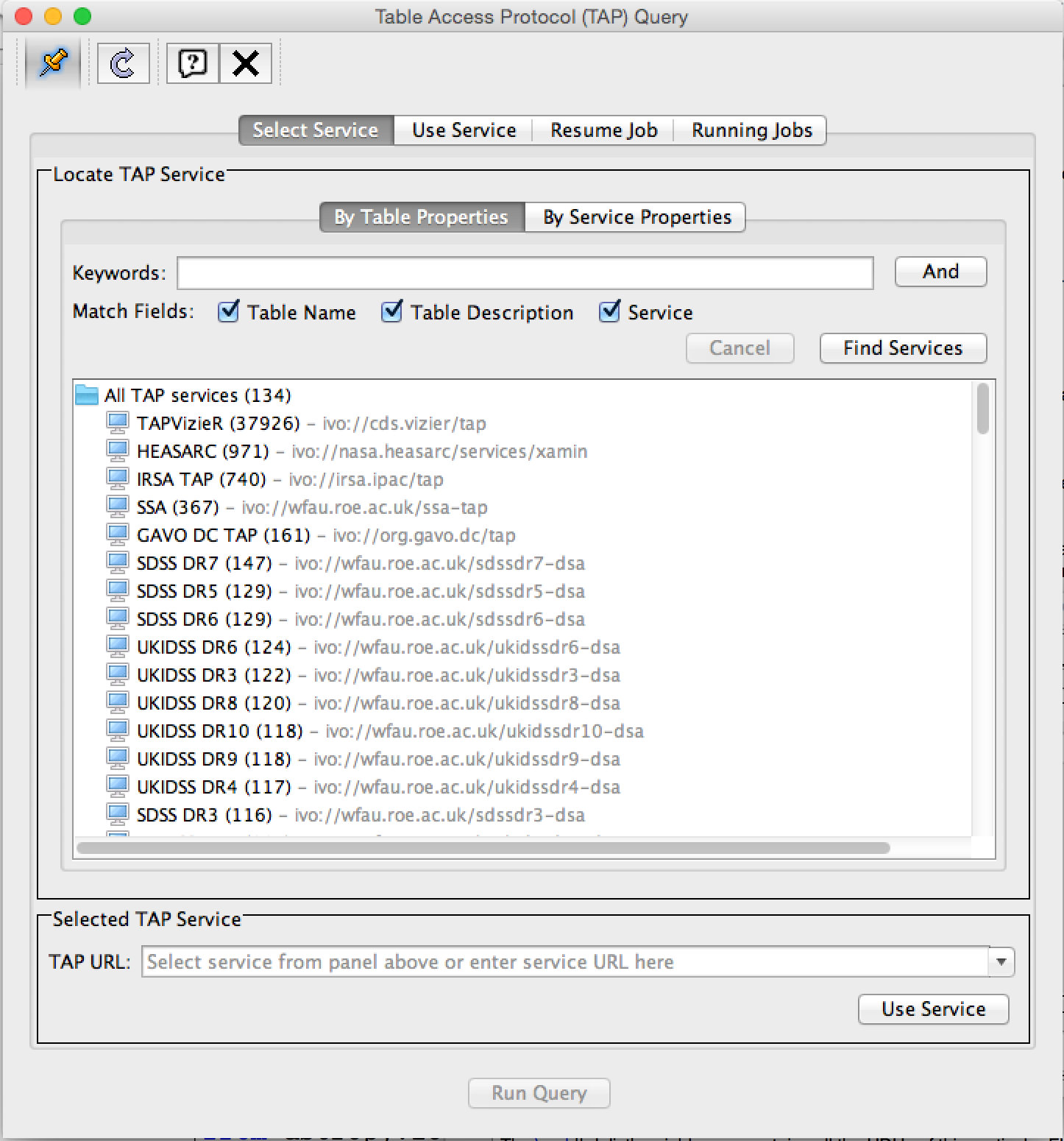}
\caption{The Table Access Protocol (TAP) window}
\label{TAPWindow}
\end{center}
\end{figure}

\begin{figure}[htbp]
\begin{center}
\includegraphics[width=\linewidth]{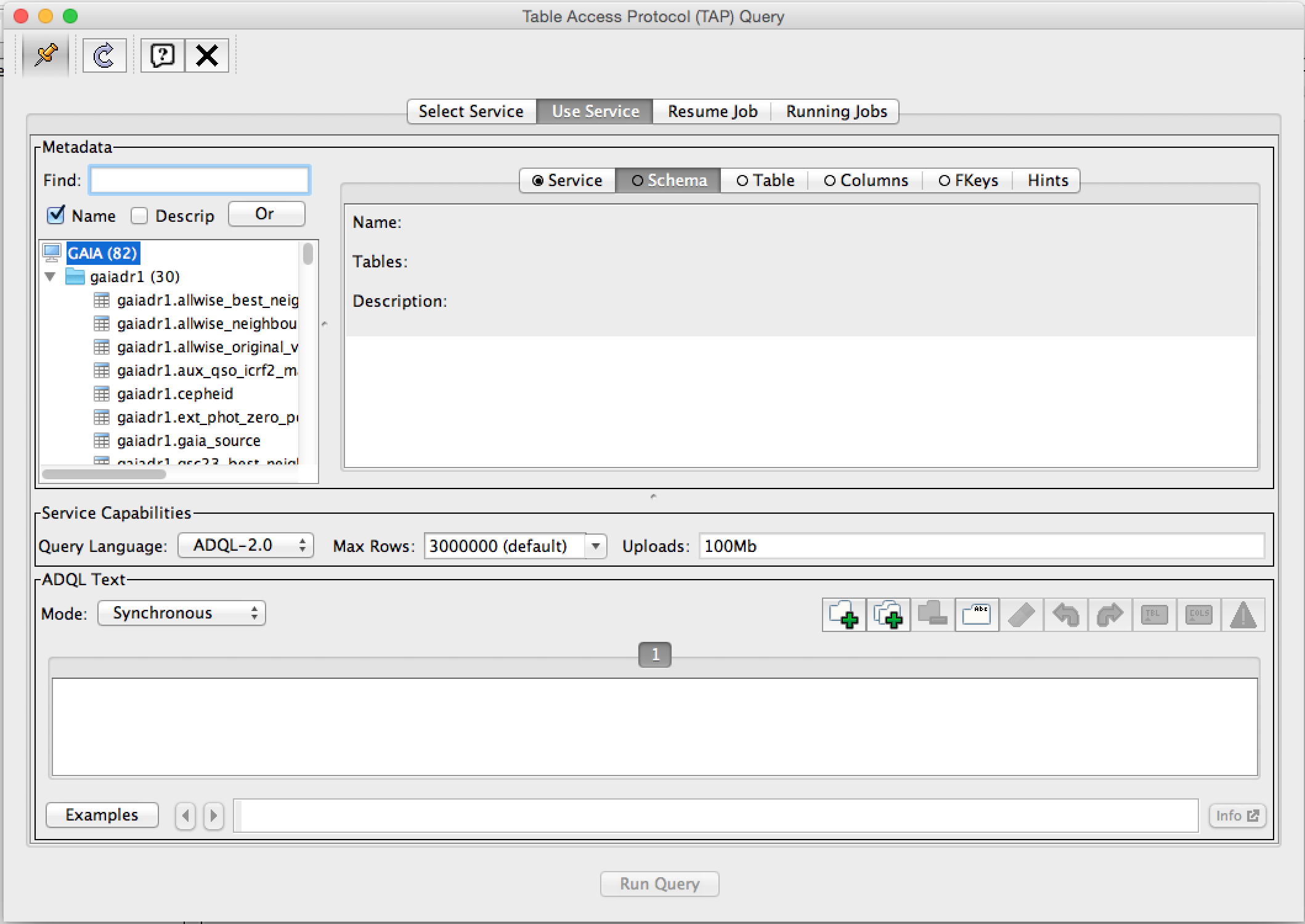}
\caption{The TAP window with the GAIA service selected}
\label{TAPWindowSelected}
\end{center}
\end{figure}

In the top menu, go to the tab ''VO'' (for ''Virtual Observatory''), and select the item ''Table Access Protocol (TAP) Query''. Alternatively, you can click on the TAP icon (cf. Fig.~\ref{TOPCATIcons}) in the main TOPCAT window. The window that should pop up can be seen in Fig.~\ref{TAPWindow}.
For data access, we will connect with a specific VO service --- that is, a data base that offers online access. We can do so either by choosing a service from the list in the window, or by using the field near the bottom to enter the URL of a specific astronomical data base service.  We will choose the ESA data base for the astrometry satellite Gaia, which is simply listed as GAIA in the TAP window. Double-click on GAIA and you will get additional information, as shown in Fig.~\ref{TAPWindowSelected}.
In the field on the left, you can see the specific tables you can access using the chosen service. In this example, in the folder gaiadr1 (that is, from the first data release) you have gaiadr1.allwise\_best\_neighbour, gaiadr1.allwise\_best\_neighbourhood, and so on (which names are only partly visible in the window as it is shown above -- expand the window by moving its boundaries if you need to!).

To learn more about any such table, specifically about the kind of information stored there, go to that table's entry in said window on the left, and double-click on it. For our example, we scroll down to gaiadr2 and double-click on gaiadr2.gaia\_source, which is the main table listing results from the second Gaia data release for the stars (or other point sources) examined, such as positions, parallaxes, proper motion and, for some of the stars, physical parameters like the effective temperature. This data release, published on 25 April 2018, revolutionized astronomy, and after one year had led to a remarkable 1200 papers based on the data --- 100 papers per month!

\begin{figure}[htbp]
\begin{center}
\includegraphics[width=\linewidth]{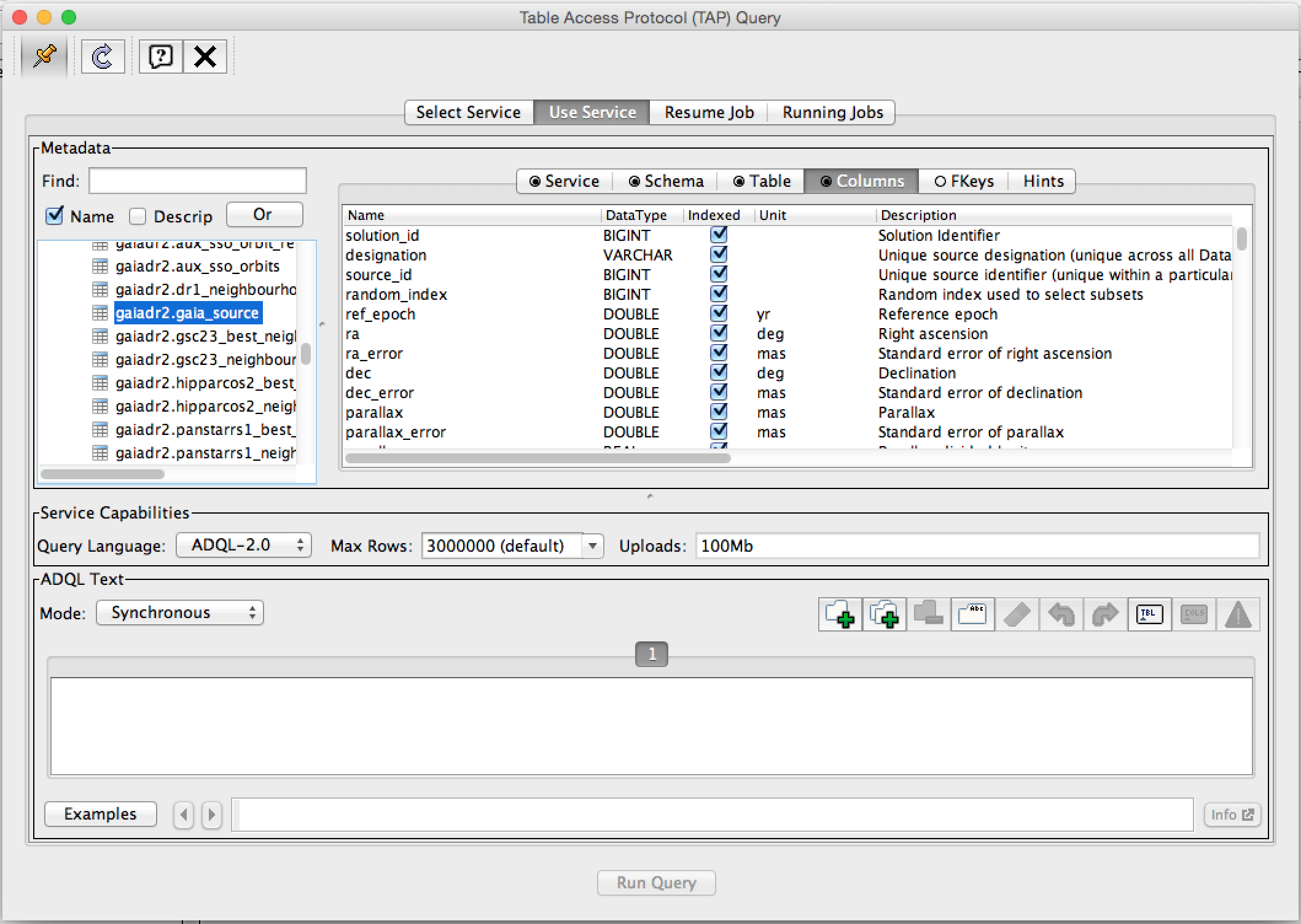}
\caption{Columns for the Gaia DR2 data set}
\label{GaiaSources}
\end{center}
\end{figure}

Let us find out the properties that the table is listing for each object. Once we have left-clicked on the specific table in the list on the left, we need only click on the  ``Columns'' tab in the sub-window on the right; TOPCAT will show us information about these properties. Picturing the table as consisting of various columns and a certain number of rows, each column corresponds to a specific property, while each row corresponds to an astronomical object included in the table. For the table gaiadr2.gaia\_source, the result looks as shown in Fig.~\ref{GaiaSources}.
For instance, the sixth line in the list on the right tells us that there is a property named ``ra'' in the table. It is stored as a DOUBLE, that is, as a particularly precise floating point number, and it has the unit ``deg'', degrees. That property (as per the description given there, as well as per universal astronomical convention) is the right ascension. So for each object in the table where that property is available, the table lists the right ascension, denoted here by 'ra'. Two below is the object's declination, in this case denoted by ``dec'', which is again a DOUBLE with units ``deg''.  

So far, none of that data is available to us to manipulate or to plot. In order to access data from that or any other VO table, we need to execute a query, asking the data base to return a specific set of data. Such queries are formulated in the Astronomical Data Query Language (ADQL), which is a close kin to the more general data base query language SQL (pronounced ``Ess-Que-Ell'' or ``Sequel,'' depending on whom you ask). Our queries go into the bottom entry field in the TAP window, which is headed ``ADQL text''. Let's talk about the basics of ADQL.

\subsection{Basic ADQL queries}
\label{BasicADQL}

Data base queries are not all that different from what you would request in natural language. ``Data base, please select for me the right ascension and declination for all objects that are listed in the gaiadr2.gaia\_source table, and return the result for me as a list'' is how we would formulate that request if we were talking to a human being. In ADQL, that same request would be
\begin{lstlisting}
SELECT ra, dec
FROM gaiadr2.gaia_source
\end{lstlisting}
where ``ra'' and ``dec'' are the table's names for these properties, standing for right ascension and declination, respectively; we looked those up in the ``columns'' list. SELECT and FROM are parts of the ADQL language. I have written them in all caps here; that is not a requirement, but makes the query more readable for us humans, as it clearly separates the language's commands from the names of properties and data bases.

If we were to run this query, it would likely take a long time; after all, Gaia DR2 provides those properties for more than 1.3 billion sources! Let us restrict our query somewhat by telling the data base that we want to look only at some specific area in the sky. Let us look at the Andromeda galaxy. In natural language, we would now ask ``Dear data base, could you please select for me right ascension and declination for those objects that are listed in the gaiadr2.gaia\_source table and that are close to the position of the Andromeda galaxy?''

What is close to the position of the Andromeda galaxy? Luckily, since that is a common astronomical question, ADQL has a function for that. Let's first see the query as a whole:
\begin{lstlisting}
SELECT ra, dec
FROM gaiadr2.gaia_source
WHERE 1=CONTAINS(POINT('ICRS',ra,dec),
                    CIRCLE('ICRS',10.684708,41.268750, 3.2))
\end{lstlisting}
The first two lines are the same as before: We select right ascension ra and declination dec from the Gaia DR2 source table. The rest of the query begins with ``WHERE'' encodes our condition. We do not want all the results for ra, dec, but only those results WHERE a specific condition is met. That condition is defined after the keyword WHERE. The CIRCLE('ICRS',10.684708,41.268750, 3.2) is a circular region in the sky --- its center point in the ICRS (``International Celestial Reference System'') coordinate system is at right ascension 10.684708 degrees and declination 41.268750 degrees, which is the center of the Andromeda galaxy's location in the sky. The circle's radius is 3.2 degrees. The POINT('ICRS', ra, dec) is a point in the sky with right ascension ra and declination dec, again referring to the ICRS coordinate system. The ``CONTAINS'' is a function that takes a point and a region (the region, in our case, is the aforementioned circle) and returns 1 if the point is contained in the region, and 0 otherwise. Taken everything together, our query should select ra and dec for all Gaia DR2 objects contained in a 3.2 degrees circle around the center point of Andromeda.

\begin{figure}[htbp]
\begin{center}
\includegraphics[width=\linewidth]{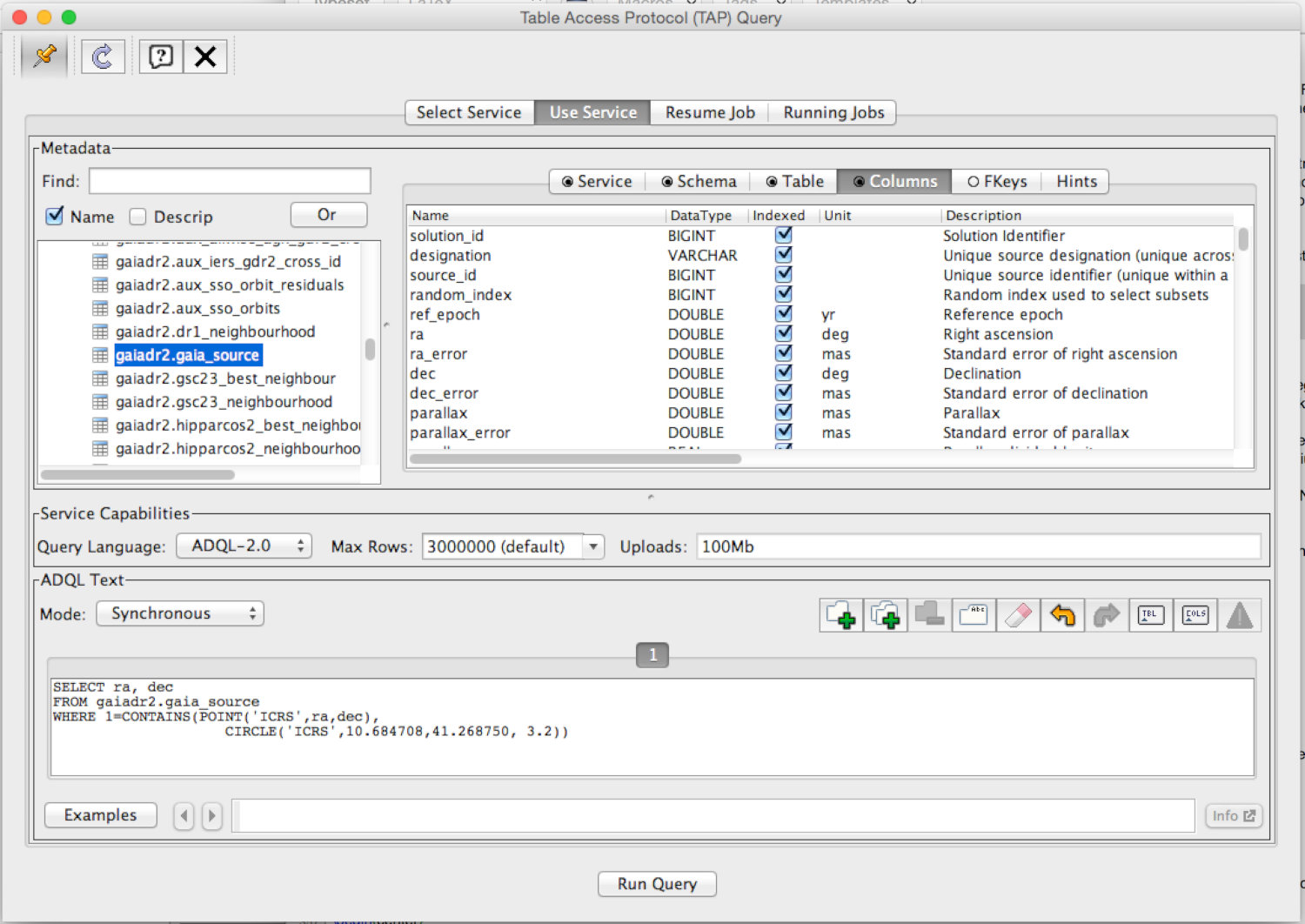}
\caption{An ADQL query entered in the TAP window}
\label{ADQLWindow}
\end{center}
\end{figure}

To execute the query, we need to write it (or paste it) in the ADQL text window, as shown in Fig.~\ref{ADQLWindow}.
If we now press the ``Run Query'' at the bottom, the query will be executed. (Or not, if we have made a syntax error --- if there is an error, we will get an error message.) During the download, there will be a temporary download window with an animated progress bar. Once the download is complete, the main TOPCAT window will push itself into the foreground, showing that the table with the results has been loaded and is ready for inspection. Our table has 419,949 rows, that is: we have retrieved ra and dec for 419,949 astronomical objects!

\begin{figure}[htbp]
\begin{center}
\includegraphics[width=\linewidth]{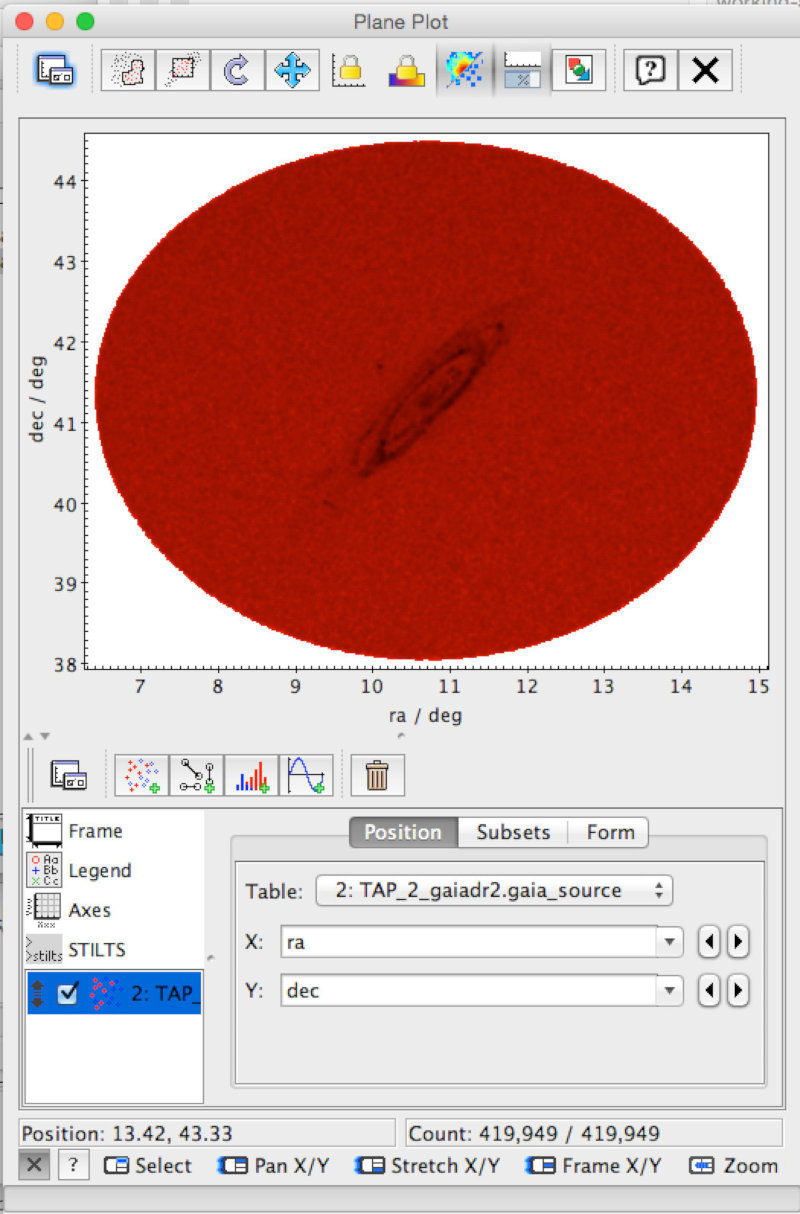}
\caption{A plane plot of the Andromeda galaxy, based on Gaia DR2}
\label{AndromedaPlot}
\end{center}
\end{figure}

Let's take a quick look at our data by clicking on the ``plane plot'' icon (cf. Fig.~\ref{TOPCATIcons})
to produce a simple plot. The result is shown in Fig.~\ref{AndromedaPlot}.
This is quite nice! Mind you, this is not an image of the Andromeda galaxy. This is a diagram showing point sources identified by Gaia DR2, in other words: this is a plot showing Andromeda as traced by separate stars identified within the galaxy! Note that TOPCAT has automatically switched to some type of density plot, where regions with more point objects in them are darker. If TOPCAT had simply plotted all the data points on top of each other in red, we would be looking at a solid red plane.

Imagine now that we want to look at the whole sky, not just at a small region, but that we still do not want to download all the Gaia DR2 data. One way out would be to just look at the brightest stars in the Gaia DR2 catalog, as the following query does:
\begin{lstlisting}
SELECT TOP 1000000 ra, dec, phot_g_mean_mag
FROM gaiadr2.gaia_source
ORDER BY phot_g_mean_mag ASC
\end{lstlisting}
Let's look at what is new with this query. First of all, we are selecting one additional property: phot\_g\_mean\_mag, which is the magnitude (that is, the brightness expressed in the usual logarithmic scale of astronomy) of the object in question in the G filter band. 

\begin{figure}[htbp]
\begin{center}
\includegraphics[width=\linewidth]{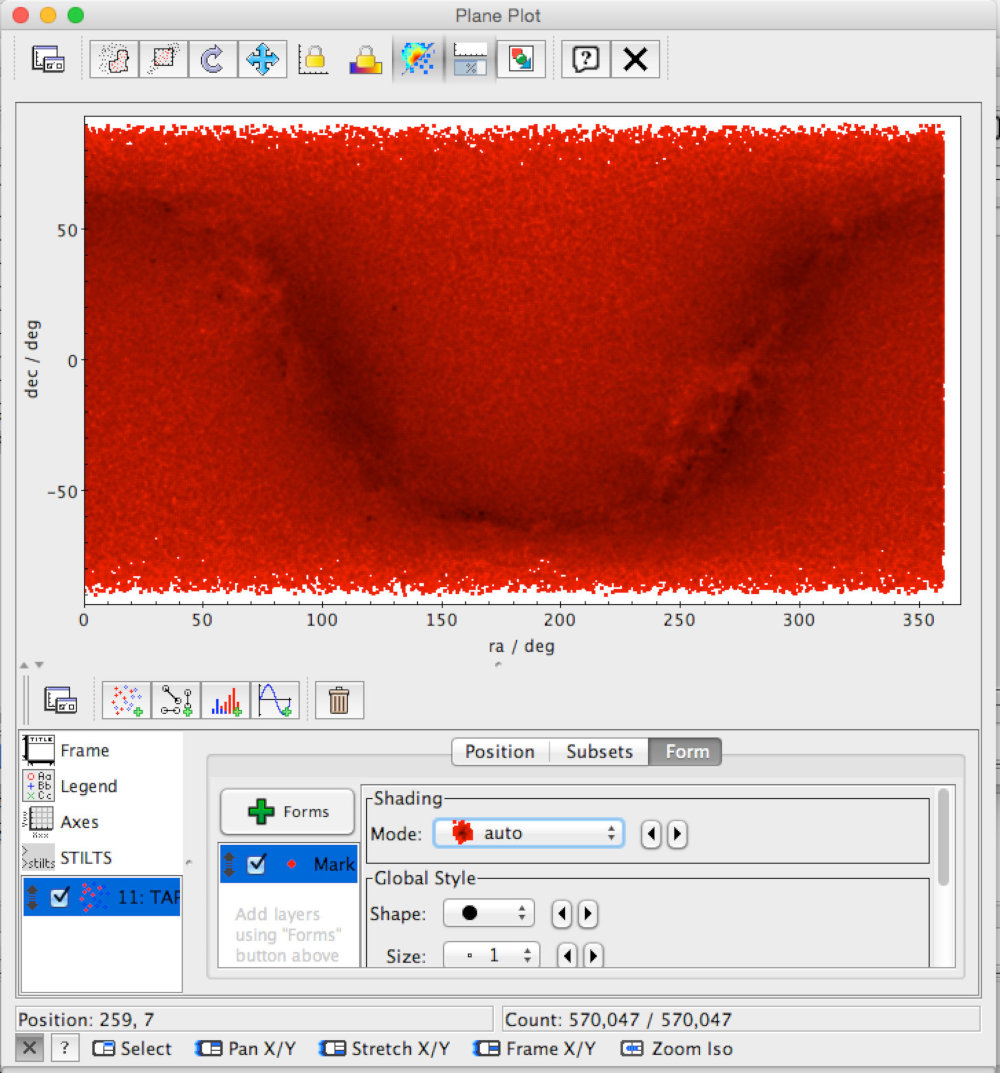}
\caption{The brightest stars from the Gaia DR2 catalog}
\label{MilkyWayPlot}
\end{center}
\end{figure}

But the SELECT has been amended as well: we are only selecting the TOP 1000000, the first million of rows of our result. Top with respect to what? That is specified in the last line, where we ask the data base to order the table so that all values of phot\_g\_mean\_mag are in ascending order (ASC; descending order would be DESC). So our TOP 1000000 selects the 1000000 objects with the smallest magnitude value, that is: the 1000000 brightest objects in the resulting table.  When we make a plane plot of those, we see that there is a curved band where there are particularly many stars, Fig.~\ref{MilkyWayPlot}.
That band is the Milky Way (in a plane ra-dec plot). Note that the brightness is reversed: Where there are many stars, the plot is dark. Those regions where parts of the Milky Way are hidden behind dark clouds, and where in consequence we see fewer stars, are brighter in this plot. If you want to see how the Milky Way is stretched as a ring across the sky, make a Sky Plot of this data.

In this case, we looked at the brightest stars. What if we do not want to single out bright stars, but instead select a representative sample of {\em all} stars? Many data bases provide auxiliary information which helps us to do just that. Consider the following query:
\begin{lstlisting}
SELECT TOP 1000000 ra, dec
FROM gaiadr2.gaia_source
ORDER BY random_index
\end{lstlisting}
The basic structure is as before: We select only the top $N$ rows, and we have ordered the table from which those top rows are selected. But this time, we order by the entry for random\_index, which is defined as follows: For a data set with $N$ entries, the column random\_index contains a random permutation of the integers from 1 to $N$. If we select the top $M$ rows from the table, ordered by the random\_index column, we should get a random sub-sample from the table. 

Note that this is not a generic ADQL feature, but instead relies on the Gaia team having supplied an extra column random\_index for the purpose. But since the need for easy extraction of a random sub-sample arises quite often, numerous catalogues provide the same service. The result of our sample extraction, shown again as a plane plot, is shown in Fig.~{MilkyWayPlot2}.
\begin{figure}[htbp]
\begin{center}
\includegraphics[width=\linewidth]{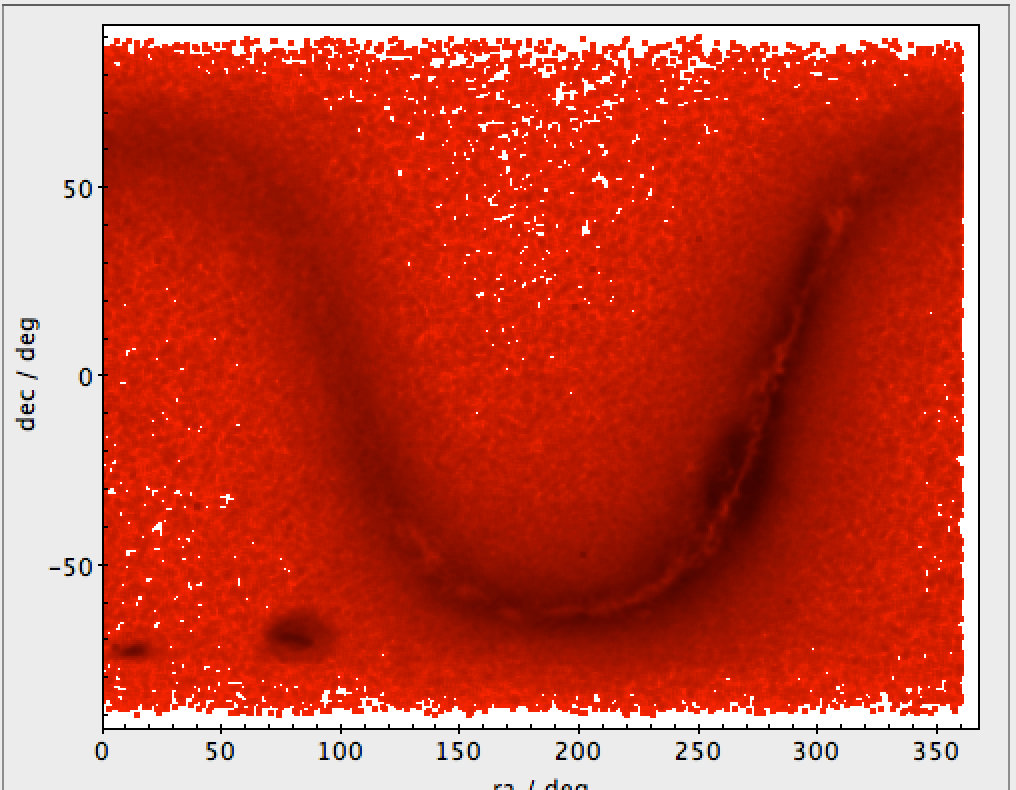}
\caption{Plot of a random sub-sample of Gaia DR2 stars}
\label{MilkyWayPlot2}
\end{center}
\end{figure}
By not focussing on the brightest stars, but on all kinds of stars, we can now make out some extra structure, namely the Large Magellanic Cloud and the Small Magellanic cloud, our nearest neighbouring galaxies, in the bottom left corner!

At other times, we need to combine data from more than one table. For instance, the GAIA service offers a table gaiadr2.vari\_cepheid containing all the stars in DR2 that the system has (so far) identified as Cepheids. The table does provide the variable star data, such as the fundamental pulsation period of the Cepheid in question. What the table does not contain is elementary astrometric information such as ra and dec.

So what do we do if we want a table listing ra, dec and pf for the DR2 Cepheids? We must somehow identify which object in the one table corresponds to which object in the other table. In ADQL, this is done via the JOIN operation. The complete query in our case is
\begin{lstlisting}
SELECT s.ra, s.dec, c.pf
FROM gaiadr2.gaia_source AS s
JOIN gaiadr2.vari_cepheid AS c
USING (source_id)
\end{lstlisting}
Begin by looking at the second row. Again, we are selecting FROM a table, namely from gaiadr2.gaia\_source, but this is followed by ``AS s''. The ``s'' is simply a name we are giving to that table. If you look in the first row, that name is added to the properties we are selecting from that particular table: ra and dec are now written as s.ra and s.dec to make clear where they come from.

The third line is ``JOIN gaiadr2.vari\_cepheid AS c'', which is the command to join that table to the first one. The second table, too, gets a short name, namely ``c''. In the first row, the property we select from the second table has a c. in front, namely c.pf for the fundamental period of the Cepheid.

How are the tables to be joined? How do we know that a row in the first table and a row in the second table refer to one and the same object? Fortunately, both tables include a unique identifier for the astronomical objects listed, namely the source\_id. The last line in the query, ``USING (source\_id)'', tells the data base that when selecting object data for one object from the first and from the second data, we should go by the source\_id. Our result includes only cases where such a match has been successful -- so in this case, our result is a table listing ra, dec and pf for all 9575 objects that are contained both in the Cepheid table and in the source table.

\subsection{Selections and subsets}
\label{TOPCATSubsets}

TOPCAT has helpful functionality that allows you to select subsets directly from a 2D plot. It happens fairly regularly that we are interested in comparing a subset of data with the rest. We encountered an example in section \ref{HighLevel}, when we used the Hertzsprung-Russell diagram to separate our sample into main sequence stars on the one hand and red giants on the other. 

For exploring the properties of those classes, we needed to look at the distribution of objects in different contexts. Our first look, in the Hertzsprung-Russell diagram, showed us there were distinct classes. But for a physical interpretation, we needed additional representations, each providing us with additional information: the combined radius histogram for both types of star showed us that when it came to the red giants, we were indeed dealing with larger-than-usual stars, and the radius-density diagram identified their unusually low density. 

Linking those different representations, and seeing the same distinction between classes in all of them, which allowed us to compare properties and to characterise the two classes of stars, was a crucial tool in that process. Our data sets for these stars are multi-dimensional; as human beings, with two-dimensional vision, we cannot explore such a multi-dimensional space at a glance without looking at different kinds of plots.

TOPCAT allows users to perform just kind of exploration by taking a diagram, defining a subset ``by eye'' directly in that diagram, and then look at that subset in other diagrams, as well.\footnote{Later on, in section \ref{Sec:Glue}, we will briefly encounter Glue, as an alternative tool for that kind of exploration, [\href{https://glueviz.org}{https://glueviz.org}].} 

As an example, let us look at open star clusters with Gaia. In order to find members of an open star cluster, it is not enough to look at the region of the sky where the star cluster is located --- our view of that region is likely to include both stars that are much closer and stars that are much further away than the cluster we are interested in, but which just happen to be visible in the same direction as our cluster. 

In the TAP window, execute the following query:
\begin{lstlisting}
SELECT ra, dec, pmra, pmdec, bp_g, phot_g_mean_mag
FROM gaiadr2.gaia_source
WHERE 1=CONTAINS(POINT('ICRS',ra,dec), CIRCLE('ICRS',12.1083,85.2550, 0.4))
\end{lstlisting}
We are getting ra and dec, but also the proper motion (that is, the motion of the stars on the celestial sphere) in the ra and dec directions, pmra and pmdec, respectively. bp\_g is the blue minus the green brightness of the object, which serves as a measure of colour, and phot\_g\_mean\_mag is the magnitude (brightness) in Gaia's broad G band, which can stand in for the object's overall brightness. 
\begin{figure}[htbp]
\begin{center}
\includegraphics[width=\linewidth]{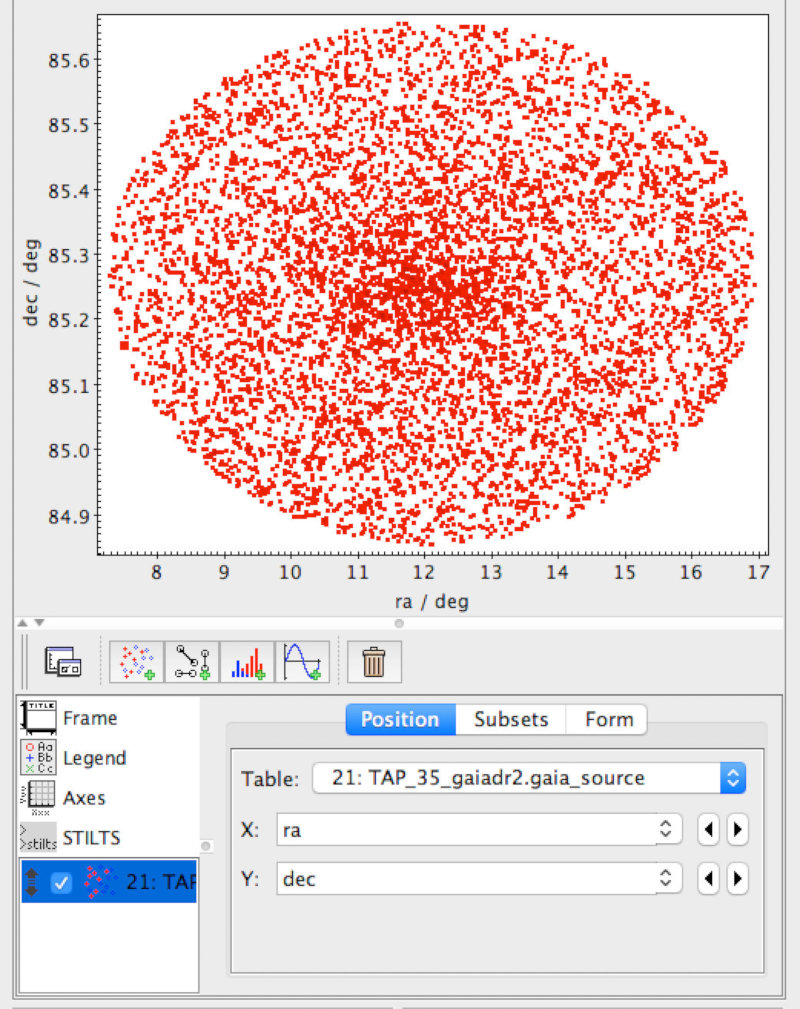}
\caption{RA-Dec plot of the region around NGC 188}
\label{NGC188Plot}
\end{center}
\end{figure}
Once the table has loaded, click the ``Plane Plot'' icon (cf. Fig.~\ref{TOPCATIcons}). The 2D plot window will open; part of it is shown in Fig.~\ref{NGC188Plot}, which is a RA-Dec plot of the stars we have retrieved. There is no straightforward way of telling which of the stars belong to our cluster, and which don't. In the bottom-right part of Fig.~\ref{NGC188Plot}, there is a line ``X: ra'' and directly below ``Y: dec.'' To the right of each line are two small arrow symbols. 
\begin{figure}[htbp]
\begin{center}
\includegraphics[width=\linewidth]{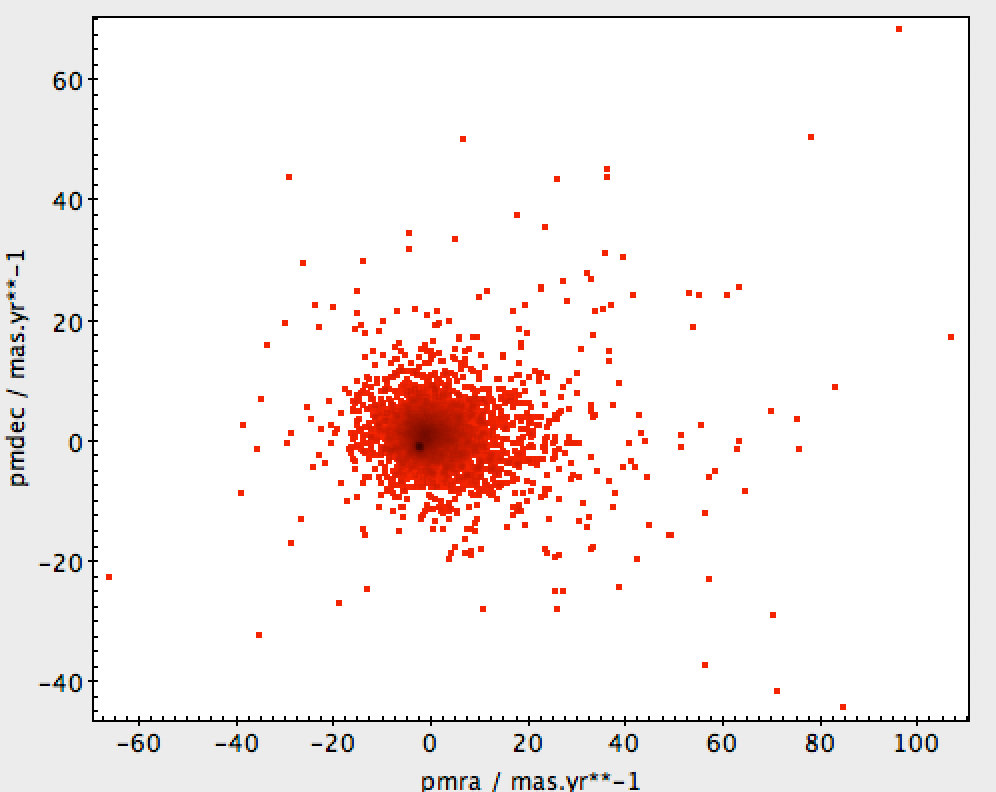}
\caption{Proper motion plot of the region around NGC 188}
\label{NGC188PM}
\end{center}
\end{figure}
Click those to choose a table column other than ra for the X coordinate, and other than dec for the Y coordinate. In particular, choose X as ``pmra'' and Y as ``pmdec''. Now you are plotting the proper motions of the stars in our table; the plot itself should now look as in Fig.~\ref{NGC188PM}.

Most of the stars are distributed symmetrically around the origin (pmra=0, pmdec=0). But there is a marked concentration, seen as a dark grey dot, that is offset to the bottom left against the overall distribution. In order to see it more clearly, let us zoom in. There are several ways of doing that. On my touchpad, swiping upwards with two fingers do the trick. (Clicking and moving my finger moves the chosen region around in the window.) A mouse wheel would do the same.
\begin{figure}[htbp]
\begin{center}
\includegraphics[width=\linewidth]{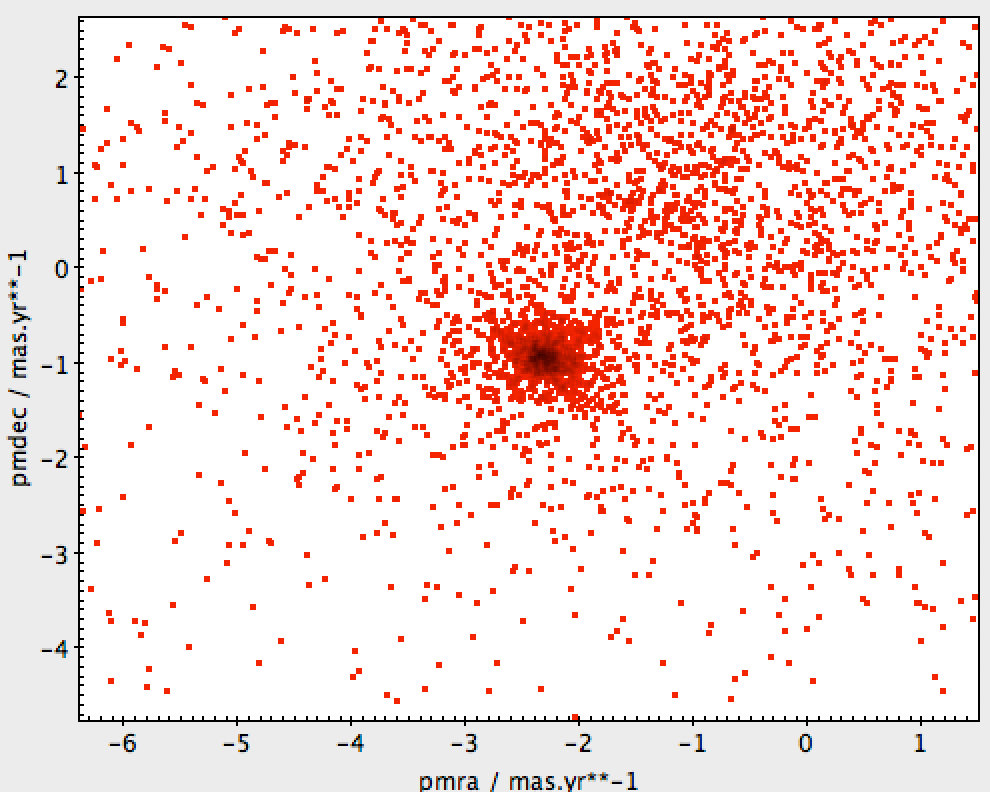}
\caption{Zoomed-in proper motion plot of the region around NGC 188}
\label{NGC188PMZoom}
\end{center}
\end{figure}
If that doesn't work, you can always click on the ``Axes'' symbol in the lower left part of the window, choose the ``Range'' tab, and adjust the range either by entering minimum and maximum values directly or by changing the sliders. Do so to get a closer look on the offset concentration, as in Fig.~\ref{NGC188PMZoom}. That concentration is our star cluster. Such star clusters form from one and the same molecular cloud, and inherit the clouds overall velocity --- with small variations, as the cluster stars themselves attract each other. That is why selecting a subset in a velocity diagram, in this case the two velocity components orthogonal to the line of sight, is a suitable way of selecting all the cluster stars, distinguishing them from other stars that might lie in the same direction, but are unlikely to move at the same velocity.

\begin{figure}[htbp]
\begin{center}
\includegraphics[width=\linewidth]{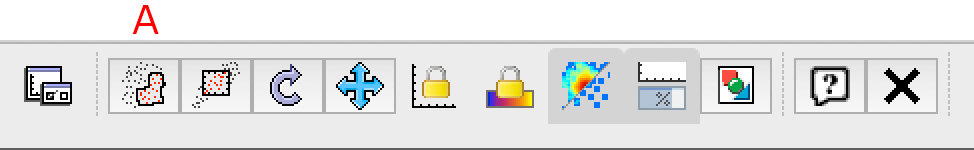}
\caption{The top row of the ``Plane Plot'' window}
\label{PlanePlotTopRow}
\end{center}
\end{figure}

\begin{figure}[htbp]
\begin{center}
\includegraphics[width=\linewidth]{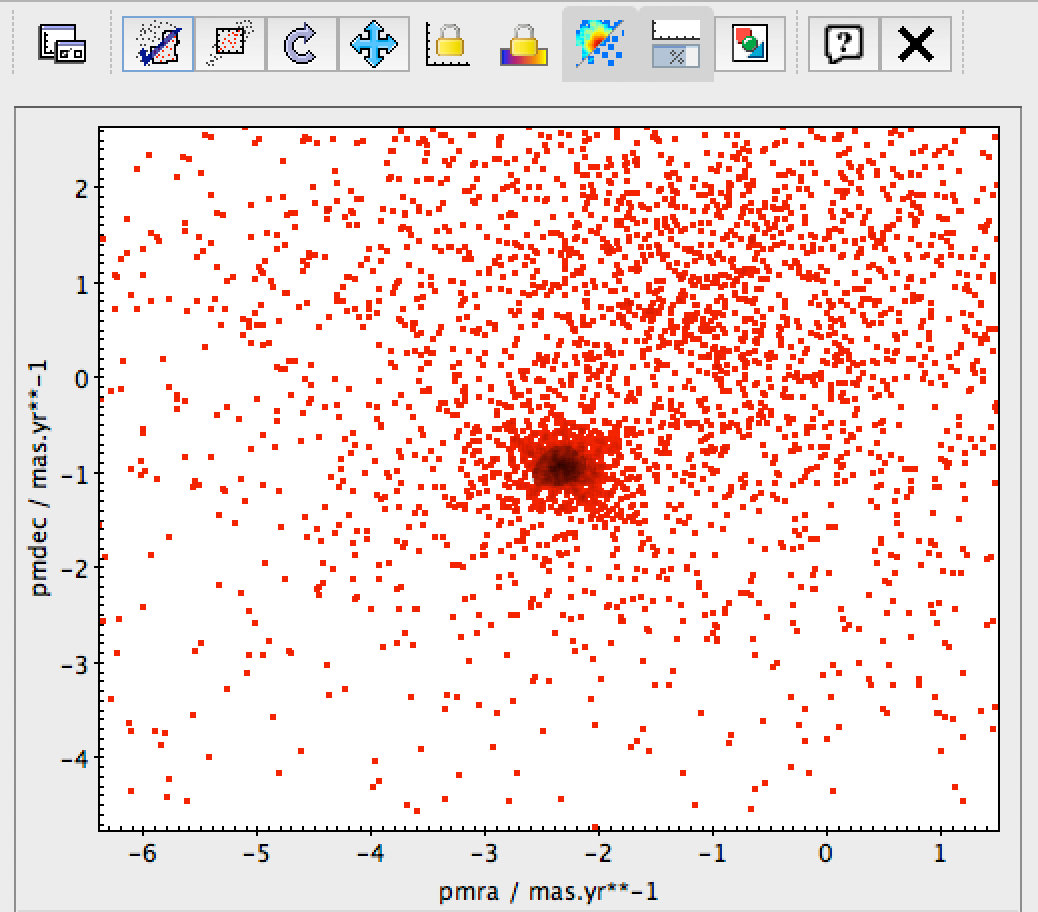}
\caption{Marked subset of the proper motion plot of NGC 188}
\label{NGC188Drawn}
\end{center}
\end{figure}

In the zoomed-in image, we can now choose a suitable subset. Look at the top row of icons in the Plane Plot window, as shown in Fig.~\ref{PlanePlotTopRow}.
The icon I have marked with the red A is the ``Draw Region'' icon. Click it, and you can draw (by holding your left mouse button pressed and dragging) a region onto the plot, delineating an area that will be filled gray as you draw. In Fig.~\ref{NGC188Drawn}, you can see the small gray region that I have drawn, right where the off-center concentration of stars is. 
You can also see that the ``Draw Region'' icon now features a check mark. Clicking on that icon and its check mark signals to TOPCAT that you have drawn your desired region, and that TOPCAT should now create a subset based on that selection.
\begin{figure}[htbp]
\begin{center}
\includegraphics[width=\linewidth]{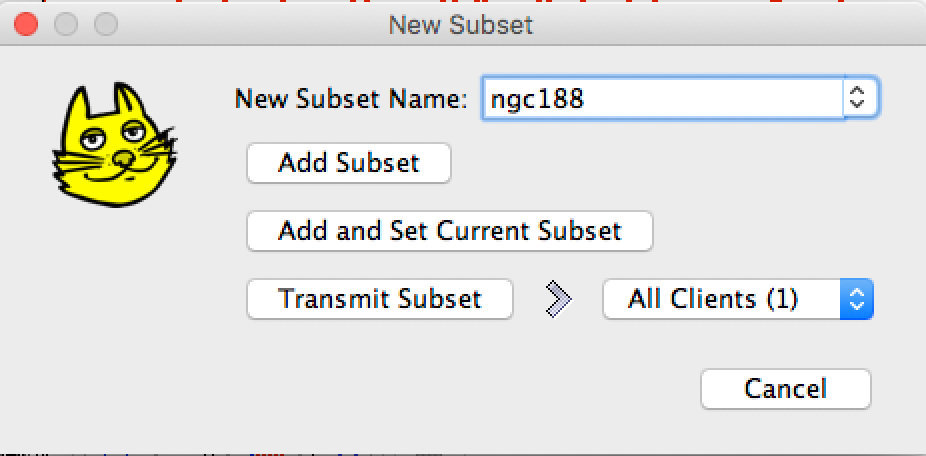}
\caption{The ``New Subset'' window}
\label{SubsetWindow}
\end{center}
\end{figure}
In response, TOPCAT will open up the ``New Subset'' window shown in Fig.~\ref{SubsetWindow}. Enter a name (as I have, in this case: ``ngc188'') and click on ``Add Subset'' to create the new subset. (If you are dissatisfied with your choice, click on ``Cancel'' and you can start over again with selecting your subset.)

\begin{figure}[htbp]
\begin{center}
\includegraphics[width=\linewidth]{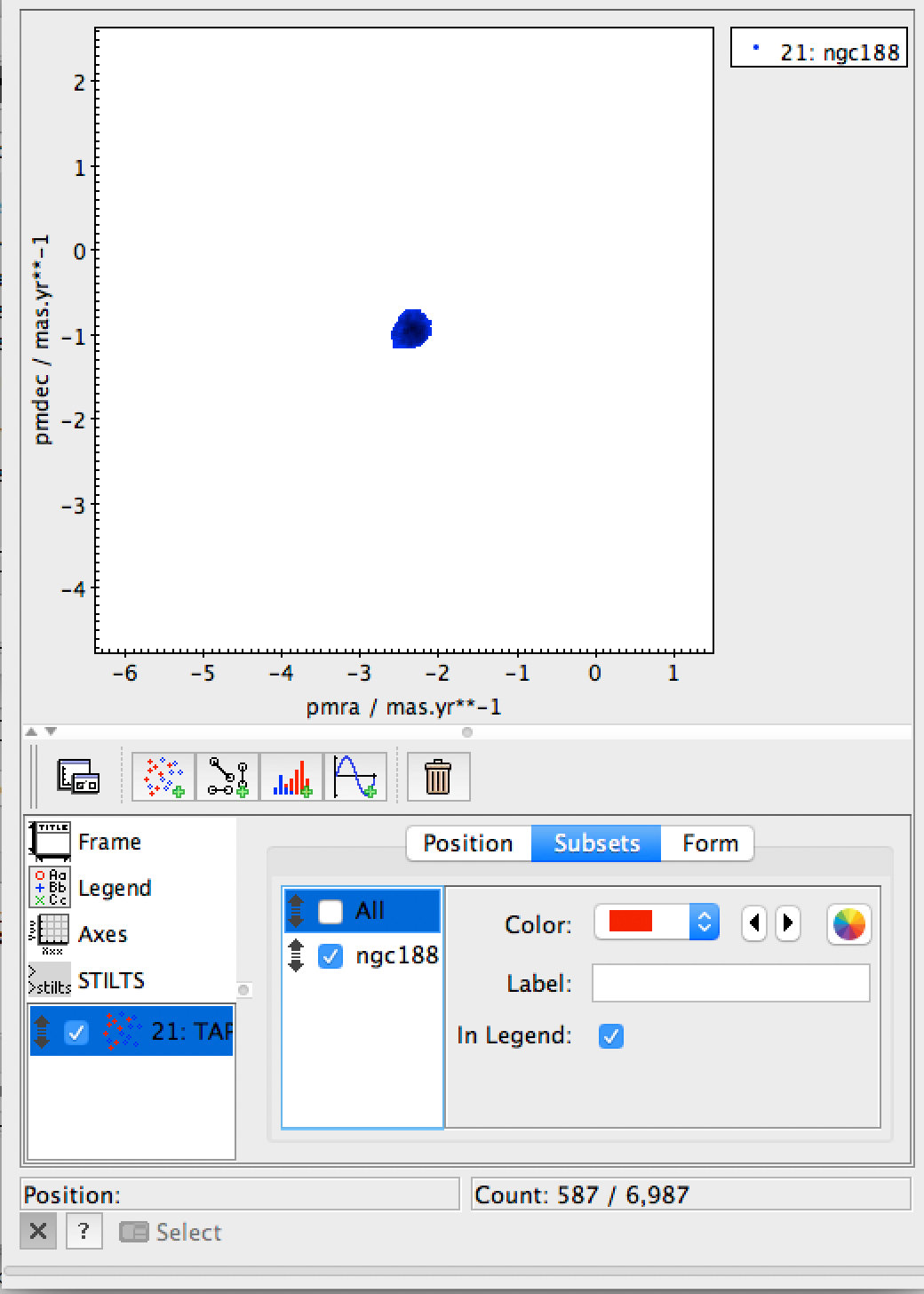}
\caption{Plane Plot with marked subset}
\label{SubsetPlot}
\end{center}
\end{figure}

Now, when you go back to the plotting window, you will see the subset marked in a different colour; in Fig.~\ref{SubsetPlot}, in blue. 
In the lower-left subwindow of Plane Plot, go to the line representing your data (in our case, the lowest, of which you can just about make out 21: TAP, marked in blue as selected in Fig.~\ref{SubsetPlot}). In the tab ``Subset'' (already selected in Fig.~\ref{SubsetPlot}), I have removed the checkmark from ``All'', so now only the subset ngc188 is visible. In that way, we can make plots or histograms for the subset only.  

\subsection{More on plotting}

\begin{figure}[htbp]
\begin{center}
\includegraphics[width=\linewidth]{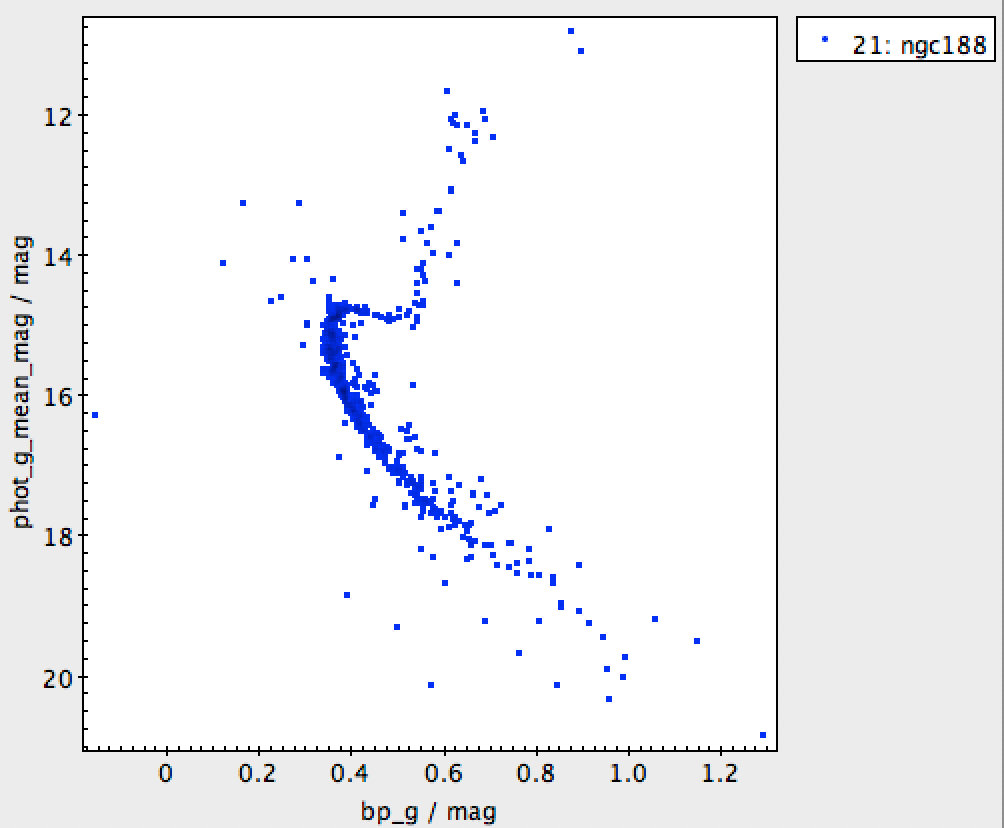}
\caption{Color-magnitude diagram of NGC 188}
\label{NGC188ColorMag}
\end{center}
\end{figure}

In the preceding section, we have already looked at basic plots, and learned how to choose custom X and Y axis values from the columns of our data set. Let us do some more plotting, and create histograms, for the subset (stars within the open cluster NGC 188) we have created in that section. To that end, let us again select the data points object in the bottom lower sub-window of Plane Plot (as in Fig.~\ref{SubsetPlot}), click on the ``Position'' tab, and select bp\_g as the quantity to be plotted along the X axis, and phot\_g\_mean\_mag for the Y axis. Next, in the bottom left sub-window of  Plane Plot, click on ``Axes'' and choose the Coords tab. Larger astronomical magnitude values correspond to lower apparent brightness, and it is customary to invert a magnitude axis, so brighter objects will be further up. In the Coords tab, we can achieve this by clicking ``Y Flip''. (In passing, note that we could flip the X axis as well, and that other options include giving the X and/or the Y axis logarithmic scaling. Also, with ``Aspect lock'' we can force an equal aspect ratio, that is, make sure that both X and Y axis are plotted to the same scale.) If you had changed the plot scale before, as I did in the previous section, you will also want to go to the ``Range'' tab and revoke any range restrictions you might have made, pulling the X and Y range sliders back to their respective boundaries. The resulting plot is shown in Fig.~\ref{NGC188ColorMag}.
The diagonal bottom-right to upper-left structure is the main sequence. Near the top, there is a turnoff point, with an S-shaped swerve to the right. There is interesting physics behind this: In our main sequence, the most luminous stars are near the upper left (blue and bright). But those stars are also the most massive, and the shortest-lived --- they spend the shortest time on the main sequence! Presumably, all the stars in our open cluster have formed at around the same time, since that is how open clusters come into existence: as the result of group star formation from one and the same giant molecular cloud; after some time, the cluster disperses. Thus what we see near the top of the main sequence of NGC 188 amounts to the shortest-lived stars having left, or being in the process of leaving, the main sequence to become red giants, moving upwards (giants are brighter!) and to the right (red giants are reddish!) as they do so. Compare those results with models of stars, which provide you with information about the lifetimes of stars as a function of their mass, and you can estimate the age of the whole star cluster from the location of that turnoff point.

\begin{figure}[htbp]
\begin{center}
\includegraphics[width=0.45\linewidth]{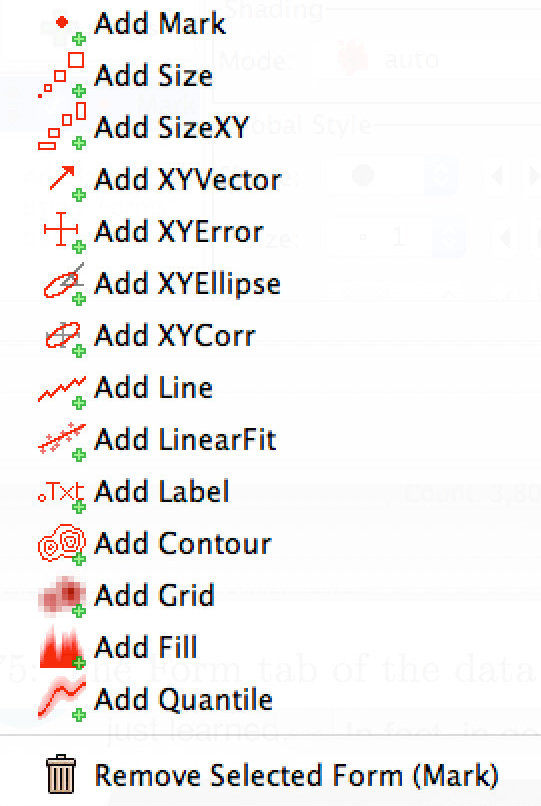}
\caption{The Form dropdown menu}
\label{FormDropdown}
\end{center}
\end{figure}
\begin{figure}[htbp]
\begin{center}
\includegraphics[width=\linewidth]{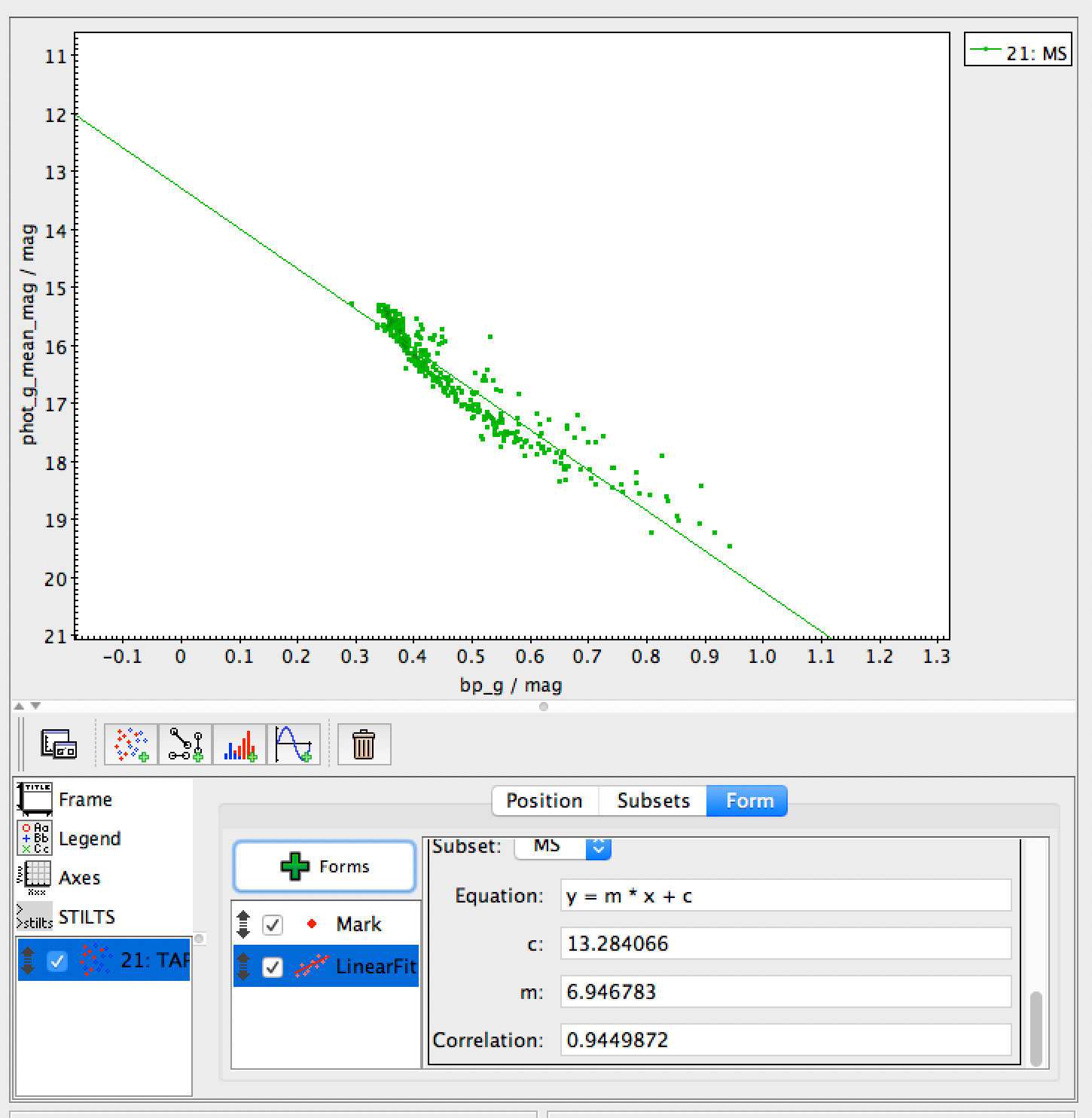}
\caption{Linear fit to the main sequence stars of NGC 188}
\label{ngc188MainFit}
\end{center}
\end{figure}

While TOPCAT does not provide for general fitting, it can be used to produce a linear fit. To this end, proceed as in section \ref{TOPCATSubsets} in order to select the cluster's main sequence stars as a subset. Display only that subset. With the data set selected, go on the ``Form'' tab, click on the big green cross to add a new form. The dropdown menu is shown in Fig.~\ref{FormDropdown}.
From that dropdown menu, select ``Add LinearFit''. TOPCAT will fit a line to your data points, and when you select the LinearFit form and scroll down, it will show you the best-fit parameters it has chosen, as well as the correlation, cf. Fig.~\ref{ngc188MainFit}.
Note that, for TOPCAT, both the data point representation (``Mark'') and the linear fit (``LinearFit'') are merely different {\em forms} of representing the same data. Other forms, including more complex ones like XY error bars or ellipses, or a text label, can be added in the same way we have added the linear fit.

We can also add an additional data set. Let us deselect the linear fit and the main sequence subset and go back to the colour-magnitude diagram for NGC 188. Following the same procedure as in section \ref{TOPCATSubsets}, select the stars in the region of the open cluster M 67, starting with the TAP query
\begin{lstlisting}
SELECT ra, dec, pmra, pmdec, bp_g, phot_g_mean_mag
FROM gaiadr2.gaia_source
WHERE 1=CONTAINS(POINT('ICRS',ra,dec), CIRCLE('ICRS',132.8250,11.800, 0.4))
\end{lstlisting}
and once more selecting a small area around the (clearly visible) concentration in the pmra-pmdec plane. In the Plane Plot for M 67, I can then add the NGC 188 data as an extra data set. There are two ways of doing this. Either you go to the icons visible directly underneath the plot sub-window. 
\begin{figure}[htbp]
\begin{center}
\includegraphics[width=\linewidth]{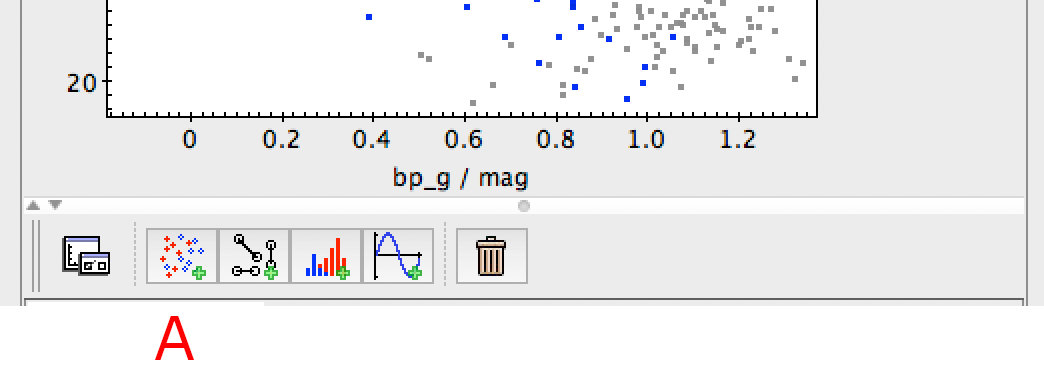}
\caption{Icons in the Plane Plot window, located directly below the displayed plot}
\label{PlotIcons}
\end{center}
\end{figure}
These icons are shown in Fig.~\ref{PlotIcons}. If you click the panel I have marked with a red A, TOPCAT will add another data set (``positional plot control'') to the plot. Click on that data set in the bottom left sub-window, and you will find you can choose which of the data sets that you have loaded into TOPCAT (cf. the table list in the main TOPCAT window) you mean to plot here. 

\begin{figure}[htbp]
\begin{center}
\includegraphics[width=\linewidth]{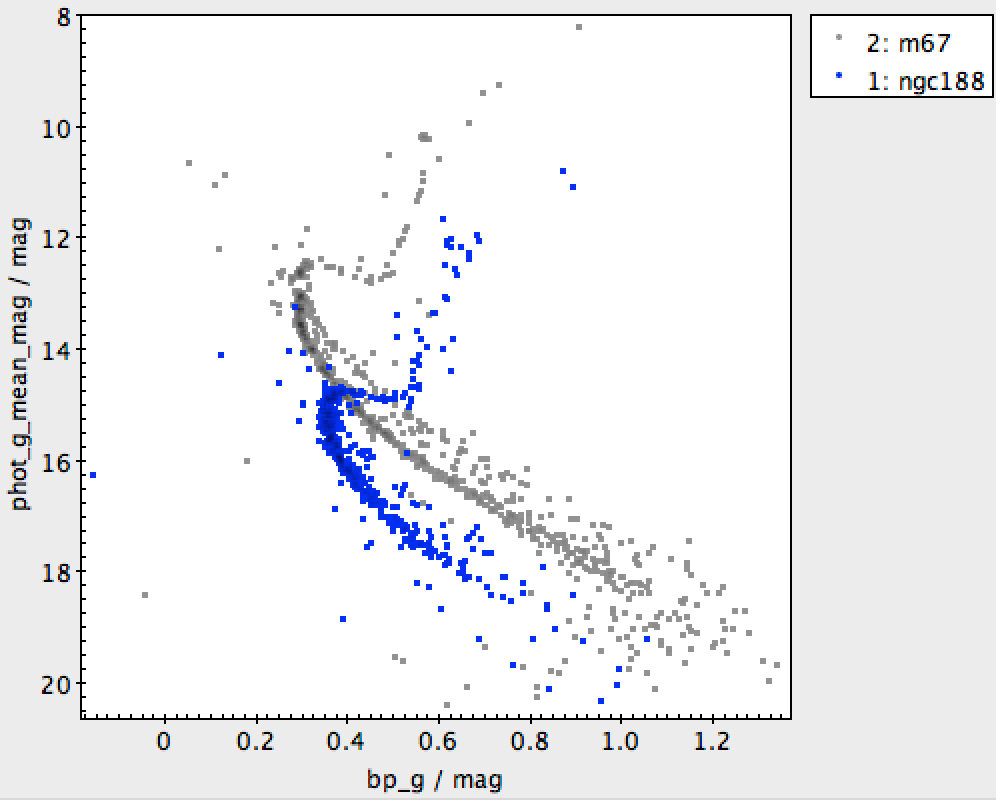}
\caption{Colour-magnitude plots for the open star clusters NGC 188 and M 67}
\label{NGC188M67Comp}
\end{center}
\end{figure}

\begin{figure}[htbp]
\begin{center}
\includegraphics[width=\linewidth]{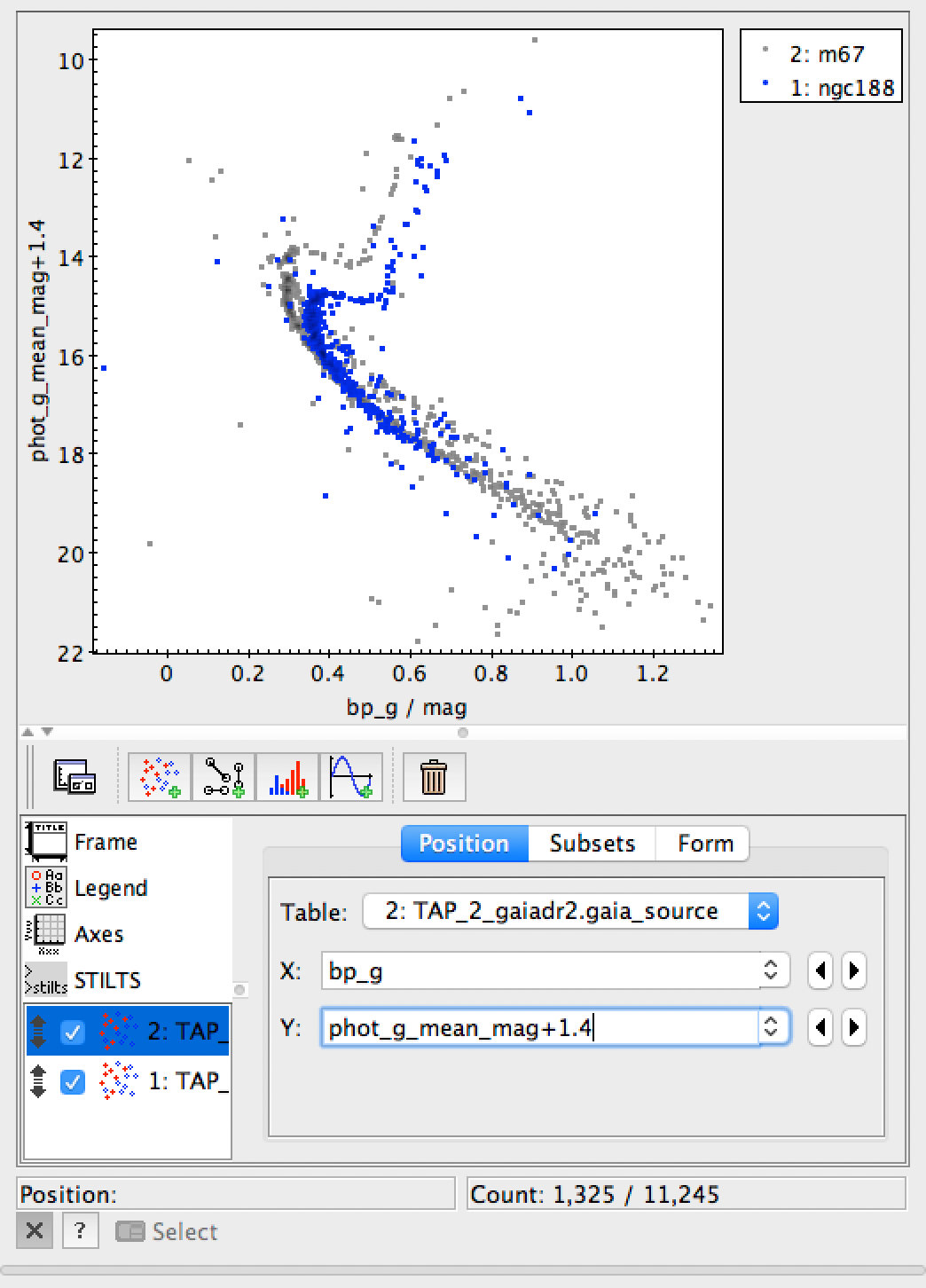}
\caption{Colour-magnitude plots for the open star clusters NGC 188 and M 67}
\label{NGC188M67CompShift}
\end{center}
\end{figure}

Alternatively you can go to the top menu that is shown while the ``Plane Plot'' window is active and, from the ``Layer'' sub-menu, select ``Add Position Control''. For the NGC 188 data, you can once more select the pmra-pmdec subset that represents the star cluster, and select only that sub-set. When you now plot the colour-magnitude diagram,  colour bp\_g against brightness phot\_g\_mean\_mag (again with the Y flip as an astronomical convention), you can see the separate plots for both of the star clusters, as in Fig.~\ref{NGC188M67Comp}.
There are two differences between the data point distribution for the two star clusters. First of all, the main sequence for M 67 is brighter than for NGC 188. Since we have no reason to assume that the light from one of the clusters is shifted to a different colour relative to the other,\footnote{This could happen: If there is more cosmic dust between us and the one cluster than between us and the second cluster, we would expect the first cluster to appear more reddish.} the simplest assumption is that this is because M 67 is closer to us than NGC 188. We can make a rough quantitative estimate of distances here, as follows. Recall the window where we chose which column to plot in the X and which in the Y direction. In that column, instead of giving the column's name, we can write a more complex expression --- such as the column name plus a constant value. By trial and error, I find that adding 1.4 to phot\_g\_mean\_mag for M 67, I can make the densest portions of the main sequence overlap, cf. Fig.~\ref{NGC188M67CompShift} (in the bottom part of the figure, you can see where I have added 1.4 to the g magnitude by hand --- if needed, you can write much more general functional expressions into that little window, involving more than one of the column names, too!).
Recall that, from the formula (\ref{AstroMagnitudeFormula}) for astronomical magnitudes, the magnitudes are related to the intensity of the light we receive from an astronomical object. That intensity is proportional to an object's intrinsic brightness and to the inverse of the square of the object's distance to us (inverse square law). When two main sequence stars have the same colour, we expect them to have the same intrinsic brightness. For such objects, the ratio of the intensity of light reaching Earth is inversely proportional to the ratio of the squares of their distances $r_1, r_2$; by (\ref{AstroMagnitudeFormula}), this means their apparent magnitudes are related as
\be
m_1 - m_2 = -2.5\cdot\log\left(\frac{r_2^2}{r_1^2}\right) = 5\cdot\log\left(\frac{r_1}{r_2}\right),
\ee
so that
\be
\frac{r_1}{r_2} = 10^{0.2(m_1-m_2)}.
\ee
With $m_1-m_2\approx 1.4$, we can estimate that NGC 188 is about twice as far away from us as M 67. According to the Simbad astronomical data base\footnote{Accessed by entering the identifiers in the web form at [\href{http://simbad.u-strasbg.fr/simbad/sim-fbasic}{http://simbad.u-strasbg.fr/simbad/sim-fbasic}], scrolling down to ``Collections of measurements'', selecting and displaying ``distance''.}, M 67 is between 0.9 and 0.99 kpc away, NGC 188 between 1.7 and 2.3 kpc, so our rough estimate is consistent with the known distance measurements. 

The shifted version of Fig.~\ref{NGC188M67CompShift} shows clearly that for NGC 188, some of the less bright stars have started to swerve off the main sequence and become red giants.  NGC 188 must be older than M 67, given that in NGC 188, less bright stars are already entering the red giants stage. Indeed, M 67 is estimated to be around 4 billion years old, NGC 188 more than five billion years (making it one of the oldest open star clusters we know).

\subsection{Histograms}
\label{TOPCAThistogram}
\begin{figure}[htbp]
\begin{center}
\includegraphics[width=\linewidth]{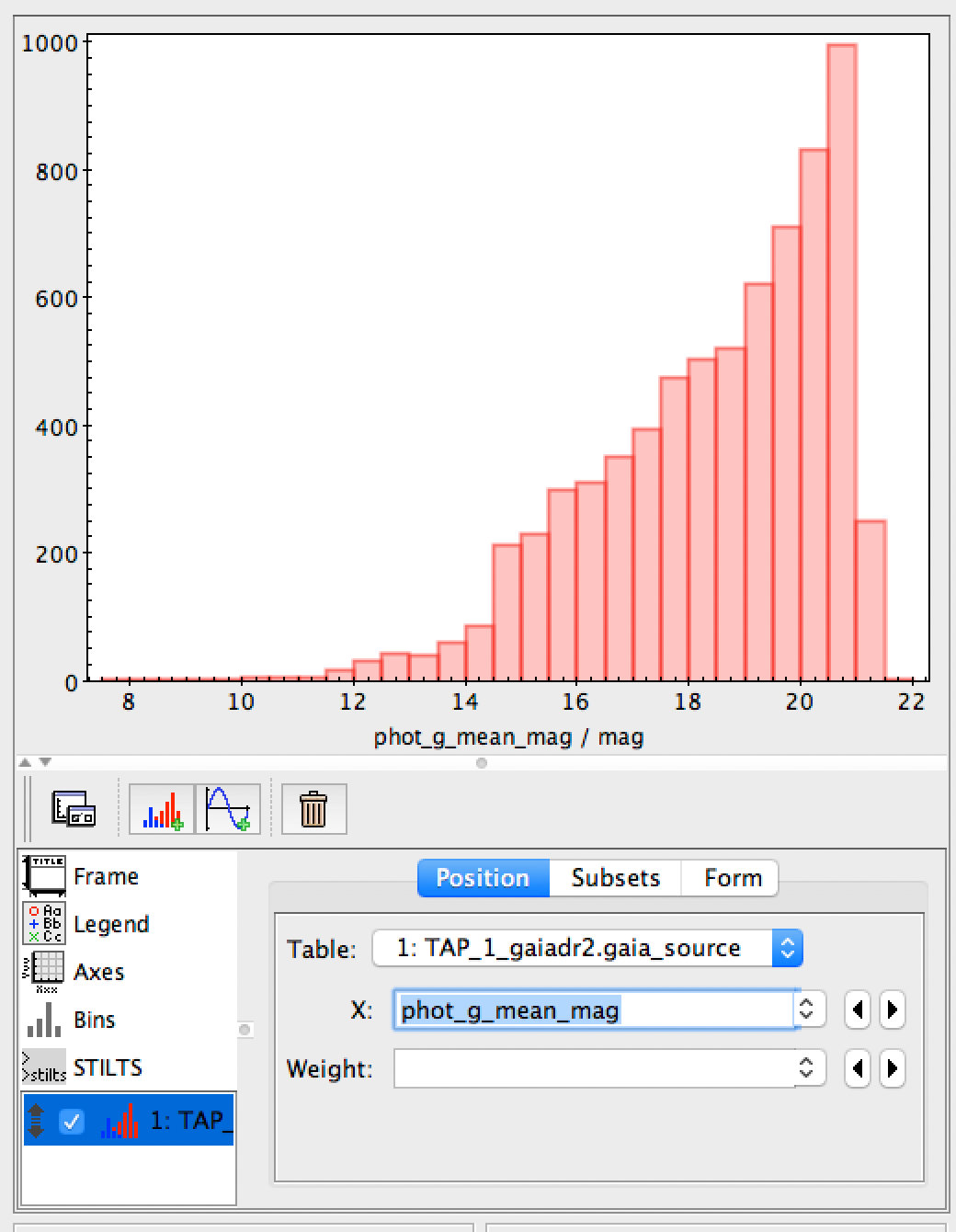}
\caption{Histogram for the g brightness in the stars in the region of NGC 188}
\label{NGC188Histogram}
\end{center}
\end{figure}

Next, let us look at some histograms. There is no separate histogram window, as you can add histogram data to an ordinary Plane Plot window, but there is a short-cut in the icons of the TOPCAT main window (cf. Fig.~\ref{TOPCATIcons}) which directly produces the Plane Plot window configured for histograms. Choose the pre-selection for, say, the NGC 188 region data set, and click on the histogram icon. A histogram based on values in the first column of your data set pops up. By selecting your histogram in the bottom left window (it should be selected by default!), you can again choose different table columns for the X axis. Choose phot\_g\_mean\_mag, for instance, and you will be rewarded with the histogram in Fig.~\ref{NGC188Histogram}.
This histogram is a combination of physics and measurement bias. In general, less luminous stars are much more common than luminous ones. Also, there are only very few stars close to us, and many more at greater distances; with increasing distance, the apparent brightness becomes less, as well. Both of these reasons are what explains the increase in the histogram from left to right. The fact that the histogram comes to a fairly abrupt end on the right is because the Gaia mission has a certain limiting magnitude beyond which stars are too faint to be included in the analysis. 

By clicking not on the data itself in the lower-left sub-window, but on ``Bins'', you can customise the histogram. Under the ``Histogram'' tab, you can use a slider to adjust bin size, or enter the size explicitly in a field. The ``Bin Phase'' slider shifts bin position. In the ``General'' tab, you can change your histogram to be cumulative, if that is what you want. Selecting subsets, or re-using subsets that have already been defined for your data set, is another possibility, as is displaying two histograms in the same window.

\subsection{A quick look at a spectrum}

Spectra play a key role in astronomy. As a special case of a Plane Plot in TOPCAT, let us download a sample galaxy spectrum from SDSS, more concretely: for data release 8, let us pick the object with the id 1237659161195249685, which at least for me is the default when I open the DR8 object explorer at [\href{http://skyserver.sdss.org/dr8/en/tools/explore/obj.asp}{http://skyserver.sdss.org/dr8/en/tools/explore/obj.asp}]. In the SpecObj submenu on the left, click on FITS to download the spectrum as a FITS file by the name of spec-1330-52822-0304.fits.

\begin{figure}[htbp]
\begin{center}
\includegraphics[width=\linewidth]{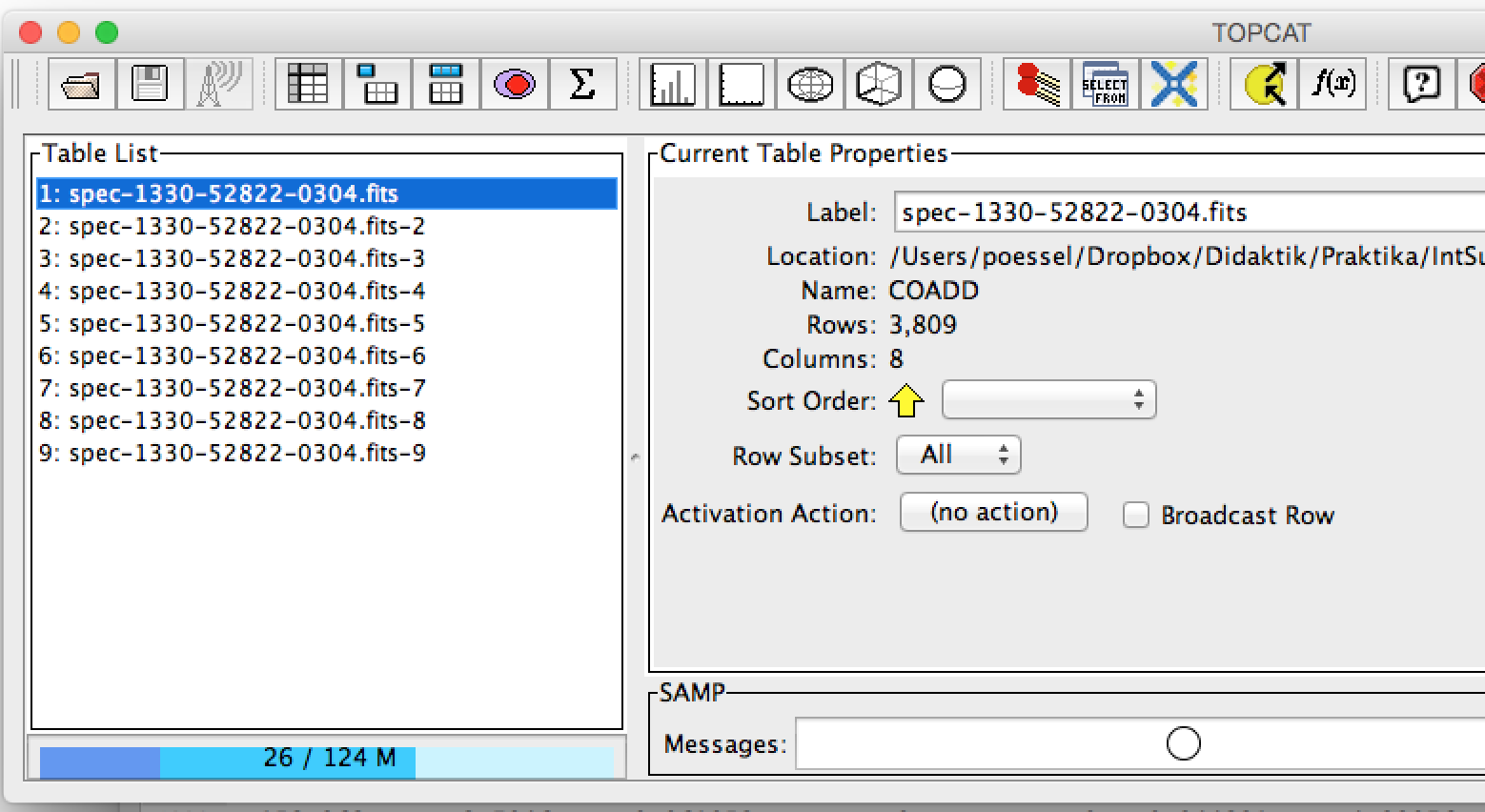}
\caption{TOPCAT main window with SDSS DR8 spectrum loaded}
\label{SpectrumLoaded}
\end{center}
\end{figure}

To take a quick first look at the spectrum, open it in TOPCAT like you would open a table (cf. section \ref{TOPCAT}). You will see that this spectrum actually opens as several tables, corresponding to different HDUs of the FITS file, as shown in Fig.~\ref{SpectrumLoaded}.
If you click on each of the tables, its name will be displayed. COADD is the spectrum we want. SPECOBJ just contains general information about the object whose spectrum this is, while SPZLINE describes the spectral lines that have been identified in the spectrum. The files starting B1 and R1are different exposures of the (overlapping) blue and red portions of the spectrum, recorded on separate chips; they have been added up and combined to give the COADD part we will use now.

Select the COADD part of the table and click the Plane Plot window icon (cf. Fig.~\ref{TOPCATIcons}) or, alternatively, choose Graphics $\to$ Plane Plot from the top menu. In the Plane Plot window, change the X and Y values -- X should be loglam, the logarithm of the wavelength, and Y should be flux. You can effect the change by clicking on the arrow buttons to the right of the field, or alternatively using the field entry as a dropdown menu. 

A spectrum is not just a collection of data points representing independent objects. It is a sample of data points from a continuous distribution curve. Thus, it makes sense to plot the data points as a connected line instead of as separate points. To this end, in the Form tab, open the ``Forms'' dropdown menu (with the big green plus sign; for the menu itself, cf. Fig.~\ref{FormDropdown}), and  create a new line plot by clicking the ``add a new line form'' button. Then unselect the existing Mark form directly above, to hide the separate data points.

\begin{figure}[htbp]
\begin{center}
\includegraphics[width=\linewidth]{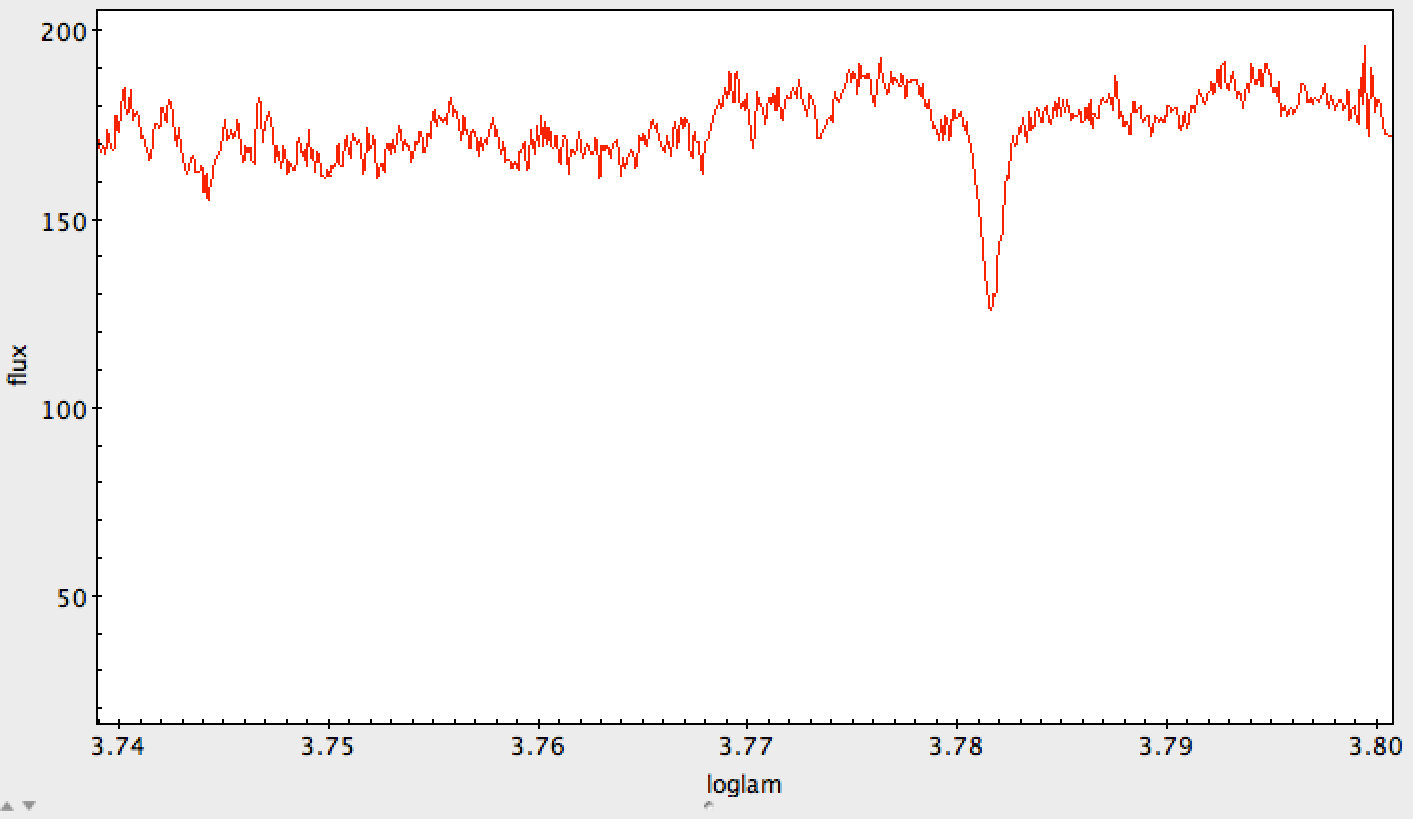}
\caption{A quick look at an SDSS spectrum, using Plane Plot}
\label{SpectrumQuickLook}
\end{center}
\end{figure}

\begin{figure}[htbp]
\begin{center}
\includegraphics[width=\linewidth]{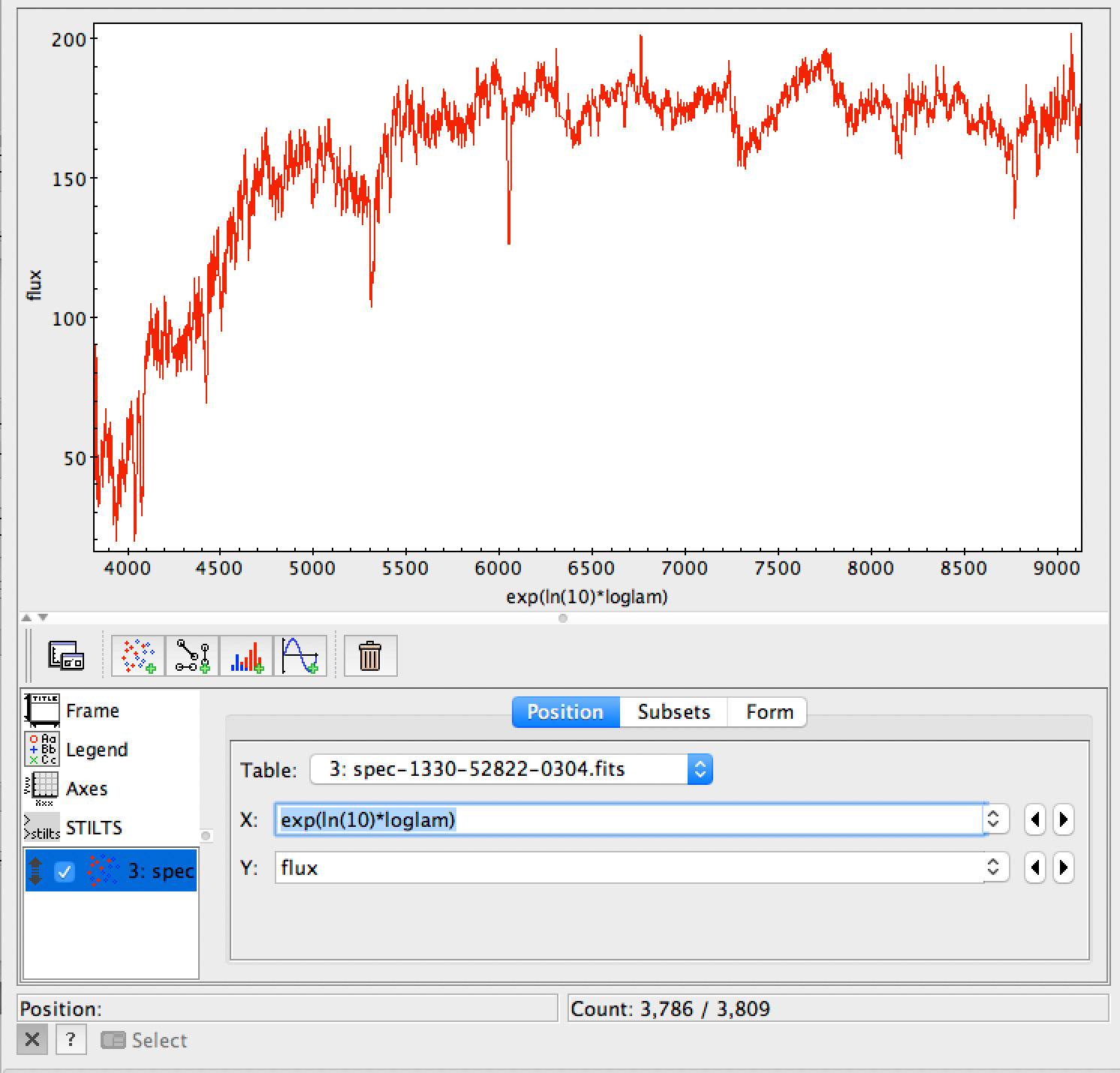}
\caption{A quick look at an SDSS spectrum, with linear wavelengths shown, using Plane Plot}
\label{SpectrumQuickLookLambda}
\end{center}
\end{figure}

Finally, click on Axes in the lower-left sub-window, go to the Range tab, and adjust the X subrange upper and lower limit, to display the details in a more restricted wavelength region. By clicking in between the two limit buttons and dragging, you can drag around the wavelength region you are looking at. All in all, this should now allow you a fairly informative quick look at your spectrum, Fig.~\ref{SpectrumQuickLook}.
What is still a bit unusual is that we are plotting the logarithm of the wavelength on the X axis. To change that, replace ``loglam'' in the X axis field by ``exp(ln(10)*loglam)'' --- now the X axis is showing the wavelength, in Angstrom (where 1\AA $= 0.1\:\mbox{nm}$), which is much better if you want to read off the positions of certain physical features, such as spectral lines. The result is shown in Fig.~\ref{SpectrumQuickLookLambda}.

As we have seen in these first sections, there are quite a lot of things you can do with application software, in this case DS9 and TOPCAT. We could have chosen differently and, for instance, introduced the Java-based image processing and analysis software ImageJ, maintained by the US National Institutes of Health (since it is also used in the life sciences), [\href{https://imagej.nih.gov/ij/}{https://imagej.nih.gov/ij/}], or its astronomical incarnation with suitable extra functionality, astroImageJ\footnote{Collins et al. 2017, \href{https://doi.org/10.3847/1538-3881/153/2/77}{doi: 10.3847/1538-3881/153/2/77}} hosted by the University of Louisville, [\href{https://www.astro.louisville.edu/software/astroimagej/}{https://www.astro.louisville.edu/software/astroimagej/}].

Up to a certain point, such application software is perfectly sufficient, and it wouldn't make much sense to re-invent the wheels it provides. This is particular true for a something that you should routinely do when you prepare to settle a new, and possibly unfamiliar data set: taking a quick look at given data and getting a feeling what that data is about. But that is not all, as we shall explore in the next sections.


\section{Getting started with Python}

Application software will only get you so far. As you follow where your research project leads, you are bound to come to the point where you need more flexibility, and more functionality, than such software can provide. The next step is to make the transition to a programming language. In astronomy, one of the most popular, if not the most popular, programming language is Python. Popularity has tangible consequences --- a popular programming language is bound to have a large community of active users, you are bound to find answers to your problems and questions on platforms such as Stackoverflow or in other forums, and specifically for a programming language that is popular in astronomy, you are bound to find that other users have written helpful modules or libraries that implement helpful astronomy-related functionality.

What follows will not replace a basic introduction to Python; before you read on, you should probably familiarise yourself with the basics of the language.\footnote{Our summer interns at Haus der Astronomie are frequently in this situation as they prepare themselves for the internship; a number of them have reported that they found [\href{https://www.codecademy.com/en/learn/python}{https://www.codecademy.com/en/learn/python}] helpful as a first introduction.} If you have written code before in another language, that is bound to help --- certain concepts, such as loops or if-conditions, are fairly universal to coding.

Once you have installed Python, you can and should work through the examples presented in the following sections. As you become more advanced, you will continue to learn by doing, by solving specific problems, by getting familiar with new techniques --- and by googling your Python questions, or parts of your error messages, which is surprisingly effective, given that there is a large Python community out there that has answered an amazing variety of questions on platforms such as Stackoverflow.

\subsection{Installing Python}

Cruel, but true: Some of the most complicated, and potentially frustrating operations come at the very beginning, as you install Python on your computer. On a Mac, things should be fairly simple. If you are working on a Linux machine, you (or whoever installed Linux on your machine!) probably know enough about what you are doing that installing Python should work. If you are on Windows, things might be more difficult.

My recommendation, in general, is to install Anaconda python, which is available for Mac, Linux, and Windows, and which can be downloaded for free at [\href{https://www.continuum.io/downloads}{https://www.continuum.io/downloads}]. Anaconda comes with many useful packages for astronomy, or science in general, installed (Astropy, Numpy, Scipy, \dots). If you are reading this as part of a course, another installation might have been recommended by your instructor --- or, a promising trend, Python might be provided to you in the form of Jupyter notebooks, accessible in your browser window with no installation required. Astrobetter, a highly useful website with helpful hints for astronomers, covering a variety of helpful issues, has a page on how to install python for astronomy.\footnote{\href{http://www.astrobetter.com/wiki/Python+Setup+for+Astronomy}{http://www.astrobetter.com/wiki/Python+Setup+for+Astronomy}} {\bf Important: For the following, I will assume that you have installed some version of Python 3.}

Anaconda also comes with a helpful programming environment called Spyder, which makes it (comparatively) easy to write, run, and debug Python code. So let's assume you have installed Python successfully, and started the Spyder software.

\subsection{Using Python in Spyder}

\begin{figure}[thbp]
\begin{center}
\includegraphics[width=\linewidth]{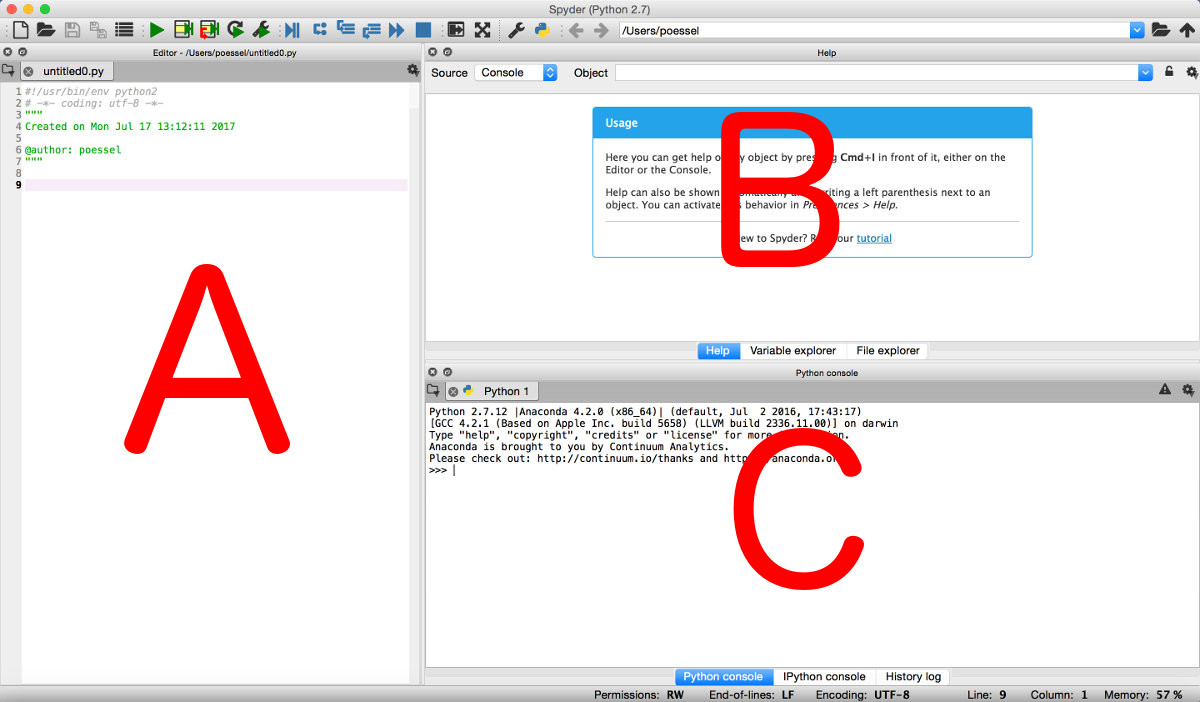}
\caption{Basic Spyder layout}
\label{SpyderDefault}
\end{center}
\end{figure}

When you open Spyder, its basic layout should look as in Fig.~\ref{SpyderDefault}.
I have added big red letters for later reference. If you are on Windows or on Linux, you might see a menu bar saying ``python file edit search'' etc. directly above the icon bar; on Mac, that same menu bar will be at the top of your screen when Spyder is active. Again, this layout might change with future versions of those operating systems, and again, we will refer to this as the top menu.

If your window arrangement should happen to look totally different, a possible remedy would be to use the top menu bar, going to View $\to$ Window Layout and clicking on ``Spyder Default Layout'' to give you the default layout. (Personally, I prefer to drag the separator at the right border of A some way to the right, enlarging the window A at the cost of the two windows B and C.)

For now, we will only use the basic functionality. In window A, we write the code we want to save, and run. You should definitely save the Python code you are writing in this window. Conversely, you can load and execute files containing code that you had written earlier.

In window C, with the tab "IPython Console" selected, we can see the results of our code. If our code prints anything, this is where we will see it. If we plot any diagrams, this is where they will be displayed. 

Window C has one additional, very practical functionality. Whenever we program something more complicated, we will write it down as a proper program/script in window A. But if we merely want to try some line of code very quickly, we can also enter it directly into window C, press return, and see the result of that particular command immediately.

For instance, if we want to know the current value of the variable \verb|a| after having run our program, simply enter \verb|a| and press return, and the value of \verb|a| will be displayed. For instance:
\begin{lstlisting}
In[0]: a=1.6
In[1]: a
Out[1]: 1.6
\end{lstlisting}
In a bit more detail, in the first row, after the prompt ``In[0]:'', I have typed ``a=1.6'' and then hit return (enter). The program has accepted my input, but does not produce any immediate output. Instead, it offers me another opportunity for input, this time numbers ``In[1]:'' -- and there, I type simply ``a'' and hit return. This time, the program does return an output value, namely the value of the variable ``a'' --- with the preface ``Out[1]:'' it returns the value 1.6. We will use this ''direct mode'' of executing a command in the following in a number of places where I am introducing some simple new concept, or command, and direct execution is the simplest way for you to see what the command does.

\begin{figure}[htbp]
\begin{center}
\includegraphics[width=\linewidth]{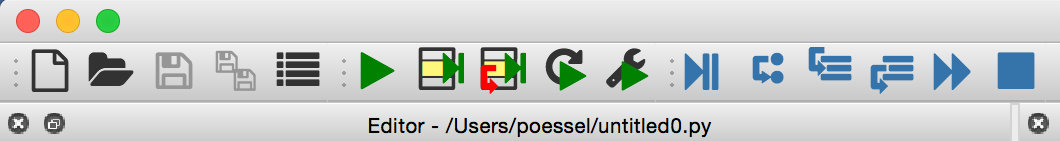}
\caption{Part of the top bar of Spyder}
\label{SpyderTopBar}
\end{center}
\end{figure}

Note that, depending on what you did before, the numbers characterising your input and output will vary. If I repeat the operation immediately, starting with the input prompt ``In[2]'' the software displays, then the whole input-output sequence would now look like this:
\begin{lstlisting}
In[2]: a=1.6
In[3]: a
Out[3]: 1.6
\end{lstlisting}
The exact number displayed for each input and possible corresponding output will change, counting up during each session. That is why, in all future examples of live interaction with the system, I will leave out the numbers altogether. Our simple sequence of defining $a$, and then retrieving its values, then looks like this:
\begin{lstlisting}
In: a=1.6
In: a
Out: 1.6
\end{lstlisting}

In order to run one of your programs that you have written in window A, press the green play arrow in the horizontal list of icons at the top. Fig.~\ref{SpyderTopBar} shows the left-hand part of that horizontal bar, to show you the green arrow, which looks like the universal play (video, song, \dots) symbol, and its neighbours.
Let us have a look at the icons. The first from the left is the ``New File'' icon, which produces a new Python script window for you to write code in. The second from the left is ``Open File'' which opens an existing file. Third from the left is the ``Save'' icon --- as everyone will tell you, save your file frequently, please! We will ignore the other icons for now and return to the green play button.

Imagine that, in window A, you have written the following program (the first lines will have been there when you created a new file):
\begin{lstlisting}
#!/usr/bin/env python
# -*- coding: utf-8 -*-
"""
Created on Mon Jul 17 13:12:11 2017

@author: poessel
"""

print( "Hello world!")
\end{lstlisting}
This is the traditional first program. Let's go through this: The first line is only important if you run the program from the command line as an executable file. It tells the computer which software to call up to run the program, in this case: Python. 

The second line is a technical one. It tells the computer that whatever text follows uses not only ASCII characters but could also contain certain special characters (such as the German Umlaute \"o, \"a, etc.). This particular set of possible characters is known as ''utf-8''. What comes next is commented out, and merely gives anyone reading the script information about when it was created, and by whom.

Only the last line, \verb|print( "Hello world!")|, is the command we want executed. So this is the program we want to run! In order to make it run, you should now press the green play button. 

If this is the first time ever you have pressed the green play button in Spyder, there will be a pop-up window asking about working directories, configuration and more. Just press OK. This window will not bother you again. 

If this is the first time you have pressed the green play button after creating a new file, the file will be called something like ``untitled1.py'', and once you press the play button, Spyder will ask you to save the file, and give it some proper name. Please do that.

Now that we're ready, press the green play button again. Now your program will run, and in the IPython Console in window C (choose the correct tab if it's not visible!), you will see its output. It will look something like this:
\begin{lstlisting}
In: runfile('/Users/poessel/pythDir/example1.py', wdir='/Users/poessel/pythDir')
Out: Hello world!
\end{lstlisting}
Under "In", spyder is telling us that it is running a particular file, in this case example1.py (as I named it earlier). The ``wdir='' is followed by the file's working directory. This is important if you need to open files, or write out your results into a file. 

If you don't specify another directory, files you are using are assumed to be in the working directory (and there will be an error message if they aren't, and the software is looking in vain). Other directories can be specified relative to the working directory. 

For instance, if you try to open a file ``figures/thisFig.png'', then ``figures'' is assumed to be a subdirectory of your working directory.

Now that you have learned about executing code that you have written in window A, and writing small snippets of code into window C, there are potential pitfalls you should know about. When you start a script from window A, or execute a small snippet in window C, the execution does not start from some blank slate. Instead, variables you have defined before are already defined, and modules you have loaded before are already loaded. 

In most cases, this ``hidden state'' is unlikely to lead to confusion, but in some cases, it can. A simple example: Your program might rely on a certain variable to have already been defined. In the situation where you write the program, and test it, running it again and again, it could be that this condition is met only because of the hidden state your program is in after having run another script, or having run the script in question before. In that case, you will get an error message as soon as you try to run that particular code from scratch. 

\begin{figure}[htbp]
\begin{center}
\includegraphics[width=\linewidth]{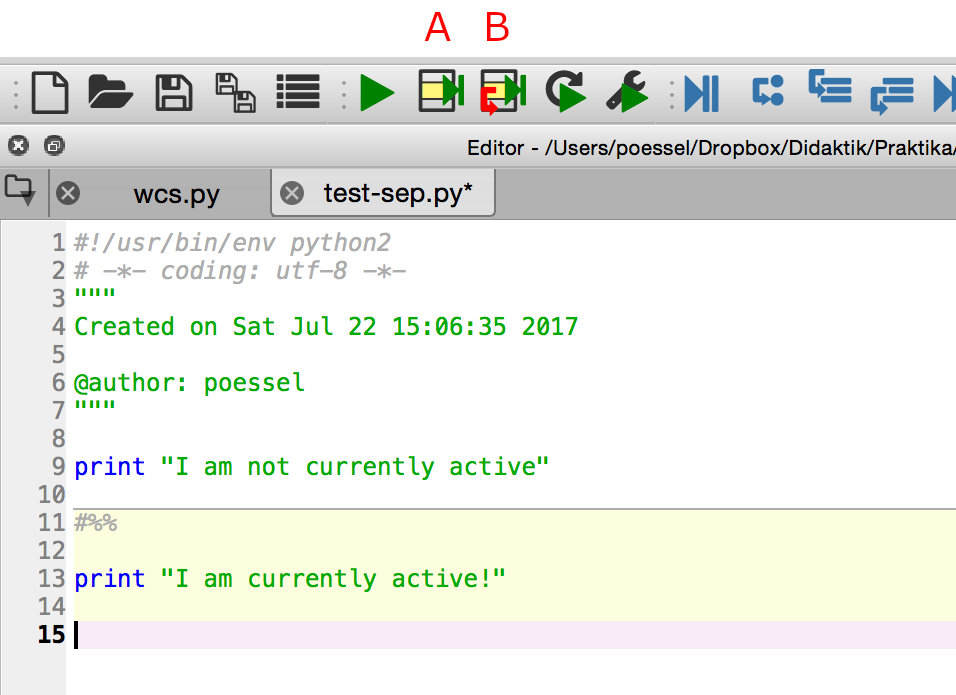}
\caption{Using hashtag-percent-percent for zoning in Spyder. The two specific play symbols have been labelled in red as A and B, by me}
\label{SpyderActiveZone}
\end{center}
\end{figure}

Moving on: a word on commenting your code. There are two ways of adding comments to your code. Use them as often as possible --- if you don't, then even you yourself might not understand what you have written, if you revisit your code after a while. Adding the hashtag symbol declares that everything from there until the end of the line is a comment:
\begin{lstlisting}
# This is a comment
This is not a comment (and will give an error message!)
\end{lstlisting}
For comments spanning several lines, we could of course add a hashtag at the beginning of each line. But there is another, more convenient way: We can use three double quotation marks in a row to mark both the beginning and the end of the commented-out section:
\begin{lstlisting}
"""
This is a commend 
which can span
several lines.
"""
\end{lstlisting}

In window A, a hashtag followed by two percent signs, \verb|#\%\%|, has an interesting effect: spyder will draw a separating line, and mark the region your cursor is in (delineated by those separating line, the beginning and the end of the file) in yellow. Fig.~\ref{SpyderActiveZone} shows an example.
In this case, the lower region, from the \verb|#\%\%| to the end of the file, is marked in yellow (except for the line the cursor is in, which is marked in pink). Note that,  to the right of the green play symbol, there are two others that show a region marked in yellow; I have labelled those A and B in Fig.~\ref{SpyderActiveZone}. If you click on the icon labelled A, Spyder will execute only the part marked in yellow. If you click on the icon labelled B, Spyder will also execute the part marked in yellow, and then move on to the next region, marking {\em that} one yellow. This is highly useful for debugging. If you are not sure what is going wrong with your code, try to separate your code into regions by using the \verb|#\%\%|. Then execute those regions one by one, seeking to understand what is happening in each.

Another application of this kind of regional execution is when you have a part of the code that takes a long time to run. As we discussed, in executing code within a region, Spyder doesn't forget what happened before -- if you loaded some modules before, they will still be loaded and accessible; if you gave certain variables certain values, these variables will still be defined in the same way. Thus, imagine that you have already executed the long-running code once. If you then want to test whatever part of your programming comes below, simply separate it from the long-running code by a separating line. That way, executing just the part of the code you are working on now will be a matter of e.g. seconds --- whereas executing the whole of the code, including the long-running bits, would take much longer.

Some brief remarks on window B: This window has two tabs. The File Explorer allows you to see what files are in your working directory. The Variable Explorer shows you the current values of the variables you have defined in your program --- very useful for a quick glance at what is going on. Select one of those variables by clicking on it, and a right mouse click on Windows, or Ctrl + mouse click on Mac will give you the options of executing specific operations on it. For list array variables, which we will encounter in section \ref{Lists}, choosing the plot or histogram option can be a good way of obtaining more information here.

\subsection{Modules}
\label{PythonModules}

Some functionality in Python comes out of the box, and is available whenever you run Python. For less common functions, you will need to import specific modules. Some modules that we will use in the following are:
\label{ModuleTable}
\begin{enumerate}
\item {\em Numpy} -- a collection of mathematical and numerical functions, from sin and cos to integration and linear algebra. We will use this instead of the python math module, which also has basic mathematical functions.\footnote{Travis E. Oliphant. A guide to NumPy, USA: Trelgol Publishing, (2006).
St\'efan van der Walt, S. Chris Colbert and Ga\"el Varoquaux. The NumPy Array: A Structure for Efficient Numerical Computation, Computing in Science \& Engineering, 13, 22-30 (2011), \href{http://dx.doi.org/10.1109/MCSE.2011.37}{DOI:10.1109/MCSE.2011.37}}
\item {\em Scipy} -- a collection of functions for scientific calculations. We will use the scipy functionality for curve fitting, for instance.\footnote{Jones E, Oliphant T E, Peterson P, et al. SciPy: Open Source Scientific Tools for Python, 2001--, [\href{http://www.scipy.org/}{http://www.scipy.org/}] [Online; accessed 2019-05-21].}
\item {\em Astropy} -- functions for astronomers, including those that let us deal with files in the fits format (as professional astronomical images usually are) and those that help us with calculating astronomical quantities.\footnote{Astropy Collaboration (2018): The Astropy Project: Building an inclusive, open-science project and status of the v2.0 core package, [\href{https://arxiv.org/abs/1801.02634}{https://arxiv.org/abs/1801.02634}]. Astropy Collaboration (2013): Astropy: A community Python package for astronomy, \href{https://doi.org/10.1051/0004-6361/201322068}{DOI:10.1051/0004-6361/201322068}}
\item {\em Matplotlib} -- our go-to library for plotting diagrams of various kinds.\footnote{John D. Hunter. Matplotlib: A 2D Graphics Environment, Computing in Science \& Engineering, 9, 90-95 (2007), \href{http://dx.doi.org/10.1109/MCSE.2007.55}{DOI:10.1109/MCSE.2007.55}}
\end{enumerate}
With Anaconda, these modules are already installed. Other, more specialised modules, you might need to install yourself. Usually, when you have found (via Google for instance) a module, it will give you some instructions on how to install the module. Installation also varies from operating system to operating system --- which is why I cannot give you a simple, general recipe here. 

Once a module is installed on your computer, you can {\em import} it, or parts of it, into the programs you write. It is customary to place all import statements at the beginning of your script, so you have an overview of what has already been imported and what hasn't.

For instance, here is a script where I import the numpy module. Once it is imported, I can use the module's functions. The name of a module's function in this context starts with the module name, then a full stop, then the specific function name. For instance, this script here will print the sine of $3.1416$:
\begin{lstlisting}
import numpy
print(numpy.sin(1.57079632679))
\end{lstlisting}
When executed, the program dutifully returns \verb|1.0| as the result (since the argument is equal to $\pi/2$, up to a numerical rounding error that is too small to influence the output).

I can also import the module under another name. Then, whenever I call a function or constant from that module, I can use the new (usually abbreviated) name. Like many other people, I habitually abbreviate numpy to ``np'' when I import it:
\begin{lstlisting}
import numpy as np
print(np.sin(1.57079632679))
\end{lstlisting}

If I only need a specific function from the module, I can import that function directly. Here, I import the sine function ``sin'' from the numpy module. Once I have done this, I do not need to prefix the function with the module name --- I can call the function directly:
\begin{lstlisting}
from numpy import sin
print(sin(1.57079632679))
\end{lstlisting}
When you import a specific function, you can also give it another name:
\begin{lstlisting}
from numpy import sin as superDuperSin
print(superDuperSin(1.57079632679))
\end{lstlisting}

Sometimes, you will not need the whole module, and instead import a submodule (that is, a predefined subset of the module's functionality). In the section where we deal with plotting diagrams, we will usually include a statement
\begin{lstlisting}
import matplotlib.pyplot as plt
\end{lstlisting}
for all our plotting and diagram needs.

\section{Basic operations with Python}

\subsection{Meet your new versatile calculator}
\label{BasicCalculations}

The least that coding can do for you is serve as a versatile calculator that, an added bonus, documents what you have been calculating. Imagine that we want to calculate the luminosity of a star with radius $R=695 500$ km and effective temperature $T_{\mathrm{eff}}=5780$ K, modelling the star as a blackbody. We use the Stefan-Boltzmann law for blackbody radiation, which states that the luminosity $L$ in that situation is given by 
\be
L = 4\pi R^2\cdot \sigma T_{\mathrm{eff}}^4,
\label{LuminosityFunc}
\ee
with the Stefan-Boltzmann constant 
$$\sigma=
5.670367\cdot 10^{-8} \frac{\mbox{W}}{\mbox{m}^2\mbox{K}^4}.
$$
If you were to type all that into a handheld calculator, you would need to keep track of all the different steps, and if worse comes to worse you might make a typo somewhere. 

In a Python script, you can define the variables involved one by one, and define the constant sigma, and then write down the formula for the result using those variables. You can easily double-check the script once you have written down. The bit of code in question would look something like the following:
\begin{lstlisting}
import numpy as np
Teff = 5780
R=695500*1000
sigma=5.670367e-8
# Now the formula:
L = 4*np.pi*R**2*sigma*Teff**4
#...and we print the result:
print(L)
\end{lstlisting}
The program dutifully returns \verb|3.68567074251e+26|, which is indeed the luminosity of that particular star, the Sun, in Watt.

Some notes on the formula: The asterisk \verb|*| is the standard sign for multiplication. The double asterisk \verb|**| stands for a power, so \verb|Teff**4| represents $T_{\mathrm{eff}}^4$. The geometric constant $\pi$ we have imported from the numpy package; hence the name \verb|np.pi|.

Note that we had to take care to use the proper units: the radius $R$ was given in kilometers; in our script, we have multiplied that number by 1000 in order to get the (SI) unit of meters. 

There is an easier way of doing this, and that is provided by astropy, as we shall see in the next section, \ref{UnitsConstants}.

The advantage of doing a calculation as a script, instead of typing it into a pocket calculator, is that you automatically document what you have done. You can go back, identify mistakes if necessary. And if you need to do the same calculation, but for values other than those you have used in the first case, you can just change the parameters and let the script run again. Or import a whole list or array of data, and apply the calculations to thousands of objects, one after the other, automatically. 

\subsection{Units and constants}
\label{UnitsConstants}
Did you ever mix up your units? If not, good for you. But even then, keeping track of your units would give you a handy cross-check for your result. If your result has the wrong units, something went wrong in the calculation.

The astropy module includes a handy selection of units and constants from physics and astronomy, and defines a straightforward way of using those units in calculations.  You can find a list of all the pre-defined units on the page [\href{http://docs.astropy.org/en/stable/units/}{http://docs.astropy.org/en/stable/units/}],
and of all the pre-defined constants on [\href{http://docs.astropy.org/en/stable/constants/}{http://docs.astropy.org/en/stable/constants/}].

Here is an astropy version of the blackbody luminosity (\ref{LuminosityFunc}) calculation:
\begin{lstlisting}
from astropy import constants as const
from astropy import units as u
Teff = 5780*u.K
R = const.R_sun
# Now the formula:
L = 4*np.pi*R**2*const.sigma_sb*Teff**4
# print the result:
print(L)
\end{lstlisting}
The output this time carries a proper physical unit, namely
\begin{lstlisting}
3.84713215453e+26 W
\end{lstlisting}
Let's see what we did there. $T_{\mathrm{eff}}$ this time was defined not as a pure number, but multiplied with the appropriate unit, namely \verb|u.K| for the unit K, Kelvin. The solar radius $R_{\odot}$ is accessible via the astropy constant module, as \verb|const.R_sun|. Similarly, the Stefan-Boltzmann constant is \verb|const.sigma_sb|.

We can also force the program to convert the result to specific, different units. For instance, the solar radius in km (instead of m) can be obtained by writing
\begin{lstlisting}
(const.R_sun).to(u.km)
\end{lstlisting}
which returns
\begin{lstlisting}
695 508 km
\end{lstlisting}
The conversion function \verb|to| is appended to the result (in this case, enclosed in parantheses); its argument is the target unit to which the expression is to be converted, in this case \verb|u.km|.

If the target unit is not equivalent to the proper physical unit of the expression, for instance if we try to ``convert'' a length to Kelvin, we get an error message like ``UnitConversionError: 'm' (length) and 'K' (temperature) are not convertible''.

\subsection{Random numbers}
\label{Random}

Sometimes, you will need random numbers -- to pick a random sample from a larger subset, for instance, or to randomly place particles at the beginning of a simulation. The random module in Python provides functions for this. Once imported with
\begin{lstlisting}
import random
\end{lstlisting}
you have a whole suite of random functions at your disposal. One is
\begin{lstlisting}
random.random()
\end{lstlisting}
which returns a random floating point number $x$ with $0\le x <1$. Scale this up and add an offset to obtain a random floating point number from any other required range of values. The function
\begin{lstlisting}
random.randint(p,q)
\end{lstlisting}
returns a random integer $i$ with $p\le i\le q$. Finally, to create a random subset of k elements from a list \verb|thisList|, call
\begin{lstlisting}
random.sample(thisList,k)
\end{lstlisting}
as in the example
\begin{lstlisting}
In: thisList = range(10)

In: random.sample(thisList,5)
Out: [5, 9, 7, 1, 3]
\end{lstlisting}
You can also draw random values from specific probability distributions. Notably,
\begin{lstlisting}
random.gauss(mu,stdDev)
\end{lstlisting}
will draw a random number from a Gaussian (normal) distribution with mean \verb|mu| and standard deviation \verb|stdDev|.

\subsection{Strings}
\label{Strings}

Strings are important whenever we are producing text output, or need to load files following a certain naming convention. If you want to concatenate strings, you can just use the plus sign:
\begin{lstlisting}
In: 'ABC' + 'def'
Out: 'ABCdef'
\end{lstlisting}

A useful tool for building strings out of numbers and other variables in Python 3 is the \verb|format| function, which works as follows: You create a string that includes a placeholder built from curly brackets. The string is followed by \verb|.format()|, where the parentheses enclose those variables that are to be substituted for the placeholders. For example, to insert an integer value into a file name, you can do this:
\begin{lstlisting}
In: thisInt = 10
In: 'egon{}.jpg'.format(thisInt)
Out: 'egon10.jpg'
\end{lstlisting}
Python sees the placeholders in the string, deduces ``ah, I need to insert a value here'' and looks for the variable whose value is to be inserted among the arguments of the format function. 

There are several useful conventions. If we want to pad our integer with a certain number of zeroes, do this:
\begin{lstlisting}
In: thisInt = 10
In: 'egon{:03d}.jpg'.format(thisInt)
Out: 'egon010.jpg'
\end{lstlisting}
This tells Python that the integer in question should always have 3 digits, and if it is too short, python should add zeroes on the left. To display floating point numbers, use \verb|{:f}|:
\begin{lstlisting}
In: thisFloat=1.51515151515151515
In: 'This is a floating point number: {:f} (see?)'.format(thisFloat)
Out: 'This is a floating point number: 1.515152 (see?)' 
\end{lstlisting}
Note that the string only includes 6 decimal places, and that the final digit has been rounded. If you want more decimal places, you can tell python like this:
\begin{lstlisting}
In: thisFloat=1.51515151515151515
In: 'This is a floating point number: {:.9f} (see?)'.format(thisFloat) 
Out: 'This is a floating point number: 1.515151515 (see?)'
\end{lstlisting}
For scientific formatting with an exponential, use \verb|{:e}|,
\begin{lstlisting}
In: thisFloat=151515.1515
In: 'This is a floating point number: {:e} (see?)'.format(thisFloat) 
Out: 'This is a floating point number: 1.515152e+05 (see?)'
\end{lstlisting}
Also, there is a placeholder  that only switches to exponentials for numbers smaller than $10^{-4}$:
\begin{lstlisting}
In: thisFloat=0.0001515
In: 'This is a floating point number: {:g} (see?)'.format(thisFloat)
Out: 'This is a floating point number: 0.0001515 (see?)'
In: thisFloat=0.00001515
In: 'This is a floating point number: {:g} (see?)'.format(thisFloat)
Out: 'This is a floating point number: 1.515e-05 (see?)'
\end{lstlisting}
All these and many more options are listed in the Python documentation, for instance for version 3.1 at [\href{https://docs.python.org/3.1/library/string.html#format-examples}{https://docs.python.org/3.1/library/string.html\#format-examples}].

From Python 3.6 onwards, there is an even simpler way of formatting numbers to yield strings, so-called ``f-strings''. Reconsider the last example. We can re-write it as
\begin{lstlisting}
In: thisFloat=0.0001515
In: f'This is a floating point number: {thisFloat:g} (see?)'
Out: 'This is a floating point number: 1.515e-05 (see?)'
\end{lstlisting}
The string now has an ``f'' in front of the quotation marks, telling Python that this is an f-string in such an f-string, expressions in curly numbers are interpreted as formatting instructions. Where previously we had \verb|{:g}| and provided the information about the variable \verb|thisFloat| via the \verb|format| function, we now have written the variable name directly into the curly brackets, as \verb|{thisFloat:g}|. The meaning of alternative formatting instructions is the same as in the previous examples we have seen for \verb|.9f|, \verb|e|, \verb|03d| and similar. 

There is one additional thing to keep in mind, since it might come in handy: Strings can be addressed as lists of characters, so lots of tricks we will talk about in section \ref{Lists} when we will have a closer look at lists are applicable to strings as well. 

For instance, if you only want to use the fourth through sixth character of a string, you would do a slice using square brackets, as with any list, like this:
\begin{lstlisting}
In: thisString='ABCDEFGHIJKLMNOP'
In: thisString[3:6]
Out: 'DEF'
\end{lstlisting}

As an astronomical example, we will automatically create a URL to download a spectrum from SDSS data release 8. Each SDSS observation of the type we are interested in took up to a few hundreds of spectra simultaneously. To this end, an aluminium plate was placed in the telescope's focal plane, with a pattern of holes made to measure for the observing field in question. In each hole, a light-conducting fiber that captures the light from a specific object, conducting it to a spectrograph. Since some plates were used more than once, we need also specify the date of the observation, using the Modified Julian Date (MJD) common in astronomy. Specifying plate number, fibre number and date picks out the spectrum of a specific object. As the SDSS software evolves, spectra are sometimes re-analysed, so we need to specify the number of the reduction run --- when the raw data was sent through a specific software pipeline to yield a reduced spectrum, suitable for astrophysics. To retrieve a specific spectrum from the SDSS server, we need combine those numbers into a suitable custom URL for download. We do this following the recipe given by SDSS, as follows:

\begin{lstlisting}[breaklines=true]
run2d=26
plate = 1324
mjd = 53088
fiberID = 456
baseURL = 'http://data.sdss3.org/sas/dr8/sdss/'
dirURL=f'spectro/redux/{run2d:d}/spectra/{plate:d}/'
fileURL="spec-{plate:d}-{mjd:d}-{fiberID:04d}"
url=baseURL+dirURL+fileURL+'.fits'
\end{lstlisting}
The resulting URL is\\{\footnotesize
\href{http://data.sdss3.org/sas/dr8/sdss/spectro/redux/26/spectra/1324/spec-1324-53088-0456.fits}{\tt http://data.sdss3.org/sas/dr8/sdss/spectro/\\redux/26/spectra/1324/spec-1324-53088-0456.fits}
}
Try and paste it into your browser, or click the link!

A simple way to actually do the download, at least on Mac or Linux computers with curl installed, would be a system call
\begin{lstlisting}
from subprocess import call
saveFileName= "spectrum.fits"

call(["curl", "-o", saveFileName, url])
\end{lstlisting}
This downloads the file to the working directory, where it will be saved as ``spectrum.fits''.

\subsection{Conditions}
\label{Conditions}
An important part of what makes coding so versatile are structures that allow you to let your program make decisions, based on the available data. For instance, you could write a script like this:
\begin{lstlisting}
a=10
if a>1:
	print("a is bigger than one!")
\end{lstlisting}
which, if you write it in window A and let it run, will return "a is bigger than one!" in window C. If, on the other hand, you set \verb|a=0| (or any other value that is not bigger than one) and run the script, it will not print anything.

Note that the actions that should happen if he condition is fulfilled, in this case the print statement, are indented (either by using tab or by putting four blank spaces in front of it). In Python, such indentation is required. This is how Python knows that these statement belong to the ``if'' block, and are only to be executed if the given condition is true.

Some decisions involve an alternative: If the condition is true, do this, if it is not true, do that other thing. This is what the if\dots else construction is for:
\begin{lstlisting}
a=0
if a>1:
	print("a is bigger than one!")
else:
	print("a is not bigger than one!")
\end{lstlisting}
Try it! There is an additional keyword called elif, which allows you to differentiate further. Check, by changing the values of \verb|a|, that this little script is indeed telling people the truth about the variable:
\begin{lstlisting}
a=0
if a>1:
	print("a is bigger than one!")
elif a==1:
	print("a is equal to one!")
else:
	print("a is smaller than one!")
\end{lstlisting}
Note that for the "a is equal to one" we have used not the equals-sign \verb|=| but a double equals sign \verb|==|. This is because 
\verb|a=1| would define a as being equal to one. The equals-sign, after all, is used to assign values to variables.

\subsection{User-defined functions}
\label{Functions}

In many situations, functions will come in handy. If I need to perform the same operation repeatedly on various variables, it makes sense to not repeat writing down all the steps of the operation again, and again, and again. Instead, we can define a function comprising these steps; whenever we need to perform the operation, we apply that function.

For instance, assume that there is a specific polynomial function
\be
\label{PolyFunc}
f(x) = 2x^2 - x,
\ee
which occurs in our analysis again and again. We need to apply this function first to a variable \verb|a|, then to a variable \verb|b|.

Writing down the polynomial explicitly each time we need it is rather cumbersome. Instead, we can define a function using \verb|def|, and apply that function twice:
\begin{lstlisting}
def polyFunc(x):
	return 2*x**2-x

a=2
b=4
print(polyFunc(a))
print(polyFunc(b))
\end{lstlisting}
Write all this down in window A and run it. In window C, you will see the results 
\begin{lstlisting}
6
28
\end{lstlisting}
--- our polynomial function applied to a and to b, respectively. Let's look a bit closer at this. The first two lines define the function, as
\begin{lstlisting}
def polyFunc(x):
	return 2*x**2-x
\end{lstlisting}
The \verb|def| keyword is followed by the name we have chosen for the function, in this case polyFunc. After the name, in parentheses, follows the list of arguments for the function. Our function will have one argument, which we have given the internal name x. Then follows the return statement, which contains what the function will report back when it is called, in our case the result of what happens when you insert the argument value into the polynomial (\ref{PolyFunc}).

The block of instructions that is called when defining a function can be rather long and complex. Here is a comparatively simple example, which returns the square root of a positive number, and zero if the argument is negative:
\begin{lstlisting}
import numpy as np

def zeroSqrt(x):
	if x<0:
		return 0
	else:
		return np.sqrt(x)
		
print(zeroSqrt(-1))
print(zeroSqrt(4))
\end{lstlisting}
If you put this in window A and run it, the result in window C will be
\begin{lstlisting}
0
2.0
\end{lstlisting}
as expected. Note the two levels of indentation here: The if and the else are indented because they are part of the function definition. The two returns are indented double, because they are subservient to the if and the else condition, respectively.

Functions can have multiple arguments. The following function will return the sum of its three arguments, specified in parentheses as x, y and z:
\begin{lstlisting}
def sumOfThreeArgs(x,y,z):
	return x+y+z
	
sumOfThreeArgs(1,20,300)
\end{lstlisting}
When executed, this little script will duly return 321. Functions can also return more complex constructs, such as lists, tuples or arrays. They can return all types of variables, or combinations thereof.

There is an alternative way of defining functions, which allows for more compact scripts at least for simple functions. It uses the keyword \verb|lambda|, which echo's mathematics' formal system of lambda calculus (which is a rather formal and abstract system for describing computations). The syntax is as follows:
\begin{lstlisting}
polFunc = lambda x: 2*x**2-x
print(polFunc(2))
\end{lstlisting}
This is a one-line-definition for our function: the keyword lambda, then the variable (or, separated by commas, variables) and to the right what would be the statements following the return keyword. Run this, and it will dutifully return the value 6.

One general remark: While you can make a function manipulate existing variables, or make them define variables that are accessible by the rest of your script afterwards, I would strongly (!) recommend that you separate your function cleanly from the rest of your script. 

Let the only information the function receives from the outside be the function arguments. Let the only information anything else receives from the function be the object after the return statement. 

For professional software, which is typically quite complex and can run into the tens of millions of lines of code, written by teams of developers, separating the code into independent sections, each with a well-defined interface to the rest of the program, is a must.

\subsection{Timing your code}
\label{Timing}

Once we get into the realm of more complicated code, there are occasions when execution time will start to matter. Programming something in one way, or another, can make the difference between waiting a few minutes for your result, or a few hours. Where the differences are that stark, you are likely to notice them right away.

Should you want to quantify runtime more precisely, you can use Python's time module. The function \verb|time()| will return the number of seconds that have passed since an operating-system-specific zero point (in UNIX, January 1, 1970). By calling the function once before and once directly after a certain point of your script, you can keep track of what takes how long, for instance:
\begin{lstlisting}
import time

start_time=time.time()
for ii in range(1000000):
	pass
end_time=time.time()

print("This took {} seconds!".format(end_time-start_time))
\end{lstlisting}
Try it once you have read the next section. The difference in speed between going through the elements of a list and performing an operation on them on the one hand, and using the numpy array functionality on the other, is quite impressive. 

\section{Taming long data sets: Lists in Python}
\label{Lists}

\subsection{A list of galaxies}

Imagine that we have 9 galaxies. For each galaxy, we know its brightness. In Python, the list of all these brightness values can be stored as an object known as, unsurprisingly, a {\em list}. Such a variable has a single name, just like a variable that only stores a single value. 

Evidently, we will need a way of defining the variable that lets Python know that it is dealing with a list, not a single value. And once the variable is defined, we will need a way of retrieving the different list entries individually. Let's see how this works using a specific example.

For instance you can define a list with the variable name \verb|galaxy_u| like this:
\begin{lstlisting}
galaxy_u = [23.4, 23.2, 26.8, 24.6, 24.5, 24.3, 23.1, 27.0, 24.0]
\end{lstlisting}
In some ways, this looks similar to the way you would write a list by hand, separating the different items using a comma. In this case, too, the comma tells the computer where the next item begins. The list as a whole is enclosed in square brackets.

If you now enter \verb|galaxy_u| in the IPython console and press return, you will obtain the whole list:
\begin{lstlisting}
In: galaxy_u
Out:
[23.4,
23.2, 
26.8, 
24.6, 
24.5, 
24.3, 
23.1, 
27.0, 
24.0]
\end{lstlisting}
What if you want to retrieve a specific item? Even though this is not shown explicitly, all the elements in this list are numbered, starting with zero. Element 0 has the value 23.4, element 1 has the value 23.2, element 2 the value 26.8 and so on. To retrieve a single element, simply add the element's number in square brackets to the variable name. Like this, as entered in the IPython console:
\begin{lstlisting}
In: galaxy_u[2]
Out: 26.8
\end{lstlisting}
Ask for the element with index 2, and you get the element in the third place of the list (since the first index is zero). Of course, you don't need to put a numerical value in there. It could be any integer variable \verb|i|, and \verb|galaxy_u[i]| would return to you the list element with the index value \verb|i|.

You can also apply some basic functions to a list. For instance, \verb|max(galaxy_u)| stands for the largest element of the list, in our case
\begin{lstlisting}
In: max(galaxy_u)
Out: 27.0
\end{lstlisting}
In the same way, you can get the smallest element:
\begin{lstlisting}
In: min(galaxy_u)
Out: 23.1
\end{lstlisting}
Another interesting property is the number of elements in the list. Use the \verb|len| function here:
\begin{lstlisting}
In: len(galaxy_u)
Out: 9
\end{lstlisting}
In this case, you can check the answer by hand: yes, this particular list has 9 entries.
 
Last but not least, using a procedure called {\em slicing}, you can obtain parts of the list. For instance, \verb|galaxy_u[2:5]| will return everything from the element with index $2$ up to and including the element with index $4=5-1$ as a smaller list:
\begin{lstlisting}
In: galaxy_u[2:5]
Out: [26.8, 24.6, 24.5]
\end{lstlisting}
Leave out the index before the colon, and your sublist will start with the first element:
\begin{lstlisting}
In: galaxy_u[:5]
Out: [23.4, 23.2, 26.8, 24.6, 24.5]
\end{lstlisting}
And, even more useful, if you leave out the index {\em after} the colon, the result will automatically include all elements up to and including the last list element:
\begin{lstlisting}
In: galaxy_u[5:]
Out: [24.3, 23.1, 27.0, 24.0]
\end{lstlisting}
You can als count off the final included element from the end, using a minus sign. For instance, this here gives you everything from the element with index 5 up to end including the next-to-last element:
\begin{lstlisting}
In: galaxy_u[5:-1]
Out: [24.3, 23.1, 27.0]
\end{lstlisting}

Finally, let us talk about various ways of changing a list. Appending an additional element to the end of the list is easy:
\begin{lstlisting}
In: galaxy_u.append(25.1)
In: galaxy_u
Out: [23.4, 23.2, 26.8, 24.6, 24.5, 24.3, 23.1, 27.0, 24.0, 25.1]
\end{lstlisting}
We can also remove the last element from a list. This is what \verb|pop| will do: apply \verb|pop| and the result will be the rightmost element of the list. But the list itself will also have been modified: the last element will have been removed, as we can see when we enter the list's name directly after the popping has been completed:
\begin{lstlisting}
In: galaxy_u.pop()
Out: 25.1
In: galaxy_u
Out: [23.4, 23.2, 26.8, 24.6, 24.5, 24.3, 23.1, 27.0, 24.0]
\end{lstlisting}

\subsection{Doing something element by element}

Oftentimes, you need to apply some function or operation to each list element separately. For instance, in the case of galaxies from the SDSS catalogue (Sloan Digital Sky Survey), the u-filter magnitude $m_u$ is related to the flux $f_u$ (energy received from the galaxy per unit frequency interval per unit time per unit receiving area) as
\be
f_u =3631 \cdot 10^{m_u/(-2.5)} \;\mbox{Jy}.
\label{fromuMagtoFlux}
\ee
I am skipping over some complications to keep things simple, and it doesn't matter if you have not encountered the unit ``Jansky,'' abbreviated Jy, before.\footnote{If you want to read up on the gory details, go to the SDSS web pages, in particular to their magnitude and flux explanations on [\href{http://www.sdss.org/dr13/algorithms/magnitudes/}{http://www.sdss.org/dr13/algorithms/magnitudes/}] and [\href{http://www.sdss.org/dr12/algorithms/fluxcal/}{http://www.sdss.org/dr12/algorithms/fluxcal/}].} What matters is that every value in our list corresponds to a u magnitude $m_u$, so for every value we want to use formula (\ref{fromuMagtoFlux}) to calculate the corresponding flux.

How would we do this in real life? Step by step. We would take the first list entry, perform the calculation described in  (\ref{fromuMagtoFlux}), and note down the result. Then we would do the same with the second list entry, then the third entry, and so on. In the end, we would have noted down a list of results. The $i$th result would be the flux for the $i$th magnitude.

If \verb|u| is some particular magnitude, then from what we have learned in section \ref{BasicCalculations}, using the \verb|numpy| package to define our mathematical functions, we know that the corresponding flux \verb|f| is given by the formula
\begin{lstlisting}
f = 3631*np.power(10,u/(-2.5))
\end{lstlisting}

There are several ways of performing this operation with all the elements of a list. The most straightforward one is close to how we would describe what we want to do in words: For each element \verb|u| in the list \verb|galaxy_u|, we want to calculate
\verb|f= 3631*np.power(10,u/(-2.5))| and then put the result in some new list, let's call it \verb|galaxy_f|. This is the actual code:
\begin{lstlisting}
import numpy as np
galaxy_f=[]
for u in galaxy_u:
	f=3631*np.power(10,u/(-2.5))
	galaxy_f.append(f)
\end{lstlisting}
The first row defines an empty list \verb|galaxy_f|, which has no elements and thus is no more than square brackets enclosing nothing whatsoever. Then, we tell the code to perform the following (indented) operation for each element \verb|u|. 

The code will repeat what is in the indented block as many times as there are elements in our list, each time on a different element, working its way systematically from the beginning to the end of the list. 

Each time we have calculated the flux \verb|f| for a particular element, we append the result to the end of our list \verb|galaxy_f|. When the code has successfully performed the operation on each element of \verb|galaxy_u|, our resulting flux list \verb|galaxy_f| is complete, and we can have a look at it in the usual way, by typing in the variable name in window C. The result is
\begin{lstlisting}
In: galaxy_f
Out: 
[1.5849889868640447e-06,
 1.9055758881669273e-06,
 6.9187278669245402e-08,
 5.2483918075784649e-07,
 5.7547471818262932e-07,
 6.9187278669245405e-07,
 2.0894224124712161e-06,
 5.7547471818262931e-08,
 9.1206596328112918e-07]
\end{lstlisting}
Looking good! There is another way of solving our problem which can come in helpful, namely using the \verb|map| function. For this, we define the operation we are interested in as a function (cf. section \ref{Functions}). \verb|map| will apply this function to each separate element of the list, collecting the results in a new list: \label{MapFunction}
\begin{lstlisting}
def flux(mag):
	return 3631*np.power(10,mag/(-2.5))
galaxy_f = map(flux, galaxy_u)
\end{lstlisting}
And there is {\em yet} another way of performing this particular task, namely creating a list from another list. The construct in question is called a {\em list comprehension}. Recall the elegant way mathematicians can define sets like $E$, the set of all even integers, like this:
$$
E = \{  \,2n\; |\; \forall n\in \mathbb{Z} \}
$$
In words, we obtain the set of all even numbers by taking the double of all elements in the set of $\mathbb{Z}$ of integers. We've neatly defined an infinite set using just a few symbols and a clever convention. List comprehensions in python work just like that (although they cannot, of course, produce infinite sets). In list comprehension form, the definition of our list of fluxes is one line:
\begin{lstlisting}
galaxy_f = [  3631*np.power(10,u/(-2.5))  for u in galaxy_u ]
\end{lstlisting}
The expression written within the square brackets is evaluated for every \verb|u| in the list \verb|galaxy_u|. If you only want results satisfying a certain condition, you can add an \verb|if| block at the end. For instance, if we only want to include galaxies for which
$u < 25$, we could write
\begin{lstlisting}
galaxy_fb = [  3631*np.power(10,u/(-2.5))  for u in galaxy_u  if u < 25]
\end{lstlisting}
which results in 
\begin{lstlisting}
In: galaxy_fb
Out: 
[1.5849889868640447e-06,
 1.9055758881669273e-06,
 5.2483918075784649e-07,
 5.7547471818262932e-07,
 6.9187278669245405e-07,
 2.0894224124712161e-06,
 9.1206596328112918e-07]
\end{lstlisting}

\subsection{Operations involving more than one list}

More often than not, when we calculate something, it involves more than one property of an object. Consider a list \verb|st_appV| of apparent magnitudes in the V band of several stars, and a list \verb|st_distPc| containing each star's distance from us in parsec:\footnote{I took these values from the Wikipedia list at the URL
[\href{https://en.wikipedia.org/wiki/List_of_nearest_bright_stars}{https://en.wikipedia.org/wiki/List\_of\_nearest\_bright\_stars}] -- they are
for Sirius, 61 Cygni A, $\tau$ Ceti, and Altair, respectively.}
\begin{lstlisting}
st_appV = [-1.46, 5.2, 3.49, 0.76]
st_distPc = [2.64, 3.5, 3.65, 5.12]
\end{lstlisting}

We want to calculate each star's absolute magnitude, using the formula relating the apparent magnitude $m$, absolute magnitude $M$, and distance $d$ as
\be
M = m - 5\cdot\log_{10}\left(\frac{d}{10\;\mbox{pc}}\right).
\label{absMag}
\ee
This time, we need to insert not one, but two properties of the star on the right-hand side: its apparent magnitude and its distance. We need to go through two lists at the same time. How do we do that?

The inelegant way would be to just go through the elements by calling them directly, using their indices. After all, the apparent magnitude \verb|st_appV[0]| and the distance \verb|st_distPc[0]| belong to the same star (in this case, to Sirius); analogously, the apparent magnitude \verb|st_appV[1]| and the distance \verb|st_distPc[1]| belong to the same star (in this case, to 61 Cygni A). Thus, we can make a for loop of indices like this:

\begin{lstlisting}
import numpy as np
st_absV=[]
for i in [0,1,2,3]:
	thisM = st_appV[i] - 5*np.log10(st_distPc[i]/10.0)
	st_absV.append(thisM)
\end{lstlisting}
Again, we begin by creating an empty list \verb|st_absV|. Then, we let \verb|i| take on all of the index values of the lists \verb|st_appV| and \verb|st_distPc|, one after the other. For each index value, we fetch the appropriate items from each list and combine them as required by the formula (\ref{absMag}).

There is one thing that is particularly awkward about this solution, and that is writing out the index values \verb|[0,1,2,3]| by hand. Once we get to longer lists, this will no longer work; also, what about cases where we do not know, beforehand, how long the lists we are processing will be?

The solution is the function \verb|range|. With a single integer argument $n$, it produces a list with $n$ values, containing integer values from 0 to $n-1$, for instance:
\begin{lstlisting}
In: range(5)
Out: [0, 1, 2, 3, 4]
\end{lstlisting}
With two arguments, you can make the list begin with a value other than zero:
\begin{lstlisting}
In: range(2,5)
Out: [2, 3, 4]
\end{lstlisting}
Combined with the length function, which gives you the number of elements in a list, we can use \verb|range| to make sure our for-loop runs over all the elements in the list \verb|st_absV|, which is of course the same number of elements as in \verb|st_distPc|. The result is
\begin{lstlisting}
import numpy as np
st_absV=[]
for i in range(len(st_appV)):
	thisM = st_appV[i] - 5*np.log10(st_distPc[i]/10.0)
	st_absV.append(thisM)
\end{lstlisting}
Using list comprehensions, we can again make this operation much shorter and simpler. In list comprehension form, we only need to write
\begin{lstlisting}
import numpy as np
st_absV = [ m - 5*np.log10(d/10.0) for m,d in zip(st_appV,st_distPc) ]
\end{lstlisting}
The difference is now there are two variables, \verb|m| and \verb|d|, in the for loop. The secret is in the \verb|zip|. Think about closing an ordinary zipper, e.g. about zipping your bag. Where before, there were two separate rows of teeth, zipping combines these rows, so in the closed zipper, there is one tooth from the left side, then one from the right side, and so on. The zip function is somewhat similar. Where, initially, we have two lists, 
\verb|st_appV| and \verb|st_distPc|, zip combines these into a single list where each entry has two values. Assume that the initial lists are
\begin{lstlisting}
lst_appV = [-1.46,5.2,3.49,0.76]
\end{lstlisting}
and
\begin{lstlisting}
lst_distPc = [2.64,3.5,3.65,5.12].
\end{lstlisting}
You should think about these lists as listing different properties of the same astronomical objects. The first list contains the visual brightness V of each object. The second list contains each object's distance from us, in parsec. So, for instance, the first object in question has visual brightness $-1.46$ and is at a distance of $2.64$ parsec from us, the second object has visual brightness $5.2$ and is at a distance of $3.5$ parsec from us, and so on. If we now apply the zip command, the two lists are combined as follows:
\begin{lstlisting}
In: zip(st_appV,st_distPc)
Out: [(-1.46, 2.64), (5.2, 3.5), (3.49, 3.65), (0.76, 5.12)]
\end{lstlisting}
Again we have a list with one entry for each object. But now that entry is itself a list-like entity: For the first object, that list-like entity contains first that object's visual brightness -1.46, and in second place the object's distance in parsec, 2.64.

In Python, the objects that look like little lists, but have round instead of square brackets, are called {\em tuples}. Lists are mutable -- as we have seen, once a list is created, you can remove items, or append items, changing the number of items in the list. Tuples, once defined, need to stay the same length. Python allows for assignments like this, mixing tuples (or lists, for that matter) and ordinary variables:
\begin{lstlisting}
m, d = (-1.46, 2.64)
\end{lstlisting}
Here, \verb|m| and \verb|d| are ordinary variables, each capable of storing a single values, while on the right-hand side, there is a tuple. Two ordinary variables on the left, a tuple with two values on the right --- that makes for an unambiguous assignment. After this assignment, the variable \verb|m| holds the value $-1.46$ while the variable \verb|d| holds the value $2.64$. 

With this background information, the list comprehension version of our calculation is straightforward to understand: With \verb|zip|, we transformed our two lists, one with all the apparent magnitudes, the other with all the distances, into a single list of tuples. Each of the tuple entries consists of two items: the apparent magnitude of a single star and its distance. 

The for loop iterates over all these tuples, one for each star, and for each star we assign to \verb|m| that star's apparent magnitude, to \verb|d| that star's distance. Then we use both \verb|m| and \verb|d| in the formula for calculating the absolute magnitude.

While our example has used two lists, the \verb|zip| works with any number of lists. For fun, we can try combining the list for the apparent magnitude, the list of distances, and the list of absolute magnitudes that we have just produced:
\begin{lstlisting}
In: zip(st_appV,st_absV,st_distPc)
Out: [(-1.46, 1.431980365650845, 2.64),
 (5.2, 7.4796597782486227, 3.5),
 (3.49, 5.6785356777176261, 3.65),
 (0.76, 2.213650195120846, 5.12)]
\end{lstlisting}
Again we have produced some kind of master list, with one entry per star, but this time each entry is a tuple with three items: first, the star's apparent magnitude, second, its absolute magnitude, and third, its distance in parsec. We can use zip to, well, zip together any number of lists --- as long as all these lists each have the same length.

\subsection{Creating lists simultaneously}

\label{firstGalaxyEx}
In the previous section, we have used tuples, and zip, to create a list that depended on two, three, or any other lists. We can use those same tools to create several lists simultaneously. The simplest application is that we have several lists referring to the same objects --- the first entry in each list refers to one specific object, the second entry in each list to the second object, and so on --- and that we want to filter these lists according to some criterion. As an example, consider these three lists that contain catalogue numbers, distances (in parsec) and redshifts $z$ for several galaxies:\footnote{
The galaxies are taken from NASA's extragalactic data base (NED), specifically its list of redshift-independent distances for galaxies, \href{https://ned.ipac.caltech.edu/Library/Distances/}{https://ned.ipac.caltech.edu/Library/Distances/}}
\begin{lstlisting}
nm = ['SDSS-II SN 21387','SDSS-II SN 13651','SDSS-II SN 03706','SDSS-II SN 10963','SDSS-II SN 03475']
dpc=[4200.,1700.,3720.,577.,1040.]
zv=[0.48,0.25,0.44,0.09,0.3]
\end{lstlisting}
On this basis, our goal is to create three new lists, call them \verb|nmN| and \verb|dpcN| and \verb|zvN|, which contain only those of the galaxies with a redshift greater than $z=0.26$.

Let us try a workable but somewhat wordy code first: iterating over the (common) index values for these lists, filtering accordingly, and filling up our new lists by appending each suitable value. This can be achieved by
\begin{lstlisting}
nmN=[]
dpcN=[]
zvN=[]
for i in range(len(nm)):
	if zv[i] > 0.26:
		nmN.append(nm[i])
		dpcN.append(dpc[i])
		zvN.append(zv[i])
\end{lstlisting}
In the end, we have three new lists, containing the values for those three of the five galaxies that meet our criterion:
\begin{lstlisting}
In: nmN
Out: ['SDSS-II SN 21387', 'SDSS-II SN 03706', 'SDSS-II SN 03475']
In: dpcN
Out: [4200.0, 3720.0, 1040.0]
In: zvN
Out: [0.48, 0.44, 0.3]
\end{lstlisting}
But once more, there is a more elegant solution using a list comprehension:
\begin{lstlisting}
nmN,dpcN,zvN = zip(*[ (n,d,z) for n,d,z in zip(nm,dpc,zv) if z> 0.26 ])
\end{lstlisting}
The expression \verb|zip(nm,dpc,zv)| again zips the three lists into a single list where each item is a tuple containing that galaxy's name, distance and redshift. For instance, the first tuple in that list is \verb|('SDSS-II SN 21387', 4200.0, 0.48)| combining the three properties of the first galaxy. For each iteration --- where we look up \verb|n,d,z| for one particular galaxy --- we do not calculate a single result, but instead tell Python to add a tuple \verb|n,d,z| to the list of results. The last part with the condition makes sure that this only happens  if our condition is met, that is, if $z>0.26$. The combination \verb|zip(* |[\dots]\verb| )| inverts the zipping. Before we applied  this combination, we were still dealing with a single list, containing one tuple per galaxy:
\begin{lstlisting}
In: [ (n,d,z) for n,d,z in zip(nm,dpc,zv) if z> 0.26 ]
Out: [('SDSS-II SN 21387', 4200.0, 0.48),
 ('SDSS-II SN 03706', 3720.0, 0.44),
 ('SDSS-II SN 03475', 1040.0, 0.3)]
\end{lstlisting}
The zip function combined with the star operation unzips this object and returns three tuples, the first of which contains all the galaxy names, the second all the galaxy distances, and the third all the galaxy redshifts:
\begin{lstlisting}
In: zip(*[ (n,d,z) for n,d,z in zip(nm,dpc,zv) if z> 0.26 ])
Out: [('SDSS-II SN 21387', 'SDSS-II SN 03706', 'SDSS-II SN 03475'),
 (4200.0, 3720.0, 1040.0),
 (0.48, 0.44, 0.3)]
 \end{lstlisting}
If we assign this expression to three variable names, separated by comma, Python automatically assigns the first tuple to the first variable, the second tuple to the second and so on:
\begin{lstlisting}
nmN,dpcN,zvN = zip(*[ (n,d,z) for n,d,z in zip(nm,dpc,zv) if z> 0.26 ])
 \end{lstlisting}
If we now call up the variables on the left-hand side, separately, we can see that each now contains one of the tuples. For instance, \verb|nmN| now contains the three names of those galaxies that fulfil our condition $z>0.26$:
\begin{lstlisting}
In: nmN
Out: ('SDSS-II SN 21387', 'SDSS-II SN 03706', 'SDSS-II SN 03475')
 \end{lstlisting}
Is it important that we now have tuples where before we had lists? For many purposes, tuples will be fine, and we can just use the resulting tuples. If we want to convert these tuples back to lists, the function \verb|list|, combined with the \verb|map| function (which we encountered in section \ref{MapFunction}, p. \pageref{MapFunction}) can help:
\begin{lstlisting}
nmN,dpcN,zvN = map(list, zip(*[ (n,d,z) for n,d,z in zip(nm,dpc,zv) if z> 0.26 ]))
 \end{lstlisting}
\verb|map| will apply the function \verb|list|, which converts a tuple into a list, to each separate tuple, and as a result, we obtain three lists, for instance
\begin{lstlisting}
In: nmN
Out: ['SDSS-II SN 21387', 'SDSS-II SN 03706', 'SDSS-II SN 03475']
\end{lstlisting}

\subsection{Numpy arrays}
Central to the numpy module is a versatile structure called an array, which is similar to a list but has several nice extra properties. For instance, if we want to create a new array out of several old ones, we can write the formula in question in exactly the same way we would write it for a simple, non-list variable. In order to define a numpy array from scratch, we define a list and transform that list into an array as follows:
\begin{lstlisting}
import numpy as np
a = np.array([1.0,2.0,3.0,4.0])
 \end{lstlisting}
If we want an array in which each value is twice than it was for our original array, simply write
\begin{lstlisting}
b=2*a
 \end{lstlisting}
and the array b will now contain the values 2,4,6 and 8. All arithmetic with numpy arrays is element-wise: Apply functions, exponentiate, add, subtract, multiply --- all this will be done as if you were to apply these operations to each element in the array, and corresponding elements in whatever additional arrays are involved. For instance, with the definitions above, 
\begin{lstlisting}
In: b+1.5*a
Out: array([  3.5,   7. ,  10.5,  14. ])
\end{lstlisting}
In order to calculate the first element of the array of results, python will take the first element of b and add 1.5 times the first element of a. This is repeated for the second elements, and so on until the operation has been performed with all elements of the arrays involved.

\subsection{Variable types, lists, arrays and speed}

We have not looked at different types of variables much, so far. We didn't have to --- Python as a programming language is ``dynamically typed,'' assigning a specific type to a variable at the moment we assign a value to that variable, with no need to declare the type beforehand. If we assign an integer to the variable \verb|a|, then that variable will be an integer variable. If we assign a string to \verb|a|, then from that moment on, \verb|a| will be a string variable (until we possibly change its type again).
\begin{lstlisting}
In: a = 'I am a string'
In: type(a)
Out: str
In: a = 1
In: type(a)
Out: int
 \end{lstlisting}
This flexibility comes at a price, in particular for list operations. One and the same list can contain an integer, a string, a floating point number, another string and so forth. Python needs to carry along the information about what is what, and about which operations can be applied to which list element. This comes at a price; list operations are slower in Python than in a language where you need to declare variable types explicitly, beforehand.

A simple one-dimensional numpy array, on the other hand, can only contain variables of the same type. (If you try to fill it with, say, a floating point number and an integer, it will choose the array type to accommodate all these values at the same time --- in this case, it would become an array of floating point numbers.) You can also force an array to have a specific type, by using the \verb|dtype| keyword. e.g. in 
\begin{lstlisting}
floatArray = np.array([1, 2, 3, 4], dtype='float32')
\end{lstlisting}

That uniformity is one reason why numpy array operations are typically much faster than list operations. The following program takes a list or array consisting of the first million integers and doubles each element:
\begin{lstlisting}
import time
import numpy as np
numberlist=[ii for ii in range(1000000)]
numberarrIt = np.array(numberlist)
numberarr = np.array(numberlist)
start_time=time.time()
for ii in range(len(numberlist)):
	numberlist[ii] = 2*numberlist[ii]
end_time=time.time()
print("List took {:.4g} seconds!".format(end_time-start_time))
start_time=time.time()
numberarr = 2*numberarr
end_time=time.time()
print("Array took {:.4g} seconds!".format(end_time-start_time))
\end{lstlisting}
On running this code, I find that the list operation takes 0.15 seconds, the array operation a mere 0.0012 seconds --- a factor hundred less! (The exact values can vary from machine to machine, and from run to run.) If you are analysing (or simulating) a lot of data, that factor hundred (or whatever it turns out to be in that particular context) can make a substantial difference.\footnote{Additional information about data types can be found in [\href{https://jakevdp.github.io/PythonDataScienceHandbook/02.01-understanding-data-types.html}{https://jakevdp.github.io/PythonDataScienceHandbook/02.01-understanding-data-types.html}].}

\subsection{Strings and base n numbers as lists}

Variables of different types can be transformed into each other. As a non-trivial example, we consider object IDs for the SDSS survey. These are long numbers to begin with; every object that has been identified in a data release of the SDSS has a unique object ID, and every object for which a spectrum has been taken has a spectral object ID, specObjID. Consider the object with the spectral object id 1490816872793270272, which is an elliptical galaxy. For a quick look, use the SDSS DR8 (data release 8) object explorer at
[\href{http://skyserver.sdss.org/dr8/en/tools/explore/obj.asp}{http://skyserver.sdss.org/dr8/en/tools/explore/obj.asp}]. Go to Search by SpecObjId in the menu on the left to call up our object and see an image, spectrum and helpful information.

The \href{http://www.sdss3.org/dr8/glossary.php#S}{DR8 glossary}, entry ``specObjID'', states that the specObjId is a 64 bit number, and that the various bits contain information, namely from left to right, starting with index zero:
\begin{center}
\begin{tabular}{c|lp{0.55\linewidth}l}
Bits & Name & Meaning\\\hline\hline
0--13 & plate & number of spectroscopic plate\\\hline
14--25 & fiberID & number of the glas fiber positioned over the object\\\hline
26--39 & MJD & modified Julian date minus 50000 when observation was made\\\hline
40--53 & rerun2d & reduction run number\\\hline
54--63 &  & only zeros -- no meaning\\
\end{tabular}
\end{center}
How can we extract the information contained in that long number? We begin by putting the specObjID into a suitable variable, and transform that to a binary number, using the function \verb|bin|:
\begin{lstlisting}
specObjID = 1490816872793270272
binVersion=bin(specObjId)
\end{lstlisting}
If you look at binVersion directly, you can see it is actually a string:
\begin{lstlisting}
In: binVersion
Out: '0b101001011000001110010000011000
0010000000000000110100000000000'
\end{lstlisting}
(You will likely see this without the line break, but that's the limitation of type-setting this script.) Let's see how we can transform this into an ordinary integer.
The `0b' in the beginning indicates that what follows is a binary number. Let's get rid of it by treating the string like a list of characters and slicing:
\begin{lstlisting}
In: bin(specObjId)[2:]
Out: '10100101100000111001000001100000
10000000000000110100000000000'
\end{lstlisting}
If you count, e.g. using the len() function, you will see those are only 61 bits. The conversion leaves out any leading zeros. To restore them, we can use the zfill function:
\begin{lstlisting}
In: bin(specObjId)[2:].zfill(64)
Out: '00010100101100000111001000001100
00010000000000000110100000000000'
\end{lstlisting}
This adds leading zeros so that, altogether, we have 64 bits. The plate number is encoded in the first 16 bits. Just as with a list, we can take the appropriate slice:
\begin{lstlisting}
In: binVersion = bin(specObjId)[2:].zfill(64)

In: binVersion[0:16]
Out: '0001010010110000'
\end{lstlisting}
Now, if we use the int function, specifying base 2 as an extra argument, we can convert this into an ordinary integer:
\begin{lstlisting}
In: int(binVersion[0:16],2)
Out: 5296
\end{lstlisting}
That is indeed the object's plate number, written as an integer, as you can confirm by looking at the object explorer. Fiber number, Julian date and the additional information contained in the object id can be extracted in an analogous fashion.

\section{Basic plotting with Python and Matplotlib}

Plots and diagrams are helpful tools for making sense of a given data set. We have already seen some examples in section \ref{HighLevel}, and seen how to make use of TOPCAT for the purpose in sections
\ref{TOPCAThistogram}. Now, it's time to learn how to do the same in Python. For this, and all our subsequent plotting needs, we use the sub-module \verb|matplotlib.pyplot|. The basic setup is very simple, as follows. We import the submodule like this:
\begin{lstlisting}
import matplotlib.pyplot as plt
\end{lstlisting}

\subsection{Plotting a function}

\begin{figure}[htbp]
\begin{center}
\includegraphics[width=\linewidth]{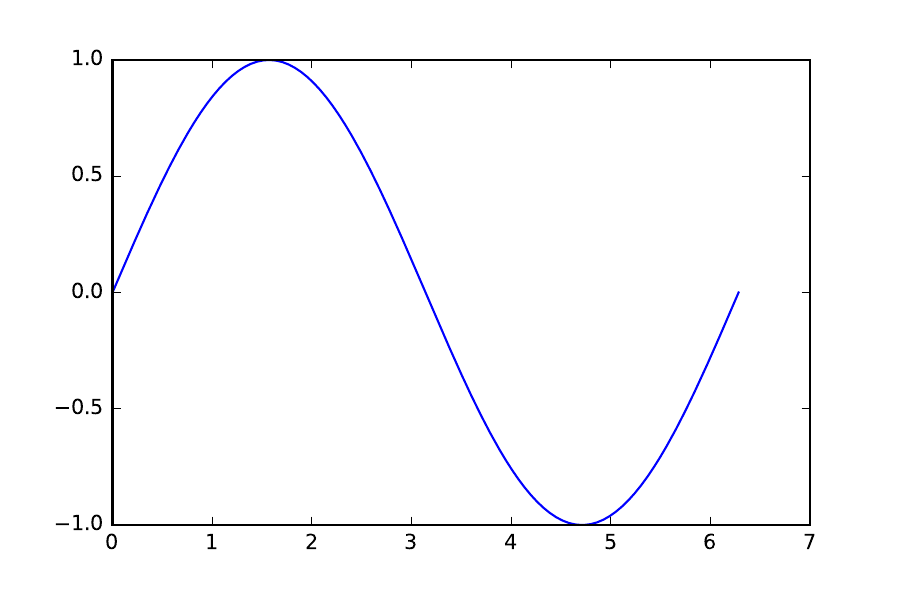}
\caption{Plotting a sine function with matplotlib}
\label{SinPlot}
\end{center}
\end{figure}

In order to plot something, we need lists of x and y values. In this piece of script, we use np.linspace to create a set of 200 points, evenly distributed between (and including) the points $0$ and $2\pi$:
\begin{lstlisting}
import numpy as np
x = np.linspace(0,2*np.pi,200)
\end{lstlisting}
Recall that the nice thing about numpy arrays is that you can write down element-wise operations just using the array variables themselves. Thus,
\begin{lstlisting}
y = np.sin(x)
\end{lstlisting}
calculates the sine function for each element of x and stores all the resulting values in the array y. The most simple plot, involving two lists or arrays, works like this:
\begin{lstlisting}
plt.clf()
plt.plot(x,y)
\end{lstlisting}
Strictly speaking, the \verb|plt.clf()| is not necessary here. It clears all figures or figure elements you might have plotted, or specified, beforehand. I usually start my figures that way, just to be on the safe side. The plot function gets two arguments in this case: a list with x values and one with y values. Both lists need to have the same length. The result is the plot shown in Fig.~\ref{SinPlot}.

\subsection{Making a plot look better}

\begin{figure}[htbp]
\begin{center}
\includegraphics[width=\linewidth]{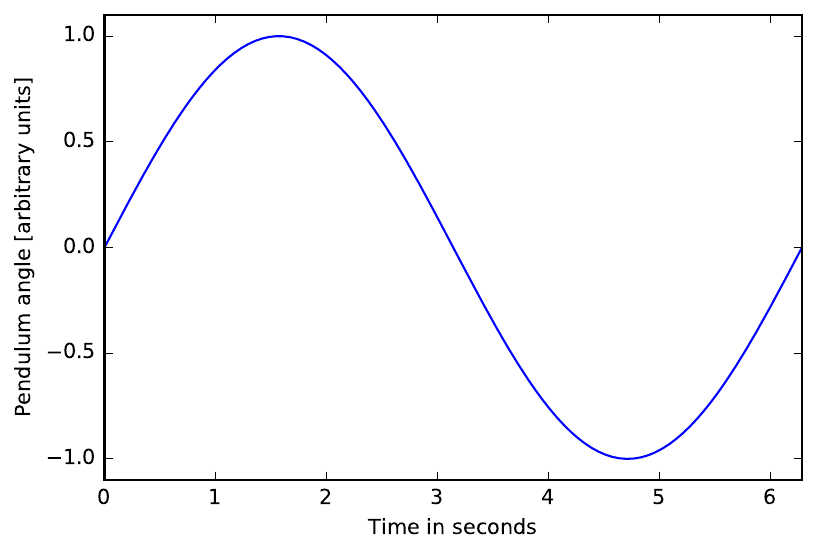}
\caption{Sine curve with custom range and proper axis names}
\label{SinPlotBetter}
\end{center}
\end{figure}

There are many ways that a plot like the last one can be made to look better. For instance, we can introduce axis names, like this:
\begin{lstlisting}
plt.clf()
plt.xlabel('Time in seconds')
plt.ylabel('Pendulum angle [arbitrary units]')
plt.plot(x,y)
\end{lstlisting}
Also, the coordinate region that is shown goes towards slightly bigger x values than necessary --- on the right-hand side, the curve is hanging in the air. In the y direction, on the other hand, we could use a bit more space, since the curve is touching the axis box, and that makes the regions near the maxima and minima less easy to see. We solve both problems by explicitly setting the xlim and ylim ranges, stating both the lowest and the highest value:
\begin{lstlisting}
plt.clf()
plt.xlabel('Time in seconds')
plt.ylabel('Pendulum angle [arbitrary units]')
plt.xlim(0,2*np.pi)
plt.ylim(-1.1,1.1)
plt.plot(x,y)
\end{lstlisting}
The resulting curve, with custom range and axis names, can be seen in Fig.~\ref{SinPlotBetter}.
That's better.

\begin{figure}[htbp]
\begin{center}
\includegraphics[width=\linewidth]{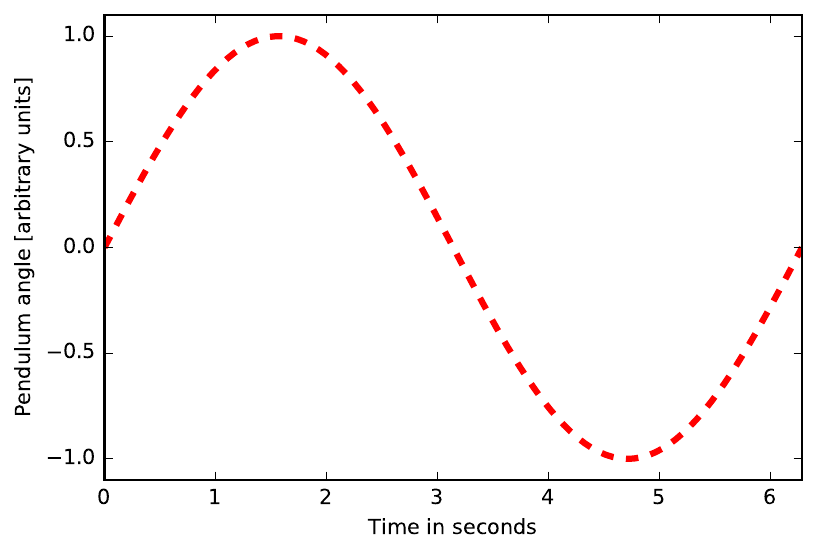}
\caption{Sine curve with different color, line width, and line style}
\label{SinPlotChanged}
\end{center}
\end{figure}

We can make a bewildering number of additional changes. (Of course, it is a sign of plotting maturity when those options are used sparingly, and only when they are in the service of helping the figure's readibility.) Here is an example where we have changed the line color, line width, and line style for our curve:\begin{lstlisting}
plt.clf()
plt.xlabel('Time in seconds')
plt.ylabel('Pendulum angle [arbitrary units]')
plt.xlim(0,2*np.pi)
plt.ylim(-1.1,1.1)
plt.plot(x,y,'r',lw=3.0,linestyle='dashed')
\end{lstlisting}
The result can be seen in Fig.~\ref{SinPlotChanged}.
The changes we have made here: specified the colour 'r', which is red; used the ``lw'' for line width option to set the line width, and set the linestyle to 'dashed' (instead of, say, 'dotted'). Colour conventions can be found on 
[\href{https://matplotlib.org/users/colors.html}{https://matplotlib.org/users/colors.html}]
--- in most situations, it suffices to know that blue, green, red, yellow, cyan, magenta and white can be called up using their initial letters, while black is 'k'.

\subsection{Annotating plots}

\begin{figure}[htbp]
\begin{center}
\includegraphics[width=\linewidth]{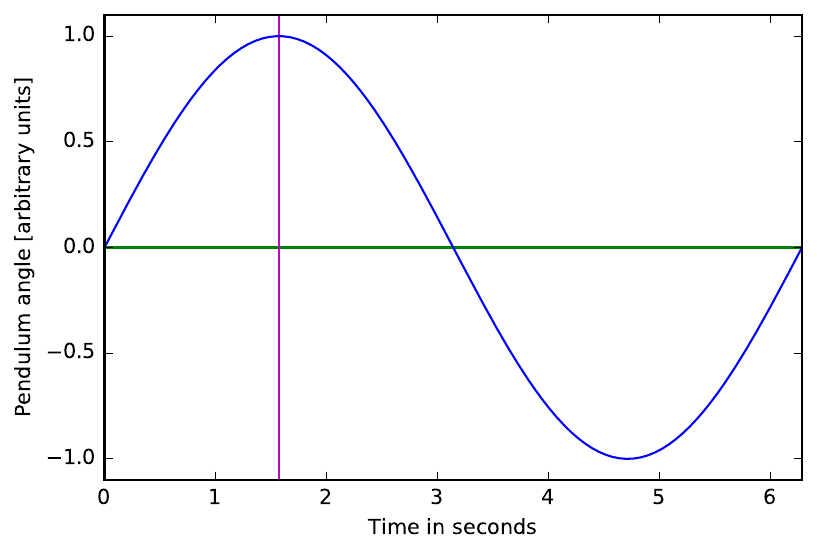}
\caption{Sine curve with horizontal and vertical lines added}
\label{SinPlotLines}
\end{center}
\end{figure}

Sometimes, we want to add straight lines to our plot, to show where certain x or y values are located. This can be done using the axvline and axhline commands, to add a vertical and horizontal line, respectively. As default argument, the vertical line takes a single x value, and the horizontal line a single y value. Here, we have given the lines two different colors, as well:
\begin{lstlisting}
plt.clf()
plt.xlabel('Time in seconds')
plt.ylabel('Pendulum angle [arbitrary units]')
plt.axhline(0.0,color='g')
plt.axvline(0.5*np.pi,color= 'm')
plt.xlim(0,2*np.pi)
plt.ylim(-1.1,1.1)
plt.plot(x,y)
\end{lstlisting}
The result is shown in Fig.~\ref{SinPlotLines}.
Note that, unlike for the function plot, where we could add the line color information by just specifying the color, for axhline and axvline we need to state explicitly \verb|color='g'|. Similar explicit keywords are needed for other specifications, e.g. for the linewidth.

\begin{figure}[htbp]
\begin{center}
\includegraphics[width=\linewidth]{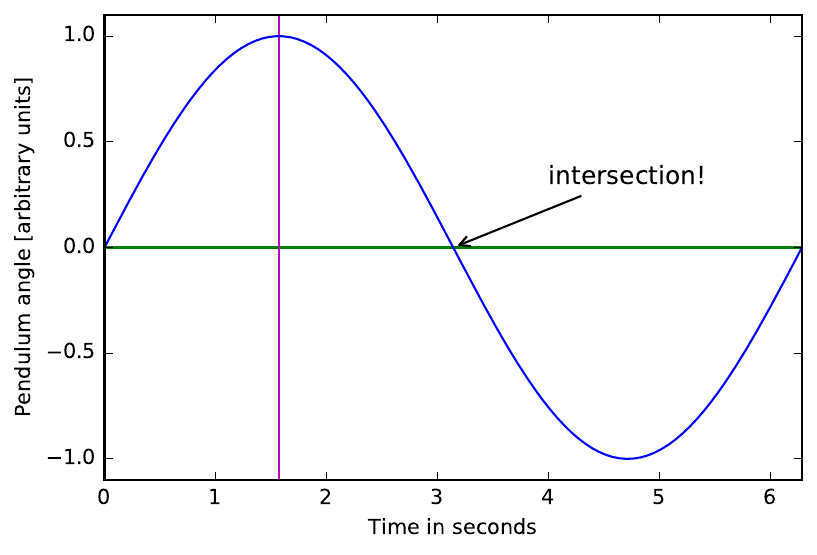}
\caption{Sine curve, with vertical and horizontal lines, and annotated}
\label{SinPlotAnnotated}
\end{center}
\end{figure}

We can also add annotations to a plot: an arrow pointing to some particular feature. The command for this is ``annotate'', and it works as follows:
\begin{lstlisting}
plt.clf()
plt.xlabel('Time in seconds')
plt.ylabel('Pendulum angle [arbitrary units]')
plt.axhline(0.0,color='g')
plt.axvline(0.5*np.pi,color= 'm')
plt.annotate('intersection!', xy=(np.pi, 0), xytext=(4, 0.3),fontsize=12)
plt.xlim(0,2*np.pi)
plt.ylim(-1.1,1.1)
plt.plot(x,y)
\end{lstlisting}
The result is shown in Fig.~\ref{SinPlotAnnotated}.
The command \verb|annotate| has added an annotation. The string argument is the text to be displayed. The xy tuple specifies the coordinate point for where the arrow points, the xytuple where the annotation text should be displayed. The fontsize option gives the size of the font used, in points; you can add this option to pretty much any function which displays text. The arrowprops option specifies the type of arrow to be used. If you leave out the arrow, you can use annotate just to place text somewhere in your diagram, without the arrow.

\subsection{Figure size}

You can change the size of your whole figure by setting the \verb|figsize| property, like this:
\begin{lstlisting}
plt.clf()
plt.figure(figsize=(4,2))
plt.xlabel('Time in seconds')
plt.ylabel('Pendulum angle [a.u.]')
plt.xlim(0,2*np.pi)
plt.ylim(-1.1,1.1)
plt.plot(x,y)
\end{lstlisting}
Both width and height are given in inches and are assigned to the figsize option. In this case, the figure would be twice as wide as it is high. 
\begin{figure}[htbp]
\begin{center}
\includegraphics[width=\linewidth]{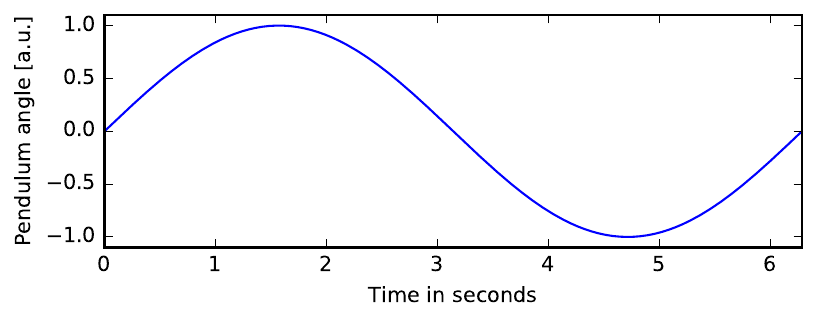}
\caption{Figure with figsize=(6,2)}
\label{FigSize}
\end{center}
\end{figure}
Fig.~\ref{FigSize} shows the same principle for \verb|figsize=(6,2)|. Note that font sizes do not scale with figure size. If you make your figsize larger, text labels and axis labels will be smaller relative to the overall size of the figure.  

\subsection{Scatter plots}

\begin{figure}[htbp]
\begin{center}
\includegraphics[width=\linewidth]{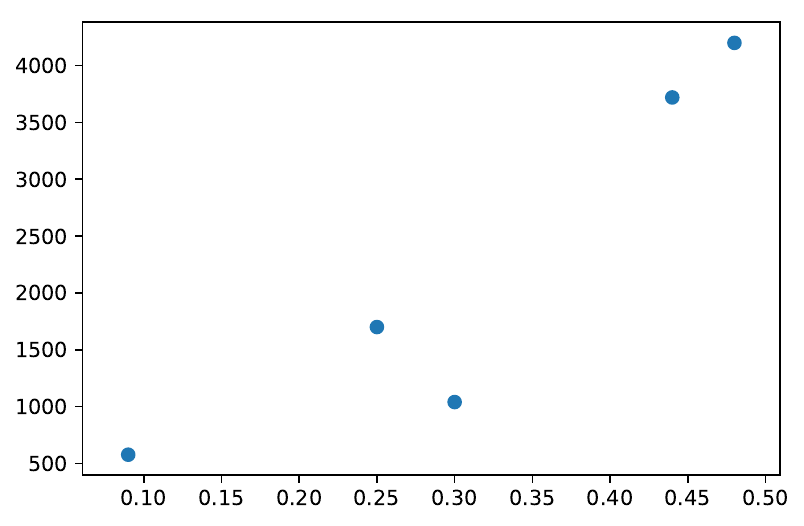}
\caption{Simple Hubble plot}
\label{HubblePlot1}
\end{center}
\end{figure}

\begin{figure}[htbp]
\begin{center}
\includegraphics[width=\linewidth]{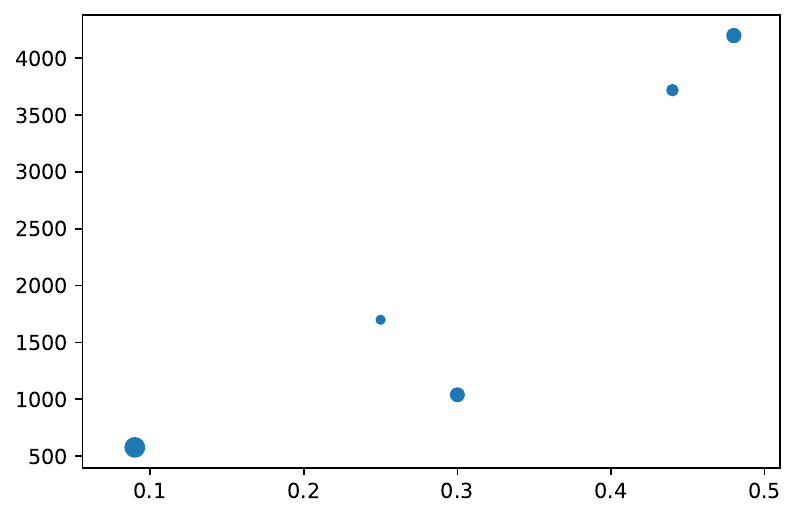}
\caption{Hubble plot with (arbitrarily) varying marker sizes}
\label{VaryingSizesExample}
\end{center}
\end{figure}

Next, let us take 5 galaxies, values for their distance to Earth in Mpc and their redshift values. (I will use the same which already made their appearance values earlier in section \ref{firstGalaxyEx}.) We begin once more by defining lists containing the galaxy's names, distances and redshifts.
\begin{lstlisting}
nm = ['SDSS-II SN 21387','SDSS-II SN 13651','SDSS-II SN 03706','SDSS-II SN 10963','SDSS-II SN 03475']
dpc=[4200.,1700.,3720.,577.,1040.]
zv=[0.48,0.25,0.44,0.09,0.3]
\end{lstlisting}
Let us create a Hubble diagram, in our case plotting redshift values on the x axis and distance values on the y axis. Since these are separate data points, it does not make sense to join them with a line. Instead, we will plot the data points as separate markers, using the scatter function:
\begin{lstlisting}
plt.clf()
plt.scatter(zv,dpc)
\end{lstlisting}
The result can be seen in Fig.~\ref{HubblePlot1}.
Again there are several possibilities to make this look nicer and more readable. For instance, you can use the option \verb|s| to change the size of your markers, or \verb|color| to change their color, and \verb|marker| to change the shape of the marker (all possible shapes can be found in [\href{https://matplotlib.org/api/markers_api.html}{https://matplotlib.org/api/markers\_api.html}]). 
\begin{figure}[htbp]
\begin{center}
\includegraphics[width=\linewidth]{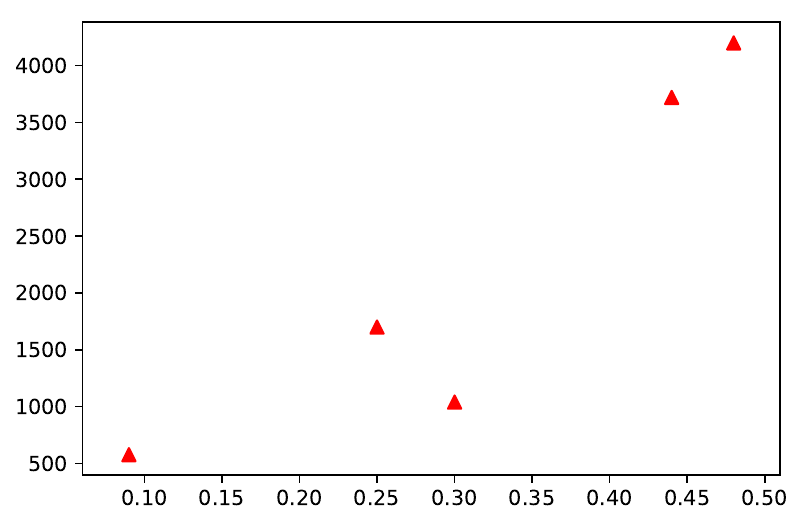}
\caption{Simple Hubble plot with custom red triangle data points}
\label{HubblePlot2}
\end{center}
\end{figure}
Fig.~\ref{HubblePlot2} is a scatter plot with red triangle markers with an area (size) of 40 square points, coded as
\begin{lstlisting}
plt.clf()
plt.scatter(zv,dpc,marker='^',s=40,color='r')
\end{lstlisting}
We can also use marker attributes such as color, size or shape to carry information. If we make the color, or the size, an array with the same length as our data set, we can set values individually for each data point. We have already seen examples for this practice in section \ref{HighLevel} in figures \ref{MassLumColor} and \ref{MassLumRadiusColor}. Here is a programming example, extending our previous diagrams, for using varying marker area (governed by the size option \verb|s|). The sizes are chosen arbitrarily and, in this particular case, convey no more information than showing you how this kind of option works:
\begin{lstlisting}
sizes = [40,15,25,80,38]

plt.clf()
plt.scatter(zv,dpc,s=sizes)
\end{lstlisting}
Remember that the values for the size correspond to the area, not the diameter. Doubling the value will not double the diameter. The resulting diagram, with each marker with its own specific size as we specified in the \verb|sizes| array, can be seen in Fig.~\ref{VaryingSizesExample}.

Last but not least, it is also possible to plot a scatter diagram using the plot function. Instead of the colour, you specify a data marker shape, in this case circles:
\begin{lstlisting}
plt.clf()
plt.plot(zv,dpc,'o')
\end{lstlisting}

\subsection{Fitting data}

The Hubble-Lema\^{\i}tre relation is supposed to be linear --- for distant galaxies, redshifts $z$ and distances $d$ are supposed to be related by 
\be
z = H_0/c\cdot d,
\ee
with $H_0$ the Hubble constant and $c$ the vacuum speed of light. Our galaxy dots, each representing a galaxy's redshift and distance, do not quite fall on a single line. So what is the line that fits these data point best?

\begin{figure}[htbp]
\begin{center}
\includegraphics[width=\linewidth]{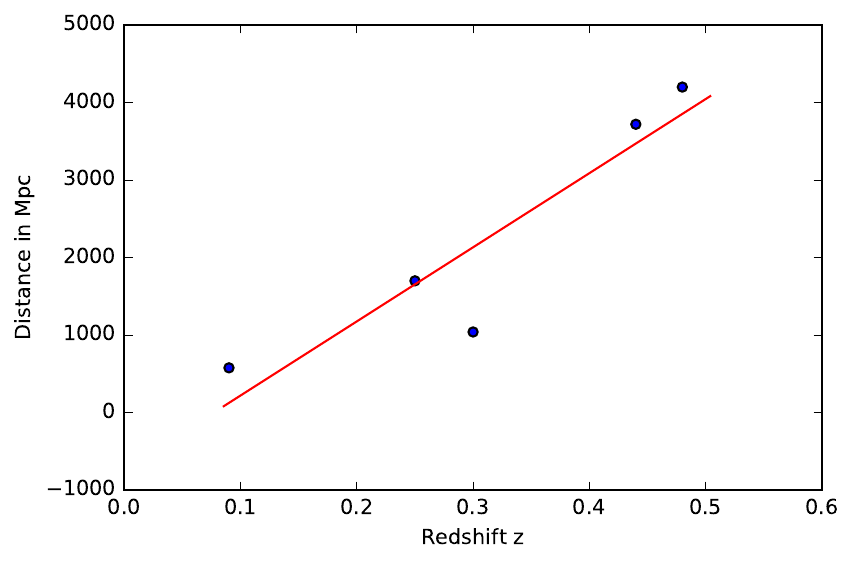}
\caption{Hubble diagram, with the best-fit line shown}
\label{HubbleFitted}
\end{center}
\end{figure}

To find out, we use the function \verb|polyfit| from the numpy module. 
\begin{lstlisting}
dpc=[4200.,1700.,3720.,577.,1040.]
zv=[0.48,0.25,0.44,0.09,0.3]

import numpy as np

popt = np.polyfit(zv, dpc,1)
\end{lstlisting}
The result is
\begin{lstlisting}
In: popt
Out: array([9560.54352268, -735.48957908])
\end{lstlisting}
We have told the \verb|polyfit| function that we want to fit the data points given by \verb|zv| as x values and \verb|dpc| as y values with a polynomial of degree 1 (the third argument we passed to the function), which is to say, with a linear function $y=a\cdot x+b$. As a result, the function has returned its best estimates for the coefficient $a$ of the first-order term, namely $a=9560,$ and for the zeroth-order (that is, constant) term $b=-735.5$ (albeit with more digits than I have reproduced here). The proportionality factor corresponds to a value for the Hubble constant of
\be
H_0 = 31.4\;\frac{\mbox{km/s}}{\mbox{Mpc}},
\ee
which is rather different from the best current values close to $70\;\mbox{km/s/Mpc}$. Can our statistical error explain the discrepancy? To find out, we re-run the fit with an additional option, namely as
\begin{lstlisting}
popt,pcov = np.polyfit(zv,dpc,1,cov=True)
perr = np.sqrt(np.diag(pcov))
\end{lstlisting}
By setting \verb|cov=True|, we tell the function to return not only its estimates, but also the associated covariance matrix. Taking the square roots of the diagonal of that matrix gives us estimates for the errors of the parameters. For the parameter $a$, we obtain $a = 9560\pm 2321$, so our Hubble constant estimate comes with an error
\be
H_0 = (31.4\pm 7.6)\;\frac{\mbox{km/s}}{\mbox{Mpc}},
\ee
so definitely not enough to explain the discrepancy. We can plot the corresponding straight line of our best fit in our diagram, cf. Fig.~\ref{HubbleFitted}.

On the other hand, from the basic definition of scale-factor expansion, we know that the relation we should be fitting is not $y=ax+b$, but rather $y=ax$, since there is no offset in the Hubble-Lema\^{\i}tre relation. Let us program this new version of our fitting function by making use of \verb|curve_fit|, which is part of the SciPy library, as follows:
\begin{lstlisting}
dpc=[4200.,1700.,3720.,577.,1040.]
zv=[0.48,0.25,0.44,0.09,0.3]

from scipy.optimize import curve_fit

def fitFunc(x,a):
	return x*a
	
popt, pcov = curve_fit(fitFunc, zv, dpc)
perr = np.sqrt(np.diag(pcov))
\end{lstlisting}
As you can see, there is now an additional step, namely that we have to define the function we want to fit to our data; I have called it \verb|fitFunc|. When we want to fit data to a more general function $f(x)$, that function will contain free parameters $a,b,c,\dots$ whose values will then need to be fixed by the fitting procedure. In defining our function, we need to make sure that $x$ is the first argument, followed by any free parameters in our function. In our specific case, we only have a single free parameter, $a$. As you can see, \verb|curve_fit| has three input slots: the first is for our pre-defined fit function, the second for the array of x data and the third for the array of y data. Even without setting an extra option, \verb|curve_fit| will return the covariance matrix, so we assign the result to the tuple \verb|popt, pcov|. In \verb|popt| we will, once more, find the best-fit values for our free parameters, arranged in an array. From the covariance matrix, the square roots of the diagonal elements, here extracted as \verb|\perr|, allow us to extract the associated statistical errors. In our particular case, the result is $a = 7597.89598735\pm 932.90257718,$ corresponding to 
$$
H_0= (39.5\pm 4.8) \frac{\mbox{Mpc}}{\mbox{km/s}}.
$$
Evidently, the discrepancy is due to something else --- to our particular selection of galaxies, perhaps, or to errors in the distances. Nothing we can resolve here, but along the way, we have learned several ways of how to fit a curve to data.

\subsection{Histograms}
\label{HistogramsPython}

\begin{figure}[htbp]
\begin{center}
\includegraphics[width=0.5\textwidth]{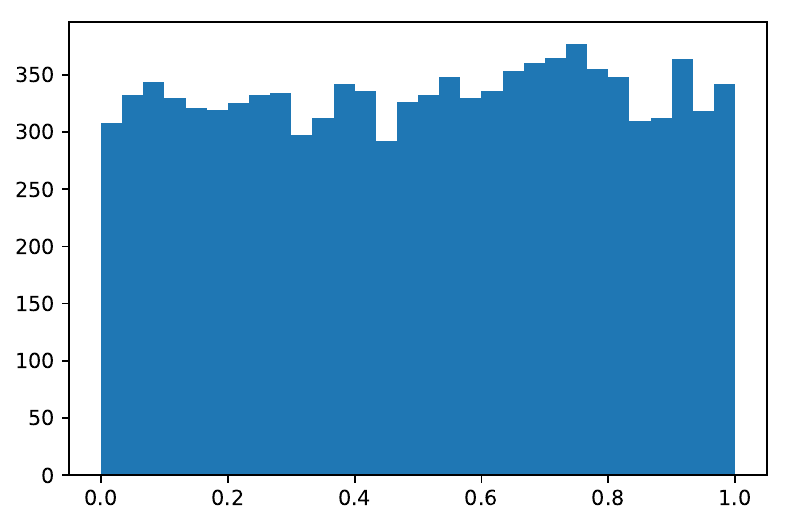}
\caption{Histogram for our array of random numbers}
\label{Histo1}
\end{center}
\end{figure}

Finally, let's make a histogram. Let's couple that with an illustration of one of the most important statistical theorems: the {\bf central limit theorem}. Let us use numpy's random function to create an array of random numbers:
\begin{lstlisting}
import numpy as np
randArray = np.random.rand(10000)
\end{lstlisting}
The simplest way of turning this into a histogram is matplotlib's hist function,
\begin{lstlisting}
plt.clf()
plt.hist(randArray,bins=30)
\end{lstlisting}
whose result can be seen in Fig.~\ref{Histo1}.
The argument ``bins=30'' indicates that we want the histogram to have 30 bins. The default bin number is 10.
The histogram has no clear structure; as we would expect, all values are of similar frequency, around the expectation value $10000/30\approx 330$, but with random fluctuations above and below that value.

\begin{figure}[htbp]
\begin{center}
\includegraphics[width=0.5\textwidth]{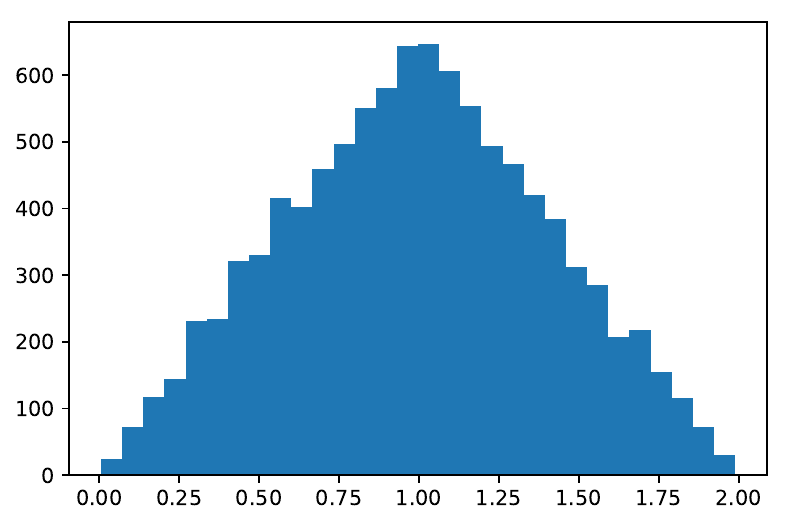}
\caption{Histogram for our array of sums of two random numbers}
\label{Histo2}
\end{center}
\end{figure}

Next, let us define our random array a bit differently. This time, we add two random arrays, so that each number in the resulting array is now the sum of two random numbers.
\begin{lstlisting}
import numpy as np
randArray = np.random.rand(10000)+np.random.rand(10000)
\end{lstlisting}
The resulting histogram can be seen in Fig.~\ref{Histo2}.
Now, the histogram has a maximum in the middle, near 1, and smaller and larger values are less common. This is straightforward to understand if you look at, say, the sums of the integers between 1 and 5. There is only one way to obtain the 10, namely 5+5. But there are several ways to obtain 5: 4+1, 3+2, 2+3, 1+4. An outcome near the middle of the set is more likely.

\begin{figure}[htbp]
\begin{center}
\includegraphics[width=0.5\textwidth]{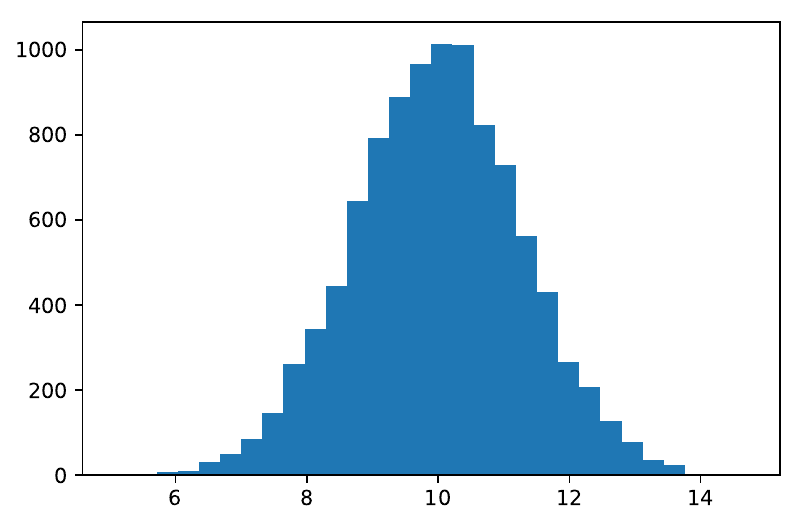}
\caption{Histogram for our array of sums of twenty random numbers}
\label{Histo3}
\end{center}
\end{figure}

We can repeat the exercise by regarding sums over more and more random numbers. Fig.~\ref{Histo3} shows the histogram for a random array where each element is the sum of twenty random numbers. 
Does the shape look familar? This is looking more and more similar to a normal (Gaussian) distribution, and that is no accident. In fact, that is what the central limit theorem says for a situation like this, where each of the quantities we document is the sum of many random variables, all drawn from the same probability distribution (which must have some additional properties such as finite variance): as the number of terms in the sum grows, the resulting distribution comes ever closer to a normal distribution. This is why normal distributions are so useful: In physics (and astronomy), there are typically many small fluctuations, many different sources of error. Those  different errors will add up to influence the sought-for result. As long as those fluctuations, and associated errors, are reasonably independent of each other, then similar to what happens in the central limit theorem, the distribution for measurement results is likely to come close to a normal distribution.

\begin{figure}[htbp]
\begin{center}
\includegraphics[width=0.5\textwidth]{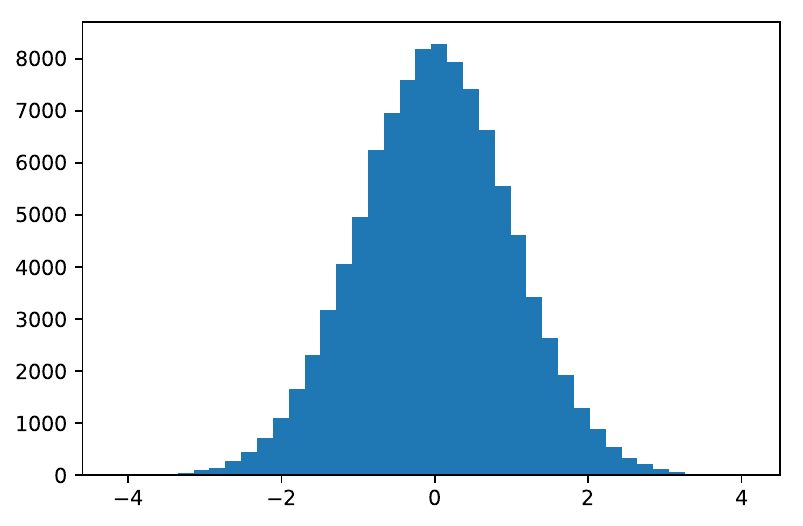}
\caption{Histogram of 100000 values randomly drawn from a normal distribution}
\label{GaussHisto}
\end{center}
\end{figure}

The \verb|np.random| sub-module also provides for specific probability distributions for us to draw samples from, and several of those functions allow us to draw samples from normal distributions. Every normal distribution is defined by its mean (where its maximum is) and its standard deviation (defining the width of the distribution). The function \verb|np.random.normal(mu, sigma, 1000)| will return an array of 1000 samples for a normal distribution whose mean is the value of the variable \verb|mu| and whose standard deviation is the value of \verb|sigma|. We choose a different way, namely the function \verb|randn|, which automatically draws the sample from the so-called standard normal distribution, which has mean 0.0 and standard deviation 1.0:
\begin{lstlisting}
import numpy as np
gaussDraw =np.random.randn(100000)
\end{lstlisting}
We can plot a histogram for our sample, choosing a suitable number of bins:
\begin{lstlisting}
plt.clf()
plt.hist(gaussDraw,bins=40)
\end{lstlisting}
The resulting histogram is shown in Fig.~\ref{GaussHisto}.
In closing, we note that it is also possible to draw two-dimensional histograms --- either as 3D objects (drawn in perspective), or else using colour to express column height. An example for the latter version can be found in section~\ref{PYVO}.

\subsection{Saving figures}

After you have drawn a figure with matplotlib, you can save it by adding the line
\begin{lstlisting}
plt.savefig('thisIsAFilename.pdf',bbox_inches='tight')
\end{lstlisting}
to your code. The string \verb|'thisIsAFilename.pdf'| may, of course, be replaced by any other string, or string-valued variable.

The file type is determined by the file extension. In this case, a pdf file is created. Other types, such as png or jpg, are possible. Resolution for pixel-based formats can be set using the option \verb|dpi=300| for 300 dots per inch, or similar. The \verb|bbox_inches='tight'| makes sure the figure fits itself nicely into the allotted space.

\subsection{Glueing data sets}
\label{Sec:Glue}

In TOPCAT, in section \ref{TOPCATSubsets}, we encountered subsets as an easy method to identify certain subsets interactively and explore their properties. To this end, the software, ``glued different diagrams together,'' identifying the different subsets in a consistent way in histograms and plots. In Python, we can program such functionality explicitly, setting criteria for the different subsets, sorting them into different arrays, and plotting the results in the histograms and diagrams we need. 

That would be a matter of coding the distinctions we make, and the visualisations. But in a number of situations, having an {\em interactive} way to select subsets from data, and then have that selection represented simultaneously in the relevant histograms and diagrams, is much easier than explicit programming. 

I will not go into this issue here, but for anyone interested, I recommend having a look at the Glue library at [\href{https://glueviz.org}{https://glueviz.org}], which provides this kind of interactive functionality.

\section{Importing table data into Python}

In all previous examples, we have used data that was written directly into the script. In most realistic cases, the data will instead be contained in a file of some type. Thus, opening files and reading in data is an important scripting skill. As we have seen (and explored with TOPCAT!) higher-level astronomical data often comes in the form of tables, where each row represents a specific object, and each column a type of property. For this kind of data, Python provides a table format, which is basically an array with additional meta-information thrown in.

\subsection{Opening a FITS table in python}
\label{FITSTablePython}

Let us begin with FITS tables. We have already encountered the FITS format as a means to encode not only images, but tables, in section \ref{TOPCATOpenTable}. Let us use open the same Galaxy Zoo FITS file table zoo2MainSpecz.fits we downloaded back then, and open it in Python:
\begin{lstlisting}
from astropy.io import fits
hdulist = fits.open('zoo2MainSpecz.fits')
\end{lstlisting}
As we have seen before, FITS files in general can have a rather complex structure. They can consist of multiple ``header/data units'', HDUs in short, where the data can be an image or a table or another type of array, and the corresponding header contains meta-information about the data. The \verb|hdulist| variable now contains all the HDUs of this particular FITS file. Call its associated function, ``index'', and you will get a brief table of contents:
\begin{lstlisting}
hdulist.info()
Filename: zoo2MainSpecz.fits
No.    Name         Type      Cards   Dimensions   Format
0    PRIMARY     PrimaryHDU      16   (19741,)     uint8   
1    Joined      BinTableHDU    480   243500R x 233C   [K, K, K, E, E, 11A, 11A, 20A, 20A, I, I, I, E, E, E, E, I, I, E, E, E, E, I, I, E, E, E, E, I, I, E, E, E, E, I, I, E, E, E, E, I, I, E, E, E, E, I, I, E, E, E, E, I, I, E, E, E, E, I, I, E, E, E, E, I, I, E, E, E, E, I, I, E, E, E, E, I, I, E, E, E, E, I, I, E, E, E, E, I, I, E, E, E, E, I, I, E, E, E, E, I, I, E, E, E, E, I, I, E, E, E, E, I, I, E, E, E, E, I, I, E, E, E, E, I, I, E, E, E, E, I, I, E, E, E, E, I, I, E, E, E, E, I, I, E, E, E, E, I, I, E, E, E, E, I, I, E, E, E, E, I, I, E, E, E, E, I, I, E, E, E, E, I, I, E, E, E, E, I, I, E, E, E, E, I, I, E, E, E, E, I, I, E, E, E, E, I, I, E, E, E, E, I, I, E, E, E, E, I, I, E, E, E, E, I, I, E, E, E, E, I, I, E, E, E, E, I, I, E, E, E, E, I]   
\end{lstlisting}
The primary HDU is not commonly used for scientific data. For us, the interesting part is the second HDU, index 1, which as you can see is a table stored in a binary format (``BinTableHDU''), with 243500 rows and 233 columns. The big list that follows lists the type of variable for each column: K are 64-bit-integers, 11A is a string with 11 characters, I a 16 bit integer, E a single precision floating point.\footnote{All the different types and their specifications can be found on [\href{http://docs.astropy.org/en/stable/io/fits/usage/table.html}{http://docs.astropy.org/en/stable/io/fits/usage/table.html}]} For more information about the columns, call the columns attribute of the table. The table, as we have seen, is the second element of the hdulist, namely \verb|hdulist[1]|:
\begin{lstlisting}
In: hdulist[1].columns
Out:
ColDefs(
    name = 'specobjid'; format = 'K'; null = -9223372036854775808
    name = 'dr8objid'; format = 'K'; null = -99
    name = 'dr7objid'; format = 'K'
    name = 'ra'; format = 'E'
    name = 'dec'; format = 'E'
    name = 'rastring'; format = '11A'
    name = 'decstring'; format = '11A'
    name = 'sample'; format = '20A'
    name = 'gz2class'; format = '20A'
    name = 'total_classifications'; format = 'I'
    name = 'total_votes'; format = 'I'
\end{lstlisting}
\dots I am not showing all of the output, but as you can see, this returns the column names as well as their types.  In order to extract the data, we will use the data attribute of the table object. Once we have opened the HDULIST and assigned it to a variable \verb|hdulist|, we can get the data via
\begin{lstlisting}
tdata = hdulist[1].data
\end{lstlisting}
To access a specific column from this table, you can use that column's name as follows:
\begin{lstlisting}
In: tdata.field('specobjid')
Out: 
array([1802674929645152256, 1992983900678285312, 1489568922213574656, 
..., 1959146978059249664,  467329305811118080,   -9999])
\end{lstlisting}
As a result, we obtain the column named 'specibjid' as an array. We can use the usual index conventions to get elements, such as
\begin{lstlisting}
In: tdata.field('specobjid')[4]
Out: 1387165355897546752
\end{lstlisting}
to access the fifth element in the list.

\subsection{Opening an ASCII table in python}
Some tables are in ascii format -- an ASCII file with elements belonging to the various column the columns separated by spaces or other symbols, or defined because each column has a pre-defined width.

As an example, download the csv (comma-separated values) version of the Galaxy Zoo Data Release table 5 we had already opened as a FITS file in the previous section, namely [\href{zooniverse-data.s3.amazonaws.com/galaxy-zoo-2/zoo2MainSpecz.csv.gz}{zooniverse-data.s3.amazonaws.com/galaxy-zoo-2/zoo2MainSpecz.csv.gz}].  Astropy has an ``ascii'' submodule to handle such files as follows:
\begin{lstlisting}
from astropy.io import ascii
tdata=ascii.read('zoo2MainSpecz.csv')
\end{lstlisting}
The resulting \verb|tdata| is a table object in Astropy. With the info attribute, you can once more get a list of all columns and their types:
\begin{lstlisting}
In: tdata.info
Out: 
<Table masked=True length=243500>
        name           dtype     n_bad
-------------------- ------- -----
           specobjid   int64        14
            dr8objid   int64      3752
            dr7objid   int64         0
                   ra  float64       0
                  dec  float64       0
            rastring   str11         0
           decstring   str11         0
              sample   str8          0
            gz2class   str8          0
\end{lstlisting}       
Again, this is only the start of a much longer list. The name and data type of the column are given; I suppose n\_bad counts empty or malformed entries, but couldn't find the proper documentation. For columns with numerical values, information like the minimum, maximum and mean value are provided, as well. Using a column name as the key will again give you data from that column:
\begin{lstlisting}
In:  tdata['specobjid']

Out: 
<MaskedColumn name='specobjid' dtype='int64' length=243500>
1802674929645152256
1992983900678285312
1489568922213574656
2924083625089591296
1387165355897546752
1833070384862226432
1809324500555163648
                ...
\end{lstlisting}                                          
where I have again shortened the output. Using a list index gives a specific entry in that column:
\begin{lstlisting}
In:  tdata['specobjid'][4]

Out: 1387165355897546752
\end{lstlisting}  

\subsection{Accessing astronomical data bases}
\label{PYVO}

In sections \ref{TOPCATVO} and \ref{BasicADQL}, we used TOPCAT to send remote queries written in the ADQL query language to Virtual Observatory (VO) services. These queries helped us to select specific data from existing catalogs. There are several ways of doing the same directly from a Python script.

For one, we can use the module \verb|pyvo| to access VO services. The module is not included in the standard Anaconda setup, though, and you will need to import it. In Linux and on a Mac, going to the command line and entering \verb|pip install pyvo| should do the trick. In Windows, your Anaconda directory should somewhere include the file pip.exe. If you open the command line tool and execute \verb|pip.exe install pivo|, that should work for you. Once you have installed pyvo, you can run queries like the following:
\begin{lstlisting}
import pyvo as vo
import matplotlib.pyplot as plt

serviceURL="http://gea.esac.esa.int/tap-server/tap"
service = vo.dal.TAPService(serviceURL)
resultset = service.search(
"""
SELECT TOP 1000000 
l,b
FROM gaiadr2.gaia_source
ORDER BY RANDOM_INDEX
""")

plt.clf()
plt.hist2d((resultset['l']+180.0) % 360,resultset['b'], bins=(200, 200), cmap=plt.cm.jet)
plt.savefig('gaia-plot.pdf',bbox_inches='tight')
\end{lstlisting}  
First, we are importing pyvo, then matplotlib. Then, we are defining a service; the URL is the service URL, where one can connect with the data base. This particular data base is the same one as the GAIA service we have used in sections \ref{TOPCATVO} and \ref{BasicADQL}. 

Then, I am performing a search using that service. Recall that a multi-line string in Python begins with a triple set of double quotation marks, \verb|"""|.  In this case, the multi-line string contains the ADQL query, using the same syntax you have learned in section \ref{BasicADQL}. This time, we are retrieving the properties galactic longitude l and galactic latitude b, referring to the coordinate system in which the Milky Way band across the sky is at latitude b=0. We are again using the RANDOM\_INDEX to retrieve a random subsample of Gaia point sources.

In the lower part, we plot a 2d-histogram, that is, a density plot for the objects we have retrieved. As you can see, we access the list of all retrieved galactic longitudes l by calling up \verb|resultset['l']|, while we get the list of latitudes b by calling up \verb|resultset['b']|. This is the same as for other tables, where we can retrieve a column by using the column name as an index. The combination \verb|(resultset['l']+180.0) % 360| 
is used to shift the galactic longitude by 180 degrees. That way, the galactic center is not around l=0, that is, at the left and right margin of the image, but in the center.\begin{figure}[htbp]
\begin{center}
\includegraphics[width=\linewidth]{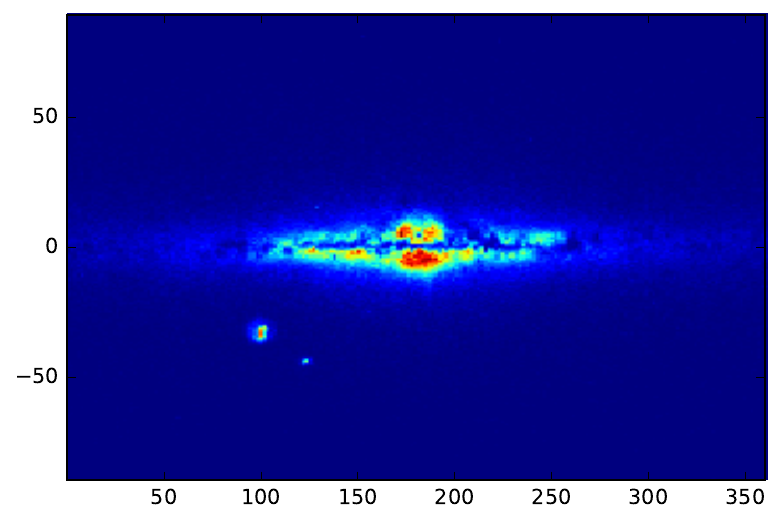}
\caption{Plot of Gaia stars, retrieved with PyVO}
\label{GaiaPlot}
\end{center}
\end{figure}
The resulting image gaia-plot.pdf can be seen in Fig.~\ref{GaiaPlot}.
Once more, the Milky Way, the Large Magellanic Cloud and the Small Magellanic Cloud are clearly visible.

Just like the TOPCAT TAP access, there is a whole world of different services to explore via this interface, from stars to exoplanets\footnote{E.g. via the API for the highly useful {\em Extrasolar Planet Encyclopaedia} website, [\href{http://exoplanet.eu/API/}{http://exoplanet.eu/API/}]} to galaxies. 

Another way of accessing various catalogues is the AstroQuery package,\footnote{A general description can be found in Ginsburg et al. 2019, ``astroquery: An Astronomical Web-Querying Package in Python,'' \href{https://arxiv.org/abs/1901.04520}{https://arXiv:1901.04520}} a collection of modules to retrieve data from various astronomical catalogues via the web. Information about the specific catalogues and the specifics of accessing them can be found in the documentation at [\href{https://astroquery.readthedocs.io}{https://astroquery.readthedocs.io}].

As a simple example, here is the same Gaia query we had already executed with \verb|pyvo|, but this time using \verb|astroquery|:
\begin{lstlisting}
from astroquery.gaia import Gaia
import matplotlib.pyplot as plt

job = Gaia.launch_job("""
SELECT TOP 1000000
l,b
FROM gaiadr2.gaia_source
ORDER BY RANDOM_INDEX
""")
resultset = job.get_results()

plt.clf()
plt.hist2d((resultset['l']+180.0) \%
    360,resultset['b'], bins=(200, 200),
    cmap=plt.cm.jet)
plt.savefig('gaia-plot2.pdf',bbox_inches='tight')
\end{lstlisting}
The resulting figure looks exactly the same as Fig.~\ref{GaiaPlot}.

\section{Astronomical image manipulation with Python}
\label{Images}
All professional astronomical images are in the FITS format, file extension .fits or .fit, which stands for ``Flexible Image Transport System''. Where your ordinary JPG file gives you 256 steps between darkest and brightest (8 bits, separately for each color RGB), FITS gives you, by default, 65536, or 16 bits. That's a lot of contrast. In section \ref{DS9}, we learned how to use application software to take a look at, and perform some measurements in, images in FITS format. Let's see how we can display and analyse FITS images in Python.

\subsection{FITS files and python}
In order to open a FITS image file in Python, we once more use the Astropy sub-module \verb|astropy.io|. Just as we did with FITS tables in section \ref{FITSTablePython}, we first load the HDU list. For convenience, we use one of the Hubble Space Telescope images we had already downloaded in section~\ref{LoadHubble}, namely the file hst\_05773\_05\_wfpc2\_f502n\_wf\_drz.fits. We load the file's HDU list like this: 
\begin{lstlisting}
from astropy.io import fits
fitsURL='hst_05773_05_wfpc2_f502n_wf_drz.fits'
hdulist=fits.open(fitsURL)
\end{lstlisting}
If we call the function info on the hdulist, we are shown the different header/data units of this fits file:
\begin{lstlisting}
In: hdulist.info()
Out:
Filename: hst_05773_05_wfpc2_f502n_wf_drz.fits
No.    Name         Type      Cards   Dimensions   Format
0    PRIMARY     PrimaryHDU     509   ()              
1    SCI         ImageHDU       103   (2150, 2150)   float32   
2    WHT         ImageHDU       124   (2150, 2150)   float32   
3    CTX         ImageHDU       123   (2150, 2150)   int32   
\end{lstlisting}
We are interested in the science image SCI, the second HDU, hence the one with index 1. The header component of the HDU variable will give you access to the header, as shown in the following listing.
\begin{lstlisting}
In: hdulist[1].header

Out: Out[279]: 
XTENSION= 'IMAGE   '           / Image extension                                
BITPIX  =                  -32 / array data type                                
NAXIS   =                    2 / number of array dimensions                     
NAXIS1  =                 2150                                                  
NAXIS2  =                 2150                                                  
PCOUNT  =                    0 / number of parameters                           
GCOUNT  =                    1 / number of groups                               
CRVAL1  =    274.7173822566667 / right ascension of reference pixel (deg)       
CRVAL2  =   -13.83106772194444 / declination of reference pixel (deg)           
CRPIX1  =               1075.0 / x-coordinate of reference pixel                
CRPIX2  =               1075.0 / y-coordinate of reference pixel  
\end{lstlisting}
Once again, I am showing only a small part of the output here. You can access specific header information by plugging in the respective keyword. This here, for instance, will give you the number of pixels along the first image dimension:
\begin{lstlisting}
In: hdulist[1].header['NAXIS1']

Out: 2150
\end{lstlisting}
An interesting piece of information is the exposure time in seconds, which is contained in the header of the primary HDU:
\begin{lstlisting}
In: hdulist[0].header['EXPTIME']

Out: 2200.0
\end{lstlisting}
The data component of the science HDU will give you the image data. Let's define
\begin{lstlisting}
imdata =  hdulist[1].data
\end{lstlisting}

\subsection{Displaying (showing) an image}
\label{MatplotlibDisplayImage}
\begin{figure}[htbp]
\begin{center}
\includegraphics[width=\linewidth]{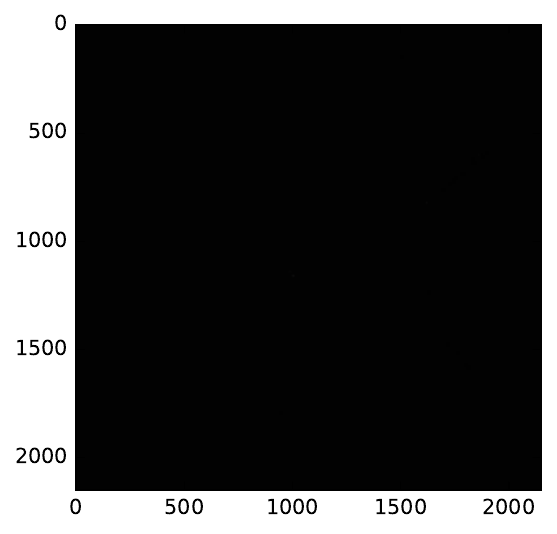}
\caption{This is not the image of a black hole}
\label{ImshowDark}
\end{center}
\end{figure}

\begin{figure}[htbp]
\begin{center}
\includegraphics[width=\linewidth]{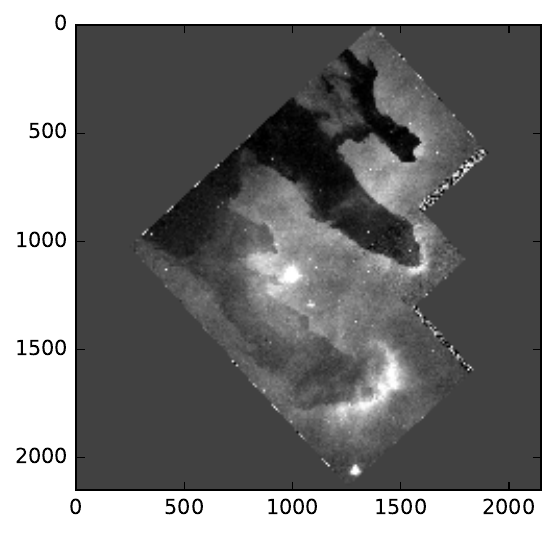}
\caption{HST image of M 16, with adapted colourmap}
\label{ImshowM16}
\end{center}
\end{figure}

Once we have the image data, we can display it, using matplotlib, using the \verb|imshow| function, as follows:
\begin{lstlisting}
plt.clf()
plt.axes().set_aspect('equal')
plt.imshow(imdata,cmap='gray')
\end{lstlisting}
The \verb|set_aspect('equal')| tells matplotlib that both x and y axes should have the same scale, as behoves two spatial directions spanning a two-dimensional plane. The cmap option tells imshow which colormap to use, in this case grayscale; \href{https://matplotlib.org/examples/color/colormaps_reference.html}{many others} are possible. 
The result is at first rather dark, as you can see in Fig.~\ref{ImshowDark}.
Just like with DS9, we somehow need to map the high contrast of the FITS image to our more modestly contrasted version. We can use the \verb|clim| option to map a more restricted range of values to our image. Let's look at the 1st and 99th percentile values of the image data (that is, the brightness value below which the darkest 1\% of the pixels fall, and the brightness value above which 1\% of the pixels fall). We can access those descriptive numbers by typing
\begin{lstlisting}
In: np.percentile(imdata,1)
Out: -0.028674498219043016

In: np.percentile(imdata,99)
Out: 0.068247721269726738
\end{lstlisting}
With the results, we can plot the image setting more suitable brightness limits (more generally, limits for our color map) like this:
\begin{lstlisting}
plt.clf()
plt.axes().set_aspect('equal')
plt.imshow(imdata,cmap='gray',clim=(-0.03,0.088))
\end{lstlisting}
The result can be seen in Fig.~\ref{ImshowM16}.

\subsection{Pixelwise operations}

The image data we have put on display is an array. Specifically, you can see this as 
\begin{lstlisting}
In: type(imdata)

Out: numpy.ndarray
\end{lstlisting}
and 
\begin{lstlisting}
In: imdata.shape

Out: (2150, 2150)
\end{lstlisting}
which shows you that the image is a Numpy array $2150\times 2150$. Just like with any other array, we can retrieve pixel-wise information. For instance, the brightness value of the pixel $x=1200, y=1400$ can be retrieved by accessing the respective element of the \verb|imdata| array, namely as
\begin{lstlisting}
In: imdata[1400][1200]

Out: 0.030607721
\end{lstlisting}
Note the counter-intuitive order --- the first square parentheses contain the y value, the second one the x value! We can change pixel values in the same way, by assigning a new value to a specific \verb|imdata[y][x]|.
Being able to read out and manipulate pixels individually gives us substantial power to analyse the image. We can use all the tools the previous sections have put at our disposal, comparing pixel values, or summing them up, or performing fits, or trying to identify point sources, or doing much more complicated types of analysis.

\begin{figure}[htbp]
\begin{center}
\includegraphics[width=\linewidth]{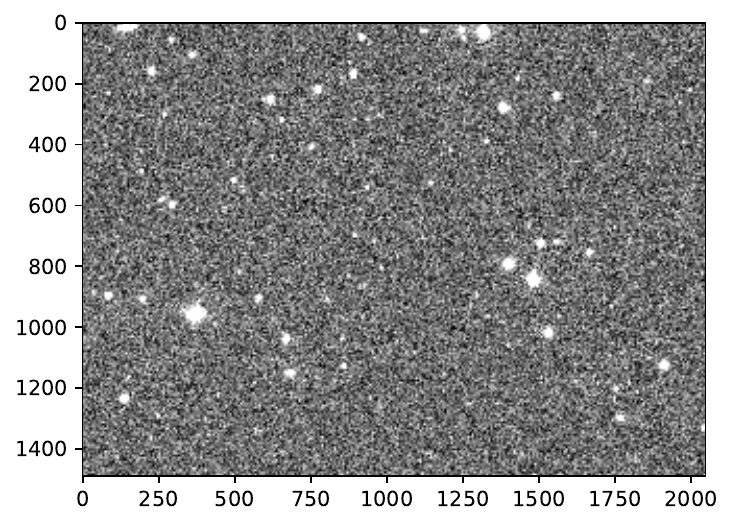}
\caption{SDSS DR9 frame displayed with imshow}
\label{PySDSS}
\end{center}
\end{figure}

Let us perform at least some basic operations on the SDSS data file frame-g-007923-5-0307.fits that we had downloaded and analysed in sections \ref{DS9DR9} and \ref{DS9Photometry}. It's usually good idea to take an overall look at the image one intends to analyse, which we have learned to do in this way:
\begin{lstlisting}
hdulistS=fits.open('frame-g-007923-5-0307.fits')
imdataS=hdulistS[0].data

lowerOne = np.percentile(imdataS,1)
upperOne = np.percentile(imdataS,99)

plt.clf()
plt.imshow(imdataS,cmap='gray',
   clim=(lowerOne,upperOne))
plt.savefig('sdss-py.pdf',bbox_inches='tight')
\end{lstlisting}
The result is shown in Fig.~\ref{PySDSS}.
Next, let us see if we can re-do the simple aperture photometry measurements. The stars we had compared in section \ref{DS9Photometry} were located at $X=1819,\; Y=1215$ and $X=1123,\;Y=32$, respectively. We can use \verb|xlim| and \verb|ylim| to zoom in onto those locations; for instance, amending the plot with
\begin{lstlisting}
centerX = 1819
centerY = 1215

plt.xlim(centerX-100,centerX+100)
plt.ylim(centerY-75,centerY+75)
\end{lstlisting}
we obtain a detailed view all around the first of the stars, 200 pixels in width, 150 in height. Next, we will visualize both the position of the star we are interested in, and the surrounding regions we will use in our aperture photometry measurements. To this end, we will add a circle to our diagram, centered on the star. The proper way of doing this is 
\begin{lstlisting}
thisCircle = plt.Circle((centerX, centerY), 10, color='r',fill=False,lw=2)
plt.gca().add_artist(thisCircle)
\end{lstlisting}
This has two parts. In the first part, we define a circle centered on $(1819, 1215)$ with radius 10, the colour red, which is not filled, just an outline with linewidth 2. In matplotlib, that circle is a so-called ``patch'', a two-dimensional artist object. Artist objects are matplotlib's way of drawing specific shapes.

In the second line, \verb|add_artist| adds this circle to the axes object of our figure. We repeat those commands with radius 20; the outer circles mark the area we will use for determining the background brightness. Note that in repeating those commands, we need not even choose a different name for the variable \verb|thisCircle|. 

\begin{figure}[htbp]
\begin{center}
\includegraphics[width=\linewidth]{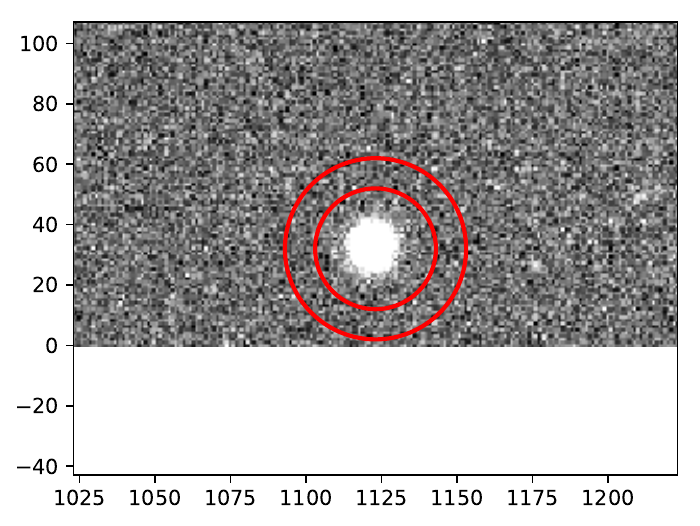}
\caption{Detailed view of the region around the first star, with an inner circle and outer annulus marked}
\label{mplCircle}
\end{center}
\end{figure}

Once we have pushed the object onto our diagram using \verb|add_artist|, the variable has done what it was meant to be, and we can use it to define, and draw, another circle. The result can be see in Fig.~\ref{mplCircle}: Our chosen location is now surrounded by two circles, drawn in red. 

That is a cosmetic marking only, though. Next, we can do aperture photometry as in section \ref{DS9Photometry}. First, we determine the sum of pixel brightness values in the outer circle, as well as the area of that outer circle in pixels. We do so by going over all pixels in a square with side length \verb|2*radius|, and including only the values of pixels whose distance is smaller than the value of \verb|radius|, in other words: those pixels that are in the circle we are interested in:
\begin{lstlisting}
radius=20
photCollector=np.array([])
for ii in range(centerX-radius, centerX+radius):
    for jj in range(centerY-radius,centerY+radius):
        distance = np.sqrt((ii-centerX)**2 + (jj-centerY)**2 )
        if distance < radius:
            photCollector= np.append(photCollector, imdataS[jj][ii])
            
C1 = np.sum(photCollector)
A1 = len(photCollector)
\end{lstlisting}
As a result, I obtain C1=107.65 and an area of A1 = 1245 pixels. For the inner circle, with radius 10 pixel, I get C2 = 101.12 with an area A2 = 305 pixels. Using the brightness formula (\ref{BrightnessFormula}),
\be
l = C_2 - \frac{C_1-C_2}{A_1-A_2}\cdot A_2,
\label{BrightnessFormula2}
\ee
we subtract from the sum of brightness values $C_2$ the sum of background brightness values (as estimated by the average of the brightness values in the annulus bounded by the inner and the outer circle). For the first star, I get the brightness $l_A=12.60$. I repeat the procedure for the second star, the one at $X=1123,\;Y=32$, and obtain C1 = 107.47 with an area of A1 = 2809 pixels, and C2 = 107.65 for an area of A1= 1245 pixels. Using once more the brightness formula (\ref{BrightnessFormula2}), the star's background-subtracted brightness is $l_B= 107.80$.

As I had already argued in section \ref{DS9Photometry}, given that the collecting area and the exposure time are the same in both cases, the ratio of our values $l_B$ and $l_A$ should be the same as the ratio of the intensities of the two stars; inserting this into equation (\ref{AstroMagnitudeFormula}) for the astronomical magnitudes, we find 
\be
m_A-m_B = -2.5\cdot\log\left(\frac{l_A}{l_B}\right) = -2.5\cdot\log\left(\frac{12.6}{107.8}\right) =2.33,
\ee
again not far from the difference in the stars' catalog g magnitudes (gmag) of 2.13. 

This is a single example of how we can use our ability to access an astronomical image pixel by pixel unlocks a treasure trove of analytical possibilities.\footnote{A more detailed guide for CCD image processing using Python is the ``CCD Data Reduction Guide'' by Matt Craig and Lauren Chambers, available at [\href{https://mwcraig.github.io/ccd-as-book/00-00-Preface.html}{https://mwcraig.github.io/ccd-as-book/00-00-Preface.html}].} If you really want to do aperture photometry on a larger scale, you would very likely {\em not} want to code your analysis from scratch --- you would use, for instance, the Astropy-affiliated package Photutils [\href{https://photutils.readthedocs.io/}{https://photutils.readthedocs.io/}] and its higher-level functions for identifying point sources and measuring their brightness. But the main point of this section was much more general: Once we have loaded FITS data, the full force of coding can be brought into play. Our aperture photometry example was comparatively simple. More complex operations are possible: We could fit a brightness profile to an extended source, or try to reconstruct the point-spread function for a point source. We could even try to have our script identify objects such as stars automatically, by checking brightness differences. The possibilities are virtually endless, definitely so in that astronomers will continue to bring to bear new analysis tools as they are being developed. Recent applications of machine learning in astronomy are a pertinent example.

\section{A simple simulation}
\label{Simulation}

So far, we have almost exclusively dealt directly with observational data, either images/spectra or higher-level table data. This is not the only kind of data that astronomers work with --- there is a whole other branch of research focused on {\em simulated} data. Such simulated data can play different roles. Some are meant to provide a point of comparison for observations. For instance, if you really want to understand all the details of the spectral line shape in a stellar spectrum, you will need access to the results of simulations for the genesis of those lines. The newest versions take into account complex effects such as the absence of local thermodynamic equilibrium (a common simplification). Compare the simulated spectra for different stellar parameter values (such as chemical abundances) with your observations, and you can deduce the properties of your star.\footnote{At the time of this writing, a sample set of tools from the group of Maria Bergemann at the Max Planck Institute for Astronomy can be found on [\href{http://nlte.mpia.de/}{http://nlte.mpia.de}].}

\begin{figure}[htbp]
\begin{center}
\includegraphics[width=\linewidth]{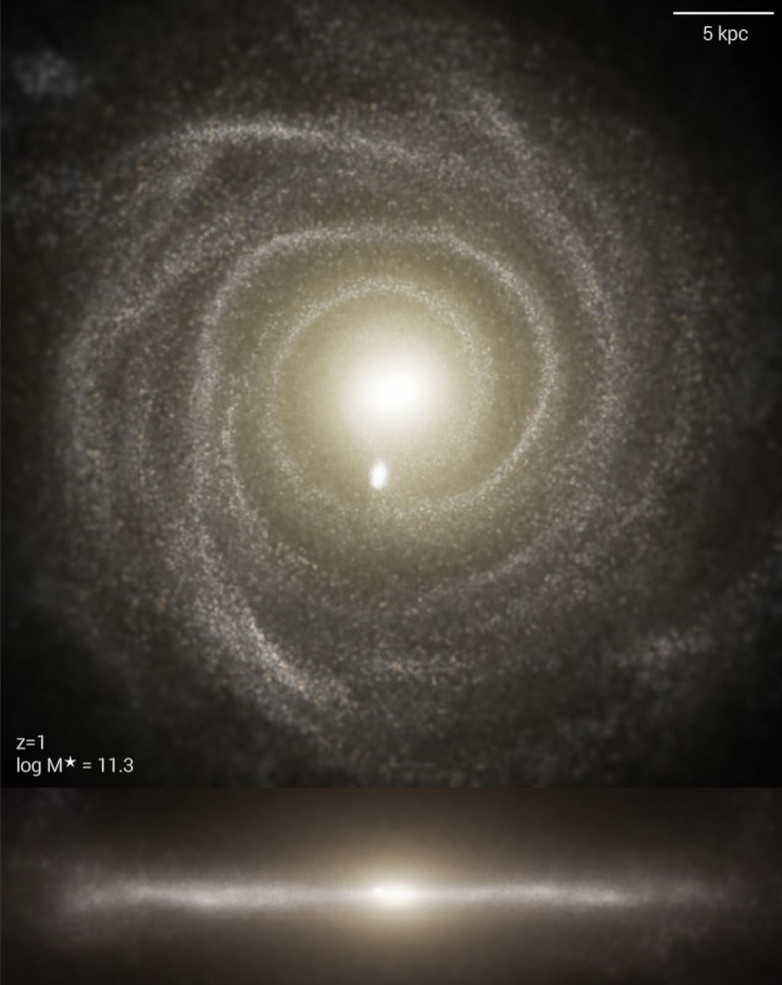}
\caption{Details of a massive disk galaxy at redshift $z=1$, simulated with the TNG50 simulation. Image: TNG collaboration}
\label{TNG}
\end{center}
\end{figure}
Other simulations are more ambitious in scope. The various runs of the IllustrisTNG\footnote{More information can be found at [\href{http://www.tng-project.org}{http://www.tng-project.org}]. } simulation, for instance, follow a cubic region within the cosmos from shortly after the Big Bang to the present. The simulation runs are of different degrees of coarseness, simulating either a very highly resolved cube with a sidelength of 50 million parsecs, a less well resolved cube of 100 Mpc or a larger, but still less resolved cube 300 Mpc a side. The simulation is mostly based on ``particles'' representing dark matter, stars, and gas, and simulates numerous physical processes including the influence of magnetic fields, the formation and evolution of stars in galaxies, the production of heavier elements in stars, and interactions with the supermassive black holes in the centres of galaxies. All in all, the IllustrisTNG simulations follow the evolution of the universe over the full 13.8 billion years, including diverse length scales from those on which the cosmos is, on average, homogeneous, down to those of the sub-structure of galaxies, cf.~Fig.~\ref{TNG}.

As I have written before, the details of such simulations call for (at least!) a whole set of lecture notes of their own. Most of that knowledge is far beyond the scope of the present text. But I firmly believe that an overview of working with astronomical data needs to include at least a brief introduction to creating your own data based on models and on the laws of physics --- that, in my mind, is an integral part of working-with-astronomical-data literacy. With this in mind, here is a simple example that illustrates at least some basic techniques, and also some elementary pitfalls, of numerical simulations. The physics in this case is simple, classical mechanics.

\subsection{Step-by-step numerical integration: Euler method}

\begin{figure}[htbp]
\begin{center}
\includegraphics[width=0.7\linewidth]{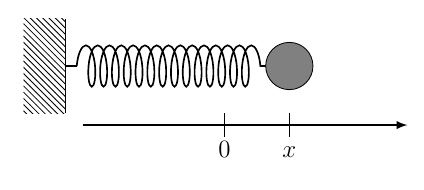}
\caption{Harmonic oscillator: particle on a spring}
\label{HarmonicSetup}
\end{center}
\end{figure}

We revert to what is probably the most basic (and arguably most useful!) elementary system in all of physics: The harmonic oscillator. The basic set-up is as shown in Fig.~\ref{HarmonicSetup}: A particle with mass $m$, which can move only in the (horizontal) x direction, is fixed to the wall with a spring. If the particle is displaced from its rest position at $x=0$, the spring exerts a force following Hooke's law, $F_x = -k\cdot x,$ with $k$ the spring constant. 
This system has the advantage that its equation of motion,
\be
m\ddot{x} = -k\cdot x,
\label{HOeom}
\ee
linking the x acceleration $\ddot{x}$ and the force using Newton's second law, is readily solved analytically, that is, in terms of a simple mathematical function. The solution is 
\be
x(t) = A\cdot\sin(\omega t),
\label{HOorbit}
\ee
where the angular frequency $\omega$ is linked to the system's oscillation period $T$ by the standard definition
\be
\omega = \frac{2\pi}{T},
\ee
and the equations of motion in this particular case demand
\be
\omega = \sqrt{\frac{k}{m}}.
\ee
Differentiating the orbit equation (\ref{HOorbit}) once with respect to time, we have
\be
\dot{x} = \omega\cdot A\cdot\cos(\omega t),
\ee
and differentiating with respect to time once more,
\be
\ddot{x} = -\omega^2\cdot A\cdot \sin(\omega t) = -\frac{k}{m}\cdot x,
\ee
which shows that our solution (\ref{HOorbit}) indeed satisfies the equation of motion (\ref{HOeom}). So far, so elementary. But now, pretend that we do not know of this simple solution. How can we simulate the system, in other words: find a solution not analytically, but numerically?

The basic idea is that, if we look at very small time interval, all of the changes during such an interval will be approximately linear. This, is, after all, the definition of a derivative: over an infinitesimally small interval $\Dd t$, the change of the function $x(t)$, namely $\Dd x$, is given by
\be
\Dd x = \dot{x}\cdot \Dd t.
\label{DerivativeDef}
\ee
Replace the infinitesimally small interval $\Dd t$ by a finite small interval $\Delta t$, and what was an equality in (\ref{DerivativeDef}) becomes an approximation, whose quality depends on the magnitude of $\Delta t$: the smaller $\Delta t$, the better the approximation. Thus, if we know the particle's x position at one time $t$, we can estimate its position at a slightly later time $t+\Delta t$ as
\be
x(t+\Delta t) = x(t) + v(t)\cdot \Delta t,
\label{xDelta}
\ee
where $v(t)$ is the particle's velocity in x direction at the time $t$. What this formula does not encode, of course, is how $v(t)$ changes over time. But for the change of $v$, we can write down a similar equation. The rate of change of the velocity, after all, is the acceleration, which by Newton's second law $F=m\ddot{x}$ is linked to the force acting on the particle. Thus, to obtain the velocity at some time $t+\Delta t$, we can use the approximation
\be
v(t+\Delta t) = v(t) + \ddot{x}(t)\cdot\Delta t = v(t) + \frac{F(t)}{m}\cdot\Delta t.
\label{vDelta}
\ee
Incidentally, in writing down the approximation equations (\ref{xDelta}) and (\ref{vDelta}), we have applied a technique that can be used much more generally, when dealing with higher-order differential equations: We have transformed a single second-order differential equation (for $\ddot{x}$) into a system of (two) first order equations, each of which we have solved approximately, in going from $t$ to $t+\Delta t$. 

Now, we can {\em discretize} the whole problem: We consider time steps $t_i$, with $i=1,\ldots,N$, and evaluate the position and the velocity of our particle at each step. We choose the times $t_i$ equidistant, with $t_{i+1}-t_i=\Delta t$ for all $i$, for some fixed, small $\Delta t$. 

(What is small, and more specifically: what is sufficiently small? That depends on the problem's characteristic time scale, and requires physical thought. With hindsight (or by looking at the analytical solution), we know that we are dealing periodic, oscillatory motion, so whatever interval $\Delta t$ we choose had better be much smaller than the system's natural period $T$. If we cannot find a physical time scale to ascribe to the system, we might need to fall back on experimentation. If the solution changes significantly when we repeat the simulation with ever smaller $\Delta t$, that is an indication we have not reached the proper resolution yet. If, on the other hand, the solution remains pretty much the same when we replace, say, $\Delta t$ by $\Delta t/2$, that is an indication that we have reached the regime where the finite size of $\Delta t$ does not exert significant influence on our result any more.)

Differential equations do not, on their own, completely determine what is happening. It is necessary to specify {\em initial conditions} in order to define a unique solution. As an example, consider what happens when you throw a ball vertically upward from the Earth's surface. The differential equations tell you how Earth's gravity will accelerate the ball. But those equations alone are not sufficient to tell you what happens. For a prediction, you will need to specify both the ball's initial position and it's initial velocity. Both those initial values are crucial in determining how the ball will move, and in particular whether it will fall back to Earth or keep going forever (when its velocity is larger than the position-dependent escape velocity).

\begin{figure}[htbp]
\begin{center}
\includegraphics[width=\linewidth]{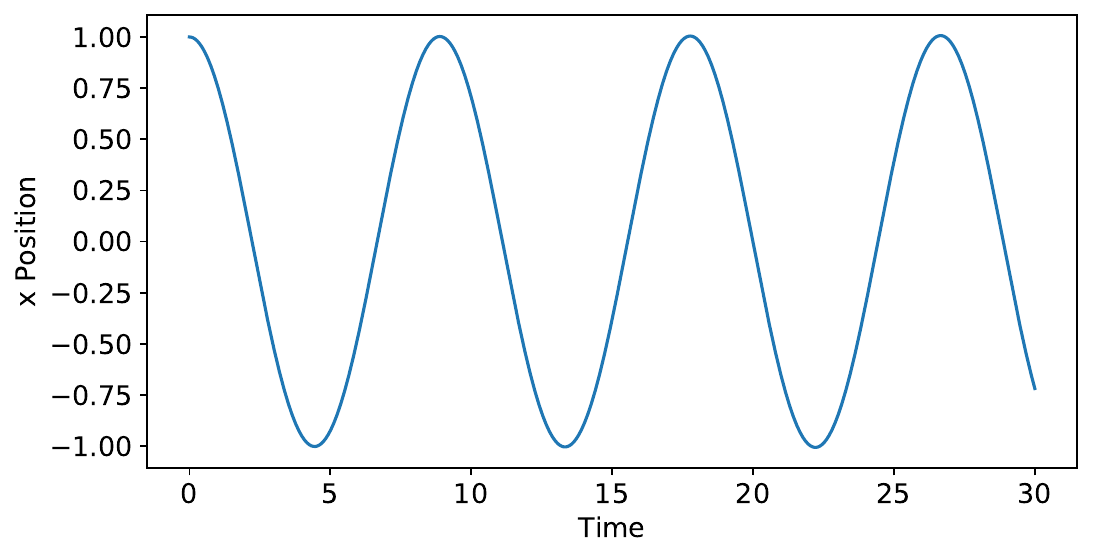}
\caption{Solution of the harmonic oscillator equation with initial position 1.0 and initial speed 0.0}
\label{HOPlot1}
\end{center}
\end{figure}

In this case, let us choose an initial position $x_0$ and initial speed $v_0$ for our particle. Let $x_i$ be the object's position at time $t_i$, $v_i$ its velocity in x direction at that time, $a_i$ its acceleration and $F_i$ the force acting on it at the time. The simple, step-wise evolution equations we have derived are 
\bea
v_{i+1} &=& v_i + a_i\cdot\Delta t = v_i + \frac{1}{m}F_i\cdot\Delta t\\[0.5em]
x_{i+1} &=& x_i + v_i\cdot\Delta t.
\eea
The process of following the evolution step by step is called {\em numerical integration}, and the simple algorithm we have given for going from one step to the next is called {\em Euler's method}. In our case, with the Hooke force (\ref{HOeom}), the force depends only on the position, so the velocity equation reads
\be
v_{ i+1}= v_i  -\frac{k}{m}x_i\cdot\Delta t.
\ee
This is readily implemented in Python, for instance in the following way, using the loop function. (The array \verb|tCollector| is only defined in preparation of plotting positions against time later on, and not used in the numerical integration itself.)
\begin{lstlisting}
k=0.5
m=1.0
numberOfSteps = 30000
DeltaT = 0.001
finalT =numberOfSteps*DeltaT

tCollector=np.linspace(0,finalT,numberOfSteps+1)

# Initial conditions:
xCollector=[1.0]
vCollector=[0]

for ii in range(numberOfSteps):
    xNew = xCollector[-1] + DeltaT*vCollector[-1]
    vNew = vCollector[-1] + DeltaT*(-k/m*xCollector[-1])
    xCollector.append(xNew)
    vCollector.append(vNew)
\end{lstlisting}
Plotting the result, as in Fig.~\ref{HOPlot1}, shows that 
we indeed obtain the proper sine shape, shifted so as to form a cosine function. (Why cosine instead of sine? Because, by choosing our initial condition to be $v=0$, we start our evolution at the maximum x value.) 

\subsection{Numerical errors}
\label{NumericalErrors}

\begin{figure}[htbp]
\begin{center}
\includegraphics[width=\linewidth]{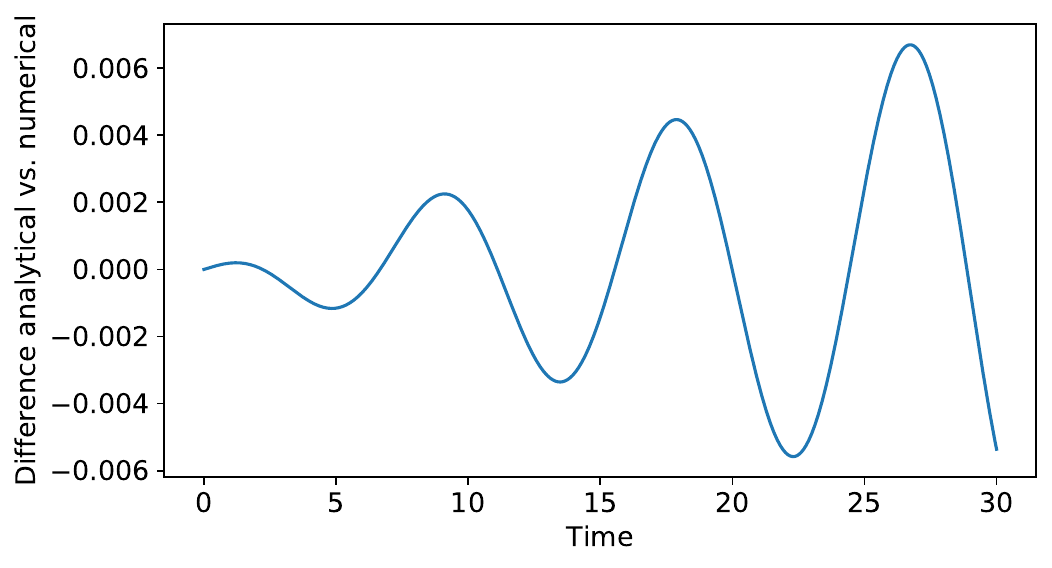}
\caption{Difference between our numerical solution and the analytical solution}
\label{HOPlotDiff}
\end{center}
\end{figure}

If I just were to plot the analytical solution and the numerical solution in the same diagram, the curves would be overlaid so closely that no difference would be visible upon direct inspection. As an alternative, Fig.~\ref{HOPlotDiff} shows the difference between the analytical and the numerical solution for our harmonic oscillator solution, at each time. 
There is good news and there is bad news. The good news is that the differences between the true solution and our simulation are very small, namely 0.6\% at most, and much better for most of the time. The bad news is that the differences are getting larger over time. If that trend continues, and
we let our simulation run much further, its deviations from the true solution will become so large as to be noticeable. Our numerical simulation is {\em unstable} in this sense. 

We can understand why that is. Imagine that, in truth, our particle reaches our maximum x value at time $t_i$. From time $t_{i-1}$ to time $t_i$, we change the x value by $\Delta x=+v_{i-1}\cdot\Delta t$. But $v_{i-1}$ is the (non-zero) velocity at the beginning of the interval. Over the time $\Delta t$, that velocity will change to zero (since velocity zero is what defines the maximum value, the turning point). Thus, the average velocity over that interval will be {\em smaller} than the initial velocity for that interval. We are overestimating the amount $\Delta x$ that x increases during that time. We pretend the object has flown with unchanged initial velocity $v_{i-1}$, whereas, in reality, it has slowed down. That means our turning-point x will be just a little further out, x a little larger, than in reality. 

There is nothing in our simulation to compensate for that larger error. The system knows nothing of its past. That slightly larger x, that slightly larger amplitude at the turning point will be carried along for the rest of our simulation. You can think about it in the following way: Because of the slightly over-large x when the turning point comes, our system, from then on, has a slightly larger potential energy than it should have. There is nothing to compensate; once the extra energy has seeped into the system through the numerical error, it will remain in the system.

Even worse: At every turning point, the same argument applies, so at every turning point, the error of having a slightly larger amplitude will increase. The errors systematically add up. {\em That} is what makes the system unstable. Errors do not compensate each other; errors just add up over time, increasing the overall deviation from the true time evolution.

Note that the systematic errors, and the instability, are a property of the algorithm we have used to simulate the time evolution. There are several different algorithms that, in the limit of infinitesimal $\Delta t$, all amount to the same integration procedure, but which differ in their stability properties and in how accurate they are for finite $\Delta t$. We will look at one of them in the next section.

\subsection{Velocity Verlet algorithm}

\begin{figure}[htbp]
\begin{center}
\includegraphics[width=\linewidth]{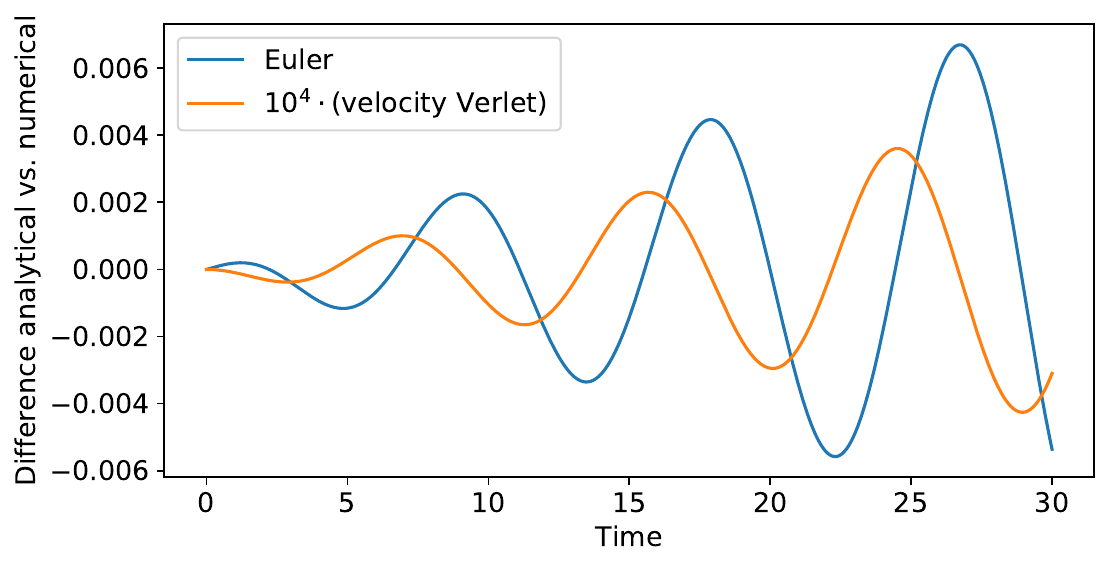}
\caption{Difference between our numerical solution and the analytical solution, once for the Euler algorithm, once for the velocity Verlet algorithm (the latter here scaled up artificially by a factor $10^4$)}
\label{VVEuler}
\end{center}
\end{figure}

One example for a better-behaved numerical integration scheme xsis the {\em velocity Verlet} algorithm, which introduces ``half-step'' velocities, in order to mitigate problems like those I described in section \ref{NumericalErrors}, as follows:
\bea
v_{i+1/2} &=& v_i + \frac{1}{2}\Delta t\cdot a_i\\[0.5em]
x_{i+1} &=& x_i + v_{i+1/2}\cdot\Delta t\\[0.5em]
v_{i+1} &=& v_{i+1/2} + \frac{1}{2}\Delta t\cdot a_{i+1},
\eea
where the $a_i$ is again calculated from the force acting in time step $i$. In this algorithm, we avoid always using the initial values of rates-of-change at each time step, leading to systematic errors; instead, the calculation of $x_{i+1}$ uses the intermediate velocity, while the transition from $v_i$ to $v_{i+1}$ proceeds in two steps, one using the acceleration value at the beginning, the second the one at the end of the interval $\Delta t$. The implementation is, again, fairly straightforward:
\begin{lstlisting}
k=0.5
m=1.0
numberOfSteps = 30000
DeltaT = 0.001
finalT =numberOfSteps*DeltaT

tCollector=np.linspace(0,finalT,numberOfSteps+1)

# Initial conditions:
xCollectorVV=[1.0]
vCollectorVV=[0]

for ii in range(numberOfSteps):
    vHalf = vCollectorVV[-1] + 0.5*DeltaT*(-k/m*xCollectorVV[-1])
    xNew = xCollectorVV[-1]  + DeltaT*vHalf
    vNew = vHalf  + 0.5*DeltaT*(-k/m*xNew)
    xCollectorVV.append(xNew)
    vCollectorVV.append(vNew)
\end{lstlisting}
The comparison between the results can be seen in Fig.~\ref{VVEuler}.
The velocity Verlet algorithm, too, appears to be unstable in the long-term, with the error increasing over time. But the simple of expedient of adding the half-step velocity has greatly improved the accuracy. After all, note that, in this diagram, I have scaled up the difference for the velocity Verlet algorithm by a whopping factor of $10^4$ to make it visible in comparison with the Euler deviations! 

This simple example shows the importance of implementing the evolution algorithm carefully; finding the best ways of doing this is a science (and possibly an art) of its own. If you decide to explore this further, you might want to look at the {\em Runge-Kutta} family of iterative methods next.

\subsection{A simple two-dimensional simulation}
\label{2D}

So far, our simple simulation was one-dimensional: motion in the x direction over time. Let us choose a two-dimensional scenario next, and one that is very important in astronomy: The motion of a test particle around a central mass under the influence of the central mass's (Newtonian) gravity, which provides a good model for the orbit of a (not too massive) planet around a star.

\begin{figure}[htbp]
\begin{center}
\includegraphics[width=0.8\linewidth]{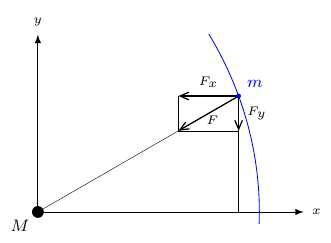}
\caption{Planet with mass $m$ orbiting in the xy plane under the gravitational influence of a central mass $M$ in the origin}
\label{2DForce}
\end{center}
\end{figure}

Let us put the central mass into the origin of our coordinate system. We treat the x and y components of the motion separately. Since we have two independent directors, we will need to treat the force as a vector as well, separating its x from its y component. Fig.~\ref{2DForce} shows the geometry of the situation.
Crucially, the triangle describing the x-y-position of the planet and the triangle describing the decomposition of the gravitational force into x and y component are similar, in the geometric sense: as the figure shows, they have the same set of three angles. Similar triangles can only differ by an overall length scale. In particular, ratios of the corresponding sides of such triangles are the same. If we abbreviate the distance of the planet from the origin (and thus from the central mass) as 
\be
r=\sqrt{x^2+y^2}
\ee
(having applied the Pythagorean theorem), then we have
\be
\frac{F_x}{F} = \frac{x}{r}
\ee
and
\be
\frac{F_y}{F} = \frac{y}{r}.
\ee
For the Newtonian gravitational force, we have
\be
F = -\frac{GMm}{r^2}.
\ee
Using $F_x = ma_x$ and $F_y=ma_y$ for the link between the force and the acceleration components, we have
\bea
v_{x,i+1/2} &=& v_{x,i} + \frac{1}{2}\Delta t\cdot a_{x,i}\\[0.5em]
x_{i+1} &=& x_i + v_{x,i+1/2}\cdot\Delta t\\[0.5em]
v_{x,i+1} &=& v_{x,i+1/2} + \frac{1}{2}\Delta t\cdot a_{x,i+1},
\eea
and analogous equations linking the $v_y$ and $y$ values. Note that, via $r$, the force at any given step, and thus the acceleration components, depend on both coordinates. 

Next, we need to choose suitable units. Let the central mass have one solar mass, $M_{\odot} = 2\cdot 10^{30}\;\mbox{kg}$. Going by Earth's orbit, a suitable unit of length is the astronomical unit (corresponding to the average Earth-Sun distance), $1\;\mbox{au} = 1.5\cdot 10^{11}\;\mbox{m}$. As our unit of time, we choose the Julian year: 365.25 standard days, abbreviated as a for the latin ``annum'' for year, related to the SI unit, the second, as $1\;\mbox{a} = 31~557~600\:\mbox{s}\sim\pi\cdot 10^7\:\mbox{s}$. The corresponding unit for speed is related to the more usual one as 
\be
1\:\frac{\mbox{km}}{\mbox{s}} = 0.21\:\frac{\mbox{au}}{\mbox{a}}
\ee
and the acceleration felt by the planet is given by
\be
F/m = -39.48\:\left(\frac{1\:\mbox{au}}{r}\right)^2\:\frac{\mbox{au}}{a^2}.
\label{FmFormulaAU}
\ee
The simulation code itself is listed here:
\begin{lstlisting}
numberOfSteps = 30000
DeltaT = 0.0001
accFac = 39.48 # Corresponding to one solar mass, in au per square year
finalT=numberOfSteps*DeltaT

tCollector=np.linspace(0,finalT,numberOfSteps+1)

# Initial conditions:
xCollector=[1.5]
vxCollector=[0]
yCollector=[0.0]
vyCollector=[2.0]

for ii in range(numberOfSteps):
    rNow   = np.sqrt(xCollector[-1]**2+yCollector[-1]**2)
    accNow = -accFac/rNow**2
    accNowx= accNow*xCollector[-1]/rNow
    accNowy= accNow*yCollector[-1]/rNow
    vxHalf = vxCollector[-1] + 0.5*DeltaT*accNowx
    vyHalf = vyCollector[-1] + 0.5*DeltaT*accNowy
    xNew = xCollector[-1]  + DeltaT*vxHalf
    yNew = yCollector[-1]  + DeltaT*vyHalf
    rNew   = np.sqrt(xNew**2 + yNew**2)
    accNew = -accFac/rNew**2
    accNewx = accNew*xNew/rNew
    accNewy = accNew*yNew/rNew
    vxNew = vxHalf  + 0.5*DeltaT*accNewx
    vyNew = vyHalf  + 0.5*DeltaT*accNewy
      
    xCollector.append(xNew)
    yCollector.append(yNew)
    vxCollector.append(vxNew)    
    vyCollector.append(vyNew)
\end{lstlisting}
For the half-step velocity, we calculate the accelerations in x and y direction, starting with the magnitude of the acceleration, which follows directly from Newton's law. Then, we evolve the position one time step further, re-calculate the acceleration for the new position, and use those to update the x and y component of the velocity for the second half of the time step. 

\begin{figure}[htbp]
\begin{center}
\includegraphics[width=\linewidth]{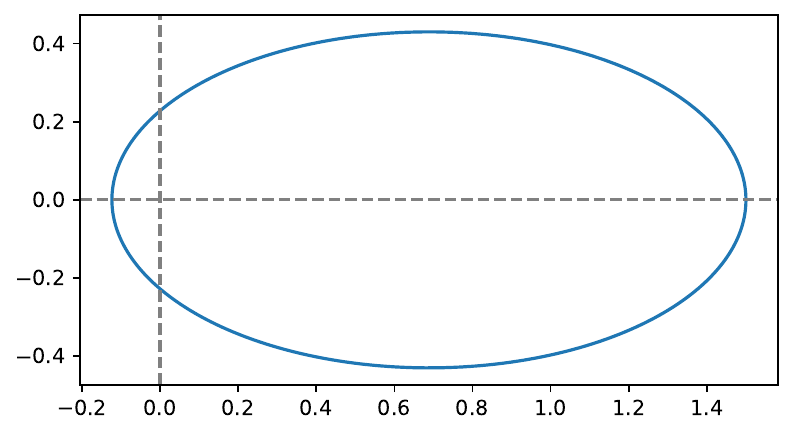}
\caption{Orbit of a planet around a central mass, simulated with the velocity Verlet algorithm}
\label{EllipticalOrbit}
\end{center}
\end{figure}

To close this section, and put our simulation to the test, let us see if we can recover Kepler's three laws of planetary motion from our simulation. We begin by plotting the shape of the orbit in Fig.~\ref{EllipticalOrbit}.
That certainly {\em looks} like an elliptical orbit, with the central mass at the origin serving as one of the focus points. But we should demonstrate this fact in a more quantitative way. Taking the minimum and maximum x value on our simulated orbit, we find that the left-most point of the orbit is at $x=-0.12335823920305192$, while the rightmost point is, by construction, at $x=1.5$ (since that is where we put the initial position of our particle, its velocity pointing straight upwards).  

\begin{figure}[htbp]
\begin{center}
\includegraphics[width=\linewidth]{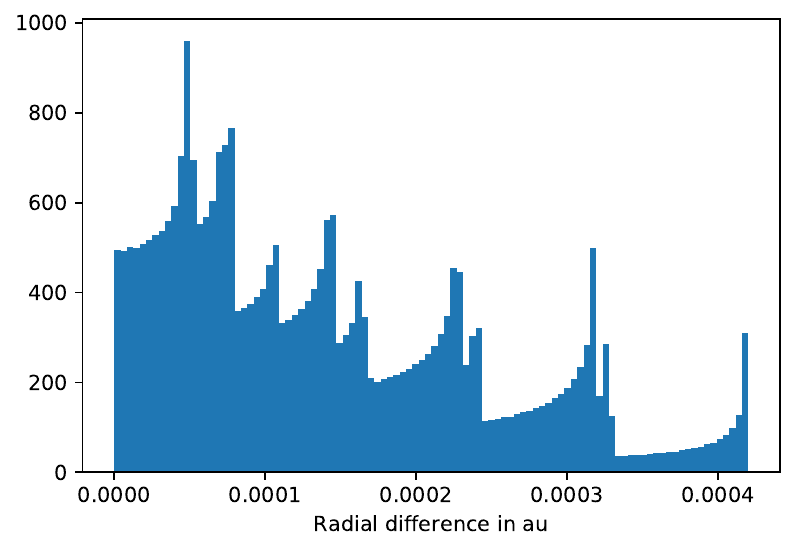}
\caption{Histogram of the square difference between fitted ellipse and simulated ellipse}
\label{EllDiffHisto}
\end{center}
\end{figure}

Thus, our simulated orbit has a major half axis of $a=1.623\:\mbox{au}$. From the minimum and maximum of the y coordinate value, we find that the minor half axis is $b=0.860\:\mbox{au}$. For an ellipse, that would correspond to an eccentricity of
\be
e = \sqrt{1-\frac{b^2}{a^2}} = 0.848.
\ee
The polar coordinate equation for an ellipse is
\be
r(\theta) = \frac{a(1-e^2)}{1+e\cos\theta}.
\label{EllipsePolarEquation}
\ee
Inserting our parameter values $a$ and $e$, we can plot this reference ellipse and compare with the simulated orbit. The resulting plot looks just like Fig.~\ref{EllipticalOrbit}, with one ellipse directly on top of the other. At least qualitatively, we have indeed confirmed Kepler's first law: the orbit of a planet orbiting a central mass is an ellipse, with the central mass in one of the focal points. More quantitatively, we can compare analytical solution and simulation directly. There are several possibilities for this; with the following piece of code, I take each simulated point, calculate the position angle $\theta$ and distance $r$ from the focus point, and compute the absolute value of the difference between the simulated value $r$ and the analytical value $r(\theta)$
given by (\ref{EllipsePolarEquation}):

\begin{lstlisting}
diffCollector=[]

for x,y in zip(xCollector,yCollector):
    r=np.sqrt(x**2+y**2)
    theta = np.arctan2(y,x)
    anr = a*(1-e**2)/(1-e*np.cos(theta))
    diffr = np.sqrt((r-anr)**2)
    diffCollector.append(diffr)
\end{lstlisting}

The histogram of the values contained in \verb|diffCollector| is shown in Fig.~\ref{EllDiffHisto}.
It has a highly structured shape, and no doubt one could learn a lot about the systematic errors involved in the simulation by understanding that cascade-like structure. Such analysis is far beyond our current scope; for our purposes, we note that this looks definitely non-random and, importantly, that the largest deviation is 4/10~000 of an astronomical unit. Given that the length scales of our orbit (half axis length, circumference) are on a scale of 1 astronomical unit, that deviation is fairly small, and our simulated planet appears to have an elliptical orbit.

\begin{figure}[htbp]
\begin{center}
\includegraphics[width=\linewidth]{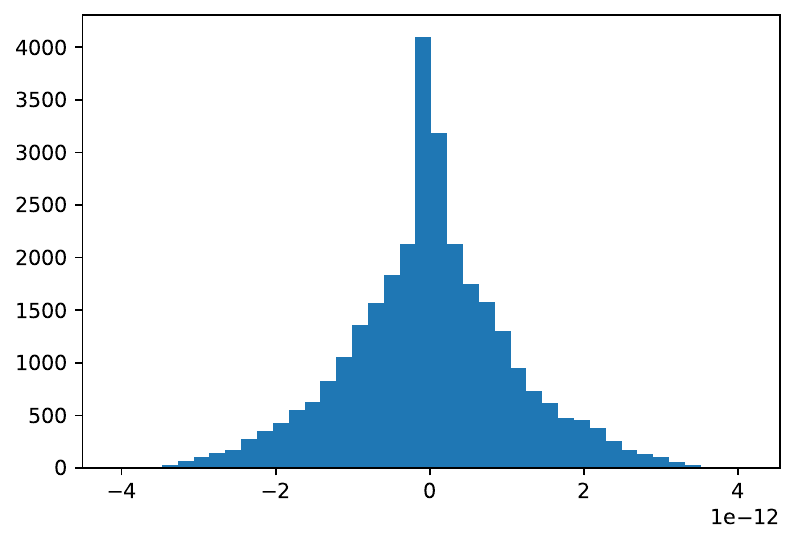}
\caption{Histogram of the relative difference from the average of the triangle areas corresponding to each time step}
\label{KeplerAreaHisto}
\end{center}
\end{figure}

Next, for Kepler's second law, which says that the connecting line between the planet and the central mass sweeps out equal areas in equal time intervals. Each of our time steps defines the same time interval, so if we calculate the triangle swept out in each time step (whose three vertices are the planet's position at the beginning and at the end of the time step, and the location of the central mass), we should always obtain the same area. For each such triangle, we know the x and y coordinates of all three vertices, and thus can calculate all the side lengths $a,b,c$ using the Pythagorean theorem. With this information, we can use Heron's formula to calculate the triangle's area as
\be
A = \sqrt{s(s-a)(s-b)(s-c)}
\ee
where $s=(a+b+c)/2$ is the triangle's semi-perimeter. (Alternatively, we can use half of the cross product of the two position vectors, now viewed as three-dimensional vectors, to obtain the same result.) The following bit of code collects the relative deviation of each such triangle area from the mean in an array \verb|relativeDiff|:
\begin{lstlisting}
areaCollector=np.array([])
for x1, x2, y1, y2 in zip(xCollector[1:], xCollector[:-1],yCollector[1:], yCollector[:-1]):
    a = np.sqrt(x1**2+y1**2)
    b = np.sqrt(x2**2+y2**2)
    c = np.sqrt((x1-x2)**2+(y1-y2)**2)
    s = 0.5*(a+b+c)
    A = np.sqrt(s*(s-a)*(s-b)*(s-c))
    areaCollector = np.append(areaCollector,A)
    
averageArea=np.average(areaCollector)
relativeDiff = (areaCollector-averageArea)/averageArea
\end{lstlisting}
A histogram of the values in \verb|relativeDiff| shows that all those areas, swept out in the same time interval, are indeed very close to their average value, as Fig.~\ref{KeplerAreaHisto} shows.
The distribution shows that we do have a strong maximum at the average area value, with small (a few parts in a trillion!) fluctuations to smaller and larger values. 

\begin{figure}[htbp]
\begin{center}
\includegraphics[width=\linewidth]{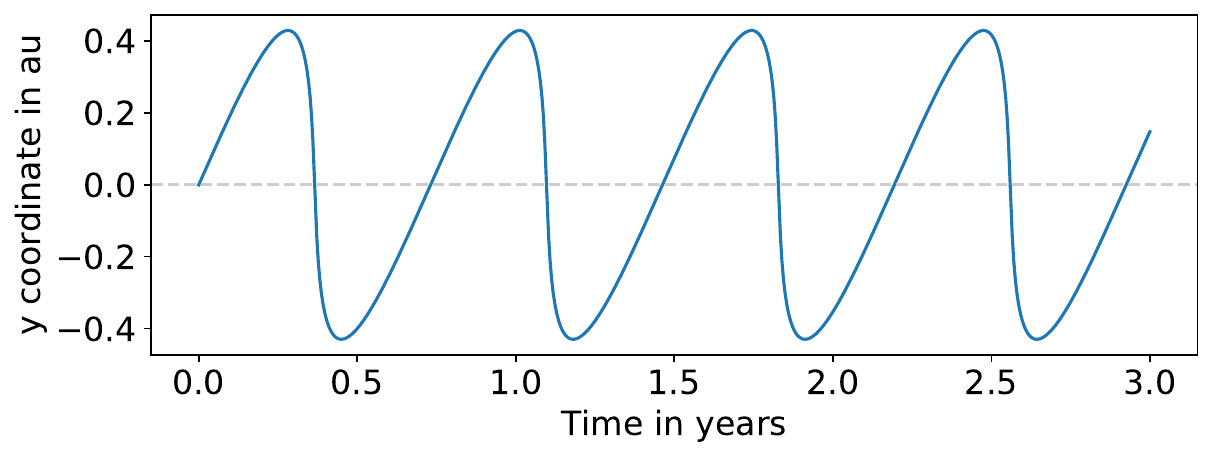}
\caption{The y coordinate of our simulated planet over time}
\label{OrbitY}
\end{center}
\end{figure}

\begin{figure}[htbp]
\begin{center}
\includegraphics[width=\linewidth]{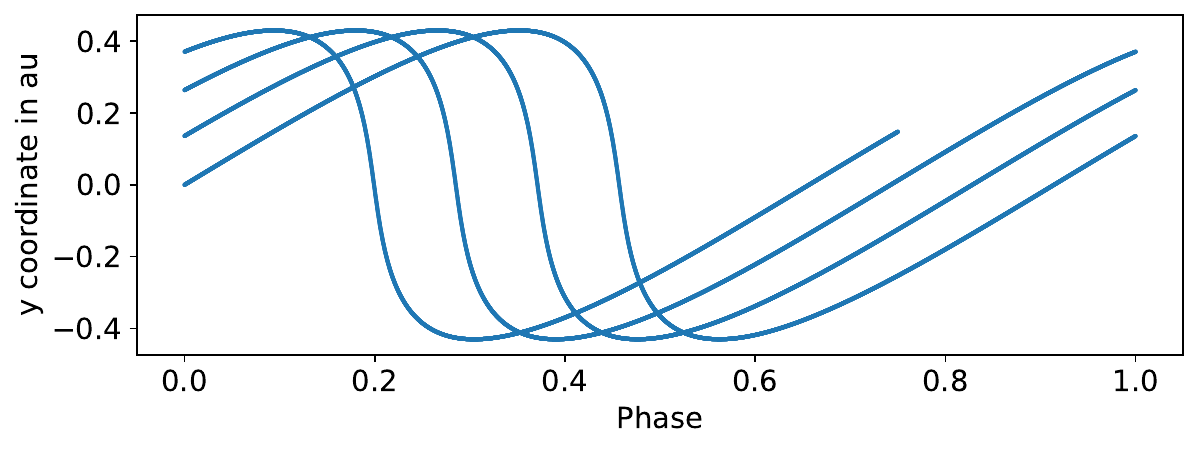}
\caption{Phase plot, with an assumed periodicity of $T=0.8\:\mbox{a}$, for our simulated orbit}
\label{Shifted08}
\end{center}
\end{figure}

Next, to Kepler's third law. In Kepler's own version, this links the orbital periods $T$ and major elliptical half axis $a$ of {\em different} planets orbiting the same central mass, stating that the ratio $a^3/T^2$ is the same for all of them. We did not simulate planets with different initial conditions (although we could), and thus will check the more advanced form of Kepler's law found by Newton, which states that
\be
\frac{a^3}{T^2} = \frac{GM}{4\pi^2}
\label{ThirdKeplerNewton}
\ee
(in the limit we have simulated, namely where the planetary mass $m$ is small against the central mass $M$). We have already estimated $a$. Let us do the same for $T$. To this end, I have plotted the y coordinate of our planet against time in Fig.~\ref{OrbitY}.
This is unsurprisingly periodical. In order to find out the period, we {\em fold} the time evolution: we {\em assume} a value for the period $T$, and define the phase $\phi$ in terms of this period as
\be
\phi = \frac{t}{T}\: \mbox{mod}\: 1.
\ee
All integer multiples of the period $T$ get mapped to 0, all times that can be written as
\be
t = (n+f)\cdot T
\ee
with integer $n$ and $0\le f<1$ get mapped to $f$. The quick-and-dirty way of finding the correct period $T$ is to vary the value by hand, and see whether or not the resulting curves coincide. From Fig.~\ref{OrbitY}, we can read off that the period is somewhat less than one year. Fig.~\ref{Shifted08} shows the resulting plot for $T=0.8\:\mbox{a}$. 
The fact that we still see separate, similar curves shows that we have not quite hit on the correct period yet. By slowly decreasing $T$, and re-plotting, I can bring those separate curves to coincidence. After a few dozen tries, the final round with a microscopic look at the steepest curve region via \verb|plt.xlim(0.475,0.525)|, I arrive at $T=0.7313\;\mbox{a}$. Note that this is very similar to how you determine the orbital of an exoplanet by folding the light curve data (from the transit method) or Doppler shift data (in the radial velocity method).
We could think about automatising this, sorting the phase values into different arrays indexed by the integer part of $t/T$, and minimising the differences between those partial curves, but for our little test, I consider the trial-and-error approach sufficient. 

Thus, our final test is to see if this $T$ value indeed satisfies equation (\ref{ThirdKeplerNewton}). The right-hand side of that equation is 
\be
1\;\frac{\mbox{au}^3}{\mbox{a}^2}
\ee
by definition, since the Earth does have a semimajor axis of length $1\;\mbox{au}$ (up to and including the sixth significant degree), and an orbital period of 1 year. By our estimates, our simulation satisfies
\be
\left(\frac{a}{1\:\mbox{au}}\right)^3\cdot\left(\frac{1\:\mbox{a}}{T}\right)^2 = 0.99991.
\ee
The two values coincide up to one part in 10~000. Our simulation reproduces Kepler's third law, as well. 

This concludes our brief excursion into the realm of simulations. We have, as ever, only scratched the surface, but along the way, you have encountered some of the pitfalls and characteristics of numerical simulations, notably the concept of numerical errors, the question of stability, and the importance of choosing a suitable algorithm. What we did not encounter was the limitation imposed by the available computing power. For our purposes, a simple script which took a few seconds to run was sufficient. As simulations become more complex, computing time increases, and can become a limiting factor. There are various ways of addressing this problem, and pushing the limit. Powerful hardware is one of them. Parallelising calculations, that is, having multiple processors (or processor cores) tackle different parts of the problem simultaneously, can be a powerful strategy for many (but not all) simulation problems. Parallelisation is achieved by linking multiple processors in a way that automatically distributes calculations among the available cores. GPUs, graphics processing units, which commonly operate at lower frequencies but consist of a greater number of cores, are frequently used for the purpose. Projects like the IllustrisTNG simulation mentioned at the beginning of this section push the envelope using an immense amount of computing power. The most calculation-intensive run of that simulation, TNG300-1, involved 24~000 cores, adding up to a total CPU time of nearly 34 million hours.\footnote{The numbers have been taken from the article by Dylan et al. 2017, [\href{http://adsabs.harvard.edu/abs/2018MNRAS.475..624N}{http://adsabs.harvard.edu/abs/2018MNRAS.475..624N}]}

We have also only reproduced the simplest type of simulation, namely an N-body simulation that follows the evolution of separate particles. Other applications in astrophysics require simulations to deal with fluid mechanics, dividing space into cells whose fluid content and properties (such as temperature, entropy and flow properties) are then followed over time.\footnote{A step-by-step to writing a hydrodynamics code can be found in [\href{https://github.com/python-hydro/how_to_write_a_hydro_code}{https://github.com/python-hydro/how\_to\_write\_a\_hydro\_code}], by Michael Zingale.}

As in the case of reducing and analysing observational data, there are packages and libraries that will help you set up specific more advanced simulations using Python.\footnote{An example is GalPy by Jo Bovy, 2014, [\href{https://arxiv.org/abs/1412.3451}{arXiv:1412.3451}].} However, should you go further in this direction, you are likely to reach a point where increased speed is so important that Python is not fast enough for what you are about to do. More advanced simulations typically use so-called compiled programming languages, such as C and its different incarnations, where the code is first translated into a ``compiled code,'' that is, into machine-readable instructions that can be carried out by the computer directly. Compiled languages are typically faster than interpreted languages like Python, where the translation into machine code happens on the fly as the program is executed. An example is the Gadget-2 code by Volker Springel et al.\footnote{See
[\href{https://ascl.net/0003.001}{https://ascl.net/0003.001}], as well as the how-to for installing and running Gadget-2 by Nathan Goldbaum on astrobites, at [\href{https://astrobites.org/2011/04/02/installing-and-running-gadget-2/}{https://astrobites.org/2011/04/02/installing-and-running-gadget-2/}].} But by the time you work at that level, you have moved far beyond the basic introduction presented in this text.

\section{Conclusion}

Working with astronomical data requires a combination of skills. If you have worked through this text, reproducing the analytical tasks on your own computer, using DS9, TOPCAT and Python, you should have acquired basic proficiency in a number of these skills, and you should now be familiar with several key tools, both conceptual (histograms! diagrams!) and practical (how to use different kinds of software to achieve specific purposes).

As we have seen on several occasions, we have only just scratched the surface. But that is perfectly fine! If you dedicate your career to research, you will continue to build on what you know. What you have learned here should allow you to start that life-long process.

You will also have seen that the knowledge needed for data analysis falls into different categories. It goes without saying that working with astronomical data requires knowledge of physics, astronomy, and mathematics, specifically statistics. When you begin working on a new topic, then likely as not in the beginning, you will not fully understand what you are doing, and how the different elements you are dealing with fit together properly. Your goal should be to, eventually, reach the stage where you do understand what is going on. Such knowledge is required if you want to understand your results, but crucially also if you want to understand the limitations of your data, and possible ways of improvement. 

As you learn to use new tools, and more advanced tools, your newly acquired knowledge might also open up new opportunities for exciting science. Applying a tool that others did not think to apply could be the key step towards finding a new result. In this respect, it pays to have an eye out for neighbouring fields (or at least sub-fields). Could what they are doing in their field help you achieve something new in yours?

Some of the knowledge you need for data analysis is a matter of convention. How to perform a certain operation in TOPCAT, or plot a certain kind of diagram, is nothing you can deduce logically from previous fundamental knowledge (although previous experience will help you find the right answer). If the software in question is well documented, it makes sense to familiarise yourself with the basics; the more common strategy is to google what you are looking for. Everybody does it. Reminding yourself how to ``matplotlib equal axis ratio'' is just one search field away. Also, if you are working in an institute, chances are there will be experienced people you can ask for help with specific problems. Last but not least, if you need to accomplish a complex task, and you happen to have a script which does something similar, it makes sense to get to understand that other script, to try out variations on what it does, and eventually to adapt it to your purposes.

Finally, there is meta-knowledge about working with astronomical data. One piece of good advice is to always look at your data in simple form before attempting complex operations on it. Make a few basic histograms and diagrams to get a feeling for your data, look at an image, or make a quick plot of a spectrum --- incidentally, you might find out that your data is somehow completely different from what you expected, and that is a good thing to know early on!

A fairly universal truth you are likely to learn early on is how easy it is to make mistakes. One reason is that, even while each of the elements of your analysis might be straightforward, code can become fairly voluminous fairly quickly once you combine all the different necessary steps. So what do you do if, at the end of your analysis, your result is surprising, possibly wrong, or even obviously wrong? Conversely, if things turn out as expected, how can you be sure that this is not the result of several mistakes cancelling each other out, or almost cancelling? Such questions would not be important if your goal was just to create a visually pleasing astronomical image, for instance. But when you are doing scientific research, you had better understand, and check, every step of what you are doing; otherwise, you cannot be sure of your result. Your analysis should contain as many cross-checks as you can come up with, and have time to implement. After every step, think about what that step is meant to do, and what you can do to check that the desired result has indeed be achieved. If you write up your research, such cross-checks and safeguards are likely to make up an important part of your description of what you did.

Also, make sure your code is comprehensible. The most important element of this is adding descriptive comments. A person reading your code should be able to follow what you are doing step by step, guided by your code, but also by your description in the comments. Remember that the person in question could be you in a few years --- it is amazing how incomprehensible ill-documented code can become once you return to it a few months or years after having written it! Meaningful variable names can help with comprehensibility, too.

Last but not least documenting your code is a matter of scientific accountability. Science should be reproducible. Your research publications should tell your colleagues what you have done, so they can build on your results, but also critically examine what you have done. In an ideal world, every scientific article would be accompanied by data files and script files; running the script on the data, you should be able to reproduce the article's results on your own computer.\footnote{Cf. Weiner et al. 2009, [\href{https://arxiv.org/abs/0903.3971}{https://arxiv.org/abs/0903.3971}].} (And, in digging into the script, you would have a complete, unambiguous representation of what the article's authors have done with their data!) We're not there yet, but why not introduce those good practices right now? Comment and document your code. Preserve the definite version of your analysis scripts together with your article. That way, you will always be able to understand what you did earlier. If colleagues ask, you can give them your script and your data, and they can check for themselves what you have done. (And yes, I know you might be worried about giving away your trade secrets. If that is the case, you could still make your scripts available after some time. The competitive edge a specific script gives you is likely to become less important over time in any case.)

If you use someone else's data, or Python module, or software, you should make sure to give them appropriate credit. Published astronomical data typically comes with accompanying publications. Citing those publications when using the data is the appropriate way of giving credit to the astronomers involved. As an example, consider the THINGS survey data used in Fig.~\ref{THINGSchannelmap}. Using this image on a slide, you would add the information ``Walter 2008,'' or ``Walter et al. 2008,'' and every astronomer would be able to use the Astrophysics Data Service ADS at \href{https://ui.adsabs.harvard.edu/}{https://ui.adsabs.harvard.edu/} to find the article. If you are writing in a medium where you can add a link, add a link. If you are writing in a more formal setting, put an entry into your reference section, and refer to that entry in your text. For specific software or larger data sets, there is often a standard paragraph of acknowledgement (``boilerplate'') you are asked to include in your text. You can find examples in the Acknowledgements section, below. Science is a networked activity, and giving credit where credit is due is an important part of good scientific practice.

Astronomical data has never been as accessible as it is now, and computing power never as cheap. Observatory and telescope archives provide images and spectra, catalogs higher-level data like never before. This is likely to get even better as new facilities come online, and new tools become available. It's an exciting time to work with astronomical data!

\section*{Acknowledgements}

I would like to thank the anonymous referee and also Wolfgang Brandner, Roland Gredel and Carolin Liefke for helpful comments, Fabian Walter for helping with the THINGS data and Catharina Hock for helpful discussions.

Working with astronomical data would not be as (comparatively) easy without the countless individuals who have invested time and effort in programming the software, packages and libraries used throughout this text. If you are using the results, please acknowledge them appropriately. A number of scientific software products have associated articles that you should cite when using those products in your scientific work (e.g. see the footnotes for the table on p.~\pageref{ModuleTable}).

SAOImageDS9 development has been made possible by funding from the Chandra X-ray Science Center (CXC) and the High Energy Astrophysics Science Archive Center (HEASARC) with additional funding from the JWST Mission office at Space Telescope Science Institute, cf. the article Joye \& Mandel 2003, [\href{http://adsabs.harvard.edu/abs/2003ASPC..295..489J}{http://adsabs.harvard.edu/abs/2003ASPC..295..489J}].

TOPCAT was developed, and keeps being developed further, by Mark Taylor (University of Bristol). A description of the application can be found in Taylor 2005, [\href{http://adsabs.harvard.edu/abs/2005ASPC..347...29T}{http://adsabs.harvard.edu/abs/2005ASPC..347...29T}].

This work makes use of observations from the LCOGT network, which is described in 
Brown et al. 2013, [\href{http://adsabs.harvard.edu/abs/2013PASP..125.1031B}{http://adsabs.harvard.edu/abs/2013PASP..125.1031B}].

Funding for SDSS-III, from which I have used some data, has been provided by the Alfred P. Sloan Foundation, the Participating Institutions, the National Science Foundation, and the U.S. Department of Energy Office of Science. The SDSS-III web site is [\href{http://www.sdss3.org/}{http://www.sdss3.org/}]. SDSS-III is managed by the Astrophysical Research Consortium for the Participating Institutions of the SDSS-III Collaboration including the University of Arizona, the Brazilian Participation Group, Brookhaven National Laboratory, Carnegie Mellon University, University of Florida, the French Participation Group, the German Participation Group, Harvard University, the Instituto de Astrofisica de Canarias, the Michigan State/Notre Dame/JINA Participation Group, Johns Hopkins University, Lawrence Berkeley National Laboratory, Max Planck Institute for Astrophysics, Max Planck Institute for Extraterrestrial Physics, New Mexico State University, New York University, Ohio State University, Pennsylvania State University, University of Portsmouth, Princeton University, the Spanish Participation Group, University of Tokyo, University of Utah, Vanderbilt University, University of Virginia, University of Washington, and Yale University. 

Some of the sample files used here are based on observations made with the NASA/ESA Hubble Space Telescope, and obtained from the Hubble Legacy Archive, which is a collaboration between the Space Telescope Science Institute (STScI/NASA), the Space Telescope European Coordinating Facility (ST-ECF/ESA) and the Canadian Astronomy Data Centre (CADC/NRC/CSA).

The astronomical data query language ADQL was developed by the Virtual Observatory Query Language Working Group, cf. the description in Ortiz et al. 2008, [\href{http://www.ivoa.net/documents/latest/ADQL.html}{http://www.ivoa.net/documents/latest/ADQL.html}].

This work has made use of data from the European Space Agency (ESA) astrometry mission
{\it Gaia} ([\url{https://www.cosmos.esa.int/gaia}]), as processed by the {\it Gaia}
Data Processing and Analysis Consortium (DPAC, see more at
[\url{https://www.cosmos.esa.int/web/gaia/dpac/consortium}]). Funding for the DPAC
has been provided by national institutions, in particular the institutions
participating in the {\it Gaia} Multilateral Agreement; cf. Gaia Collaboration 2016, [\href{http://adsabs.harvard.edu/abs/2016A&A...595A...1G}{http://adsabs.harvard.edu/abs/2016A\&A...595A...1G}], and the description in Gaia Collaboration 2018, [\href{http://adsabs.harvard.edu/abs/2018arXiv180409365G}{http://adsabs.harvard.edu/abs/2018arXiv180409365G}].
\end{document}